\documentclass[a4paper,onecolumn,11pt]{quantumarticle}
\pdfoutput=1

\usepackage[utf8]{inputenc}
\usepackage[english]{babel}
\usepackage[T1]{fontenc}

\usepackage{amsmath}
\usepackage{amssymb}
\usepackage[numbers,sort&compress]{natbib}
\usepackage{amsthm}
\usepackage{braket}
\usepackage{graphicx}
\usepackage{hyperref}
\usepackage[dvipsnames]{xcolor}
\usepackage{dsfont}
\usepackage{verbatim}
\usepackage{amscd}
\usepackage[all,cmtip]{xy}
\usepackage{multirow}

\newcommand{\eX}{\dimexpr\fontcharht\font`X\relax}

\definecolor{darkblue}{RGB}{0,0,127} 
\definecolor{darkgreen}{RGB}{0,180,0}
\hypersetup{
	pdftitle={Symmetry-enriched topological order in tensor networks: Defects, gauging and anyon condensation},
	pdfauthor={Dominic J. Williamson, Nick Bultinck, Frank Verstraete}
}

\newcommand{\dom}[1]{\textcolor{darkgreen}{#1}}
\newcommand{\cut}[1]{\textcolor{red}{#1}}
\newcommand{\tofix}[1]{\textcolor{red}{#1}}
\newcommand{\red}[1]{\textcolor{red}{#1}}
\newcommand{\blue}[1]{\textcolor{blue}{#1}}

\graphicspath{{./Figures/}}

\usepackage{tikz,pgfplots}\pgfplotsset{compat=newest}
\usetikzlibrary{external,calc,decorations.pathreplacing,decorations.markings,decorations.pathmorphing,arrows.meta,shapes.geometric}
\newlength\figureheight
\newlength\figurewidth

\newcommand{\includeTikz}[2]{
\includegraphics{#1}
}

\newcommand{\includeTikzrm}[2]{
\tikzset{external/remake next}
\tikzsetnextfilename{#1}#2}

\newcommand{\openone}{\ensuremath{\mathds{1}}}
\newcommand{\g}[1]{{\ensuremath{\text{\bf{#1}}}}}
\newcommand{\defect}[2]{\ensuremath{{{#1}_{\g{#2}}}}}
\newcommand{\act}[1]{\ensuremath{{}^{\g{#1}}}}
\newcommand{\factsys}[2]{\ensuremath{{\eta_{#1}({#2})}}}
\newcommand{\splitact}[2]{\ensuremath{U_{\g{#1}}({#2})}}
\newcommand{\tube}[2]{\ensuremath{\mathcal{T}^{#2}_{#1}}}
\newcommand{\f}[4]{\ensuremath{{[F^{#1}_{#2}]_{#3}^{#4}}}}
\newcommand{\finv}[4]{\ensuremath{{[(F^{#1}_{#2})^{-1}]_{#3}^{#4}}}}
\newcommand{\shortf}[4]{\ensuremath{{F^{#1}_{#2 #3 #4}}}}
\newcommand{\rmatrix}[2]{\ensuremath{{R^{#1}_{#2}}}}
\newcommand{\mpo}[2]{\ensuremath{\text{MPO}^{#1}_{#2}}}
\newcommand{\qd}{\ensuremath{\mathcal{D}}}
\newcommand{\pepsbd}{\ensuremath{\mathsf{D}}}
\newcommand{\mpobd}{\ensuremath{{\chi}}}
\newcommand{\cat}{\ensuremath{\mathcal{C}}}
\newcommand{\catD}{\ensuremath{\mathcal{C}_\g{1}}}
\newcommand{\aban}{\ensuremath{\mathcal{A}}}
\newcommand{\ici}[1]{\ensuremath{\underline{#1}}}
\newcommand{\sector}[1]{\ensuremath{\lfloor{#1}\rfloor}}
\newcommand{\fs}[1]{\ensuremath{\varkappa_{#1}}}
\newcommand{\rbend}[2]{\ensuremath{B^{#1}_{#2}}}
\newcommand{\lbend}[2]{\ensuremath{A^{#1}_{#2}}}
\newcommand{\splittingspace}[2]{\ensuremath{{V}^{#1}_{#2}}}
\newcommand{\double}[1]{\ensuremath{\mathcal{Z}({#1})}}
\newcommand{\gdouble}[1]{\ensuremath{\mathcal{Z}_{\G}({#1})}}
\newcommand{\rdouble}[2]{\ensuremath{\mathcal{Z}_{#2}({#1})}}
\newcommand{\vecg}[2]{\ensuremath{\text{Vec}_{#1}^{#2}}}
\newcommand{\slant}[1]{\ensuremath{{\alpha^{(#1)}}}}
\newcommand{\dw}[2]{\ensuremath{\mathcal{B}^{\g{#1}}_{{#2}}}}
\newcommand{\tdw}[2]{\ensuremath{\widetilde{\mathcal{B}}^{\g{#1}}_{{#2}}}}
\newcommand{\irrep}[3]{\ensuremath{\chi^{#2}_{{#1}}(\g{#3})}}
\newcommand{\cirrep}[3]{\ensuremath{\conj{\chi}^{#2}_{{#1}}(\g{#3})}}
\newcommand{\irrp}[2]{\ensuremath{\chi^{#2}_{{#1}}}}
\newcommand{\gauged}[2]{\ensuremath{{[#1,#2]}}}
\newcommand{\gaugedii}[2]{\ensuremath{\ici{{(#1,#2)}}}}
\newcommand{\norm}[1]{\ensuremath{|{#1}|}}
\newcommand{\mpotensor}[2]{\ensuremath{B_{#1}^{#2}}}
\newcommand{\std}[1]{{$({#1}+1)$D}}
\newcommand{\autoequiv}[1]{\ensuremath{\mathsf{Z}_{\g{#1}}}}
\newcommand{\vac}{\ensuremath{0}}
\newcommand{\orbit}[1]{{\ensuremath{[{#1}]}}}
\newcommand{\pirrep}[3]{\ensuremath{\pi^{#2}_{{#1}}(\g{#3})}}
\newcommand{\globalu}[1]{\ensuremath{\text{\bf{U}}_{\g{#1}}}}
\newcommand{\localu}[1]{\ensuremath{U_{\g{#1}}}}
\newcommand{\entropy}[1]{\ensuremath{H_{#1}}}
\newcommand{\fusiontube}[2]{\ensuremath{\mathcal{V}_{#2}^{#1}}}
\newcommand{\ufactsys}[2]{\ensuremath{{\omega_{#1}({#2})}}}
\newcommand{\ctl}[2]{\ensuremath{{CL}^{-1}_{{#1, #2}}}}
\newcommand{\ctr}[2]{\ensuremath{{CR}^{-1}_{{#1, #2}}}}

\newcommand{\fref}[1]{Fig.~\ref{Fig:#1}}

\newcommand{\peps}{\ensuremath{T}}
\newcommand{\pepspert}[1]{\ensuremath{V_{#1}}}
\newcommand{\hilbert}{\ensuremath{\mathbb{H}}}

\newcommand{\rg}[1]{\ensuremath{\mathcal{L}_{(#1)}}}

\newcommand{\G}{\ensuremath{\mathcal{G}}}
\newcommand{\repg}{\ensuremath{\text{Rep}(\G)}}
\newcommand{\rep}[1]{\ensuremath{\text{Rep}(#1)}}
\newcommand{\Uone}{\ensuremath{\mathsf{U}(1)}}
\newcommand{\coho}[1]{\ensuremath{\mathcal{H}^{#1}(\G,\Uone)}}
\newcommand{\cohom}[2]{\ensuremath{\mathcal{H}^{#1}(#2,\Uone)}}
\newcommand{\cohomo}[3]{\ensuremath{\mathcal{H}^{#1}(#2,#3)}}
\newcommand{\twistedcoho}{\ensuremath{\mathcal{H}_\rho^{2}(\G,\Uone)}}
\newcommand{\disent}[1]{\ensuremath{\mathcal{D}_{#1}}}
\newcommand{\ghz}[1]{\ensuremath{\ket{\text{GHZ}_{#1}}}}

\newcommand{\trans}[1]{\ensuremath{\tau_{#1}}}
\newcommand{\dehn}[1]{\ensuremath{{T}_{#1}}}
\newcommand{\centralizer}[1]{\ensuremath{{Z}_{#1}}}
\newcommand{\sso}[2]{\ensuremath{\mathcal{S}_{#1}({#2})}}
\newcommand{\order}[1]{\ensuremath{n_{#1}}}
\newcommand{\crossing}[2]{\ensuremath{{M^{(#1)}_{#2}}}}
\newcommand{\conjclass}[1]{\ensuremath{{C}_{#1}}}

\newcommand{\N}{\ensuremath{\mathbb{Z}^+}}
\newcommand{\z}{\ensuremath{\mathbb{Z}}}
\newcommand{\zt}{\ensuremath{\mathbb{Z}_2}}
\newcommand{\zn}{\ensuremath{\mathbb{Z}_n}}
\newcommand{\zN}{\ensuremath{\mathbb{Z}_N}}
\newcommand{\conj}[1]{\ensuremath{{\overline{#1}}}}
\newcommand{\proj}[1]{\ensuremath{P}_{#1}}
\newcommand{\hilb}{\ensuremath{\mathbb{H}}}
\newcommand{\transp}{\ensuremath{\text{T}}}
\newcommand{\inv}[1]{\ensuremath{{#1}^{\tiny{-1}}}}
\newcommand{\vect}[1]{\ensuremath{{\underline{\mathbf{#1}}}}}
\newcommand{\basis}[1]{\ensuremath{\hat{e}_{#1}}}
\newcommand{\tr}{\ensuremath{\text{Tr}}}
\newcommand{\sdim}{\ensuremath{d}}
\newcommand{\symtyp}{\ensuremath{T}}

\newcommand{\drawtube}[4]{
\begin{tikzpicture}[scale=.1]
\def\dx{-1};
\def\ddx{1.5};
\def\dy{.5};
\def\p{5.5};
\filldraw[UFCBackground](0,0)--(20,0)--(20,20)--(0,20)--(0,0);
\draw[anyon](13,9)--(20,9) node[pos=0.55,above] {$#3$};
\draw[anyon](7,11)--(13,9) node[pos=0.45,below] {$#2$};
\draw[anyon](0,11)--(7,11) node[pos=0.55,above] {$#1$};
\draw[anyon](10,0)--(13,9)  node[pos=0.3,right] {$#4$};
\draw[anyon](7,11)--(10,20) node[pos=0.55,right] {$#4$};
\draw[draw=black!29,dashed](0,0)--(20,0) (20,20)--(0,20);
\filldraw[draw=black!29,fill=black!29](\p,0)--(\p+\dx,-\dy)--(\p+\dx,\dy)--(\p,0);
\filldraw[draw=black!29,fill=black!29](\p,20)--(\p+\dx,20-\dy)--(\p+\dx,20+\dy)--(\p,20);
\filldraw[draw=black!29,fill=black!29](\p+\ddx,0)--(\p+\dx+\ddx,-\dy)--(\p+\dx+\ddx,\dy)--(\p+\ddx,0);
\filldraw[draw=black!29,fill=black!29](\p+\ddx,20)--(\p+\dx+\ddx,20-\dy)--(\p+\dx+\ddx,20+\dy)--(\p+\ddx,20);
	\end{tikzpicture}
	}
	
\newcommand{\drawsettube}[4]{
\begin{tikzpicture}[scale=.1]
\def\dx{-1};
\def\ddx{1.5};
\def\dy{.5};
\def\p{5.5};
\filldraw[UFCBackground](0,0)--(20,0)--(20,20)--(0,20)--(0,0);
\draw[anyon](13,9)--(20,9) node[pos=0.55,above] {$#3$};
\draw[anyon](7,11)--(13,9) node[pos=0.25,below] {$#2$};
\draw[anyon](0,11)--(7,11) node[pos=0.55,above] {$#1$};
\draw[anyon](10,0)--(13,9)  node[pos=0.3,right] {$#4$};
\draw[anyon](7,11)--(10,20) node[pos=0.55,right] {$#4$};
\draw[draw=black!29,dashed](0,0)--(20,0) (20,20)--(0,20);
\filldraw[draw=black!29,fill=black!29](\p,0)--(\p+\dx,-\dy)--(\p+\dx,\dy)--(\p,0);
\filldraw[draw=black!29,fill=black!29](\p,20)--(\p+\dx,20-\dy)--(\p+\dx,20+\dy)--(\p,20);
\filldraw[draw=black!29,fill=black!29](\p+\ddx,0)--(\p+\dx+\ddx,-\dy)--(\p+\dx+\ddx,\dy)--(\p+\ddx,0);
\filldraw[draw=black!29,fill=black!29](\p+\ddx,20)--(\p+\dx+\ddx,20-\dy)--(\p+\dx+\ddx,20+\dy)--(\p+\ddx,20);
	\end{tikzpicture}
	}
	
\newcommand{\drawisingtube}[4]{
\begin{tikzpicture}[scale=.1]
\def\dx{-1};
\def\ddx{1.5};
\def\dy{.5};
\def\p{5.5};
\filldraw[UFCBackground](0,0)--(20,0)--(20,20)--(0,20)--(0,0);
\draw[#3](20,9)--(13,9);
\draw[#2](13,9)--(7,11);
\draw[#1](7,11)--(0,11);
\draw[#4](13,9)--(10,0);
\draw[#4](10,20)--(7,11);
\draw[draw=black!29,dashed](0,0)--(20,0) (20,20)--(0,20);
\filldraw[draw=black!29,fill=black!29](\p,0)--(\p+\dx,-\dy)--(\p+\dx,\dy)--(\p,0);
\filldraw[draw=black!29,fill=black!29](\p,20)--(\p+\dx,20-\dy)--(\p+\dx,20+\dy)--(\p,20);
\filldraw[draw=black!29,fill=black!29](\p+\ddx,0)--(\p+\dx+\ddx,-\dy)--(\p+\dx+\ddx,\dy)--(\p+\ddx,0);
\filldraw[draw=black!29,fill=black!29](\p+\ddx,20)--(\p+\dx+\ddx,20-\dy)--(\p+\dx+\ddx,20+\dy)--(\p+\ddx,20);
	\end{tikzpicture}
	}
	
\newcommand{\drawtorus}[4]{
\begin{tikzpicture}[scale=.1]
\def\dx{-1};
\def\ddx{1.5};
\def\dy{.5};
\def\p{5.5};
\def\pp{14};
\filldraw[UFCBackground](0,0)--(20,0)--(20,20)--(0,20)--(0,0);
\draw[draw=black!29,dashed](0,0)--(20,0)--(20,20)--(0,20)--(0,0);
\draw[anyon](13,9)--(20,9) node[pos=0.55,above] {$#3$};
\draw[anyon](7,11)--(13,9) node[pos=0.45,below] {$#2$};
\draw[anyon](0,11)--(7,11) node[pos=0.55,above] {$#1$};
\draw[anyon](10,0)--(13,9)  node[pos=0.3,right] {$#4$};
\draw[anyon](7,11)--(10,20) node[pos=0.55,right] {$#4$};
\filldraw[draw=black!29,fill=black!29](\p,0)--(\p+\dx,-\dy)--(\p+\dx,\dy)--(\p,0);
\filldraw[draw=black!29,fill=black!29](\p,20)--(\p+\dx,20-\dy)--(\p+\dx,20+\dy)--(\p,20);
\filldraw[draw=black!29,fill=black!29](\p+\ddx,0)--(\p+\dx+\ddx,-\dy)--(\p+\dx+\ddx,\dy)--(\p+\ddx,0);
\filldraw[draw=black!29,fill=black!29](\p+\ddx,20)--(\p+\dx+\ddx,20-\dy)--(\p+\dx+\ddx,20+\dy)--(\p+\ddx,20);
\begin{scope}[yshift=19.25cm,rotate=-90]
\filldraw[draw=black!29,fill=black!29](\pp,0)--(\pp-\dx,-\dy)--(\pp-\dx,\dy)--(\pp,0);
\filldraw[draw=black!29,fill=black!29](\pp,20)--(\pp-\dx,20-\dy)--(\pp-\dx,20+\dy)--(\pp,20);
\end{scope}
	\end{tikzpicture}
	}
	
\newcommand{\drawisingtorus}[4]{
\begin{tikzpicture}[scale=.1]
\def\dx{-1};
\def\ddx{1.5};
\def\dy{.5};
\def\p{5.5};
\def\pp{14};
\filldraw[UFCBackground](0,0)--(20,0)--(20,20)--(0,20)--(0,0);
\draw[draw=black!29,dashed](0,0)--(20,0)--(20,20)--(0,20)--(0,0);
\draw[#3](20,9)--(13,9);
\draw[#2](13,9)--(7,11);
\draw[#1](7,11)--(0,11);
\draw[#4](13,9)--(10,0);
\draw[#4](10,20)--(7,11);
\filldraw[draw=black!29,fill=black!29](\p,0)--(\p+\dx,-\dy)--(\p+\dx,\dy)--(\p,0);
\filldraw[draw=black!29,fill=black!29](\p,20)--(\p+\dx,20-\dy)--(\p+\dx,20+\dy)--(\p,20);
\filldraw[draw=black!29,fill=black!29](\p+\ddx,0)--(\p+\dx+\ddx,-\dy)--(\p+\dx+\ddx,\dy)--(\p+\ddx,0);
\filldraw[draw=black!29,fill=black!29](\p+\ddx,20)--(\p+\dx+\ddx,20-\dy)--(\p+\dx+\ddx,20+\dy)--(\p+\ddx,20);
\begin{scope}[yshift=19.25cm,rotate=-90]
\filldraw[draw=black!29,fill=black!29](\pp,0)--(\pp-\dx,-\dy)--(\pp-\dx,\dy)--(\pp,0);
\filldraw[draw=black!29,fill=black!29](\pp,20)--(\pp-\dx,20-\dy)--(\pp-\dx,20+\dy)--(\pp,20);
\end{scope}
	\end{tikzpicture}
	}

\definecolor{tensorblue}{rgb}{0.8,0.8,1}
\definecolor{tensorred}{rgb}{1,0.5,0.5}
\definecolor{tensorpurp}{rgb}{1,0.5,1}

\tikzset{nonesty/.style={fill=none,draw=none}}
\tikzset{ten/.style={fill=tensorblue}}
\tikzset{tenred/.style={fill=tensorred}}
\tikzset{tengreen/.style={fill=green!50!black!50}}
\tikzset{tenpurp/.style={fill=tensorpurp}}
\tikzset{tengrey/.style={fill=black!20}}
\tikzset{tenorange/.style={fill=orange!30}}
\tikzset{u/.style={fill=blue!20,draw=black}}
\tikzset{w/.style={fill=green!50!black!50,draw=black}}

\tikzset{anyon/.style={line width=1pt,draw=black,postaction={on each segment={mid arrow=black}}}}
\tikzset{ vacuum/.style={line width=1pt, draw=black, loosely dotted}}
\tikzset{psi/.style={line width=.5pt,decorate, decoration={snake, amplitude=1.5pt, segment length=5pt}, draw=black}}
\tikzset{sigma/.style={line width=1pt, draw=black}}
\tikzset{UFCBackground/.style={draw=black!10!white!80,fill=black!10!white!80}}

\tikzset{external/system call={pdflatex \tikzexternalcheckshellescape -halt-on-error -interaction=batchmode -jobname "\image" "\texsource" ; rm "\image".log ; rm "\image".dpth ; rm "\image"Notes.bib ; rm "\image".md5 }}

\tikzset{
  on each segment/.style={
    decorate,
    decoration={
      show path construction,
      moveto code={},
      lineto code={
        \path [#1]
        (\tikzinputsegmentfirst) -- (\tikzinputsegmentlast);
      },
      curveto code={
        \path [#1] (\tikzinputsegmentfirst)
        .. controls
        (\tikzinputsegmentsupporta) and (\tikzinputsegmentsupportb)
        ..
        (\tikzinputsegmentlast);
      },
      closepath code={
        \path [#1]
        (\tikzinputsegmentfirst) -- (\tikzinputsegmentlast);
      },
    },
  },
  mid arrow/.style={postaction={decorate,decoration={
        markings,
        mark=at position .7 with {\arrow[#1]{Stealth}}
      }}},
}

\makeatletter
\def\l@subsubsection#1#2{}
\makeatother

\begin{document}

\title{Symmetry-enriched topological order in tensor networks: Defects, gauging and anyon condensation}

\author{Dominic J.~Williamson}
\affiliation{Vienna Center for Quantum Technology, University of Vienna, Boltzmanngasse 5, 1090 Vienna, Austria}
\orcid{0000-0002-8029-6408}
\author{Nick Bultinck}
\affiliation{Department of Physics and Astronomy, Ghent University, Krijgslaan 281 S9, B-9000 Ghent, Belgium}
\orcid{0000-0002-2781-4085}
\author{Frank Verstraete}
\affiliation{Department of Physics and Astronomy, Ghent University, Krijgslaan 281 S9, B-9000 Ghent, Belgium}
\orcid{0000-0003-0270-5592}

\maketitle

\begin{abstract}
We study symmetry-enriched topological order in two-dimensional tensor network states by using graded matrix product operator algebras to represent symmetry-induced domain walls. 
A close connection to the theory of graded unitary fusion categories is established. 
Tensor network representations of the topological defect superselection sectors are constructed for all domain walls. 
The emergent symmetry-enriched topological order is extracted from these representations, including the symmetry action on the underlying anyons. 
Dual phase transitions, induced by gauging a global symmetry, and condensation of a bosonic subtheory, are analyzed and the relationship between topological orders on either side of the transition is derived. 
Several examples are worked through explicitly. 
\end{abstract}

\tableofcontents

  \section{ Introduction}
  \label{setsection:intro}

Symmetry plays a fundamental role in the classification of phases of matter. 
From spontaneous symmetry breaking in the Landau-Ginzburg theory of second-order phase transitions~\cite{landau1965course} to the homeomorphism invariance of  topological quantum field theories (TQFTs)~\cite{segal1988definition,witten1988topological,atiyah1988topological}. 
Even in the absence of a global symmetry, equivalence classes of gapped Hamiltonians under adiabatic deformation break up into inequivalent topological phases~\cite{wegner1971duality,kosterlitz1973ordering,PhysRevB.40.7387,einarsson,doi:10.1142/S0217979290000139}. These phases have topological symmetries inherited from the TQFTs that describe their low energy behavior, which are often realized by nontrivial string operators.  
Enforcing a global symmetry drastically refines the classification of phases that is found.  Even the trivial phase splits into a set of symmetry-protected topological phases, which cannot be adiabatically connected while preserving the symmetry~\cite{PhysRevLett.50.1153,gu2009tensor,pollmann2010entanglement,chen2013symmetry}. 
Nontrivial topological phases also split into symmetry-enriched topological phases which are distinguished by the interplay of the global symmetry with their topological superselection sectors~\cite{turaev2000homotopy,kirillov2004g,kitaev2006anyons,drinfeld2010braided,etingof2009fusion,bombin2010topological,hung2013quantized,mesaros2013classification,Barkeshli2013a,barkeshli2014symmetry,tarantino2015symmetry,teo2015theory,PhysRevB.94.235136,cheng2016exactly}.

The interplay of a global symmetry with topological degrees of freedom has also received attention from the quantum information community due to  its relevance for topological quantum codes and computation~\cite{beigi2011quantum,bombin2010topological,PhysRevLett.111.220402,yoshida2015gapped,PhysRevA.91.012305,PhysRevLett.110.170503,doi:10.1063/1.4939783,PhysRevB.91.245131,bridgeman2017tensor,webster2017locality}. 
Transversal gates on topological codes are generated by the action of on-site symmetries on superselection sectors, and symmetry defects have been used to alter the code properties of a topological order. 
In the quantum information approach to many-body condensed matter systems, tensor networks~\cite{VerstraeteMurgCirac2008,Orus2014,TNReview} are used to describe ground states and excitations. 
These provide efficient representations that faithfully capture the local entanglement structure of the many-body states. 
In one spatial dimension, matrix product states (MPS) have been used to classify gapped phases by studying the way a global symmetry acts on the auxiliary entanglement degrees of freedom~\cite{PhysRevLett.69.2863,PhysRevLett.75.3537,0295-5075-43-4-457,PhysRevLett.59.799,Fannes92,0295-5075-24-4-010,MPSrepresentations,1Done,SchuchGarciaCirac11,Cirac2017100}.  
In two spatial dimensions, projected entangled pair states (PEPS) have been used to study topological phases~\cite{doi:10.1143PTP.105.409,peps,Ginjectivity,Buerschaper14,MPOpaper,Bultinck2017183,ribbons,williamson2016fermionic,NewFermionPEPSPaper2017,bridgeman2017tensor}. A classification of nonchiral phases has been obtained from the algebra of matrix product operator (MPO) symmetries on the entanglement degrees of freedom of a local PEPS tensor that satisfy a pulling through equation. The resulting topological phases are given by Morita equivalence~\cite{muger2003subfactors} classes of MPO symmetry algebras. 
 This class of tensor networks includes all the string-net states~\cite{levin2005string}, for which explicit PEPS and MPO representations have been found~\cite{stringnet1,stringnet2,MPOpaper}. 
Chiral phases have also been studied~\cite{PhysRevLett.114.106803,PhysRevB.92.205307,PhysRevLett.111.236805}, however a complete coherent theoretical framework is lacking.

In this work, we study tensor network representations of symmetry-enriched nonchiral topological phases.
We derive results that allow us to incorporate the full theory of non-anomalous symmetry-enriched topological (SET) order~\cite{Barkeshli2013a,barkeshli2014symmetry,tarantino2015symmetry,teo2015theory,PhysRevB.94.235136,cheng2016exactly} into the tensor network formalism of Refs.~\cite{williamson2014matrix,Bultinck2017183}. To achieve this we introduce general tools to calculate emergent SET order from categorical input data~\cite{etingof2009fusion,gelaki2009centers} which are not specific to tensor networks. 
We focus on the interplay of a global symmetry with an MPO symmetry algebra on the entanglement degrees of freedom of a PEPS. To this end we consider the generalized symmetry-enriched pulling through equation introduced in Ref.~\cite{williamson2014matrix}.  
The resulting symmetry-enriched tensor networks are classified by graded Morita equivalence classes of MPO symmetry algebras. 
We demonstrate that this class of tensor networks includes the recently introduced symmetry-enriched string-net models~\cite{PhysRevB.94.235136,cheng2016exactly,chang2015enriching} by constructing the local PEPS and MPO tensors. 
We generalize Ocneanu's tube algebra~\cite{ocneanu1994chirality,tubealgebra,evans1998quantum} to include nontrivial defect sectors, from which we construct the superselection sectors of the emergent SET order. 
We explain how the physical data of the SET can be extracted from this construction. 
We calculate the effects of gauging the global symmetry~\cite{wilson1974confinement,kogut1975hamiltonian,levin2012braiding,Gaugingpaper}, and the relation of the SET to the resulting topological phase. We also find an anyon condensation phase transition~\cite{PhysRevB.79.045316} --- dual to gauging --- that is induced by breaking a graded MPO symmetry algebra, and calculate the relation of the topological order to the resulting SET. 
The nonchiral case of many of the general results shown in Ref.~\cite{barkeshli2014symmetry} follow directly from our construction. 
The mathematical content of our work includes an explicit construction of the automorphism induced on the Drinfeld center (double) by a given extension of a UFC~\cite{etingof2009fusion}, which forms part of our more general construction of the defect tube algebra and graded Drinfeld center from a graded UFC~\cite{gelaki2009centers}. 
Several examples are provided to illustrate various aspects of our formalism. 

The paper is organized as follows: 
In Section~\ref{setsection:example1}, we present a detailed analysis of the electromagnetic duality symmetry-enriched toric code: including a summary of relevant background results from Refs.~\cite{MPOpaper} and~\cite{Bultinck2017183}, a $\zt$ gauging procedure that maps it to the doubled Ising model, and a dual $\rep{\zt}$ anyon condensation phase transition. 
In Section~\ref{set:stringnetexample}, we write down explicit tensor network representations of the symmetry-enriched string-net ground states and their $\G$-graded MPO symmetry algebras, and show that they satisfy the symmetry-enriched pulling through equation. 
In Section~\ref{set:toptubealgsection}, we summarize the results of Ref.~\cite{Bultinck2017183} including a derivation of Ocneanu's tube algebra from the MPO symmetry algebra of a tensor network, a construction of the emergent topological superselection sectors from the tube algebra, and the extraction of physical data from superselection sectors thus constructed. 
In Section~\ref{setsection:setmpos}, we describe the theory of $\G$-graded matrix product operator algebras in terms of $\G$-extensions of an underlying MPO algebra, we draw an analogy to the theory of $\G$-graded unitary fusion categories, and define the symmetry-enriched pulling through equation for a tensor network. 
In Section~\ref{setsection:dubes}, we generalize the tube algebra to include nontrivial $\G$-defect sectors, which are constructed from a $\G$-graded MPO symmetry algebra of a tensor network, we describe a construction of the emergent SET order, and the extraction of its physical data. 
In Section~\ref{setsection:gauging}, we calculate the effect that gauging a global $\G$ symmetry has upon a $\G$-graded MPO symmetry algebra, derive the relationship between the emergent SET and the topological order that results from gauging, and demonstrate the gauging procedure explicitly for SPT phases. 
In Section~\ref{setsection:condensation}, we describe the $\repg$ anyon condensation phase transitions that are induced by $\G$-graded MPO algebra symmetry breaking, these phase transitions are dual to gauging a global $\G$ symmetry, we derive the relation between the emergent topological order and the SET that results from the condensation.  
In Section~\ref{setsection:examples2}, we present several examples that demonstrate different aspects of our formalism, including the interplay between Morita equivalence and anyon condensation phase transitions induced by MPO algebra symmetry breaking. 
In Section~\ref{setsection:conclusion}, we present conclusions and future directions. 
In Appendix~\ref{setappendix:gauging}, we describe how to sequentially gauge a global symmetry group, possessing a nontrivial normal subgroup, on the lattice.

\section{A motivating example}
\label{setsection:example1}

We begin with a concrete example that demonstrates the methods developed throughout the paper. In this example we study the relation between the \emph{doubled Ising model}~\cite{levin2005string} and the \emph{toric code}~\cite{qdouble} enriched by $\zt$ electromagnetic duality symmetry~\cite{bombin2010topological,PhysRevB.94.235136,cheng2016exactly}. These models are related by the dual processes of gauging the $\zt$ electromagnetic duality symmetry or condensing a $\text{Rep}(\zt)$ subtheory, as follows
\begin{align}
\xymatrixcolsep{7pc}\xymatrix{
\text{Toric code } \ar@<+4pt>[r]^-{\text{Add } \zt \text{ defects}}  & \txt{ $\zt$ symmetry-enriched \\ toric code }\text{ } \ar@<+4pt>[r]^-{\text{Gauge } \zt \text{ symmetry}}  \ar@<+4pt>[l]^-{\text{Confine } \zt \text{ defects}}  & \txt{ Doubled \\  Ising model.} \ar@<+4pt>[l]^-{\text{Condense Rep}(\zt)}
}
\end{align}
Each of the steps in this diagram can be calculated concretely in the language of tensor networks. To achieve this we summarize the mathematical objects, known as \emph{unitary fusion categories}, from which the toric code and doubled Ising model are built~\cite{etingof2005fusion,etingof2015tensor}. 
Using this data, we recount the tensor network description of the toric code and doubled Ising model string-net ground states~\cite{stringnet1,stringnet2} and the algebras of matrix product operator symmetries they posses~\cite{MPOpaper}. 
We present an extension of the fixed point tensor network construction to the recently defined class of symmetry-enriched string-net models~\cite{PhysRevB.94.235136,cheng2016exactly}, and demonstrate the interplay of the MPO symmetries with the physical symmetry. This extension is based upon the notion of graded unitary fusion categories~\cite{kirillov2004g,turaev2000homotopy,etingof2009fusion}. By viewing the Ising fusion category as a $\zt$-graded fusion category the authors of Refs.~\cite{PhysRevB.94.235136,cheng2016exactly} were able to construct a local commuting projector Hamiltonian, in two spatial dimensions, with a $\zt$-enriched toric code topological order where an on-site group action implements the electromagnetic duality of the toric code anyons. 

To calculate the action of the symmetry explicitly with the tensor network approach we build upon the anyon ansatz presented in Ref.~\cite{Bultinck2017183}, which is summarized in Section~\ref{set:toptubealgsection}. 
In that work, a second algebra --- corresponding to \emph{Ocneanu's tube algebra}~\cite{ocneanu1994chirality,tubealgebra,evans1998quantum} --- was derived from the matrix product operator symmetry algebra of the local tensors. By diagonalizing this tube algebra into irreducible blocks, the topological sectors are constructed. From the projectors onto these blocks, the full set of topological data --- characterizing the emergent anyonic excitations --- can be extracted. This set of data forms a mathematical object known as a \emph{modular tensor category}~\cite{bakalov2001lectures,etingof2005fusion,kitaev2006anyons,drinfeld2010braided,etingof2015tensor} which includes an underlying set of anyons, their fusion rules and $F$-symbols, $R$ matrices which describe braiding processes, as well as the gauge invariant Frobenius-Schur indicators, and the modular $S$ and $T$ matrices. 

We extend the tube algebra to a set of $|\G|$ \emph{defect tube algebras} whose irreducible blocks describe the definite topological $\G$-defect sectors. 
Similar to the anyon tubes, all topological data of the emergent symmetry-enriched theory can be extracted from the defect tubes. This data constitutes a \emph{$\G$-crossed modular tensor category}~\cite{kirillov2004g,turaev2000homotopy,etingof2009fusion,barkeshli2014symmetry,gelaki2009centers}, and includes the underlying set of topological defects, their fusion rules and $F$-symbols, the $\G$ action on the set of defects, the $\G$-crossed $R$ matrices which describe the braiding of defects, the projective representation carried by each defect and its 2-cocycle $\factsys{\defect{a}{g}}{\g{h},\g{k}}$, and the projective representation on the fusion and splitting spaces $\splitact{\g{g}}{\defect{a}{h},\defect{b}{k};\defect{c}{hk}}$, as well as the Frobenius-Schur indicators, and the $\G$-crossed modular $S$ and $T$ matrices. 

Rather than calculating the topological defect sectors of the EM duality enriched toric code directly, we develop an extremely simple recipe to find the effect of $\rep{\zt}$ \emph{anyon condensation} on the tubes of the doubled Ising model. This approach yields the pair of inequivalent topological defect sectors, known as~\cite{bombin2010topological} $\sigma_+$, $\sigma_-$. We also describe a recipe to calculate the effects of \emph{gauging} the $\zt$ EM duality symmetry on the defect tubes to recover the tubes of the doubled Ising model. 
We derive relations between the topological data of the models related by the dual gauging and condensation procedures. 
We remark that related dualities have previously been studied using the concept of bond algebras~\cite{Nussinov06102009,NUSSINOV2009977,doi:10.1080/00018732.2011.619814}.

\subsection{Background: Constructing topological sectors in a tensor network}
\label{set:tubebackground}

\subsubsection{Toric code}

The toric code is based on a very simple unitary fusion category $\cat=\vecg{\zt}{}$ built from the algebra $\mathbb{C}[\zt]$. We denote the simple objects $\{\vac,\psi\}$ with abelian multiplication
\begin{align}
\label{set:tcalgebra}
\vac\times a = a, \quad  \psi\times\psi=\vac,
\end{align}
for $a\in\{\vac,\psi\}$. 

The $F$-symbols are trivial, containing only the fusion constraints
\begin{align}
\shortf{a b c}{d}{e}{f}=\delta_{ab}^e\delta_{ec}^d \delta_{bc}^f \delta_{af}^d ,
\end{align}
which are defined by
\begin{align}\label{set:fusiondelta}
\delta^c_{a b} :=\left\{\begin{array}{ll}
1 & \text{when $c$ appears in the fusion product } a\times b
\\
0 & \text{otherwise.}
\end{array}\right.
\end{align}

A tensor network representation of the toric code ground state on a trivalent lattice can be constructed from these $F$-symbols, see Eq.~\eqref{set:TCtensors}. This is a specific instance of the more general tensor network representation of string-net ground states~\cite{stringnet1,stringnet2,MPOpaper}. This tensor network specifies a state in a Hilbert space where the qubit degrees of freedom that usually live on edges are instead defined by the diagonal subspace of a pair of qubits that live on the neighbouring vertices 
\begin{align}
\mathbb{C}[\zt]_e \cong \text{span} \{\ket{00},\ket{11} \} \subseteq \mathbb{C}[\zt]_{u} \otimes \mathbb{C}[\zt]_{v},
\end{align}
where $\partial e = \{u,v\}$. This convention is used as it avoids the need to define separate tensors on the edges of the tensor network. 

This tensor network has a $\zt$ MPO symmetry specified by the tensor in  Eq.~\eqref{set:TCtensors}.
\begin{align}
\begin{array}{c}
	\includeTikz{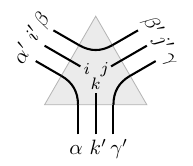}{
	\begin{tikzpicture}[scale=.1]
	\node[transform shape,draw=black!29,fill=black!10!white!80,regular polygon,regular polygon sides=3,minimum size=20cm] at (0,0){};
	\begin{scope}
	\draw[draw=black,line width=1pt,postaction=decorate] (0,-10) -- (0,-3);
	\draw[draw=black,line width=1pt] (-3,-10) -- (-3,-5) ;
	\draw[draw=black,line width=1pt] (3,-10) -- (3,-5);
	\draw[draw=black,line width=1pt,rounded corners=1mm] (-3,-5) to [bend right=30] ({3*cos(240)+5*sin(240)},{3*sin(240)-5*cos(240)});
	\node at (.5,-12.5){$\alpha\ k'\, \gamma'$};
	\end{scope}
	\begin{scope}[rotate=120]
	\draw[draw=black,line width=1pt,postaction=decorate] (0,-3) --  (0,-10);
	\draw[draw=black,line width=1pt] (-3,-10) -- (-3,-5) ;
	\draw[draw=black,line width=1pt] (3,-10) -- (3,-5);
	\draw[draw=black,line width=1pt,rounded corners=1mm] (-3,-5) to [bend right=30] ({3*cos(240)+5*sin(240)},{3*sin(240)-5*cos(240)});
	\node[rotate=300] at (-.5,-12.5){$\beta' \, j'\, \gamma$};
	\end{scope}
	\begin{scope}[rotate=-120]
	\draw[draw=black,line width=1pt,postaction=decorate]  (0,-3) -- (0,-10);
	\draw[draw=black,line width=1pt] (-3,-10) -- (-3,-5) ;
	\draw[draw=black,line width=1pt] (3,-10) -- (3,-5);
	\draw[draw=black,line width=1pt,rounded corners=1mm] (-3,-5) to [bend right=30] ({3*cos(240)+5*sin(240)},{3*sin(240)-5*cos(240)});
	\node[rotate=60] at (0,-12.5){$\alpha' \ i'\, \beta$};
	\end{scope}
			\node at (-1.5,1){\footnotesize $i$};
			\node at (1.5,1){\footnotesize $j$};
			\node at (0,-1.5){\footnotesize $k$};
	\end{tikzpicture}
	}
	\end{array}
&=\delta_{i i'}\delta_{j j'}\delta_{k k'}\delta_{\alpha\alpha'}\delta_{\beta \beta'}\delta_{\gamma\gamma'} \shortf{\alpha i j}{\gamma}{\beta}{k}
\, ,
\label{set:TCtensors}
\\
\begin{array}{c}
\includeTikz{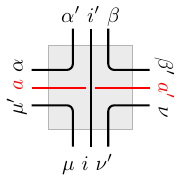}{
\begin{tikzpicture}[scale=.1]
\begin{scope}
\node[transform shape,draw=black!29,fill=black!10!white!80,regular polygon,regular polygon sides=4,minimum size=20cm] at (0,-15){};
\draw[draw=red,line width=1pt,postaction=decorate] (-10,-15) -- (10,-15);
\node[transform shape,draw=black!10!white!80,fill=black!10!white!80,regular polygon,regular polygon sides=4,minimum size=2cm] at (0,-15){};
\draw[draw=black,line width=1pt,postaction=decorate] (0,-25) -- (0,-5);
\draw[draw=black,line width=1pt, rounded corners=1mm] (-10,-12) -- (-3,-12) -- (-3,-5)
										(-10,-18) -- (-3,-18) -- (-3,-25)
										(10,-18) -- (3,-18) -- (3,-25)
										(10,-12 )-- (3,-12) -- (3,-5);
\end{scope}
\node at (-.5,-28){$\mu \ i\ \nu'$};
\node at (0,-3){$\alpha' \ i'\ \beta$};
\node[rotate=90] at (-12.5,-15){$\mu' \ \red{a}\ \alpha$};
\node[rotate=-90] at (12.5,-15){$\beta' \ \red{a'}\ \nu$};
\end{tikzpicture}
}
\end{array}
&=\delta_{\alpha\alpha'}\delta_{\beta \beta'}\delta_{\mu\mu'}\delta_{\nu\nu'}\delta_{i i'}\delta_{a a'} \shortf{a \mu i}{\beta}{\alpha}{\nu}
\, ,
\end{align}
where all labels lie in $\zt$. Here $a\in\zt\cong\{\vac,\psi\}$ controls which element of $\zt$ the MPO represents 
\begin{align}
\mpo{}{a}:= 
%
,
\end{align}
where the vertical dashed lines denote periodic boundary conditions and
\begin{align}
\begin{array}{c}
\includeTikz{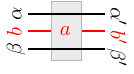}{
\begin{tikzpicture}[scale=.1]
\begin{scope}
\draw[draw=red,line width=1pt,postaction=decorate] (-20,-15) -- (-7,-15);
\filldraw[draw=black!29,fill=black!10!white!80] (-16,-10) -- (-11,-10) -- (-11,-20) -- (-16,-20) -- cycle;
\draw[draw=black,line width=1pt, rounded corners=1mm] (-20,-12) -- (-7,-12)
										(-20,-18) -- (-7,-18) ;										
\node[anchor=east] at (-11.5,-15){$ \red{a}$};																	
\end{scope}
\node[rotate=90] at (-22,-15){$\beta \ \red{b}\ \alpha$};
\node[rotate=-90] at (-5,-16){$\alpha' \, \red{b'}\, \beta'$};
\end{tikzpicture}
}
\end{array}
:= \delta_{\alpha \alpha'}\delta_{\beta \beta'} \delta_{b,b'}\delta_{b,a}
.
\end{align}
We will also make use of the following short hand notation for the projector onto the $a$ block
\begin{align}
	\raisebox{.2cm}{$
	\begin{array}{c}
	\includeTikz{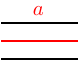}{
	\begin{tikzpicture}[scale=.1]
	\begin{scope}
	\draw[draw=red,line width=1pt,postaction=decorate] (-20,-15) -- (-7,-15);
	\draw[draw=black,line width=1pt, rounded corners=1mm] (-20,-12) -- (-7,-12)
											(-20,-18) -- (-7,-18) ;							
	\node[anchor=east] at (-11.5,-10){$ \red{a}$};																	
	\end{scope}
	\end{tikzpicture}
	}
	\end{array} 
	$}
:=
	\begin{array}{c}
	\includeTikz{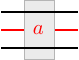}{
	\begin{tikzpicture}[scale=.1]
	\begin{scope}
	\draw[draw=red,line width=1pt,postaction=decorate] (-20,-15) -- (-7,-15);
	\filldraw[draw=black!29,fill=black!10!white!80] (-16,-10) -- (-11,-10) -- (-11,-20) -- (-16,-20) -- cycle;
	\draw[draw=black,line width=1pt, rounded corners=1mm] (-20,-12) -- (-7,-12)
											(-20,-18) -- (-7,-18) ;									
	\node[anchor=east] at (-11.5,-15){$ \red{a}$};																	
	\end{scope}
	\end{tikzpicture}
	}
	\end{array}
.
\end{align}
These MPOs reproduce the fusion algebra in Eq.~\eqref{set:tcalgebra}
\begin{align}
\mpo{}{\vac}\mpo{}{a}=\mpo{}{a}, && \mpo{}{\psi}\mpo{}{\psi}=\mpo{}{\vac}.
\end{align}
In fact, the individual MPO tensors satisfy the fusion rules locally, up to a fusion/splitting tensor shown below
\begin{align}
	\begin{array}{c}
	\includeTikz{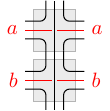}{
	\begin{tikzpicture}[scale=.05]
	\begin{scope}[yshift=17cm]
	\begin{scope}
	\node[transform shape,draw=black!29,fill=black!10!white!80,regular polygon,regular polygon sides=4,minimum size=20cm] at (0,-15){};
	\draw[draw=red,line width=.5pt,postaction=decorate] (-10,-15) -- (10,-15);
	\node[transform shape,draw=black!10!white!80,fill=black!10!white!80,regular polygon,regular polygon sides=4,minimum size=2cm] at (0,-15){};
	\end{scope}
	\begin{scope}
	\draw[draw=black,line width=.5pt,postaction=decorate] (0,-22) -- (0,-5);
	\draw[draw=black,line width=.5pt, rounded corners=1mm] (-10,-12) -- (-3,-12) -- (-3,-5)
											(-10,-18) -- (-3,-18) -- (-3,-22)
											(10,-18) -- (3,-18) -- (3,-22)
											(10,-12 )-- (3,-12) -- (3,-5);
	\end{scope}
	\end{scope}
	\begin{scope}
	\node[transform shape,draw=black!29,fill=black!10!white!80,regular polygon,regular polygon sides=4,minimum size=20cm] at (0,-15){};
	\draw[draw=red,line width=.5pt,postaction=decorate] (-10,-15) -- (10,-15);
	\node[transform shape,draw=black!10!white!80,fill=black!10!white!80,regular polygon,regular polygon sides=4,minimum size=2cm] at (0,-15){};
	\end{scope}
	\begin{scope}
	\draw[draw=black,line width=.5pt,postaction=decorate] (0,-25) -- (0,-5);
	\draw[draw=black,line width=.5pt, rounded corners=1mm] (-10,-12) -- (-3,-12) -- (-3,-5)
											(-10,-18) -- (-3,-18) -- (-3,-25)
											(10,-18) -- (3,-18) -- (3,-25)
											(10,-12 )-- (3,-12) -- (3,-5);
	\end{scope}
	\node[anchor=east] at (-10,2){$ \red{a}$};
	\node[anchor=east] at (-10,-15){$ \red{b}$};
	\node[anchor=west] at (10,2){$ \red{a}$};
	\node[anchor=west] at (10,-15){$ \red{b}$};
	\end{tikzpicture}
	}
	\end{array}
	\, ,
    \label{eq:sec21zipper}
\end{align}
where
\begin{align}
\begin{array}{c}
	\includeTikz{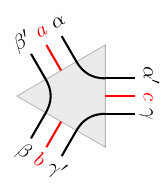}{
	\begin{tikzpicture}[scale=.1]
	\begin{scope}[rotate=90]
	\node[transform shape,draw=black!29,fill=black!10!white!80,regular polygon,regular polygon sides=3,minimum size=20cm] at (0,0){};
	\begin{scope}
	\draw[draw=red,line width=1pt,postaction=decorate](0,-5) -- (0,-10);
	\draw[draw=black,line width=1pt] (-3,-10) -- (-3,-5) ;
	\draw[draw=black,line width=1pt] (3,-10) -- (3,-5);
	\draw[draw=black,line width=1pt,rounded corners=1mm] (-3,-5) to [bend right=30] ({3*cos(240)+5*sin(240)},{3*sin(240)-5*cos(240)});
	\node[rotate=-90] at (0.5,-12.5){$\alpha' \ \red{c}\ \gamma$};
	\end{scope}
	\begin{scope}[rotate=120]
	\draw[draw=red,line width=1pt,postaction=decorate] (0,-10) -- (0,-5);
	\draw[draw=black,line width=1pt] (-3,-10) -- (-3,-5) ;
	\draw[draw=black,line width=1pt] (3,-10) -- (3,-5);
	\draw[draw=black,line width=1pt,rounded corners=1mm] (-3,-5) to [bend right=30] ({3*cos(240)+5*sin(240)},{3*sin(240)-5*cos(240)});
	\node[rotate=30] at (0,-12.5){$\beta' \ \red{a} \ \alpha$};
	\end{scope}
	\begin{scope}[rotate=-120]
	\draw[draw=red,line width=1pt,postaction=decorate]  (0,-10) -- (0,-5);
	\draw[draw=black,line width=1pt] (-3,-10) -- (-3,-5) ;
	\draw[draw=black,line width=1pt] (3,-10) -- (3,-5);
	\draw[draw=black,line width=1pt,rounded corners=1mm] (-3,-5) to [bend right=30] ({3*cos(240)+5*sin(240)},{3*sin(240)-5*cos(240)});
	\node[rotate=330] at (0.5,-12.5){$\beta \ \red{b} \ \gamma'$};
	\end{scope}
	\end{scope}
	\end{tikzpicture}
	}
	\end{array}
&=\delta_{\alpha\alpha'}\delta_{\beta \beta'}\delta_{\gamma\gamma'} \shortf{a b  \gamma }{\alpha}{c}{\beta}
.
\end{align}
This MPO representation of $\zt$ falls into the framework developed in Refs.~\cite{Ginjectivity,Bultinck2017183}.

The local order of fusion/splitting does not matter in this case
\begin{align}
	%
	,
    \label{eq:sec21FMove}
\end{align}
as the $F$-symbols are trivial and simply enforce the fusion constraint $\delta_{bc}^{f} \delta_{af}^{d}$.

The toric code tensor network and its $\zt$ MPO symmetry satisfy the pulling through equation (left), as well as a constraint that allows one to remove a contractible loop of the MPO from the tensor network (right)
	\begin{align}
	\raisebox{-.45cm}{$
	\begin{array}{c}
	\includeTikz{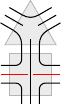}{
	\begin{tikzpicture}[scale=.05]
	\node[transform shape,draw=black!29,fill=black!10!white!80,regular polygon,regular polygon sides=3,minimum size=20cm] at (0,0){};
	\begin{scope}
	\node[transform shape,draw=black!29,fill=black!10!white!80,regular polygon,regular polygon sides=4,minimum size=20cm] at (0,-15){};
	\draw[draw=red,line width=.5pt,postaction=decorate] (-10,-15) -- (10,-15);
	\node[transform shape,draw=black!10!white!80,fill=black!10!white!80,regular polygon,regular polygon sides=4,minimum size=2cm] at (0,-15){};
	\end{scope}
	\begin{scope}
	\draw[draw=black,line width=.5pt,postaction=decorate] (0,-25) -- (0,-3);
	\draw[draw=black,line width=.5pt, rounded corners=1mm] (-10,-12) -- (-3,-12) -- (-3,-5)
											(-10,-18) -- (-3,-18) -- (-3,-25)
											(10,-18) -- (3,-18) -- (3,-25)
											(10,-12 )-- (3,-12) -- (3,-5);
	\draw[draw=black,line width=.5pt,rounded corners=1mm] (-3,-5) to [bend right=30] ({3*cos(240)+5*sin(240)},{3*sin(240)-5*cos(240)});
	\end{scope}
	\begin{scope}[rotate=120]
	\draw[draw=black,line width=.5pt,postaction=decorate] (0,-3) --  (0,-10);
	\draw[draw=black,line width=.5pt] (-3,-10) -- (-3,-5) ;
	\draw[draw=black,line width=.5pt] (3,-10) -- (3,-5);
	\draw[draw=black,line width=.5pt,rounded corners=1mm] (-3,-5) to [bend right=30] ({3*cos(240)+5*sin(240)},{3*sin(240)-5*cos(240)});
	\end{scope}
	\begin{scope}[rotate=-120]
	\draw[draw=black,line width=.5pt,postaction=decorate] (0,-3) --  (0,-10);
	\draw[draw=black,line width=.5pt] (-3,-10) -- (-3,-5) ;
	\draw[draw=black,line width=.5pt] (3,-10) -- (3,-5);
	\draw[draw=black,line width=.5pt,rounded corners=1mm] (-3,-5) to [bend right=30] ({3*cos(240)+5*sin(240)},{3*sin(240)-5*cos(240)});
	\end{scope}
	\end{tikzpicture}
	}
	\end{array}
	$}
	,
	&
	\begin{array}{c}
	\includeTikz{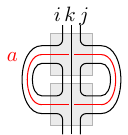}{
	\begin{tikzpicture}[scale=.05]
	\node at (0,15){$i\,k\,j$};
	\node at (-20,1){$\red{a}$};
	\begin{scope}[yshift=17cm]
	\node[transform shape,draw=black!29,fill=black!10!white!80,regular polygon,regular polygon sides=4,minimum size=20cm] at (0,-15){};
	\draw[draw=red,line width=.5pt,postaction=decorate] (10,-15) -- (-10,-15);
	\node[transform shape,draw=black!10!white!80,fill=black!10!white!80,regular polygon,regular polygon sides=4,minimum size=2cm] at (0,-15){};
	\draw[draw=black,line width=.5pt,postaction=decorate] (0,-22) -- (0,-5);
	\draw[draw=black,line width=.5pt, rounded corners=1mm] (-10,-12) -- (-3,-12) -- (-3,-5)
											(-10,-18) -- (-3,-18) -- (-3,-22)
											(10,-18) -- (3,-18) -- (3,-22)
											(10,-12 )-- (3,-12) -- (3,-5);
	\end{scope}
	\begin{scope}
	\node[transform shape,draw=black!29,fill=black!10!white!80,regular polygon,regular polygon sides=4,minimum size=20cm] at (0,-15){};
	\draw[draw=red,line width=.5pt,postaction=decorate] (-10,-15) -- (10,-15);
	\node[transform shape,draw=black!10!white!80,fill=black!10!white!80,regular polygon,regular polygon sides=4,minimum size=2cm] at (0,-15){};
	\draw[draw=black,line width=.5pt,postaction=decorate] (0,-25) -- (0,-5);
	\draw[draw=black,line width=.5pt, rounded corners=1mm] (-10,-12) -- (-3,-12) -- (-3,-5)
											(-10,-18) -- (-3,-18) -- (-3,-25)
											(10,-18) -- (3,-18) -- (3,-25)
											(10,-12 )-- (3,-12) -- (3,-5);
	\end{scope}
	\draw[draw=black,line width=.5pt, rounded corners=1mm] 
	(10,-12) to [bend right=90] (10,-12+11)
	(10,-18) to [bend right=90] (10,-18+23)
	(-10,-12) to [bend left=90] (-10,-12+11)
	(-10,-18) to [bend left=90] (-10,-18+23);
	\draw[draw=red,line width=.5pt] 
	(10,-15) to [bend right=90] (10,-15+17)
	(-10,-15) to [bend left=90] (-10,-15+17);
	\end{tikzpicture}
	}
	\end{array}
	&=
	\delta_{ij}^{k}
	\begin{array}{c}
	\includeTikz{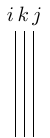}{
	\begin{tikzpicture}[scale=.05]
	\node at (0,0){$i\,k\,j$};
	\begin{scope}
	\draw[draw=black,line width=.5pt,postaction=decorate] (0,-41) -- (0,-5);
	\draw[draw=black,line width=.5pt] (-3,-41) -- (-3,-5) ;
	\draw[draw=black,line width=.5pt] (3,-41) -- (3,-5);
	\end{scope}
	\end{tikzpicture}
	}
	\end{array}
	.
    \label{eq:sec21PullThru}
	\end{align}
Furthermore the toric code tensor network is MPO-injective, in the terminology of Ref.~\cite{MPOpaper}, with respect to the MPO projector onto the symmetric subspace: $\frac{1}{2}(\mpo{}{\vac}+\mpo{}{\psi})$. 

The framework developed in Refs.~\cite{MPOpaper,Bultinck2017183,1751-8121-49-35-354001} demonstrated that calculations involving an algebra of MPO symmetries on the virtual level of an MPO-injective tensor network follow the same rules as the planar isotopy invariant, diagrammatic calculus for unitary fusion categories described in Refs.~\cite{JOYAL199155,JOYAL199320,kitaev2006anyons} and elsewhere. 
In other words, the properties of an MPO algebra on the virtual level of an MPO-injective tensor network can be abstracted to the diagrammatic calculus for anyons (without braiding). 

To see this from another perspective, recall that the vector space associated to a manifold by the Turaev-Viro \emph{topological quantum field theory} (TQFT)~\cite{turaev1992state,barrett1996invariants} --- for a given unitary fusion category --- is defined to be the space of allowed pictures modulo local relations~\cite{walker1991witten,freedman2008picture} (this is isomorphic to the ground space of the string net model on that manifold). 
In this way a tensor network on the sphere satisfying MPO-injectivity, together with its MPO symmetry algebra, can be thought of as an explicit representation of all pictures that can be reduced to the empty diagram via local relations.  
In this representation the local relations are implemented by the pulling through equation, together with the local conditions satisfied by the MPOs. 
We remark that one can construct the full set of different ground states by closing the tensor network using different MPOs, as described below.

The upshot of the previous paragraph is: for the purposes of calculations involving elements of the MPO algebra within an MPO-injective tensor network, we can work at the level of the diagrammatic UFC calculus~\cite{JOYAL199155,JOYAL199320,kitaev2006anyons}. The MPO-injective tensor network itself can be thought of as a sort of background, and MPOs on the virtual level will become loops on top of this background. For example
\begingroup
\renewcommand*{\arraystretch}{1}
\begin{align}

\end{align}
\endgroup
where the $\psi$ loop is depicted as a squiggly line in the diagrammatic calculus. As demonstrated by this example local relations in the diagrammatic calculus correspond to equalities in the tensor network representation. For this reason we will work with the vector space of diagrams modulo local relations, hence the equivalence above becomes an equality.

As was mentioned above, on a topologically nontrivial manifold the vector space of the Turaev-Viro TQFT may have dimension greater than one. Equivalently the string-net model may have a ground state degeneracy~\cite{koenig2010quantum}. 
This appears in the tensor network formalism when one considers closing a disc to form a nontrivial topology. It was shown in Ref.~\cite{MPOpaper} that the MPO-injectivity property is stable under concatenation. That is, one can contract arbitrarily many MPO-injective tensors on a disc and the resulting tensor will maintain the MPO-injectivity property. 
The resulting tensor network is a PEPS, and hence comes equipped with a frustration free, local, parent Hamiltonian for which it is a ground state~\cite{GarciaVerstraeteWolfCirac08}. 
It was also shown in Ref.~\cite{MPOpaper}, that the ground space of this Hamiltonian on a disc is spanned by the tensor network on the disc contracted with an arbitrary boundary tensor, see Eq.~\eqref{set:puncturedsphere}, LHS. 
We remark that the local relations in the bulk give rise to an action of the MPO algebra on the boundary tensor which project it onto the symmetric subspace. 
A generalization of this story also holds in higher spatial dimensions~\cite{walker1991witten,higherdto}.

The local tensor in Eq.~\eqref{set:TCtensors} can be used to build a unique state on a closed manifold. 
However there may be linearly independent states on the  same manifold that are constructed from the same tensor on any disc, but also contain virtual level MPOs that are free to move through the tensor network.  As a matter of terminology, we will refer to a spanning set of ground states of the local parent Hamiltonian on a given manifold as the set of tensor networks assigned to that manifold by the local tensor. 

To form a closed manifold one can consider adding handles to a disc. The simplest choice is to close the disc into a sphere. Even in this case one finds a closure tensor, located on a single link, that is free to move throughout the tensor network~\cite{MPOpaper}. The different choices for this closure tensor pick out different ground states in an unstable TQFT. Here unstable means that the TQFT has a ground state degeneracy on the sphere. For a stable TQFT, which is always the case when the MPO algebra yields a unitary fusion category, there is a single ground state on the sphere and the closure tensor is trivial. Since an unstable TQFT can be written as a sum of stable TQFTs we will only consider the stable case from here on. 

The next example we consider is attaching a 1-handle to a disc to form a cylinder. Before the closure, the class of states we are considering corresponds to the tensor network with arbitrary boundary conditions $A$, depicted below on the left. After taking the closure we are guaranteed that any disc must look like the original tensor network. Hence the only boundary conditions along the attaching edge that survive are the ones that can be freely deformed through the tensor network. This implies that these boundary conditions must be a linear combination of the irreducible MPOs, shown below on the right. This argument was given explicitly in Ref.~\cite{MPOpaper}. 
\begin{align}
\label{set:puncturedsphere}
\begin{array}{c}
	\includeTikz{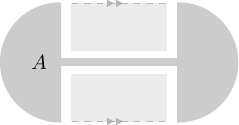}{
	\begin{tikzpicture}[scale=.1]
	\def\dx{-1};
	\def\ddx{1.5};
	\def\dy{.5};
	\def\p{9};
	\filldraw[UFCBackground](2,0)--(18,0)--(18,8)--(2,8)--cycle (18,20)--(2,20)--(2,12)--(18,12)--cycle;
	\filldraw[line width=1pt,draw=black!20,fill=black!20] (0,9.5)--(20,9.5)--(20,10.5)--(0,10.5) -- cycle;
	\draw[draw=black!29,dashed](2,0)--(18,0) (2,20)--(18,20);
	\filldraw[draw=black!29,fill=black!29](\p,0)--(\p+\dx,-\dy)--(\p+\dx,\dy)--(\p,0);
	\filldraw[draw=black!29,fill=black!29](\p,20)--(\p+\dx,20-\dy)--(\p+\dx,20+\dy)--(\p,20);
	\filldraw[draw=black!29,fill=black!29](\p+\ddx,0)--(\p+\dx+\ddx,-\dy)--(\p+\dx+\ddx,\dy)--(\p+\ddx,0);
	\filldraw[draw=black!29,fill=black!29](\p+\ddx,20)--(\p+\dx+\ddx,20-\dy)--(\p+\dx+\ddx,20+\dy)--(\p+\ddx,20);
	\filldraw[line width=1pt,draw=black!20,fill=black!20] (20,0)--(20,20) arc (90:-90:10) -- cycle;
	\filldraw[line width=1pt,draw=black!20,fill=black!20] (0,20)--(0,0) arc (270:90:10) -- cycle;
	\node[anchor=east] at (-1,10){$A$};
		\end{tikzpicture}
		}
\end{array}
= 
\sum_{p \mu} c_p
\begin{array}{c}
	\includeTikz{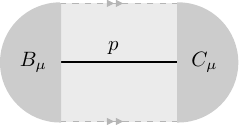}{
	\begin{tikzpicture}[scale=.1]
	\def\dx{-1};
	\def\ddx{1.5};
	\def\dy{.5};
	\def\p{9};
	\def\y{12.5};
	\def\x{1};
	\filldraw[UFCBackground](0,0)--(20,0)--(20,20)--(0,20)--(0,0);
	\draw[line width=1pt,draw=black](20,10)--(0,10) node[pos=0.55,above] {$p$};
	\draw[draw=black!29,dashed](0,0)--(20,0) (20,20)--(0,20);
	\filldraw[draw=black!29,fill=black!29](\p,0)--(\p+\dx,-\dy)--(\p+\dx,\dy)--(\p,0);
	\filldraw[draw=black!29,fill=black!29](\p,20)--(\p+\dx,20-\dy)--(\p+\dx,20+\dy)--(\p,20);
	\filldraw[draw=black!29,fill=black!29](\p+\ddx,0)--(\p+\dx+\ddx,-\dy)--(\p+\dx+\ddx,\dy)--(\p+\ddx,0);
	\filldraw[draw=black!29,fill=black!29](\p+\ddx,20)--(\p+\dx+\ddx,20-\dy)--(\p+\dx+\ddx,20+\dy)--(\p+\ddx,20);
	\filldraw[line width=1pt,draw=black!20,fill=black!20] (20,0)--(20,20) arc (90:-90:10) -- cycle;
	\filldraw[line width=1pt,draw=black!20,fill=black!20] (0,20)--(0,0) arc (270:90:10) -- cycle;
	\node[anchor=east] at (-1,10){$B_\mu$};
	\node[anchor=west] at (21,10){$C_\mu$};
		\end{tikzpicture}
		}
\end{array}
,
\end{align}
where the dotted lines indicate a periodic identification of the boundaries and the solid line connecting the caps indicates that they are part of the same tensor. 
Hence all boundary conditions on the cylinder can be divided into an arbitrary tensor $A':=\sum_\mu B_\mu \otimes C_\mu$, acting on the open left and right boundaries, joined by a linear combination of MPOs along the closure edge. More specifically, the virtual indices of the MPO are contracted with indices of $A'$. 

Similarly, when closing a disc to form a torus, the most general boundary condition is given by a linear combination of MPOs along  the cycles of the torus. Furthermore the tensor at their intersection point can be resolved into a linear combination of fusion and splitting vertices, shown below on the right. 
\begin{align}
\label{set:torusclosure}
\begin{array}{c}
	\includeTikz{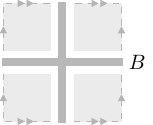}{
	\begin{tikzpicture}[scale=.1]
	\def\dx{-1};
	\def\ddx{1.5};
	\def\dy{.5};
	\def\p{16};
	\filldraw[UFCBackground] (0,0)--(8,0)--(8,8)--(0,8)--cycle (12,0)--(20,0)--(20,8)--(12,8)--cycle  
	(20,20)--(12,20)--(12,12)--(20,12)--cycle (0,12)--(8,12)--(8,20)--(0,20)--cycle;
	\filldraw[line width=1pt,draw=black!28,fill=black!28] (0,9.5)--(20,9.5)--(20,10.5)--(0,10.5) -- cycle;
	\filldraw[line width=1pt,draw=black!28,fill=black!28] (9.5,0)--(9.5,20)--(10.5,20)--(10.5,0) -- cycle;
	\draw[draw=black!29,dashed](0,8)--(0,0)--(8,0) (0,12)--(0,20)--(8,20) (12,0)--(20,0)--(20,8) (20,12)--(20,20)--(12,20);
	\filldraw[draw=black!29,fill=black!29](\p,0)--(\p+\dx,-\dy)--(\p+\dx,\dy)--(\p,0);
	\filldraw[draw=black!29,fill=black!29](\p,20)--(\p+\dx,20-\dy)--(\p+\dx,20+\dy)--(\p,20);
	\filldraw[draw=black!29,fill=black!29](\p+\ddx,0)--(\p+\dx+\ddx,-\dy)--(\p+\dx+\ddx,\dy)--(\p+\ddx,0);
	\filldraw[draw=black!29,fill=black!29](\p+\ddx,20)--(\p+\dx+\ddx,20-\dy)--(\p+\dx+\ddx,20+\dy)--(\p+\ddx,20);
\begin{scope}[xshift=-12.5cm]
	\filldraw[draw=black!29,fill=black!29](\p,0)--(\p+\dx,-\dy)--(\p+\dx,\dy)--(\p,0);
	\filldraw[draw=black!29,fill=black!29](\p,20)--(\p+\dx,20-\dy)--(\p+\dx,20+\dy)--(\p,20);
	\filldraw[draw=black!29,fill=black!29](\p+\ddx,0)--(\p+\dx+\ddx,-\dy)--(\p+\dx+\ddx,\dy)--(\p+\ddx,0);
	\filldraw[draw=black!29,fill=black!29](\p+\ddx,20)--(\p+\dx+\ddx,20-\dy)--(\p+\dx+\ddx,20+\dy)--(\p+\ddx,20);
\end{scope}
\begin{scope}[yshift=20.5cm,rotate=-90]
	\filldraw[draw=black!29,fill=black!29](\p,0)--(\p-\dx,-\dy)--(\p-\dx,\dy)--(\p,0);
	\filldraw[draw=black!29,fill=black!29](\p,20)--(\p-\dx,20-\dy)--(\p-\dx,20+\dy)--(\p,20);
\begin{scope}[xshift=-11.5cm]
	\filldraw[draw=black!29,fill=black!29](\p,0)--(\p-\dx,-\dy)--(\p-\dx,\dy)--(\p,0);
	\filldraw[draw=black!29,fill=black!29](\p,20)--(\p-\dx,20-\dy)--(\p-\dx,20+\dy)--(\p,20);
\end{scope}
\end{scope}
	\node[anchor=west] at (20,10){$B$};
		\end{tikzpicture}
		}
\end{array}
=
\sum_{xyz} c_{xy}^{z}
\begin{array}{c}
\includeTikz{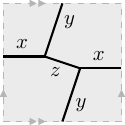}{
	\begin{tikzpicture}[scale=.1]
	\def\dx{-1};
	\def\ddx{1.5};
	\def\dy{.5};
	\def\p{5.5};
	\def\pp{14};
	\filldraw[UFCBackground](0,0)--(20,0)--(20,20)--(0,20)--(0,0);
	\draw[draw=black!29,dashed](0,0)--(20,0)--(20,20)--(0,20)--(0,0);
	\draw[line width=1pt,draw=black](20,9)--(13,9) node[pos=0.55,above] {$x$};
	\draw[line width=1pt,draw=black](13,9)--(7,11) node[pos=0.7,below] {$z$};
	\draw[line width=1pt,draw=black](7,11)--(0,11) node[pos=0.55,above] {$x$};
	\draw[line width=1pt,draw=black](13,9)--(10,0)  node[pos=0.7,right] {$y$};
	\draw[line width=1pt,draw=black](10,20)--(7,11) node[pos=0.35,right] {$y$};
	\filldraw[draw=black!29,fill=black!29](\p,0)--(\p+\dx,-\dy)--(\p+\dx,\dy)--(\p,0);
	\filldraw[draw=black!29,fill=black!29](\p,20)--(\p+\dx,20-\dy)--(\p+\dx,20+\dy)--(\p,20);
	\filldraw[draw=black!29,fill=black!29](\p+\ddx,0)--(\p+\dx+\ddx,-\dy)--(\p+\dx+\ddx,\dy)--(\p+\ddx,0);
	\filldraw[draw=black!29,fill=black!29](\p+\ddx,20)--(\p+\dx+\ddx,20-\dy)--(\p+\dx+\ddx,20+\dy)--(\p+\ddx,20);
	\begin{scope}[yshift=19.25cm,rotate=-90]
	\filldraw[draw=black!29,fill=black!29](\pp,0)--(\pp-\dx,-\dy)--(\pp-\dx,\dy)--(\pp,0);
	\filldraw[draw=black!29,fill=black!29](\pp,20)--(\pp-\dx,20-\dy)--(\pp-\dx,20+\dy)--(\pp,20);
	\end{scope}
	\end{tikzpicture}
	}
\end{array}
=
\sum_{xyz} c_{xy}^{z} \ket{xyz}
.
\end{align}

For the toric code, a basis for the ground space on the torus is given by
\begin{align}
&\ket{ \vac \vac \vac}=
\begin{array}{c}
\includeTikz{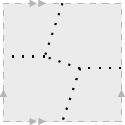}{\drawisingtorus{vacuum}{vacuum}{vacuum}{vacuum}}
  \end{array} \, ,
\quad
&\ket{\vac \psi\psi}=
\begin{array}{c}
\includeTikz{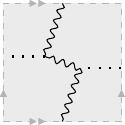}{\drawisingtorus{vacuum}{psi}{vacuum}{psi}}
  \end{array}\, ,
\nonumber 
\\
&\ket{\psi \vac \psi}=
\begin{array}{c}
\includeTikz{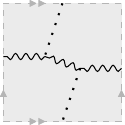}{\drawisingtorus{psi}{psi}{psi}{vacuum}}
  \end{array}\, ,
\quad
&\ket{\psi\psi \vac}=
\begin{array}{c}
\includeTikz{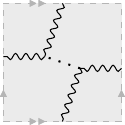}{\drawisingtorus{psi}{vacuum}{psi}{psi}}
  \end{array}\,
  .
\end{align}

To find the superselection sectors within this tensor network, we consider a well separated pair of pointlike excitations on a sphere. This sphere can be stretched out to a cylinder with an excitation at each end. Away from these excitation points the tensor network locally looks the same, hence the whole tensor network is of the form in Eq.~\eqref{set:puncturedsphere} with $A'$ now also containing physical indices located around each excitation.   
The local relations in the bulk of the tensor network on the cylinder give rise to an MPO action on each boundary. 
\begin{align}
\label{set:TCcylinderrelation}
\begin{array}{c}
	\includeTikz{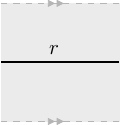}{
	\begin{tikzpicture}[scale=.1]
	\def\dx{-1};
	\def\ddx{1.5};
	\def\dy{.5};
	\def\p{9};
	\filldraw[UFCBackground](0,0)--(20,0)--(20,20)--(0,20)--(0,0);
	\draw[line width=1pt,draw=black](20,10)--(0,10) node[pos=0.55,above] {$r$};
	\draw[draw=black!29,dashed](0,0)--(20,0) (20,20)--(0,20);
	\filldraw[draw=black!29,fill=black!29](\p,0)--(\p+\dx,-\dy)--(\p+\dx,\dy)--(\p,0);
	\filldraw[draw=black!29,fill=black!29](\p,20)--(\p+\dx,20-\dy)--(\p+\dx,20+\dy)--(\p,20);
	\filldraw[draw=black!29,fill=black!29](\p+\ddx,0)--(\p+\dx+\ddx,-\dy)--(\p+\dx+\ddx,\dy)--(\p+\ddx,0);
	\filldraw[draw=black!29,fill=black!29](\p+\ddx,20)--(\p+\dx+\ddx,20-\dy)--(\p+\dx+\ddx,20+\dy)--(\p+\ddx,20);
		\end{tikzpicture}
		}
\end{array}
=
\begin{array}{c}
	\includeTikz{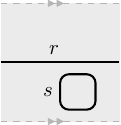}{
	\begin{tikzpicture}[scale=.1]
	\def\dx{-1};
	\def\ddx{1.5};
	\def\dy{.5};
	\def\p{9};
	\filldraw[UFCBackground](0,0)--(20,0)--(20,20)--(0,20)--(0,0);
	\draw[line width=1pt,draw=black](20,10)--(0,10) node[pos=0.55,above] {$r$};
	\draw[draw=black!29,dashed](0,0)--(20,0) (20,20)--(0,20);
	\filldraw[draw=black!29,fill=black!29](\p,0)--(\p+\dx,-\dy)--(\p+\dx,\dy)--(\p,0);
	\filldraw[draw=black!29,fill=black!29](\p,20)--(\p+\dx,20-\dy)--(\p+\dx,20+\dy)--(\p,20);
	\filldraw[draw=black!29,fill=black!29](\p+\ddx,0)--(\p+\dx+\ddx,-\dy)--(\p+\dx+\ddx,\dy)--(\p+\ddx,0);
	\filldraw[draw=black!29,fill=black!29](\p+\ddx,20)--(\p+\dx+\ddx,20-\dy)--(\p+\dx+\ddx,20+\dy)--(\p+\ddx,20);
	\draw[draw=black,line width=1pt, rounded corners=1.5mm] (16,8)--(10,8)--(10,2)--(16,2)--cycle;
	\node[anchor=east] at (10,5){$s$};
		\end{tikzpicture}
		}
\end{array}
=
\begin{array}{c}
	\includeTikz{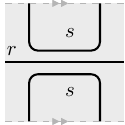}{
	\begin{tikzpicture}[scale=.1]
	\def\dx{-1};
	\def\ddx{1.5};
	\def\dy{.5};
	\def\p{9};
	\filldraw[UFCBackground](0,0)--(20,0)--(20,20)--(0,20)--(0,0);
	\draw[line width=1pt,draw=black](20,10)--(0,10);
	\draw[draw=black,line width=1pt, rounded corners=1.5mm] (16,0)--(16,8)--(4,8)--(4,0);
	\draw[draw=black,line width=1pt, rounded corners=1.5mm] (16,20)--(16,12)--(4,12)--(4,20);
	\node[anchor=east] at (13,5){$s$};
	\node[anchor=east] at (13,15){$s$};
	\node[anchor=west] at (-1,12){$r$};
	\draw[draw=black!29,dashed](0,0)--(20,0) (20,20)--(0,20);
	\filldraw[draw=black!29,fill=black!29](\p,0)--(\p+\dx,-\dy)--(\p+\dx,\dy)--(\p,0);
	\filldraw[draw=black!29,fill=black!29](\p,20)--(\p+\dx,20-\dy)--(\p+\dx,20+\dy)--(\p,20);
	\filldraw[draw=black!29,fill=black!29](\p+\ddx,0)--(\p+\dx+\ddx,-\dy)--(\p+\dx+\ddx,\dy)--(\p+\ddx,0);
	\filldraw[draw=black!29,fill=black!29](\p+\ddx,20)--(\p+\dx+\ddx,20-\dy)--(\p+\dx+\ddx,20+\dy)--(\p+\ddx,20);
		\draw[line width=1pt,draw=black](20,10)--(0,10);
		\end{tikzpicture}
		}
\end{array}
&=\sum_{q} \delta_{rs}^{q}
\begin{array}{c}
	\includeTikz{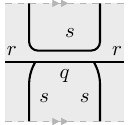}{
	\begin{tikzpicture}[scale=.1]
	\def\dx{-1};
	\def\ddx{1.5};
	\def\dy{.5};
	\def\p{9};
	\filldraw[UFCBackground](0,0)--(20,0)--(20,20)--(0,20)--(0,0);
	\begin{scope}[rotate=180,xshift=-20cm,yshift=-20cm]
	\draw[draw=black,line width=1pt, rounded corners=1.5mm] (16,20)--(16,12)--(15,10) (5,10)--(4,12)--(4,20);
	\draw[draw=black,line width=1pt, rounded corners=1.5mm] (16,0)--(16,8)--(4,8)--(4,0);
	\end{scope}
	\node[anchor=east] at (13,15){$s$};
	\node[anchor=west] at (-1,12){$r$};
	\node[anchor=east] at (21,12){$r$};
	\node[anchor=north] at (10,10){$q$};
	\node[anchor=east] at (15.5,4){$s$};
	\node[anchor=west] at (4.5,4){$s$};
	\draw[draw=black!29,dashed](0,0)--(20,0) (20,20)--(0,20);
	\filldraw[draw=black!29,fill=black!29](\p,0)--(\p+\dx,-\dy)--(\p+\dx,\dy)--(\p,0);
	\filldraw[draw=black!29,fill=black!29](\p,20)--(\p+\dx,20-\dy)--(\p+\dx,20+\dy)--(\p,20);
	\filldraw[draw=black!29,fill=black!29](\p+\ddx,0)--(\p+\dx+\ddx,-\dy)--(\p+\dx+\ddx,\dy)--(\p+\ddx,0);
	\filldraw[draw=black!29,fill=black!29](\p+\ddx,20)--(\p+\dx+\ddx,20-\dy)--(\p+\dx+\ddx,20+\dy)--(\p+\ddx,20);
		\draw[line width=1pt,draw=black](20,10)--(0,10);
		\end{tikzpicture}
		}
\end{array}
\nonumber \\
&=\sum_{q r} \delta_{rs}^{q} \delta_{sq}^{p}
\begin{array}{c}
	\includeTikz{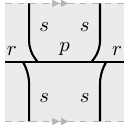}{
	\begin{tikzpicture}[scale=.1]
	\def\dx{-1};
	\def\ddx{1.5};
	\def\dy{.5};
	\def\p{9};
	\filldraw[UFCBackground](0,0)--(20,0)--(20,20)--(0,20)--(0,0);
	\begin{scope}[rotate=180,xshift=-20cm,yshift=-20cm]
	\draw[draw=black,line width=1pt, rounded corners=1.5mm] (16,0)--(16,8)--(14.5,10) (5.5,10)--(4,8)--(4,0);
	\draw[draw=black,line width=1pt, rounded corners=1.5mm] (16,20)--(16,12)--(17,10) (3,10)--(4,12)--(4,20);
	\end{scope}
	\node[anchor=east] at (15.5,16){$s$};
	\node[anchor=west] at (4.5,16){$s$};
	\node[anchor=east] at (15.5,4){$s$};
	\node[anchor=west] at (4.5,4){$s$};
	\node[anchor=west] at (-1,12){$r$};
	\node[anchor=east] at (21,12){$r$};
	\node[anchor=south] at (10,10){$p$};
	\draw[draw=black!29,dashed](0,0)--(20,0) (20,20)--(0,20);
	\filldraw[draw=black!29,fill=black!29](\p,0)--(\p+\dx,-\dy)--(\p+\dx,\dy)--(\p,0);
	\filldraw[draw=black!29,fill=black!29](\p,20)--(\p+\dx,20-\dy)--(\p+\dx,20+\dy)--(\p,20);
	\filldraw[draw=black!29,fill=black!29](\p+\ddx,0)--(\p+\dx+\ddx,-\dy)--(\p+\dx+\ddx,\dy)--(\p+\ddx,0);
	\filldraw[draw=black!29,fill=black!29](\p+\ddx,20)--(\p+\dx+\ddx,20-\dy)--(\p+\dx+\ddx,20+\dy)--(\p+\ddx,20);
		\draw[line width=1pt,draw=black](20,10)--(0,10);
		\end{tikzpicture}
		}
\end{array}
\, .
\end{align}
This defines an algebra acting on each of the punctures. In fact, this is Ocneanu's celebrated \emph{tube algebra}. 
We remark that the tube algebra contains a copy of the fusion algebra but is generally distinct from the fusion algebra. 
The irreducible superselection sectors within a puncture are given by the irreducible blocks of the tube algebra. 
Projecting onto a superselection sector fixes a definite anyon flux through the puncture. 
A particular sector can be constructed by projecting onto the corresponding block of the tube algebra using the \emph{irreducible central idempotent} (ICI) that acts as the identity within that block.

We use the following labeling convention for basis elements of the tube algebra
  \begin{align}
  \label{set:tctubeconvention}
\tube{p q r}{s} = 
\begin{array}{c}
	\includeTikz{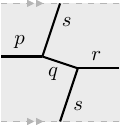}{
	\begin{tikzpicture}[scale=.1]
	\def\dx{-1};
	\def\ddx{1.5};
	\def\dy{.5};
	\def\p{5.5};
	\filldraw[UFCBackground](0,0)--(20,0)--(20,20)--(0,20)--(0,0);
	\draw[line width=1pt,draw=black](20,9)--(13,9) node[pos=0.55,above] {$r$};
	\draw[line width=1pt,draw=black](13,9)--(7,11) node[pos=0.7,below] {$q$};
	\draw[line width=1pt,draw=black](7,11)--(0,11) node[pos=0.55,above] {$p$};
	\draw[line width=1pt,draw=black](13,9)--(10,0)  node[pos=0.7,right] {$s$};
	\draw[line width=1pt,draw=black](10,20)--(7,11) node[pos=0.35,right] {$s$};
	\draw[draw=black!29,dashed](0,0)--(20,0) (20,20)--(0,20);
	\filldraw[draw=black!29,fill=black!29](\p,0)--(\p+\dx,-\dy)--(\p+\dx,\dy)--(\p,0);
	\filldraw[draw=black!29,fill=black!29](\p,20)--(\p+\dx,20-\dy)--(\p+\dx,20+\dy)--(\p,20);
	\filldraw[draw=black!29,fill=black!29](\p+\ddx,0)--(\p+\dx+\ddx,-\dy)--(\p+\dx+\ddx,\dy)--(\p+\ddx,0);
	\filldraw[draw=black!29,fill=black!29](\p+\ddx,20)--(\p+\dx+\ddx,20-\dy)--(\p+\dx+\ddx,20+\dy)--(\p+\ddx,20);
		\end{tikzpicture}
	}
  \end{array}
  \, ,
  \end{align}
  which correspond to the minimal tensor networks:
  \begin{align} 
  %
	,
  \end{align}
note $\tube{pqr}{s}$ is only nonzero when $\delta_{qs}^{r}\delta_{sq}^{p}=1$,  hence the dimension of the algebra is given by 
$$\sum\limits_{pqrs} \delta_{qs}^{r} \delta_{sq}^{p} < |\cat|^4 \, ,$$ 
where $|\cat|$ is the number of simple objects in the fusion algebra, in this case $|\cat|=2$. As noted above, the tube algebra contains a copy of the input fusion algebra which is generated by $\tube{\vac s \vac}{s}$.

    Multiplication of tubes is defined by the stacking operation
    \begin{align}
    \tube{p q r}{s} \tube{p'q'r'}{s'} = 
    \delta_{r p'}
	\begin{array}{c}
	\includeTikz{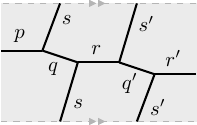}{
	\begin{tikzpicture}[scale=.1]
	\def\dx{-1};
	\def\ddx{1.5};
	\def\dy{.5};
	\def\p{16};
	\def\dz{1}
	\filldraw[UFCBackground](0,0)--(33,0)--(33,20)--(0,20)--(0,0);
	\draw[line width=1pt,draw=black](20,9+\dz)--(13,9+\dz) node[pos=0.55,above] {$r$};
	\draw[line width=1pt,draw=black](13,9+\dz)--(7,11+\dz) node[pos=0.7,below] {$q$};
	\draw[line width=1pt,draw=black](7,11+\dz)--(0,11+\dz) node[pos=0.55,above] {$p$};
	\draw[line width=1pt,draw=black](13,9+\dz)--(10,0)  node[pos=0.7,right] {$s$};
	\draw[line width=1pt,draw=black](10,20)--(7,11+\dz) node[pos=0.35,right] {$s$};
	\draw[draw=black!29,dashed](0,0)--(33,0) (33,20)--(0,20);
	\filldraw[draw=black!29,fill=black!29](\p,0)--(\p+\dx,-\dy)--(\p+\dx,\dy)--(\p,0);
	\filldraw[draw=black!29,fill=black!29](\p,20)--(\p+\dx,20-\dy)--(\p+\dx,20+\dy)--(\p,20);
	\filldraw[draw=black!29,fill=black!29](\p+\ddx,0)--(\p+\dx+\ddx,-\dy)--(\p+\dx+\ddx,\dy)--(\p+\ddx,0);
	\filldraw[draw=black!29,fill=black!29](\p+\ddx,20)--(\p+\dx+\ddx,20-\dy)--(\p+\dx+\ddx,20+\dy)--(\p+\ddx,20);
	\begin{scope}[xshift=13cm]
	\draw[line width=1pt,draw=black](20,9-\dz)--(13,9-\dz) node[pos=0.55,above] {$r'$};
	\draw[line width=1pt,draw=black](13,9-\dz)--(7,11-\dz) node[pos=0.7,below] {$q'$};
	\draw[line width=1pt,draw=black](13,9-\dz)--(10,0)  node[pos=0.7,right] {$s'$};
	\draw[line width=1pt,draw=black](10,20)--(7,11-\dz) node[pos=0.35,right] {$s'$};
	\end{scope}
		\end{tikzpicture}
	}
  \end{array}
  = 
\delta_{r p'} \sum_{q'' s''} \shortf{sqs'}{q' }{r}{ q''} \shortf{q'' s' s}{p }{q }{s''} \shortf{s' s q''}{r' }{q'}{ s''} \  \tube{pq'' r'}{s''} 
\, ,
    \end{align}
    where the only tubes $\tube{pq'' r'}{s''} $ that appear on the RHS with  nonzero coefficients satisfy $\delta_{s s'}^{s''}$. 
     Hermitian conjugation is given by
    \begin{align}
   (\tube{p q r}{s})^\dagger = 
	\begin{array}{c}
	\includeTikz{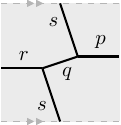}{
	\begin{tikzpicture}[scale=.1]
	\def\dx{-1};
	\def\ddx{1.5};
	\def\dy{.5};
	\def\p{5.5};
	\filldraw[UFCBackground](0,0)--(-20,0)--(-20,20)--(0,20)--(0,0);
	\draw[line width=1pt,draw=black](-20,9)--(-13,9) node[pos=0.55,above] {$r$};
	\draw[line width=1pt,draw=black](-13,9)--(-7,11) node[pos=0.7,below] {$q$};
	\draw[line width=1pt,draw=black](-7,11)--(0,11) node[pos=0.55,above] {$p$};
	\draw[line width=1pt,draw=black](-13,9)--(-10,0)  node[pos=0.7,left] {$s$};
	\draw[line width=1pt,draw=black](-10,20)--(-7,11) node[pos=0.35,left] {$s$};
	\begin{scope}[xshift=-20cm]
	\draw[draw=black!29,dashed](0,0)--(20,0) (20,20)--(0,20);
	\filldraw[draw=black!29,fill=black!29](\p,0)--(\p+\dx,-\dy)--(\p+\dx,\dy)--(\p,0);
	\filldraw[draw=black!29,fill=black!29](\p,20)--(\p+\dx,20-\dy)--(\p+\dx,20+\dy)--(\p,20);
	\filldraw[draw=black!29,fill=black!29](\p+\ddx,0)--(\p+\dx+\ddx,-\dy)--(\p+\dx+\ddx,\dy)--(\p+\ddx,0);
	\filldraw[draw=black!29,fill=black!29](\p+\ddx,20)--(\p+\dx+\ddx,20-\dy)--(\p+\dx+\ddx,20+\dy)--(\p+\ddx,20);
	\end{scope}
		\end{tikzpicture}
	}
  \end{array}
  =
\sum_{q'}  \shortf{srs}{p}{q}{q'} \tube{r q' p}{s}
\, .
    \end{align}   
    
    With this notation Eq.~\eqref{set:TCcylinderrelation} becomes 
    \begin{align}
\tube{rrr}{\vac}
&=
\sum_{p q}
(\tube{p q r}{s})^\dagger
\tube{ppp}{\vac}
\tube{p q r}{s}
.
\end{align}
    
    We have verified that the tube algebra is closed under multiplication and Hermitian conjugation. Hence the tube algebra is a $C^*$ algebra, spanned by the tubes in Eq.~\eqref{set:tctubeconvention} and can be block diagonalized.

The toric code tube algebra is spanned by the following basis elements  
  \begin{align}
&\tube{\vac \vac \vac}{\vac} = 
\begin{array}{c}
\includeTikz{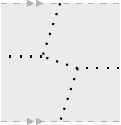}{\drawisingtube{vacuum}{vacuum}{vacuum}{vacuum}}
  \end{array}\, ,
  \quad
&  \tube{\vac \psi \vac}{\psi} = 
\begin{array}{c}
\includeTikz{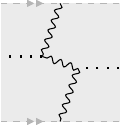}{\drawisingtube{vacuum}{psi}{vacuum}{psi}}
  \end{array}\, ,
\nonumber \\
&  \tube{\psi \psi \psi }{\vac} = 
\begin{array}{c}
\includeTikz{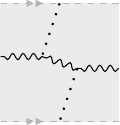}{\drawisingtube{psi}{psi}{psi}{vacuum}}
  \end{array}\, ,
  \quad
&  \tube{\psi \vac \psi}{\psi} = 
\begin{array}{c}
\includeTikz{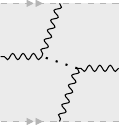}{\drawisingtube{psi}{vacuum}{psi}{psi}}
  \end{array}\, .
\label{set:TCtubepictures}
  \end{align}

  We remark that this tube algebra splits into two copies of $\mathbb{C}[\zt]$ and hence is diagonalized by a set of projectors onto the irreducible representations (irreps) of each $\zt$. 
  These projectors are summarized in the following table
  \begin{align}
\label{set:tctubes}
\begin{tabular}{| c | c c c c |} 
\hline 
Anyons & \tube{\vac \vac \vac}{\vac} & \tube{\vac \psi \vac}{\psi } & \tube{\psi\psi\psi}{\vac} & \tube{\psi \vac \psi}{\psi} 
\\ \hline
$\vac$ &  $1$ & $1$ & &
\\
$e$ & $1$ & $-1$  & &
\\
$m$ & & & $1$ & $1$ 
\\
$em$ & & & $1$ & $-1$ 
\\
\hline
\end{tabular}
\end{align}
which contains the coefficients of each ICI. In this table --- and for all superselection sector tables throughout the paper --- we suppress a normalization by the quantum dimension squared $\qd^{2}$, which gives $2$ in this case. 
Hence the ICIs of the toric code are given by
\begin{align}
\ici{\vac}=\frac{1}{2}( \tube{\vac  \vac \vac}{\vac} +\tube{\vac \psi \vac}{\psi})\, , &&  \ici{e}=\frac{1}{2}( \tube{\vac \vac \vac}{\vac} -\tube{\vac \psi \vac}{\psi})\, , \nonumber \\
\ici{m}=\frac{1}{2}(\tube{\psi \psi \psi}{\vac} +\tube{ \psi \vac \psi}{\psi})\, ,  && \ici{em}=\frac{1}{2}( \tube{\psi \psi \psi}{\vac} - \tube{ \psi \vac \psi}{\psi}) \, .
\end{align}
We remark that the vacuum $\ici{\vac}$ superselection sector is always given by the projector onto the symmetric subspace of the fusion algebra generated by $\tube{\vac s \vac}{s}$. 
To justify the labeling of the other superselection sectors, we need a method to extract the physical data characterizing the anyon theory from the ICIs. 
The $S$ and $T$ matrices of the emergent topological theory are independent of any gauge choice in defining the input fusion category. 
These invariants prove very useful for the classification and identification of non-chiral theories, which are precisely those arising from the double construction we have described. In the current example $S$ and $T$ turn out to uniquely specify the anyon theory amongst other possibilities with the same number of anyons, however they are known to be incomplete invariants in general~\cite{mignard2017modular}. 

On the tube algebra $S$ and $T$ are defined as
\begin{align}
S( \tube{p q r}{s}) &:= \delta_{p r}
\begin{array}{c}
\includeTikz{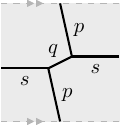}{
\begin{tikzpicture}[scale=.1]
\def\dx{-1};
\def\ddx{1.5};
\def\dy{.5};
\def\p{5.5};
\filldraw[UFCBackground](0,0)--(20,0)--(20,20)--(0,20)--(0,0);
\draw[line width=1pt,draw=black](10,0)--(8,9) node[pos=0.5,right] {$p$};
\draw[line width=1pt,draw=black](8,9)--(12,11) node[pos=0.2,above] {$q$};
\draw[line width=1pt,draw=black](12,11)--(10,20) node[pos=0.5,right] {$p$};
\draw[line width=1pt,draw=black](0,9)--(8,9)  node[pos=0.5,below] {$s$};
\draw[line width=1pt,draw=black](20,11)--(12,11) node[pos=0.5,below] {$s$};
\draw[draw=black!29,dashed](0,0)--(20,0) (20,20)--(0,20);
\filldraw[draw=black!29,fill=black!29](\p,0)--(\p+\dx,-\dy)--(\p+\dx,\dy)--(\p,0);
\filldraw[draw=black!29,fill=black!29](\p,20)--(\p+\dx,20-\dy)--(\p+\dx,20+\dy)--(\p,20);
\filldraw[draw=black!29,fill=black!29](\p+\ddx,0)--(\p+\dx+\ddx,-\dy)--(\p+\dx+\ddx,\dy)--(\p+\ddx,0);
\filldraw[draw=black!29,fill=black!29](\p+\ddx,20)--(\p+\dx+\ddx,20-\dy)--(\p+\dx+\ddx,20+\dy)--(\p+\ddx,20);
	\end{tikzpicture}
	}
  \end{array}
  = \delta_{p r} \sum_{q'} \shortf{sps}{p}{q'}{q} \tube{s q' s}{p}
  \, ,
   &
  T (\tube{pqr}{s}) &:= \tube{p 1 p}{p} \tube{pqr}{s} 
  \, ,
  \\
  S(\ici{a} ) &= \sum_{\ici{b}}  S_{ab}\, \ici{b}
  \, ,
    &
    T (\ici{a}) &= \theta_a\,  \ici{a}
    \, ,
\end{align}
in writing the second formula for $S$ we have assumed that all idempotents have dimension one, which is true for the toric code example. The general formula is given in Section~\ref{set:stringnetexample}, Eq.~\eqref{set:snegsandt}. 
We remark that $S$ corresponds to changing basis on the torus from a definite flux through the $y$-cycle to a definite flux through the $x$-cycle. Similarly $T$ corresponds to a Dehn twist along the $y$-cycle of the torus with definite flux through the $y$-cycle, or equivalently a $2\pi$ rotation of a superselection sector.

For toric code $S$ and $T$ are given by
  \begin{align}
\begin{tabular}{ |c | c c c c |} 
\hline
$\qd^2 S$ & $\vac$ & $e$ & $m$  & $em$
\\ \hline
$\vac$ & $1$ & $1$ & $1$ & $1$
\\
$e$ & $1$ & $1$  &  $-1$ & $-1$
\\
$m$ & $1$ & $-1$ & $1$ & $-1$ 
\\
$em$ & $1$ & $-1$ & $-1$ & $1$ 
\\ \hline
\end{tabular}
&\quad &
\begin{tabular}{ | c | c c c c |} 
\hline 
$T$ & $\vac$ & $e$ & $m$ & $em$
\\ \hline
$\vac$ & $1$ & & &
\\
$e$ & & $1$ & &
\\
$m$ & && $1$ &
\\
$em$ & & & & $-1$
\\
\hline
\end{tabular}
\end{align}
We remark the normalization of the $S$ matrix by $\qd^2$ refers to the quantum dimension of the input fusion category which satisfies $\qd ^2=\qd_\text{out}$. In this example the normalization is $\qd^2=2$.  
The $T$ matrix uniquely identifies the $em$ sector as it is the only fermion in the theory. However one can see that the choice of $e$ and $m$ sectors is arbitrary as the data is invariant under swapping $e$ with $m$.  This is an EM duality $\zt$ symmetry of the toric code. In the next section we will construct the symmetry enriched toric code, including the EM duality defects. 

In fact we can also extract the full set of gauge-variant data, the $F$ and $R$ symbols,  that define the emergent topological theory from the ICIs. The ICIs can be used to construct a realization of the emergent theory, which may be of use for topological quantum computation in wavefunctions that admit a tensor network description. 

Throughout this subsection all explanations were given with the toric code in mind and hence many subtleties have been left out. In particular we neglected orientation dependence, factors of the quantum dimension, pivotal phases, Frobenius-Schur indicators and possible fusion degeneracy. These subtleties are covered in Section~\ref{set:stringnetexample}.

\subsubsection{Doubled Ising model}
\label{set:isingexample}

The Ising fusion category is based on an algebra of three objects which we denote $\{\vac,\psi,\sigma\}$ with commutative multiplication rules 
\begin{align}
\vac\times a = a, \quad 
\psi\times \psi = \vac, \quad 
\sigma \times \psi = \sigma, \quad
\sigma \times \sigma = \vac + \psi,
\end{align}
for $a\in\{\vac,\psi,\sigma\}$. The quantum dimensions are $d_\vac=1,d_\psi=1,d_\sigma=\sqrt{2}$, hence $\qd^2=4$. The nontrivial $F$-symbols are given by
\begin{align}
\label{set:isingfsymbols}
\f{\sigma \sigma \sigma}{\sigma }{i}{j} = 
\frac{1}{\sqrt{2}}
\left[
\begin{matrix}
1 & 1 \\
1 & -1
\end{matrix}
\right],\quad
\f{\sigma \psi \sigma}{\psi}{\sigma}{\sigma}=\f{\psi \sigma \psi}{\sigma}{\sigma}{\sigma}=-1 ,
\end{align}
for $i,j\in\{\vac,\psi\}$. 
The remaining $F$-symbols are either $0$ or $1$, which is completely determined by the convention that $\f{a b c}{d}{e}{f}$ contains the constraint $\delta^e_{ a b} \delta^d_{ e c} \delta^f_{ b c} \delta^d_{ a f}$, defined in Eq.~\eqref{set:fusiondelta}. One can check that these $F$-symbols satisfy the pentagon equation.  
We remark that $\vecg{\zt}{}$ is contained as a subcategory of Ising, generated by the objects $\{\vac,\psi\}$. This observation is the basis of the construction in Section~\ref{set:settceg}.

The tensor network and MPO algebra for the Ising model are defined in a similar fashion to the toric code example. Both examples fall under the general case that covers all string-net models~\cite{MPOpaper,Bultinck2017183}, which is described in Section~\ref{set:stringnetexample}. Here we will skip directly to the tube algebra. 
The construction of the tube algebra proceeds similarly to the toric code, but one must also deal with the appearance of quantum dimensions. 
For an explanation of the explicit details we have skipped over see Section~\ref{set:stringnetexample}.

The Ising tube algebra is twelve dimensional and is spanned by the tubes in Eq.~\eqref{set:TCtubepictures}, together with the following
  \begin{align}
&\tube{\sigma\sigma\sigma}{\vac} = 
\begin{array}{c}
\includeTikz{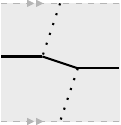}{\drawisingtube{sigma}{sigma}{sigma}{vacuum}}
  \end{array}\, ,
  \quad
\tube{\sigma\sigma\sigma}{\psi} = 
\begin{array}{c}
\includeTikz{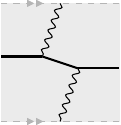}{\drawisingtube{sigma}{sigma}{sigma}{psi}}
  \end{array}\, ,
  \nonumber \\
&\tube{\sigma \vac \sigma}{\sigma} = 
\begin{array}{c}
\includeTikz{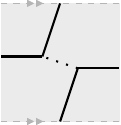}{\drawisingtube{sigma}{vacuum}{sigma}{sigma}}
  \end{array}\, ,
  \quad
  \tube{\sigma\psi\sigma}{\sigma} = 
\begin{array}{c}
\includeTikz{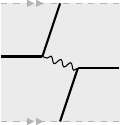}{\drawisingtube{sigma}{psi}{sigma}{sigma}}
  \end{array}\, ,
  \nonumber \\
&\tube{\vac\sigma\vac}{\sigma} = 
\begin{array}{c}
\includeTikz{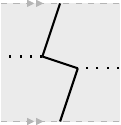}{\drawisingtube{vacuum}{sigma}{vacuum}{sigma}}
  \end{array}\, ,
  \quad
  \tube{\psi\sigma\psi}{\sigma} = 
\begin{array}{c}
\includeTikz{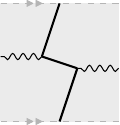}{\drawisingtube{psi}{sigma}{psi}{sigma}}
  \end{array}\, ,
  \nonumber \\
&\tube{\psi\sigma \vac}{\sigma} = 
\begin{array}{c}
\includeTikz{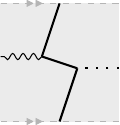}{\drawisingtube{psi}{sigma}{vacuum}{sigma}}
  \end{array}\, ,
  \quad
  \tube{\vac\sigma\psi}{\sigma} = 
\begin{array}{c}
\includeTikz{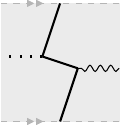}{\drawisingtube{vacuum}{sigma}{psi}{sigma}}
  \end{array}\, 
.
  \end{align}

The coefficients of the ICIs that determine the superselection sectors of the tensor network are given by 
\begin{align}
\label{set:isingtable}
	\begin{tabular}{ | c | c c c c c c c c c c | } 
\hline
Anyons & \tube{\vac\vac\vac}{\vac} & \tube{\vac\psi \vac}{\psi } & \tube{\psi\psi\psi}{\vac} & \tube{\psi \vac\psi}{\psi} & \tube{\sigma \sigma \sigma}{\vac} & \tube{\sigma \sigma \sigma}{ \psi} & \tube{\vac \sigma \vac}{\sigma} & \tube{\psi \sigma \psi}{\sigma}  & \tube{\sigma \vac \sigma}{\sigma} & \tube{\sigma\psi\sigma}{\sigma} 
\\ \hline
$\vac$ & $1$ & $1$ & & & & &  $\sqrt{2}$ & & &
 \\ 
 $\psi\conj{\psi}$ & $1$ & $1$ &  & & & & $-\sqrt{2}$ & & & 
  \\ 
${\psi}$& & & $1$ & $-1$ & & &  &  $-i\sqrt{2}$  & &
 \\ 
$\conj{\psi}$  & & & $1$ & $-1$ & & &  & $i\sqrt{2}$  & & 
 \\ 
 $\sigma\conj{\sigma}$ & $2$ & $-2$ & $2$ & $2$ & & &  & & &
 \\ 
$\sigma$ & & & & & $1$ & $-i$ &  & & $e^{-\frac{\pi i}{8}}$ & $e^{\frac{3 \pi i }{8}}$ 
 \\ 
$\sigma\conj{\psi}$  & & & & & $1$ & $-i$ & & & $e^{\frac{7 \pi i}{8}}$ & $e^{-\frac{5 \pi i }{8}}$ 
 \\ 
$\conj{\sigma}$ & & & & & $1$ & $i$ &  & & $e^{\frac{\pi i}{8}}$ & $e^{-\frac{3 \pi i }{8}}$ 
 \\ 
$\psi\conj{\sigma}$ & & & & & $1$ & $i$ & & & $e^{-\frac{7 \pi i}{8}}$ & $e^{\frac{5 \pi i }{8}}$ 
\\
\hline
	\end{tabular}
\end{align}
where we have suppressed a normalization by $\qd^2=4$.  
To be explicit, the table should be read as follows
\begin{align}
\ici{a_i} = \frac{1}{\qd^2} \sum_{j} t_{i}^{j} (\tube{pqr}{s})_j ,
\end{align}
where $a_i$ is the $i$-th anyon row label,  $(\tube{pqr}{s})_j$ is the $j$-th tube column label, and  $t_{i}^{j}$ is the $(i,j)$-th table entry. 
We remark that our normalization convention for the tubes $\tube{pqr}{s}$ differs slightly from that in Ref.~\cite{Bultinck2017183} by a factor $\sqrt{d_s}$, after adjusting for this difference our results match those reported in Ref.~\cite{Bultinck2017183}. 

In the Ising tube algebra the $\ici{\sigma \conj{\sigma}}$ block is two dimensional. The full block, including off diagonal elements, is  given by 
\begin{align}
\ici{\sigma \conj{\sigma}}_{00} &= \frac{1}{2}(\tube{\vac\vac\vac}{\vac}-\tube{\vac\psi\vac}{\psi}) & \ici{\sigma \conj{\sigma}}_{01} &= \frac{1}{\sqrt{2}} \tube{\vac\sigma \psi}{\sigma}
\nonumber \\
\ici{\sigma \conj{\sigma}}_{10} &= \frac{1}{\sqrt{2}}  \tube{\psi\sigma \vac}{\sigma} &  \ici{\sigma \conj{\sigma}}_{11} &=  \frac{1}{2}(\tube{\psi \psi \psi}{\vac}+\tube{\psi \vac \psi}{\psi}) .
\end{align}

To justify the labeling of the particles we have calculated the $S$ and $T$ matrices from the ICIs using the formula in Section~\ref{set:stringnetexample}, Eq.~\eqref{set:snegsandt}. Note that since $\ici{\sigma\conj{\sigma}}$ is a two dimensional block, the $S$ matrix is defined in terms of its action on the normalized ICI $\frac{1}{2}\ici{\sigma\conj{\sigma}}$.
  \begin{align}
  \qd^2 S =& \
\begin{tabular}{ |c | c c c | c | c | c c c|} 
\cline{1-4} \cline{6-9}
 & \vac & $\psi$ & $\sigma$  & & & \conj{\vac} & $\conj{\psi}$ & $\conj{\sigma}$
\\ \cline{1-4} \cline{6-9}
$\vac$ & $1$ & $1$ & $\sqrt{2}$ & \multirow{3}{*}{$\bigotimes$} &
$\vac$ & $1$ & $1$ & $\sqrt{2}$
\\
$\psi$ & $1$ & $1$  &  $-\sqrt{2}$ & &
$\conj{\psi}$ & $1$ & $1$  &  $-\sqrt{2}$
\\
$\sigma$ & $\sqrt{2}$ & $-\sqrt{2}$ & $0$ & &
$\conj{\sigma}$ &  $\sqrt{2}$ & $-\sqrt{2}$ & $0$ 
\\ \cline{1-4} \cline{6-9}
\end{tabular}
\\
T=& \ 
\begin{tabular}{ |c | c c c | c | c | c c c|} 
\cline{1-4} \cline{6-9}
 & \vac & $\psi$ & $\sigma$  & & & \conj{\vac} & $\conj{\psi}$ & $\conj{\sigma}$
\\ \cline{1-4} \cline{6-9}
$\vac$ & $1$ &  &  & \multirow{3}{*}{$\bigotimes$} &
$\vac$ & $1$  & &
\\
$\psi$ &  & $-1$  &  & &
$\conj{\psi}$ & & $-1$  &
\\
$\sigma$ & & & $e^{\frac{\pi i}{8}}$ & &
$\conj{\sigma}$ & & & $e^{-\frac{\pi i}{8}}$
\\ \cline{1-4} \cline{6-9}
\end{tabular}
\end{align}
We remark the $S$ and $T$ matrices have a tensor product structure since the input Ising fusion category admits a modular braiding. This implies~\cite{DrinfeldCenter} $\double{\text{Ising}}\cong \text{Ising}^{(1)} \boxtimes \conj{\text{Ising}}^{(1)}$, where $\text{Ising}^{(1)}$ denotes the Ising theory equipped with a modular braiding following the convention of Ref.~\cite{barkeshli2014symmetry}.

\subsection{Symmetry-enriched tensor networks from anyon condensation}

\subsubsection{$\zt$-paramagnet from the toric code}

To demonstrate the condensation procedure we note the toric code trivially has a $\zt$-grading $\cat_{\g{1}} \oplus \cat_{\g{x}}=\{\vac\}\oplus\{\psi\}$, where we use the following presentation  $\zt\cong \left\langle \g{x}\ | \ \g{x}^2= {\g{1}}\right\rangle$. We make use of the notation $\psi_{\g{x}}$ to indicate $\psi\in\cat_{\g{x}}$ and $\sector{\psi}=\g{x}$ to indicate the sector containing $\psi$. In this example each sector contains only a single simple object, so there is no need to keep track of the sector label separately. However, we will explicitly keep track of the sectors to prepare the reader for more complicated examples.

The tensors for the symmetry-enriched model~\cite{PhysRevB.94.235136,cheng2016exactly} are given by a simple modification of the string-net tensors, which amounts to promoting the sector label of the virtual indices to a physical degree of freedom. 
\begin{align}
	\begin{array}{c}
	\includeTikz{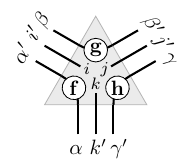}{
	\begin{tikzpicture}[scale=.1]
	\node[transform shape,draw=black!29,fill=black!10!white!80,regular polygon,regular polygon sides=3,minimum size=20cm] at (0,0){};
	\begin{scope}
	\draw[draw=black,line width=1pt,postaction=decorate] (0,-10) -- (0,-3);
	\draw[draw=black,line width=1pt] (-3,-10) -- (-3,-5) ;
	\draw[draw=black,line width=1pt] (3,-10) -- (3,-5);
	\draw[draw=black,line width=1pt,rounded corners=1mm] (-3,-5) to [bend right=30] ({3*cos(240)+5*sin(240)},{3*sin(240)-5*cos(240)});
	\node at (.5,-12.5){$\alpha\ k'\, \gamma'$};
	\end{scope}
	\begin{scope}[rotate=120]
	\draw[draw=black,line width=1pt,postaction=decorate] (0,-3) --  (0,-10);
	\draw[draw=black,line width=1pt] (-3,-10) -- (-3,-5) ;
	\draw[draw=black,line width=1pt] (3,-10) -- (3,-5);
	\draw[draw=black,line width=1pt,rounded corners=1mm] (-3,-5) to [bend right=30] ({3*cos(240)+5*sin(240)},{3*sin(240)-5*cos(240)});
	\node[rotate=300] at (-.5,-12.5){$\beta' \, j'\, \gamma$};
	\end{scope}
	\begin{scope}[rotate=-120]
	\draw[draw=black,line width=1pt,postaction=decorate]  (0,-3) -- (0,-10);
	\draw[draw=black,line width=1pt] (-3,-10) -- (-3,-5) ;
	\draw[draw=black,line width=1pt] (3,-10) -- (3,-5);
	\draw[draw=black,line width=1pt,rounded corners=1mm] (-3,-5) to [bend right=30] ({3*cos(240)+5*sin(240)},{3*sin(240)-5*cos(240)});
	\node[rotate=60] at (0,-12.5){$\alpha' \ i'\, \beta$};
	\end{scope}
	\draw[draw=black,fill=white] (0,4.25) circle (2cm);
	\node at (0,4.25){$\g{g}$};
	\begin{scope}[rotate=120]
	\draw[draw=black,fill=white] (0,4.25) circle (2cm);
	\node at (0,4.45){$\g{f}$};
	\end{scope}
		\begin{scope}[rotate=-120]
	\draw[draw=black,fill=white] (0,4.25) circle (2cm);
	\node at (0,4.25){$\g{h}$};
	\end{scope}
		\node at (-1.5,1){\footnotesize $i$};
		\node at (1.5,1){\footnotesize $j$};
		\node at (0,-1.5){\footnotesize $k$};
	\end{tikzpicture}
	}
	\end{array}
&=\delta_{\g{f} \sector{\alpha}} \delta_{\g{g} \sector{\beta}} \delta_{\g{h} \sector{\gamma}}
\delta_{i i'} \delta_{j j'} \delta_{k k'}
 \delta_{\alpha\alpha'}\delta_{\beta \beta'}\delta_{\gamma\gamma'} \shortf{\alpha i j}{\gamma}{\beta}{k}
,
\end{align}

The set of MPOs for the SET string-net tensor are still given by Eq.~\eqref{set:TCtensors}, we simply reinterpret them in light of the additional  $\G$-grading structure. 
Hence the $\g{1}$ sector of the MPO algebra is given by $\mpo{}{\vac}$ and the $\g{x}$ sector is given by $\mpo{}{\psi}$. 
Since we are using the same MPOs as in Section~\ref{set:tubebackground} they still satisfy the conditions introduced in Eqs.~\eqref{eq:sec21zipper},~\eqref{eq:sec21FMove},~\eqref{eq:sec21PullThru}. 

Eq.~\eqref{eq:sec21PullThru} must be modified in accordance with the $\G$-grading. That is, pulling through an MPO in sector $\g{g}$ results in a $\g{g}$ action on the physical indices
	\begin{align}
	\label{set:tcsetpullthru}
	\raisebox{-.9cm}{$

	$}
	\, ,
\end{align}
where
\begin{align}
			\raisebox{-.15cm}{$
			\begin{array}{c}
	\includeTikz{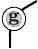}{
	\begin{tikzpicture}[scale=.1]
	\begin{scope}[rotate=-120]
		\begin{scope}[rotate=-120]
	\draw[draw=black,line width=1pt,rounded corners=1mm] (-3,-5) to [bend right=30] ({3*cos(240)+5*sin(240)},{3*sin(240)-5*cos(240)});
		\draw[draw=black,line width=1pt, rounded corners=1mm] (-3,-7) -- (-3,-5);
	\end{scope}
		\draw[draw=black,fill=white] (0,4.25) circle (2cm);
		\begin{scope}[rotate=120]
	\draw[draw=black,line width=1pt, rounded corners=1mm] (3,-7) -- (3,-5);
	\end{scope}
			\draw[line width=.3pt ,draw=black!40,fill=black!20]  (0,4.25) circle (1.3cm);
			\node at (0,4.25){\small $\g{g}$};
	\end{scope}
	\end{tikzpicture}
	}
	\end{array}
	$}
	:=
		\raisebox{-.15cm}{$
		\begin{array}{c}
	\includeTikz{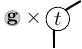}{
	\begin{tikzpicture}[scale=.1]
	\begin{scope}[rotate=-120]
		\begin{scope}[rotate=-120]
	\draw[draw=black,line width=1pt,rounded corners=1mm] (-3,-5) to [bend right=30] ({3*cos(240)+5*sin(240)},{3*sin(240)-5*cos(240)});
		\draw[draw=black,line width=1pt, rounded corners=1mm] (-3,-7) -- (-3,-5);
	\end{scope}
		\draw[draw=black,fill=white] (0,4.25) circle (2cm);
		\begin{scope}[rotate=120]
	\draw[draw=black,line width=1pt, rounded corners=1mm] (3,-7) -- (3,-5);
	\end{scope}
			\node at (0,4.25){$t$};
	\end{scope}
		\draw[line width=.3pt ,draw=black!40,fill=black!20] (-4,-2) circle (1.3cm);
	\node at (-4,-2.2){\small $\g{g}$};
		\node[anchor=west] at (-3,-2){$\times$};
	\end{tikzpicture}
	}
	\end{array}
	$}
	=
		\raisebox{-.15cm}{$
			\begin{array}{c}
	\includeTikz{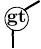}{
	\begin{tikzpicture}[scale=.1]
	\begin{scope}[rotate=-120]
		\begin{scope}[rotate=-120]
	\draw[draw=black,line width=1pt,rounded corners=1mm] (-3,-5) to [bend right=30] ({3*cos(240)+5*sin(240)},{3*sin(240)-5*cos(240)});
		\draw[draw=black,line width=1pt, rounded corners=1mm] (-3,-7) -- (-3,-5);
	\end{scope}
		\draw[draw=black,fill=white] (0,4.25) circle (2cm);
		\begin{scope}[rotate=120]
	\draw[draw=black,line width=1pt, rounded corners=1mm] (3,-7) -- (3,-5);
	\end{scope}
			\node at (-.03,4.25){\small $\g{gt}$};
	\end{scope}
	\end{tikzpicture}
	}
	\end{array}
	$}
	.
	\end{align}

Similar to the purely topological case, the calculations involving MPOs within a tensor network that satisfies the SET pulling though condition can be abstracted to a diagrammatic calculus. This calculus additionally involves discs that are labeled by group elements, which represent the application of the on-site symmetry to the tensor network within the labeled region. For example 
\begin{align}
\begin{array}{c}
\label{set:abstractingtopictures2}
	\includeTikz{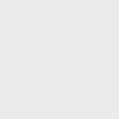}{
	\begin{tikzpicture}[scale=.1]
	\filldraw[UFCBackground] (0,0) --(20,0)--(20,20)--(0,20)--cycle;
	\end{tikzpicture}
	}
\end{array}
\sim
\begin{array}{c}
	\includeTikz{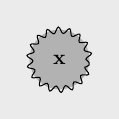}{
	\begin{tikzpicture}[scale=.1]
	\filldraw[UFCBackground] (0,0) --(20,0)--(20,20)--(0,20)--cycle;
			\filldraw[line width=.3pt ,draw=black!30,fill=black!30] (10,10) circle (5cm);
	\draw[psi] (10,9.5) circle (5cm);
	\node at (10,10){ $\g{x}$};
	\end{tikzpicture}
	}
\end{array}
\, , \
\text{ alternatively, } \
\begin{array}{c}
	\includeTikz{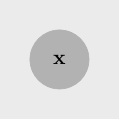}{
	\begin{tikzpicture}[scale=.1]
	\filldraw[UFCBackground] (0,0) --(20,0)--(20,20)--(0,20)--cycle;
		\filldraw[line width=.3pt ,draw=black!30,fill=black!30] (10,10) circle (5cm);
	\node at (10,10){ $\g{x}$};
	\end{tikzpicture}
	}
\end{array}
\sim
\begin{array}{c}
	\includeTikz{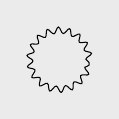}{
	\begin{tikzpicture}[scale=.1]
	\filldraw[UFCBackground] (0,0) --(20,0)--(20,20)--(0,20)--cycle;
	\draw[psi] (10,9.5) circle (5cm);
	\end{tikzpicture}
	}
\end{array}
\, ,
\end{align}
where the $\g{1}$ label, corresponding to the action of the trivial element, remains implicit. This corresponds to a diagrammatic calculus for a 2-category~\cite{barkeshli2016reflection,cui2017state} where the set of objects is $\G$, the morphisms from $\g{g}$ to $\g{h}$ are given by the objects in $\cat_{\conj{\g{g}} \g{h}}$, and the 2-morphisms are given by the 1-morphisms of $\cat_{\G}$.

The emergent SET order of the tensor network is described by a unitary $\G$-crossed braided fusion category. This mathematical structure captures the topological properties of extrinsic monodromy defects as they braid and fuse. The inequivalent monodromy defects for a given $\g{g}\in\G$ correspond to the superselection sectors at the end point of a $\g{g}$ domain wall. 
The defect superselection sectors can be calculated using a modification of the tube algebra, which is described in Section~\ref{setsection:dubes}. 

The SET is intimately related to a topological order obtained by gauging the $\G$ symmetry~\cite{gelaki2009centers,barkeshli2014symmetry,teo2015theory}, which can be done directly on the lattice~\cite{Gaugingpaper,williamson2014matrix,PhysRevB}. The superselection sectors of the gauged topological order can be derived from the defect superselection sectors and the action of the group upon them. 
Alternatively, the defect sectors of the SET can be found via a $\repg$ condensation~\cite{PhysRevB.79.045316} procedure on the gauged topological order. 

Here we demonstrate the effect of condensing Rep$(\zt)$ on the toric code tubes. 
This is implemented by dividing the tube algebra into $\g{1}$ and $\g{x}$ sectors, according to the group element $\sector{p}=\sector{r}$ in $\tube{pqr}{s}$. Next we simply drop the basis elements $\tube{pqr}{s}$ with $\sector{s}=\g{x}$, corresponding to a $\psi$ line wrapping the tube, from the tube algebra. 
The ICIs in the $\g{1}$ sector correspond to anyons or $\g{1}$-defects, while those in the $\g{x}$ sector correspond to $\g{x}$-defects. 
Since the defect tube algebra in each sector becomes one dimensional, we find only the vacuum anyon and a single $\g{x}$-defect, $\Delta_{\g{x}}$. 
\begin{align}
&\begin{tabular}{ |c | c c c c |} 
\hline 
Anyons & \tube{\vac\vac\vac}{\vac} & {\tube{\vac\psi \vac}{\psi }} & \tube{\psi\psi\psi}{\vac} & {\tube{\psi \vac\psi}{\psi}}
\\ \hline
$\vac$ &  $1$ & {$1$} & &
\\
$e$ & $1$ & {$-1$}  & &
\\
$m$ & &  & $1$ & {$1$}
\\
$em$ & & \makebox(1,-8){\red{\rule[16.75ex]{.5pt}{6\normalbaselineskip}}} 
& $1$ & {$-1$}\makebox(-13,-8){\red{\rule[16.75ex]{.5pt}{6\normalbaselineskip}}}
\\
\hline
\end{tabular}
\nonumber \\
& \qquad \text{\large $\mapsto$ } \qquad
\begin{tabular}{ | c | c  | c | c | c  | c | c | c | } 
\cline{1-5} \cline{7-8}
\multirow{2}{*}{Sectors} &\multirow{2}{*}{Defects} & Condensed & \multirow{2}{*}{$\tube{\vac\vac\vac}{\vac}$}  & \multirow{2}{*}{$\tube{\psi\psi\psi}{\vac}$  } & & \multicolumn{2}{|c|}{$ \zt$-action}
\\ \cline{7-8}
&&anyons&& & & {\tube{\vac\psi \vac}{\psi }} & {\tube{\psi \vac\psi}{\psi}}
\\ \cline{1-5} \cline{7-8}
\multirow{2}{*}{$\g{1}$ } & \multirow{2}{*}{$\vac$} & $\vac$ &  $1$  & & & \multirow{2}{*}{$1$} & 
\\
 && $e$ & $1$ & & & &
\\ \cline{1-5} \cline{7-8}
\multirow{2}{*}{$\g{x}$ } &\multirow{2}{*}{$\Delta_{\g{x}}$} & $m$ & & $1$  & & & \multirow{2}{*}{$1$}
\\
&& $em$ & &  $1$ & & &
\\ \cline{1-5} \cline{7-8}
\end{tabular}
\end{align}
In the right table we retain the implicit normalization of $\qd^2=2$,  from the toric code, in the coefficients of the ICIs. However, we also have multiple contributions to each defect ICI from the different anyon ICIs that condense into it. In this case the factors cancel and we simply have 
\begin{align}
\ici{\vac}=\tube{\vac\vac\vac}{\vac} & & \ici{\Delta_{\g{x}}}=\tube{\psi \psi \psi}{\vac}
.
\end{align}
From the table we see that the Rep$(\zt)$ boson which condensed was the $e$ particle of the toric code. By picking a different tensor network representation of the toric code one can also realize the equivalent condensation of $m$. The result is a trivial topological order with a single $\zt$ defect. Since the underlying topological order is trivial, the model is an SPT which is classified by the cohomology class of the associator for the defects. This label lies in $\coho{3}$ and for this example it is trivial. 
Hence we have found a trivial $\zt$-paramagnet as expected. 

The tubes that were dropped to make the defect tube algebra can now be used to realize the action of the symmetry on the defect sectors. This is achieved by projecting the collection of discarded tubes onto each defect ICI. Hence we find the action $\tube{\vac \psi \vac}{\psi}$ on the vacuum, as expected, and $\tube{\psi \vac \psi}{\psi}$ acting on the $\Delta_{\g{x}}$ defect. These actions generate linear representations of $\zt$, note the identity in each sector is given by the corresponding defect ICI. 

With the representation of the symmetry for each defect we can also calculate the effect of gauging the $\zt$ symmetry. Specifically there is a simple recipe to calculate the full set of anyon ICIs of the gauged theory. 
For each defect the post-gauging anyon ICIs are given by the projectors onto the irreps of the group action on that defect. 
In the toric code example this implies
\begin{align}
\ici{\vac} \
 \mapsto 
\left. 
\begin{array}{c c c}
\frac{1}{2}(\tube{\vac\vac\vac}{\vac} + \tube{\vac\psi \vac}{\psi}) & = & \ici{\vac}
\\
\frac{1}{2}(\tube{\vac\vac\vac}{\vac} - \tube{\vac\psi \vac}{\psi}) & =  & \ici{e}
\end{array}
\right. &
&
\ici{\Delta_{\g{x}}} \
 \mapsto 
\left. 
\begin{array}{c c c}
\frac{1}{2}(\tube{\psi \psi\psi}{\vac} + \tube{\psi \vac \psi}{\psi}) & = & \ici{m}
\\
\frac{1}{2}(\tube{\psi \psi\psi}{\vac} - \tube{\psi \vac \psi}{\psi}) & = & \ici{em}
\end{array}
\right. 
.
\end{align}

We remark that there are much simpler ways to analyze the condensation / gauging procedure for the toric code, we have included it here to demonstrate our general approach on a familiar example.

\subsubsection{Symmetry-enriched toric code from the doubled Ising model}
\label{set:settceg}

The Ising fusion category has a $\zt$-graded structure $\cat_{\g{1}}\oplus\cat_{\g{x}}=\{\vac,\psi\}\oplus\{\sigma\}$. In this case there are two simple objects in the trivial sector, so it is necessary to keep track of the object and sector index separately.

A tensor network representation of the recently introduced SET string-net models~\cite{cheng2016exactly,PhysRevB.94.235136} is given by the following tensors
\begin{align}

&= 
\begin{array}{c}
 \delta_{\g{f} \sector{\alpha}} \delta_{\g{g} \sector{\beta}} \delta_{\g{h} \sector{\gamma}}
 \delta_{ii'}
 \\
\delta_{jj'}\delta_{kk'}\delta_{\alpha\alpha'}\delta_{\beta \beta'}\delta_{\gamma\gamma'} 
\end{array}
&= 
\begin{array}{c}
 \delta_{\g{f} \sector{\alpha}} \delta_{\g{g} \sector{\beta}} \delta_{\g{h} \sector{\gamma}}
 \delta_{ii'}
 \\
\delta_{jj'}\delta_{kk'}\delta_{\alpha\alpha'}\delta_{\beta \beta'}\delta_{\gamma\gamma'} 
\end{array}
 \frac{\shortf{\alpha i j}{\gamma}{\beta}{k}^*}{\sqrt{d_{\beta} d_{k}}}
\end{align}
where the $F$-symbols are those of the input $\G$-graded fusion category. For the current example they are given in Eq.~\eqref{set:isingfsymbols}. We remark that the tensors are related by a combination of reflection and complex conjugation. 

The MPOs for this tensor network are the same as those for the usual string-net~\cite{levin2005string}, which are described in Section~\ref{set:stringnetexample}. 
However, the MPOs in the $\g{g}$-sector now represent $\g{g}$-domain walls. The multiplication of these MPOs is $\zt$-graded, and in particular $\{ \frac{1}{2}( \mpo{}{\vac}+\mpo{}{\psi}),\frac{1}{\sqrt{2}} \mpo{}{\sigma}\}$ forms a representation of $\zt$. 

These domain walls require a group action to be moved through the lattice. 
This is captured by the symmetry-enriched pulling through equation 
	\begin{align}
	\raisebox{-.9cm}{$

	$}.
	\label{set:defectisingpullthru}
\end{align}
We have used the notation defined in Eq.~\eqref{set:tcsetpullthru} for the group action on the physical indices. The equation on the right is  obtained from the left via complex conjugation and reflection. All other pulling through configurations  can be derived from Eq.~\eqref{set:defectisingpullthru} by using properties of the string-net MPO given in Section~\ref{set:stringnetexample}.

The defect tube algebra is spanned by the subset of Ising tubes $\tube{pqr}{s}$, given in Section~\ref{set:isingexample}, with $s\in\{ \vac, \psi\}$. 
The algebra is $\zt$-graded, with the $\g{1}$ sector containing tubes with $\sector{p}=\sector{q}=\g{1}$, and the $\g{x}$ sector containing those with $\sector{p}=\sector{q}=\g{x}$. 
The ICIs can be found by condensing the $\rep{\zt}$ boson in doubled Ising, corresponding to the $\zt$ grading of the input Ising fusion category. 
First the --- possibly non-central --- irreducible idempotents of the defect tube algebra are derived from those of the doubled Ising model, given in Eq.~\eqref{set:isingtable}, by simply dropping the elements $\tube{pqr}{s}$ with $s=\sigma$. 
In this case all the resulting idempotents turn out to be central in the defect tube algebra. In particular the two non-central irreducible idempotents in the ${\sigma\conj{\sigma}}$ block of doubled Ising split into distinct sectors after condensation. 
The results are summarized in the following tables.
\begin{align}
\label{set:defectisingexample}
\begin{tabular}{ |c | c | c | c c c c | c c  | } 
\cline{1-9} 
\multirow{2}{*}{Sectors} & \multirow{2}{*}{Defects} & Condensed    & \multirow{2}{*}{\tube{\vac\vac\vac}{\vac}} & \multirow{2}{*}{\tube{\vac\psi \vac}{\psi }} & \multirow{2}{*}{\tube{\psi\psi\psi}{\vac}} & \multirow{2}{*}{\tube{\psi \vac\psi}{\psi}} & \multirow{2}{*}{\tube{\sigma \sigma \sigma}{\vac}} & \multirow{2}{*}{\tube{\sigma \sigma \sigma}{ \psi}} 
\\ 
&&anyons&& &&&& 
\\ \cline{1-9}
\multirow{6}{*}{$\g{1}$ }  & \multirow{2}{*}{$\vac$} & $\vac$ & $1$ & $1$ & & & & 
 \\ 
 & & $\psi\conj{\psi}$ & $1$ & $1$ &  & & & 
 \\ \cline{2-3} 
& \multirow{2}{*}{$em$} & ${\psi}$ & & & $1$ & $-1$ & & 
 \\ 
& & $\conj{\psi}$ & & & $1$ & $-1$ & & 
 \\ \cline{2-3}  
& $e$ & $\sigma\conj{\sigma}_{00}$ & $2$ & $-2$ & &  & & 
\\ \cline{2-2} 
& $m$ & $\sigma\conj{\sigma}_{11}$ &  & & $2$ & $2$ & & 
 \\ \cline{1-9}
\multirow{4}{*}{$\g{x}$ } & \multirow{2}{*}{$\sigma_+$} & $\sigma$ & & & & & $1$ & $-i$ 
 \\ 
& & $\sigma\conj{\psi}$  & & & & & $1$ & $-i$ 
 \\ \cline{2-3}   
& \multirow{2}{*}{$\sigma_-$} & $\conj{\sigma}$ & & & & & $1$ & $i$ 
 \\ 
& & $\psi\conj{\sigma}$ & & & & & $1$ & $i$
\\ \cline{1-9}
	\end{tabular}
\end{align}
\begin{align}
\begin{tabular}{ |c | c | c |  c c  c c | c c |} 
\cline{1-9}
\multirow{2}{*}{Sectors} & \multirow{2}{*}{Defects} & Condensed    &  \multicolumn{6}{c|}{$\zt$-action}
\\ \cline{4-9}
&&anyons& \tube{\vac \sigma \vac}{\sigma} & \tube{\psi \sigma \psi}{\sigma} &  \tube{\vac \sigma \psi}{\sigma} &  \tube{\psi \sigma \vac}{\sigma} & \tube{\sigma \vac \sigma}{\sigma} & \tube{\sigma\psi\sigma}{\sigma} 
\\ \cline{1-9} 
\multirow{6}{*}{$\g{1}$ }  & \multirow{2}{*}{$\vac$} & $\vac$ & \multirow{2}{*}{$\sqrt{2}$} && & &&
 \\ 
 & & $\psi\conj{\psi}$ &  &&& &&
 \\ \cline{2-3} 
& \multirow{2}{*}{$em$} & ${\psi}$ &  & \multirow{2}{*}{$ \sqrt{2}$} 
&& &&
 \\ 
& & $\conj{\psi}$ & &&& &&
 \\ \cline{2-3}   
& $e$ & $\sigma\conj{\sigma}_{00}$ & &&  & $\sqrt{2}$  & &
\\ \cline{2-2} 
& $m$ & $\sigma\conj{\sigma}_{11}$ &  &&  $\sqrt{2}$ & &&
 \\ \cline{1-9} 
\multirow{4}{*}{$\g{x}$ } & \multirow{2}{*}{$\sigma_+$} & $\sigma$ &   && && \multirow{2}{*}{$e^{-\frac{\pi i}{4}}$} & \multirow{2}{*}{$e^{\frac{ \pi i }{4}}$}  
 \\ 
& & $\sigma\conj{\psi}$  &  &&& &&
 \\ \cline{2-3}   
& \multirow{2}{*}{$\sigma_-$} & $\conj{\sigma}$ & && && \multirow{2}{*}{$e^{\frac{\pi i}{4}}$} & \multirow{2}{*}{$e^{-\frac{ \pi i }{4}}$}  
 \\ 
& & $\psi\conj{\sigma}$ & &&& &&
\\ \cline{1-9} 
	\end{tabular}
\end{align}
The table above carries over the implicit normalization of $\qd^2=4$ from Eq.~\eqref{set:isingtable}, but there are also two contributions to each of the condensed defects, except $e$ and $m$. For example we have $\ici{\vac}=\frac{1}{2}(\tube{\vac\vac\vac}{\vac}+\tube{\vac\psi \vac}{\psi})$.

Each defect sector, $\ici{a}$, carries a $\zt$ action, $\dw{g}{a}$. This is found by projecting the Ising tubes that were thrown away onto the ICI of each sector and normalizing the results so that they satisfy the unitarity condition $(\dw{g}{a})^\dagger \dw{g}{a}=\ici{a}$. 
\begin{align}
K\, (c_0 \tube{\vac \sigma \vac}{\sigma} +c_1 \tube{\psi \sigma \psi}{\sigma} +c_2  \tube{\psi \sigma \vac}{\sigma} +c_3 \tube{\vac \sigma \psi}{\sigma} +c_4 \tube{\sigma \vac \sigma}{\sigma} +c_5 \tube{\sigma\psi\sigma}{\sigma} )  \, \ici{a}
= \dw{x}{a}
,
\end{align}
for arbitrary constants $c_i\in\mathbb{C}$ and some positive constant normalization $K$ that depends on the $c_i$.  We remark that certain specific choices of $c_i$ can result in a zero projection, but any generic choice will lead to a nonzero result. 
There is a gauge freedom in the definition of $\dw{x}{a}$ corresponding to multiplication by a coboundary. In particular, one can always choose  $\dw{1}{a}=
\ici{a}$ for all $\ici{a}$. 

The resulting $\dw{x}{a}$ for the symmetry-enriched toric code are contained on the right of Eq.~\eqref{set:defectisingexample}, note there is an implicit normalization by $\qd_\g{1}^2=2$ there. 
We remark that the tubes which were thrown away correspond to an $\g{x}$ domain wall, hence projecting them onto a defect captures the action of the $\g{x}$ domain wall. 

The nontrivial group element $\g{x}$ may permute each defect $\ici{a}$, resulting in a defect $\act{x}\ici{a}$,  this permutation is found by considering
\begin{align}
\dw{x}{a} (\dw{x}{a} )^\dagger = \act{x}\ici{a} 
.
\end{align}
In our example we find the only nontrivial permutation corresponds to the EM duality $\act{x}\ici{e}=\ici{m}, \act{x}\ici{m}=\ici{e}$ as expected. This is summarized in the following table
\begin{align}
\begin{tabular}{ | c | c c c c | c c| } 
\hline
 $\rho_{\g{x}}$  & $\act{x}\vac$ & $\act{x}em$ & $\act{x}e$ & $\act{x}m$ & $\act{x}\sigma_+$ & $\act{x}\sigma_-$
\\ \hline
$\vac$ & $1$ & & & & &
\\
$em$ & & $1$ & & & &
\\
$e$ & & & & $1$  & &
\\
$m$ &&  & $1$ & & &
\\  \hline
$\sigma_+$ &  & & & & $1$ &
\\
$\sigma_-$ &  & & & & & $1$
\\ \hline
\end{tabular}
\end{align}

For each defect, the action of the subgroup that does not permute it may form a projective representation. For our example we find 
\begin{align}
(\dw{x}{\vac})^2= \dw{1}{\vac}  \, , && (\dw{x}{em})^2= - \dw{1}{em} \, , && (\dw{x}{\sigma_+})^2= e^{-\frac{\pi i}{4}}\dw{1}{\sigma_+}  \, , && (\dw{x}{\sigma_-})^2= e^{\frac{\pi i}{4}} \dw{1}{\sigma_-} 
.
\end{align}
Note, the symmetry groups preserving $\ici{e}$ and $\ici{m}$ are trivial. In this case the projective phases of the $\g{x}$-defect sectors can be trivialized by a gauge transformation. 

The $\dw{h}{a_{\g{g}}}$ domain walls determine a basis of minimally entangled states for each symmetry twisted sector on the torus. The state $\ket{\dw{h}{a_{\g{g}}}}$ is given by a linear combination of states derived from basis elements of the tube algebra
\begin{align}
\ket{p_{\g{g}} \, s_{\g{h}}  \, q_{\g{g\conj{h}}}}
= \begin{array}{c}
\includeTikz{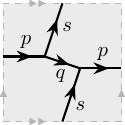}{\drawtorus{p}{q}{p}{s}}
\end{array}
.
\end{align}
 The coefficient of each state is determined by those of $\dw{h}{a_{\g{g}}}$ expanded in the tube basis.  For a more detailed description see Eq.~\eqref{set:defdomainwallstate}. 
For this state to be nonzero we must have $\act{h}{a}=a$, as the periodic boundary conditions imply that the following tubes yield the same state
\begin{align}
\dw{h}{a_{\g{g}}}= \dw{h}{a_{\g{g}}} \, \ici{a} \, \sim \,  \ici{a} \, \dw{h}{a_{\g{g}}}  = \delta_{\act{h}{a}\, a} \dw{h}{a_{\g{g}}}
.
\end{align}
Hence the dimension of the $(\g{g},\g{h})$-twisted sector is equal to the number of $\g{h}$-invariant $\g{g}$-defects.

The $\dw{h}{a_{\g{g}}}$  domain walls, satisfying $\act{h}{a}=a$, give rise to $\G$-crossed $S$ and $T$ matrices. These matrices are defined by their action on tube basis elements, which is described in Eq.~\eqref{set:snegsandt}. They satisfy 
\begin{align}
\frac{1}{ D_a} S(\dw{h}{a_{\g{g}}}) &= \sum_{b_{\g{h}} } \frac{ S_{a_{\g{g}} b_{\g{h}}} }{ D_b} \, \dw{{\conj{\text{g}}}}{b_{\g{h}}}
\, ,    &
    T (\dw{h}{a_{\g{g}}} ) &=  \theta_a \dw{gh}{a_\g{g}}
    ,
\\
   S_{\defect{a}{g} \defect{b}{h}} &=  \frac{D_b \tr[ ( \dw{{\conj{\text{g}}}}{b_{\g{h}}})^\dagger S(\dw{h}{a_{\g{g}}}) ]}{D_a \tr[ ( \dw{{\conj{\text{g}}}}{b_{\g{h}}})^\dagger \dw{{\conj{\text{g}}}}{b_{\g{h}}} ]} \, ,
    &
 \theta_a \delta_{a b} &= \frac{D_b \tr[ ( \dw{{\conj{\text{g}}}}{b_{\g{h}}})^\dagger T(\dw{h}{a_{\g{g}}}) ]}{D_a \tr[ ( \dw{{\conj{\text{g}}}}{b_{\g{h}}})^\dagger \dw{{\conj{\text{g}}}}{b_{\g{h}}} ]}
    \, ,
\end{align}
where we have assumed that the defect ICIs are one dimensional for the top left equality for $S$, which is true for the current example. We remark that the trace used above is the conventional matrix trace applied to elements of the tube algebra. 

For the current example the $\G$-crossed modular matrices are given by 
\begin{align}
&\begin{tabular}{ | c c | c c c c | c c| c c | c c | } 
\hline
 \multicolumn{2}{|c|}{\multirow{2}{*}{$\qd_\g{1}^2 S$}}  
  &  \multicolumn{4}{c|}{$(\g{1},\g{1})$} &  \multicolumn{2}{c|}{$(\g{1},\g{x})$}  &  \multicolumn{2}{c|}{$(\g{x},\g{1})$} &  \multicolumn{2}{c|}{$(\g{x},\g{x})$}
 \\
&& $\vac$ & $em$ & $e$ & $m$ & $\dw{x}{\vac}$ & $\dw{x}{em}$ & $\sigma_+$ & $\sigma_-$  & $\dw{x}{\sigma_+}$ & $\dw{x}{\sigma_-}$
\\ \hline
\multirow{4}{*}{$(\g{1},\g{1})$}  &$\vac$ & $1$ & $1$ & $1$ & $1$ & & & & & &
\\
&$em$ & $1$ & $-1$ & $-1$ & $1$ & & & & & &
\\
&$e$ & $1$ & $1$ & $-1$ & $-1$  & & & & & &
\\
&$m$ &$1$ & $-1$ & $1$ & $-1$ & & & & & &
\\  \hline
\multirow{2}{*}{$(\g{1},\g{x})$} & $\dw{x}{\vac}$ &  & & & &  & & $\sqrt{2}$ & $\sqrt{2}$ & &
\\
& $\dw{x}{em}$ &  & & & & &  & $i \sqrt{2}$ & $-i \sqrt{2}$ & &
\\  \hline
\multirow{2}{*}{$(\g{x},\g{1})$} & $\sigma_+$ &  & & & & $\sqrt{2}$ & $-i \sqrt{2}$ & & & &
\\
& $\sigma_-$ &  & & & & $\sqrt{2}$  & $i \sqrt{2}$ & & & &
\\  \hline
\multirow{2}{*}{$(\g{x},\g{x})$} & $\dw{x}{\sigma_+}$ &  & & & & & & & & & $e^{\frac{\pi i }{4} }$ 
\\
& $\dw{x}{\sigma_-}$ &  & & & & &  & & &  $e^{-\frac{\pi i }{4} }$  &
\\ \hline
\end{tabular}
 \\
&\begin{tabular}{ | c | c c c c | c c| c c | c c | } 
\hline
 $T$  & $\vac$ & $em$ & $e$ & $m$ &  $\dw{x}{\vac}$ & $\dw{x}{em}$ &  $\sigma_+$ & $\sigma_-$ & $\dw{x}{\sigma_+}$ & $\dw{x}{\sigma_-}$
\\ \hline
$\vac$ & $1$ & & & & & & & & &
\\
$em$ & & $-1$ & & & & & & & &
\\
$e$ & & &  $1$ &  & & & & & &
\\
$m$ &&  && $1$ & & & & & &
\\  \hline
$\dw{x}{\vac}$&  & & & & $1$   & & & & &
\\
$\dw{x}{em}$ &  & & & & & $-1$   & & & &
\\  \hline
$\sigma_+$ &  & & & &  & & & & $1$ &
\\
$\sigma_-$ &  & & & & &  & & & & $1$
\\  \hline
$\dw{x}{\sigma_+}$  &  & & & & &  & $e^{\frac{\pi i}{4}}$  & & &
\\
$\dw{x}{\sigma_-}$  &  & & & & &  &  & $e^{-\frac{\pi i}{4}}$ & &
\\ \hline
\end{tabular}
\end{align}
We remark that these matrices are not gauge invariant quantities. However, they do contain gauge invariant data, in particular the $S$ and $T$ matrices of the underlying toric code, and the quantum dimensions of the defects. 
Unitarity of the $S$-matrix implies that the number of $\g{h}$-invariant $\g{g}$-defects equals the number of $\g{g}$-invariant $\g{h}$-defects, in particular the number of $\g{g}$-defects equals the number of $\g{g}$-invariant anyons~\cite{barkeshli2014symmetry}. It also allows one to use the $\G$-crossed Verlinde formula to calculate the fusion rules~\cite{barkeshli2014symmetry}.

The ICIs of the gauged model can be recovered from the full set of domain walls, the projective representations they form, and their permutation action on the defects. First, a set of idempotents $\gaugedii{a}{\mu}$ are constructed by forming projectors onto the projective irreps of the stabilizer group of each defect. 
The ICIs $\ici{\gauged{a}{\mu}} $ are then found by taking a sum over projectors in the orbit of each defect
\begin{align}
\gaugedii{a}{\mu} =\frac{d_\mu}{|\centralizer{a}|} \sum_{\g{h}\in\centralizer{a} } \cirrep{a}{\mu}{h} \, \dw{h} {a}
\, ,
&&
\ici{\gauged{a}{\mu}} =\sum_{\g{k}\in \G / \centralizer{a}} \gaugedii{\act{k}a}{\act{k} \mu}
\end{align}

For example 
\begin{align}
\ici{\sigma_+} \
 \mapsto 
\left. 
\begin{array}{c c c}
\frac{1}{2}(\dw{1}{\sigma_+} +  e^{\frac{\pi i}{8}} \dw{\g{x}}{\sigma_+}) & = & \ici{\sigma}
\\
\frac{1}{2}(\dw{1}{\sigma_+} - e^{\frac{\pi i}{8}} \dw{\g{x}}{\sigma_+}) & =  & \ici{\sigma\conj{\psi}}
\end{array}
\right.
\, ,
 &
&
\begin{array}{c }
\ici{e}
\\
\ici{m}
\end{array}
 \mapsto  \ \dw{1}{e} + \dw{1}{m} \ = \   \ici{\sigma\conj{\sigma}} 
\, .
\end{align}
note the projective irreps $\irrp{\sigma_+}{\pm}$ of the representation $\{ \dw{1}{\sigma_+}, \dw{x}{\sigma_+}\}$ are given by ${\irrep{\sigma_+}{\pm}{1}=1,}$ ${\irrep{\sigma_+}{\pm}{x} = \pm e^{-\frac{\pi i}{8}}}$. 

The full set of gauged ICIs is once again given by Eq.~\eqref{set:isingtable}, and the resulting topological order is the doubled Ising model. 
 In this case, knowing the topological order post-gauging is enough to diagnose that the SET must be the EM duality enriched toric code $\{\vac, e, m, em\}\oplus \{ \sigma_+, \sigma_-\}$ where $\sigma_\pm$ have trivial FS indicator. 

 The $S$ and $T$ matrix of the gauged theory can be derived from their $\G$-crossed counter parts~\cite{barkeshli2014symmetry} by using the the gauged ICIs, see Section~\ref{setsection:gauging}, 
\begin{align}
\label{set:isingegSTgauging}
  S^{\double{\text{Ising}}}_{\gauged{a_{\g{g}}}{\mu} \gauged{b_{\g{h}}}{\nu}} &=  \frac{1}{|\G|} 
  \sum_{\substack{\g r \in \G/\centralizer{a}\\ \g s \in \G/\centralizer{b}}} 
  \act{r}\cirrep{a}{\mu}{\act{s}h}\, \act{s} \irrep{b}{\nu}{\act{r}\conj{\text{g}}}  \, S_{\act{r}a_{\g{g}} \act{s}b_{\g{h}}}
  \, ,
    &
    T^{\double{\text{Ising}}}_{\gauged{a_{\g{g}}}{\mu} \gauged{b_{\g{h}}}{\nu}} &=   \delta_{ {\gauged{a_{\g{g}}}{\mu} \gauged{b_{\g{h}}}{\nu}}}
    \frac{\irrep{a}{\mu}{g}}{\irrep{a}{\mu}{1}} \theta_a
    ,
\end{align}
where $S_{a_{\g{g}} b_{\g{h}}}$ 
denotes an element of the SET $S$ matrix, and $\theta_a$ an element of the SET $T$ matrix. 
We have also used the notation 
\begin{align}
\act{r}\irrp{a}{\mu }:= \irrp{\act{r} a}{\act{r} \mu},
\end{align}
to denote the action of an element $\g r\in \G / \centralizer{a}$, which  permutes $a$,  on a character of $\centralizer{a}$. Note the permutation action induces a fixed choice of irrep $\act{r}\mu$ on $\act{r} a$ which specifies the aforementioned character. 
By using Eq.~\eqref{set:isingegSTgauging}, one can diagnose the gauged theory once the $\G$-crossed $S$ and $T$ matrices have been calculated, without needing to carry out the gauging procedure explicitly. In the current example we recover the $S$ and $T$ matrices of the doubled Ising model.

We remark that one can go further, and use the defect ICIs and domain walls to construct a realization of the full emergent U$\G$xBFC, including the full gauge variant, $\{N,F,R,\rho,U,\eta\}$ data. For the purposes of diagnosing the SET this does not appear to be particularly useful, but we anticipate that it may have applications to analyzing topological quantum computation~\cite{WalterOgburn1999,qdouble,freedman2003topological,nayak2008non,wang2010topological} with defects in a tensor network.

\section{Symmetry-enriched string-net tensor networks}
\label{set:stringnetexample}

In this section we present the PEPS and MPO tensors for the class of symmetry-enriched string-net examples~\cite{PhysRevB.94.235136,cheng2016exactly}. These include the conventional string-nets as a subcase~\cite{levin2005string}. We first introduce the unitary fusion category data~\cite{etingof2005fusion,etingof2015tensor} that is the input for the string-net construction. We then present the MPOs that are used in both the string-nets and their symmetry-enriched counterparts. The local tensors for the (symetry-enriched) string nets are then given, along with the (symmetry-enriched) pulling through equation that they satisfy. The global symmetry of the symmetry-enriched string-net tensors is explicitly gauged, which is shown to recover the regular string-net tensor.

\subsection{Input (graded) unitary fusion category data}

We give a pedestrian description of the algebraic data~\cite{moore1988polynomial,Moore1989} of a unitary fusion category (UFC) $\cat$ that is taken as input to the string-net construction. 
The skeleton of this data is given by a fusion algebra of a finite set of simple objects $a\in\cat$
\begin{align}
a \times b = \sum_{c\in\cat} N_{ab}^{c}  c
\, ,
\end{align}
where $N_{ab}^{c}\in\mathbb{N}$ are a set of fusion coefficients. We define $\delta_{ab}^{c}\in\zt$ to be $0$ when $N_{ab}^{c}=0$ and $1$ otherwise. 

For a $\G$-graded UFC $\cat_\G$ the simple objects are organized into sectors $\cat_{\g{g}}$ and the fusion respects the $\G$-grading
\begin{align}
\cat_\G = \bigoplus_{\g g \in \G} \cat_\g{g} \, ,
&&
N_{\defect{a}{g}\defect{b}{h}}^{\defect{c}{k}} = \delta_{\g{k},\g{gh}} N_{\defect{a}{g}\defect{b}{h}}^{\defect{c}{k}} 
\, ,
\end{align}
where $\defect{a}{g}$ is shorthand for $a\in\cat_\g{g}$. 

There is a distinguished unit element $\vac$, corresponding to the vacuum in a physical theory, that satisfies $N_{\vac a}^{b}=\delta_{a,b}=N_{a \vac }^{b}$. 
Each simple object $a$ has a unique conjugate simple object $\conj{a}$ that satisfies $N_{a b}^{\vac}=\delta_{\conj{a},b}=N_{b a }^{\vac}$, and $\conj{\conj{a}}=a$. For a graded fusion category $\vac\in\cat_\g{1}$ and $\conj{\defect{a}{g}}\in\cat_{\conj{\g{g}}}$. 

Each simple object $a$ is assigned a quantum dimension $d_a$, given by the Perron-Frobenius eigenvalue of the matrix obtained by fixing $a$ in $N_{ab}^{c}$. These quantum dimensions satisfy
\begin{align}
d_a \, d_b = \sum_c N_{ab}^{c} \, d_c
\, ,
&&
d_\vac =1
&&
\text{and}
&&
d_{a}=d_{\conj{a}}
\, . 
\end{align}
The total quantum dimension $\qd$ of a UFC $\cat$ is defined by 
\begin{align}
\qd^2 = \sum_a d_a^2 
\, .
\end{align}
For $\G$-graded UFCs the quantum dimension of $\cat_\g{g}$ is defined by 
\begin{align}
\qd^2_\g{g} = \sum_\defect{a}{g} d_\defect{a}{g}^2 
\, ,
\end{align}
 they satisfy $\qd_\g{g}=\qd_\g{1}$ for all $\g{g} \in\G$, and hence $\qd_\G^2=|\G|\qd_\g{1}^2$. 
 
The following sum is a projector known as the $\omega_0$-loop of $\cat$  
\begin{align}
\omega_0 := \sum_{a \in \cat} \frac{d_a}{\qd^2} a 
\, .
\end{align}
For a $\G$-graded UFC the $\omega_0$-loop of $\cat_\g{1}$ plays the role of the unit element in a representation of $\G$ given by 
\begin{align}
\pi_\g{g}:=\sum_{a\in\cat_\g{g}} \frac{d_\defect{a}{g}}{\qd^2} \, \defect{a}{g}
\, ,
\end{align}
hence the $\omega_0$-loop of $\cat_\G$ is the projector onto the symmetric subspace of this representation. 

In a UFC the fusion coefficients are promoted to $N_{ab}^{c}$-dimensional vector spaces over $\mathbb{C}$. These vector spaces $V_{ab}^c$ (and their duals $V^{ab}_c$) appear on fusion (and splitting) vertices. Basis states for $V^{ab}_c$ are denoted
\begin{align}
	\begin{array}{c}
	\includeTikz{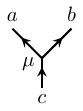}{
	\begin{tikzpicture}[scale=.5]
	\begin{scope}[xshift=0cm,yshift=0cm,decoration={markings,mark=at position 0.7 with {\arrow{stealth}}}]
	\draw[line width=1pt,draw=black,postaction=decorate] (0,0)--(0,1);
	\draw[line width=1pt,draw=black,postaction=decorate](0,1)-- (-1,2);
	\draw[line width=1pt,draw=black,postaction=decorate]   (0,1)--(1,2);
	\node[anchor=south] at (-1,2){$ a$};
	\node[anchor=south] at (1,2){$ b$};	
	\node[anchor=north] at (0,0){$ c$};
	\node[anchor=east] at (0,.8){$ \mu$};
	\end{scope}
	\end{tikzpicture}
	}
	\end{array}
	\, ,
\end{align}
where $\mu\in\{1,\dots,N_{ab}^{c}\}$. 

The fusion algebra is associative and the fusion coefficients satisfy
\begin{align}
\sum_{e\in\cat} N_{ab}^{e} N_{ec}^{d}  =  \sum_{f\in\cat} N_{bc}^{f} N_{af}^{d}
\, .
\end{align}
However, the composition of fusion or splitting vertices is not strictly associative. Rather, two different orders of fusion or splitting are related by an associator matrix known as the $F$-symbol.

At this point,  for clarity of exposition, we restrict our discussion to multiplicity free UFCs. That is, we assume $N_{ab}^{c}=\delta_{ab}^{c}\in \zt$. 
It is straightforward, although cumbersome, to extend the whole construction to include fusion multiplicity, but we will not cover that here. 

The $F$-symbol associators satisfy 
\begin{align}
	\begin{array}{c}
	\includeTikz{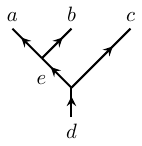}{
	\begin{tikzpicture}[scale=.5]
	\begin{scope}[xshift=0cm,yshift=0cm,decoration={markings,mark=at position 0.7 with {\arrow{stealth}}}]
	\draw[line width=1pt,draw=black,postaction=decorate] (0,0)--(0,1);
	\draw[line width=1pt,draw=black,postaction=decorate](0,1)-- (-1,2);
	\draw[line width=1pt,draw=black,postaction=decorate] (-1,2)-- (-2,3);
	\draw[line width=1pt,draw=black,postaction=decorate]   (0,1)--(2,3);
	\draw[line width=1pt,draw=black,postaction=decorate] (-1,2)-- (0,3);
	\node[anchor=south] at (-2,3){$ a$};
	\node[anchor=south] at (0,3){$ b$};	
	\node[anchor=south] at (2,3){$ c$};	
	\node[anchor=east] at (-.6,1.25){$ e$};	
	\node[anchor=north] at (0,0){$ d$};
	\end{scope}
	\end{tikzpicture}
	}
	\end{array}
	&
= \sum_{f} \f{abc}{d}{e}{f}
	\begin{array}{c}
	\includeTikz{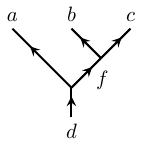}{
	\begin{tikzpicture}[scale=.5]
	\begin{scope}[xshift=0cm,yshift=0cm,decoration={markings,mark=at position 0.7 with {\arrow{stealth}}}]
	\draw[line width=1pt,draw=black,postaction=decorate] (0,0)--(0,1);
	\draw[line width=1pt,draw=black,postaction=decorate](0,1)-- (1,2);
	\draw[line width=1pt,draw=black,postaction=decorate] (0,1)-- (-2,3);
	\draw[line width=1pt,draw=black,postaction=decorate]   (1,2)--(2,3);
	\draw[line width=1pt,draw=black,postaction=decorate] (1,2)-- (0,3);
	\node[anchor=south] at (-2,3){$ a$};
	\node[anchor=south] at (0,3){$ b$};	
	\node[anchor=south] at (2,3){$ c$};	
	\node[anchor=east] at (1.5,1.25){$ f$};	
	\node[anchor=north] at (0,0){$ d$};
	\end{scope}
	\end{tikzpicture}
	}
	\end{array}
	\, .
\end{align}
The $F$-symbols are only defined upon the allowed splitting vertices, we use the convention that they are $0$ outside of this subspace
\begin{align}
 \f{abc}{d}{e}{f} = \delta_{ab}^{e} \delta_{ec}^{d} \delta_{bc}^{f} \delta_{af}^{d}  \f{abc}{d}{e}{f}
 \, .
\end{align}

The $F$-symbols are invertible matrices on the splitting spaces for which they are defined
\begin{align}
\sum_f \f{abc}{d}{e}{f} \finv{abc}{d}{f}{\widetilde e}  = \delta_{e \widetilde{e}} \, \delta_{ab}^{e} \, \delta_{ec}^{d}
\, .
\end{align}

The quantum dimension and Frobenius-Shur indicator can be extracted from the $F$-symbol
\begin{align}
d_a &= \frac{1}{| \shortf{a \conj{a} a}{a}{\vac}{\vac} | }  \, ,
&
 \fs{a} &= \frac{ \shortf{a \conj{a} a}{a}{\vac}{\vac} }{| \shortf{a \conj{a} a}{a}{\vac}{\vac} |} \, .
\end{align} 

Since the associativity constraint of an algebra was relaxed to a nontrivial associator, we find a higher order constraint on the composition of associators known as the pentagon equation. The pentagon equation can be derived by equating two inequivalent paths of associators that relate a pair of splitting diagrams on four objects. 
\begin{align}

\end{align}
Which can be written in terms of matrix elements as
\begin{align}
	\shortf{pcd}{e}{q}{r}\shortf{abr}{e}{p}{s}=\sum_{x}\shortf{abc}{q}{p}{x}\shortf{axd}{e}{q}{s}\shortf{bcd}{s}{x}{r}
	\, ,
	&&
	\text{where}
	&&
\shortf{abc}{d}{e}{f} := \f{abc}{d}{e}{f} 
\, .
\end{align}
A result known as Maclane's coherence theorem~\cite{mac2013categories} implies that the pentagon equation alone is enough to ensure that any path of associators between a pair of splitting diagrams are consistent. 
Another fundamental result known as Ocneanu rigidity~\cite{etingof2005fusion,kitaev2006anyons}, implies that a continuous deformation of a solution to the pentagon equation can be absorbed into a gauge transformation of the associator. 

The fusion categories that lead to Hermitian string-net Hamiltonians have unitary $F$-symbols on the relevant splitting subspace
\begin{align}
\finv{abc}{d}{f}{ e}=[(F^{abc}_{d})^\dagger]_{f}^{ e} = \left( \shortf{abc}{d}{ e}{f} \right) ^*
\, ,
&&
\sum_f \shortf{abc}{d}{e}{f} \left( \shortf{abc}{d}{\widetilde e}{f} \right) ^* = \delta_{e \widetilde{e}} \, \delta_{ab}^{e} \, \delta_{ec}^{d}
\, .
\end{align}

\subsection{Matrix product operator tensors and their fusion}

The string-net examples considered here are defined on oriented lattices, which give rise to both left and right handed MPO tensors. These tensors are related by a combined reflection and complex conjugation. A factor of the quantum dimension must be included for each variable that is summed over to achieve a correct overall normalization. The variables being summed over appear as closed loops in the tensor network diagrams and we use the convention established in Ref.~\cite{MPOpaper},  that such a sum over a variable $s$ is weighted by the quantum dimensions $d_s$. We remark that the weighting by quantum dimension can also be incorporated into the local tensors directly, by making use of the planar geometry, which was done in Refs.~\cite{stringnet1,stringnet2,Bultinck2017183}. We do not take this approach here.  

The MPOs described in this section are for both the string-nets and their symmetry-enriched variants. The only difference in the symmetry-enriched case is the inclusion of a $\G$-grading on the objects of the input fusion category $\cat_\G$. 

The left and right handed MPO tensors are given by
\begin{align}
\begin{array}{c}
	\includeTikz{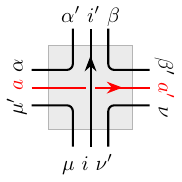}{
	\begin{tikzpicture}[scale=.1]
	\begin{scope}[decoration={markings,mark=at position 0.77 with {\arrow{Stealth}}}] 
	\node[transform shape,draw=black!29,fill=black!10!white!80,regular polygon,regular polygon sides=4,minimum size=20cm] at (0,-15){};
	\draw[draw=red,line width=1pt,postaction=decorate] (-10,-15) -- (10,-15);
	\node[transform shape,draw=black!10!white!80,fill=black!10!white!80,regular polygon,regular polygon sides=4,minimum size=2cm] at (0,-15){};
	\draw[draw=black,line width=1pt,postaction=decorate] (0,-25) -- (0,-5);
	\draw[draw=black,line width=1pt, rounded corners=1mm] (-10,-12) -- (-3,-12) -- (-3,-5)
											(-10,-18) -- (-3,-18) -- (-3,-25)
											(10,-18) -- (3,-18) -- (3,-25)
											(10,-12 )-- (3,-12) -- (3,-5);
	\node at (-.5,-28){$\mu \ i\ \nu'$};
	\node at (0,-3){$\alpha' \ i'\ \beta$};
	\node[rotate=90] at (-12.5,-15){$\mu' \ \red{a}\ \alpha$};
	\node[rotate=-90] at (12.5,-15){$\beta' \ \red{a'}\ \nu$};
	\end{scope}
	\end{tikzpicture}
	}
	\end{array}
&=
\begin{array}{c}
\delta_{\alpha\alpha'}\delta_{\beta \beta'}\delta_{\mu\mu'}\delta_{\nu\nu'}
\\
\delta_{i i'}\delta_{a a'}
\end{array}
\frac{\shortf{a \mu i}{\beta}{\alpha}{\nu}}{\sqrt{d_{\alpha} d_{\nu}}}
\, ,
\\
	\begin{array}{c}
	\includeTikz{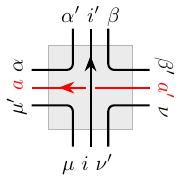}{
	\begin{tikzpicture}[scale=.1]
	\begin{scope}[decoration={markings,mark=at position 0.77 with {\arrow{Stealth}}}] 
	\node[transform shape,draw=black!29,fill=black!10!white!80,regular polygon,regular polygon sides=4,minimum size=20cm] at (0,-15){};
	\draw[draw=red,line width=1pt,postaction=decorate] (10,-15) -- (-10,-15);
	\node[transform shape,draw=black!10!white!80,fill=black!10!white!80,regular polygon,regular polygon sides=4,minimum size=2cm] at (0,-15){};
	\draw[draw=black,line width=1pt,postaction=decorate] (0,-25) -- (0,-5);
	\draw[draw=black,line width=1pt, rounded corners=1mm] (-10,-12) -- (-3,-12) -- (-3,-5)
											(-10,-18) -- (-3,-18) -- (-3,-25)
											(10,-18) -- (3,-18) -- (3,-25)
											(10,-12 )-- (3,-12) -- (3,-5);
											\node at (-.5,-28){$\mu \ i\ \nu'$};
	\node at (0,-3){$\alpha' \ i'\ \beta$};
	\node[rotate=90] at (-12.5,-15){$\mu' \ \red{a}\ \alpha$};
	\node[rotate=-90] at (12.5,-15){$\beta' \ \red{a'}\ \nu$};
	\end{scope}
	\end{tikzpicture}
	}
	\end{array}
&=
\begin{array}{c}
\delta_{\alpha\alpha'}\delta_{\beta \beta'}\delta_{\mu\mu'}\delta_{\nu\nu'}
\\
\delta_{i i'}\delta_{a a'}
\end{array}
 \frac{\shortf{a \alpha i}{\nu}{\mu}{\beta}^*}{\sqrt{d_{\mu} d_{\beta}}}
 \, ,
\end{align}
where the indices take values in the set of simple objects of the input category, and $d$ and $F$ are also taken from the input category. Each index of the tensor is $|\cat|^3$ dimensional, and red is used to indicate the virtual indices. The MPO tensors have some redundancy, as they are only nonzero on the subspace of the bond indices that satisfy $\delta_{\beta'a'}^{\nu}$ and $\delta_{\alpha a}^{\mu'}$, they are injective once projected onto this subspace. Here $a$ corresponds to the block label of the MPO. 

These MPO tensors satisfy an equation that corresponds to dragging an MPO string off a tensor network bond, this leads to a bubble popping equation that replaces a loop of MPO $a$ with the quantum dimension $d_a$ 
\begin{align}
\label{set:draggingMPOoffvirtualindex}
\begin{array}{c}
	\includeTikz{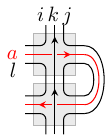}{
	\begin{tikzpicture}[scale=.05]
	\node at (0,15){$i\,k\,j$};
	\node[anchor=east] at (-10,2){$\red{a}$};
	\node[anchor=east] at (-10.5,-3.5){$l$};
	\begin{scope}[yshift=17cm,decoration={markings,mark=at position 0.77 with {\arrow{stealth}}}] 
	\node[transform shape,draw=black!29,fill=black!10!white!80,regular polygon,regular polygon sides=4,minimum size=20cm] at (0,-15){};
	\draw[draw=red,line width=.5pt,postaction=decorate] (-10,-15) -- (10,-15);
	\node[transform shape,draw=black!10!white!80,fill=black!10!white!80,regular polygon,regular polygon sides=4,minimum size=2cm] at (0,-15){};
	\draw[draw=black,line width=.5pt,postaction=decorate] (0,-22) -- (0,-5);
	\draw[draw=black,line width=.5pt, rounded corners=1mm] (-10,-12) -- (-3,-12) -- (-3,-5)
											(-10,-18) -- (-3,-18) -- (-3,-22)
											(10,-18) -- (3,-18) -- (3,-22)
											(10,-12 )-- (3,-12) -- (3,-5);
	\end{scope}
	\begin{scope}[decoration={markings,mark=at position 0.77 with {\arrow{stealth}}}] 
	\node[transform shape,draw=black!29,fill=black!10!white!80,regular polygon,regular polygon sides=4,minimum size=20cm] at (0,-15){};
	\draw[draw=red,line width=.5pt,postaction=decorate] (10,-15) -- (-10,-15);
	\node[transform shape,draw=black!10!white!80,fill=black!10!white!80,regular polygon,regular polygon sides=4,minimum size=2cm] at (0,-15){};
	\draw[draw=black,line width=.5pt,postaction=decorate] (0,-25) -- (0,-5);
	\draw[draw=black,line width=.5pt, rounded corners=1mm] (-10,-12) -- (-3,-12) -- (-3,-5)
											(-10,-18) -- (-3,-18) -- (-3,-25)
											(10,-18) -- (3,-18) -- (3,-25)
											(10,-12 )-- (3,-12) -- (3,-5);
	\end{scope}
	\draw[draw=black,line width=.5pt, rounded corners=1mm] 
	(10,-12) to [bend right=90] (10,-12+11)
	(10,-18) to [bend right=90] (10,-18+23);
	\draw[draw=red,line width=.5pt] 
	(10,-15) to [bend right=90] (10,-15+17);
	\end{tikzpicture}
	}
	\end{array}
	=	\frac{1}{d_i}
	\delta_{i i'}
	\delta_{i k}^{j} 	\delta_{a l }^{i'}
	\begin{array}{c}
	\includeTikz{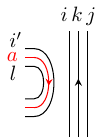}{
	\begin{tikzpicture}[scale=.05]
	\def\l{3};
	\node at (0,0){$i\,k\,j$};
	\begin{scope}[decoration={markings,mark=at position 0.55 with {\arrow{stealth}}}]
	\draw[draw=black,line width=.5pt,postaction=decorate] (0,-41) -- (0,-5);
	\draw[draw=black,line width=.5pt] (-3,-41) -- (-3,-5) ;
	\draw[draw=black,line width=.5pt] (3,-41) -- (3,-5);
	\end{scope}
	\begin{scope}[xshift=-25cm,yshift=-16cm]
		\draw[draw=black,line width=.5pt, rounded corners=1mm] 
	(10,-12) to [bend right=90] (10,-12+11)
	(10,-18) to [bend right=90] (10,-18+23)
	(10,-12) --	(10-\l,-12) 
	(10,-12+11) -- (10-\l,-12+11)
	(10,-18) -- (10-\l,-18)
	(10,-18+23) -- (10-\l,-18+23);
	\begin{scope}[decoration={markings,mark=at position 0.55 with {\arrow{stealth}}}] 
	\draw[draw=red,line width=.5pt,postaction=decorate] 
	(10,-15+17) to [bend left=90] (10,-15);
	\draw[draw=red,line width=.5pt] 
	(10,-15+17)  -- 	(10-\l,-15+17) 
	(10,-15) -- (10-\l,-15);
	\end{scope}
	\node[anchor=east] at (10-\l,2){$\red{a}$};
	\node[anchor=east] at (11.5-\l,8){$i'$};
	\node[anchor=east] at (9.5-\l,-3.5){$l$};
	\end{scope}
	\end{tikzpicture}
	}
	\end{array}
	=
	d_a \delta_{i k}^{j}
	\begin{array}{c}
	\includeTikz{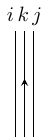}{
	\begin{tikzpicture}[scale=.05]
	\node at (0,0){$i\,k\,j$};
	\begin{scope}[decoration={markings,mark=at position 0.55 with {\arrow{stealth}}}]
	\draw[draw=black,line width=.5pt,postaction=decorate] (0,-41) -- (0,-5);
	\draw[draw=black,line width=.5pt] (-3,-41) -- (-3,-5) ;
	\draw[draw=black,line width=.5pt] (3,-41) -- (3,-5);
	\end{scope}
	\end{tikzpicture}
	}
	\end{array}
	\, .
\end{align}
The factor of $\frac{1}{d_i}$ in the left equation corrects for an excess quantum dimension that arises when two separate loops are merged. This can be seen by imagining the equation as part of a larger, closed,  tensor network diagram in which the $\delta_{i i'}$ condition becomes redundant. 
Reflecting and conjugating the above equations, leads to analogous relations with reversed orientations.

The matrices constructed by contracting a periodic chain of $L$  local tensors, with block label $a$ and virtual index pointing left to right, are denoted $\mpo{L}{a}$. These matrices form an MPO representation of the input fusion algebra 
\begin{align}
\mpo{L}{a} \, \mpo{L}{b} = \sum_{c\in\cat} N_{ab}^{c} \,  \mpo{L}{c} \, .
\end{align}
In the symmetry-enriched case this algebra is $\G$-graded. 

For open boundary conditions the MPOs only obey the algebra multiplication up to boundary fusion and splitting tensors 
\begin{align}
\begin{array}{c}
	\includeTikz{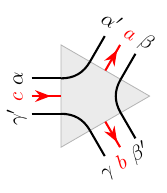}{
	\begin{tikzpicture}[scale=.1]
	\begin{scope}[rotate=-90]
	\node[transform shape,draw=black!29,fill=black!10!white!80,regular polygon,regular polygon sides=3,minimum size=20cm] at (0,0){};
	\begin{scope}[decoration={markings,mark=at position 0.65 with {\arrow{Stealth}}}]
	\draw[draw=red,line width=1pt,postaction=decorate] (0,-10) -- (0,-5);
	\draw[draw=black,line width=1pt] (-3,-10) -- (-3,-5) ;
	\draw[draw=black,line width=1pt] (3,-10) -- (3,-5);
	\draw[draw=black,line width=1pt,rounded corners=1mm] (-3,-5) to [bend right=30] ({3*cos(240)+5*sin(240)},{3*sin(240)-5*cos(240)});
	\node[rotate=90] at (.5,-12.5){$\gamma'\ \red{c}\ \alpha$};
	\end{scope}
	\begin{scope}[rotate=120,decoration={markings,mark=at position 0.8 with {\arrow{Stealth}}}]
	\draw[draw=red,line width=1pt,postaction=decorate] (0,-5) --  (0,-10);
	\draw[draw=black,line width=1pt] (-3,-10) -- (-3,-5) ;
	\draw[draw=black,line width=1pt] (3,-10) -- (3,-5);
	\draw[draw=black,line width=1pt,rounded corners=1mm] (-3,-5) to [bend right=30] ({3*cos(240)+5*sin(240)},{3*sin(240)-5*cos(240)});
	\node[rotate=30] at (-.5,-12.5){$\gamma \ \red{b}\ \beta'$};
	\end{scope}
	\begin{scope}[rotate=-120,decoration={markings,mark=at position 0.8 with {\arrow{Stealth}}}]
	\draw[draw=red,line width=1pt,postaction=decorate]  (0,-5) -- (0,-10);
	\draw[draw=black,line width=1pt] (-3,-10) -- (-3,-5) ;
	\draw[draw=black,line width=1pt] (3,-10) -- (3,-5);
	\draw[draw=black,line width=1pt,rounded corners=1mm] (-3,-5) to [bend right=30] ({3*cos(240)+5*sin(240)},{3*sin(240)-5*cos(240)});
	\node[rotate=330] at (0,-12.5){$\alpha' \ \red{a}\ \beta$};
	\end{scope}
	\end{scope}
	\end{tikzpicture}
	}
	\end{array}
&=\delta_{\alpha\alpha'}\delta_{\beta \beta'}\delta_{\gamma\gamma'} \frac{d_a^{\frac{1}{4}}d_b^{\frac{1}{4}}  \shortf{a b \gamma}{\alpha}{c}{\beta}}{d_c^{\frac{1}{4}} \sqrt{d_{\beta}}}
\, ,
\\
	\begin{array}{c}
	\includeTikz{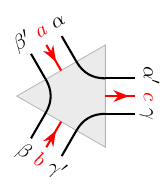}{
	\begin{tikzpicture}[scale=.1]
	\begin{scope}[rotate=90]
	\node[transform shape,draw=black!29,fill=black!10!white!80,regular polygon,regular polygon sides=3,minimum size=20cm] at (0,0){};
	\begin{scope}[decoration={markings,mark=at position 0.8 with {\arrow{Stealth}}}]
	\draw[draw=red,line width=1pt,postaction=decorate](0,-5) -- (0,-10);
	\draw[draw=black,line width=1pt] (-3,-10) -- (-3,-5) ;
	\draw[draw=black,line width=1pt] (3,-10) -- (3,-5);
	\draw[draw=black,line width=1pt,rounded corners=1mm] (-3,-5) to [bend right=30] ({3*cos(240)+5*sin(240)},{3*sin(240)-5*cos(240)});
	\node[rotate=-90] at (0.5,-12.5){$\alpha' \ \red{c}\ \gamma$};
	\end{scope}
	\begin{scope}[rotate=120,decoration={markings,mark=at position 0.65 with {\arrow{Stealth}}}]
	\draw[draw=red,line width=1pt,postaction=decorate] (0,-10) -- (0,-5);
	\draw[draw=black,line width=1pt] (-3,-10) -- (-3,-5) ;
	\draw[draw=black,line width=1pt] (3,-10) -- (3,-5);
	\draw[draw=black,line width=1pt,rounded corners=1mm] (-3,-5) to [bend right=30] ({3*cos(240)+5*sin(240)},{3*sin(240)-5*cos(240)});
	\node[rotate=30] at (0,-12.5){$\beta' \ \red{a} \ \alpha$};
	\end{scope}
	\begin{scope}[rotate=-120,decoration={markings,mark=at position 0.65 with {\arrow{Stealth}}}]
	\draw[draw=red,line width=1pt,postaction=decorate]  (0,-10) -- (0,-5);
	\draw[draw=black,line width=1pt] (-3,-10) -- (-3,-5) ;
	\draw[draw=black,line width=1pt] (3,-10) -- (3,-5);
	\draw[draw=black,line width=1pt,rounded corners=1mm] (-3,-5) to [bend right=30] ({3*cos(240)+5*sin(240)},{3*sin(240)-5*cos(240)});
	\node[rotate=330] at (0.5,-12.5){$\beta \ \red{b} \ \gamma'$};
	\end{scope}
	\end{scope}
	\end{tikzpicture}
	}
	\end{array}
&=\delta_{\alpha\alpha'}\delta_{\beta \beta'}\delta_{\gamma\gamma'} \frac{d_a^{\frac{1}{4}}d_b^{\frac{1}{4}} \shortf{a b  \gamma }{\alpha}{c}{\beta}^*}{d_c^{\frac{1}{4}}\sqrt{d_{\beta}}}
\, .
\end{align}
These tensors are again related by a combined reflection and complex conjugation. 
We have chosen a particular normalization of the fusion and splitting tensors such that the following equations hold
\begin{align}
\label{set:SNMPOsplittingeqn}
	
&=
\delta_{c,c'}
\delta_{ab}^{c}
\delta_{c \beta}^{\alpha}
\sqrt{\frac{d_a d_b}{d_c}}
	\begin{array}{c}
	\includeTikz{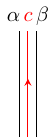}{
	\begin{tikzpicture}[scale=.05]
	\node at (0,0){$\alpha\,\red{c}\,\beta$};
	\begin{scope}[decoration={markings,mark=at position 0.55 with {\arrow{stealth}}}]
	\draw[draw=red,line width=.5pt,postaction=decorate] (0,-41) -- (0,-5);
	\draw[draw=black,line width=.5pt] (-3,-41) -- (-3,-5) ;
	\draw[draw=black,line width=.5pt] (3,-41) -- (3,-5);
	\end{scope}
	\end{tikzpicture}
	}
	\end{array}
	\, ,
\end{align}
where the equation below reduces to bubble popping for $b=\conj{a}$ and $c=\vac$. 

While the multiplication of closed MPOs is clearly associative, as they are matrices, for open boundary conditions the multiplication is not strictly associative. Rather, two different orders of multiplication are related by an associator, that is given by the $F$-symbol of the input category	
\begin{align}
	%
	\, .
\end{align}
An analogous equation for the fusion tensors is obtained by a combined reflection and complex-conjugation. 
The resulting $F$ symbols satisfy the pentagon equation, as the input was a consistent UFC. 
Since we are working with the tensor product of tensors over $\mathbb{C}$ there is no difference between taking a left or right trace, which is analogous to the \emph{spherical} property of UFCs.

In our examples we assume that the $F$ symbols have been brought into a gauge where
\begin{align}
\shortf{\vac bc}{d}{e}{f} = \shortf{a\vac c}{d}{e}{f} = \shortf{ab\vac}{d}{e}{f} = 1
\, .
\end{align}
We remark that, in general, these elements may be arbitrary complex phases.  
This constraint is equivalent to the following \emph{triangle equations}, which allow a vacuum line to be added or removed at fusion and splitting vertices
\begin{align}

	$}
	\, .
\end{align}
In the above equations we have used the following states to terminate a vacuum line
\begin{align}
	\begin{array}{c}
	\includeTikz{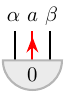}{
	\begin{tikzpicture}[scale=.1]
	\begin{scope}[rotate=180]
		\begin{scope}[decoration={markings,mark=at position 0.8 with {\arrow{Stealth}}}]
	\draw[draw=red,line width=1pt,postaction=decorate](0,-5) -- (0,-10);
	\draw[draw=black,line width=1pt] (-3,-10) -- (-3,-5) ;
	\draw[draw=black,line width=1pt] (3,-10) -- (3,-5);
	\node at (0,-12.5){$\alpha \ a\ \beta$};
		\filldraw[line width=1pt,draw=black!29,fill=black!10!white!80] (-5,-5)--(5,-5) arc (0:180:5) -- cycle;
			\node[anchor=north] at (0,-5){ $\vac$};
	\end{scope}
	\end{scope}
	\end{tikzpicture}
	}
	\end{array}
&:= \delta_{a \vac} \delta_{\alpha \beta} \sqrt{d_\beta}
\, ,
&
	\begin{array}{c}
	\includeTikz{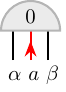}{
	\begin{tikzpicture}[scale=.1]
		\begin{scope}[decoration={markings,mark=at position 0.65 with {\arrow{Stealth}}}]
	\draw[draw=red,line width=1pt,postaction=decorate] (0,-10) -- (0,-5);
	\draw[draw=black,line width=1pt] (-3,-10) -- (-3,-5) ;
	\draw[draw=black,line width=1pt] (3,-10) -- (3,-5);
	\node at (.5,-12.5){$\alpha\ a\ \beta$};
		\filldraw[line width=1pt,draw=black!29,fill=black!10!white!80] (-5,-5)--(5,-5) arc (0:180:5) -- cycle;
			\node[anchor=south] at (0,-5){ $\vac$};
	\end{scope}
	\end{tikzpicture}
	}
	\end{array}
&:= \delta_{a \vac} \delta_{\alpha \beta} \sqrt{d_\beta}
\, . 
\end{align}

Terminating a vacuum line that splits into (results from the fusion of) a particle-antiparticle pair leads to an orientation reversing ``flag'' matrix, corresponding to a cup (cap) 
\begin{align}
\label{set:flags}
	\begin{array}{c}
	\includeTikz{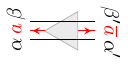}{
	\begin{tikzpicture}[scale=.05]
	\node[transform shape,draw=black!29,fill=black!10!white!80,regular polygon,regular polygon sides=3,minimum size=15cm,rotate=-30] at (0,0){};
	\begin{scope}[xshift=-2.5cm,rotate=-90,decoration={markings,mark=at position 0.8 with {\arrow{stealth}}}]
	\draw[draw=red,line width=.5pt,postaction=decorate]  (0,-5) -- (0,-10);
		\draw[draw=red,line width=.5pt,postaction=decorate]  (0,6.25) -- (0,12);
	\draw[draw=black,line width=.5pt] (-3,-10) -- (-3,12) ;
	\draw[draw=black,line width=.5pt] (3,-10) -- (3,12);
	\node[anchor=south,rotate=90] at (-.5,-9){$\alpha\, \red{a}\, \beta$};
	\end{scope}
		\begin{scope}[rotate=90,decoration={markings,mark=at position 0.8 with {\arrow{stealth}}}]
	\node[anchor=south,rotate=-90] at (-.75,-9){$\beta' \red{\conj{a}}\, \alpha'$};
	\end{scope}
	\end{tikzpicture}
	}
	\end{array}
&:=
	\begin{array}{c}
	\includeTikz{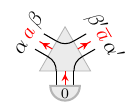}{
	\begin{tikzpicture}[scale=.05]
	\begin{scope}[decoration={markings,mark=at position 0.8 with {\arrow{stealth}}}]
	\node[transform shape,draw=black!29,fill=black!10!white!80,regular polygon,regular polygon sides=3,minimum size=20cm] at (0,0){};
	\draw[draw=red,line width=.5pt,postaction=decorate] (0,-10) -- (0,-5);
	\draw[draw=black,line width=.5pt] (-3,-10) -- (-3,-5) ;
	\draw[draw=black,line width=.5pt] (3,-10) -- (3,-5);
	\begin{scope}[decoration={markings,mark=at position 0.65 with {\arrow{stealth}}}]
	\draw[draw=black,line width=.5pt,rounded corners=1mm] (-3,-5) to [bend right=30] ({3*cos(240)+5*sin(240)},{3*sin(240)-5*cos(240)});
	\end{scope}
			\filldraw[line width=1pt,draw=black!29,fill=black!10!white!80] (-5,-10)--(5,-10) arc (0:-180:5) -- cycle;
			\node[anchor=north] at (0,-8.5){ \scriptsize $\vac$};
	\begin{scope}[rotate=120,decoration={markings,mark=at position 0.8 with {\arrow{stealth}}}]
	\draw[draw=red,line width=.5pt,postaction=decorate] (0,-5) --  (0,-10);
	\draw[draw=black,line width=.5pt] (-3,-10) -- (-3,-5) ;
	\draw[draw=black,line width=.5pt] (3,-10) -- (3,-5);
	\draw[draw=black,line width=.5pt,rounded corners=1mm] (-3,-5) to [bend right=30] ({3*cos(240)+5*sin(240)},{3*sin(240)-5*cos(240)});
			\node[anchor=south,rotate=-60] at (-1,-9){$\beta' \red{\conj{a}}\, \alpha'$};
	\end{scope}
	\begin{scope}[rotate=-120,decoration={markings,mark=at position 0.8 with {\arrow{stealth}}}]
	\draw[draw=red,line width=.5pt,postaction=decorate]  (0,-5) -- (0,-10);
	\draw[draw=black,line width=.5pt] (-3,-10) -- (-3,-5) ;
	\draw[draw=black,line width=.5pt] (3,-10) -- (3,-5);
	\draw[draw=black,line width=.5pt,rounded corners=1mm] (-3,-5) to [bend right=30] ({3*cos(240)+5*sin(240)},{3*sin(240)-5*cos(240)});
	\node[anchor=south,rotate=60] at (-.5,-9){$\alpha\, \red{a}\, \beta$};
	\end{scope}
	\end{scope}
	\end{tikzpicture}
	}
	\end{array}%
	\, .
\end{align}
The absolute orientation chosen for the flag matrices is not important, as they already define their oppositely oriented variants via transposition and swapping $a\leftrightarrow\conj{a}$.

The closed MPOs satisfy $ \mpo{\dagger}{a}=\mpo{}{\conj{a}}=\mpo{-}{a}$, where $-$ indicates orientation reversal. 
The orientation reversing flag matrices implement this at the level of the local tensors 
\begin{align}
\label{set:orientationreversalofSNlocaltensor}
	%
	\, .
\end{align}
From the above we see that the relative orientation of the flags does matter. Changing this orientation leads to the FS indicator $\fs{a}\in\Uone$, which is a piece  of gauge invariant physical data --- given by $\fs{a}=\pm1$ --- for $a=\conj{a}$
\begin{align}
	%
	\, .
\end{align}

The orientation reversing flag matrices can be used to derive a relation between the fusion and splitting spaces
\begin{align}
\label{set:SNexamplelinebendingab}
	%
	\, ,
\end{align}
with the normalization we have chosen these maps are unitary. They can be used to show a version of the \emph{pivotal} identity satisfied by UFCs. 
For the case of no fusion multiplicity these maps are simply phases 
\begin{align}
\lbend{ab}{c} &= \sqrt{\frac{d_a d_b}{d_c}} \shortf{\conj{a} a b}{b}{\vac}{c}^* \, ,
& 
\rbend{ab}{c} &= \sqrt{\frac{ d_a d_b}{d_c}} \shortf{a b \conj{b}}{a}{c}{\vac} \, . 
\end{align}

Furthermore, the flag matrices allow one to derive a version of Eqs.~\eqref{set:draggingMPOoffvirtualindex}, and \eqref{set:SNMPOsplittingeqn} with the orientation of only the red lines reversed.

\subsection{String-net tensors and the pulling through equation}

The string-net models are defined on a trivalent lattice that is dual to a triangulation endowed with branching structure. A PEPS representation of the ground state is specified by the local tensors 
\begin{align}
\label{set:stringnetfixptpepstensors}

	\frac{ d_i^\frac{1}{4} d_j^{\frac{1}{4}} \shortf{\alpha i j}{\gamma}{\beta}{k}^*}{d_k^\frac{1}{4} \sqrt{d_{\beta} }}
\, ,
\end{align}
which are related by a combined reflection and complex conjugation.  
Each of the indices is $|\cat|^3$ dimensional, the physical indices are depicted within the tensors.  In our representation of the string-nets each edge degree of freedom appears twice on the adjacent triangles. This convention, although redundant, is used for convenience and is equivalent to other conventions where the redundancy is removed.

The string-net PEPS was found to be MPO-injective in Ref.~\cite{MPOpaper} with respect to the MPO defined by the same UFC $\cat$. The pulling through equations for the string-nets are given by
\begin{align}
	\raisebox{-.9cm}{$

	$}\, .
\end{align}
We emphasize the inclusion of a factor of the quantum dimension $d_a$ on the index that is summed over, which appears as a closed loop. These equations are implied by the pentagon equation for the $F$-symbol used to define both the PEPS and the MPO tensors. The two equations are related by a combined reflection and complex conjugation.  All other pulling through equations, involving different combinations of indices and orientations,  can be derived from the above by using Eqs.~\eqref{set:draggingMPOoffvirtualindex}, and \eqref{set:orientationreversalofSNlocaltensor}.

\subsection{Symmetry-enriched string-net tensors, pulling through, and gauging}

For the $\G$ symmetry-enriched string-nets the simple objects of the input UFC are equipped with a $\G$-grading. We use the shorthand notation $\sector{a}=\g{g}$ to denote the sector $a\in\cat_\g{g}$, i.e. $\sector{\defect{a}{g}}=\g{g}$. 
The PEPS tensors only differ from the ungraded string-nets via the inclusion of the sector information of the virtual indices in the physical index 
\begin{align}
	\begin{array}{c}
	\includeTikz{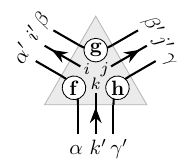}{}
	\end{array}
&= 
\begin{array}{c}
 \delta_{\g{f} \sector{\alpha}} \delta_{\g{g} \sector{\beta}} \delta_{\g{h} \sector{\gamma}}
 \delta_{ii'}
 \\
\delta_{jj'}\delta_{kk'}\delta_{\alpha\alpha'}\delta_{\beta \beta'}\delta_{\gamma\gamma'} 
\end{array}
	\frac{d_i^\frac{1}{4} d_j^{\frac{1}{4}} \shortf{\alpha i j}{\gamma}{\beta}{k}}{d_{k}^\frac{1}{4}\sqrt{d_{\beta}}}
\, ,
\\
	\begin{array}{c}
	\includeTikz{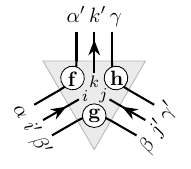}{}
	\end{array}
&= 
\begin{array}{c}
 \delta_{\g{f} \sector{\alpha}} \delta_{\g{g} \sector{\beta}} \delta_{\g{h} \sector{\gamma}}
 \delta_{ii'}
 \\
\delta_{jj'}\delta_{kk'}\delta_{\alpha\alpha'}\delta_{\beta \beta'}\delta_{\gamma\gamma'} 
\end{array}
	 \frac{d_i^\frac{1}{4} d_j^{\frac{1}{4}} \shortf{\alpha i j}{\gamma}{\beta}{k}^*}{ d_{k}^\frac{1}{4} \sqrt{d_{\beta}}}
 \, .
\end{align}
The internal tensor that copies the sector information is depicted as a circle. Again the tensors are related by combined reflection and complex conjugation. Similar tensors were described in Refs.~\cite{barkeshli2016reflection,cui2017state,luo2016structure}. 

These tensors are MPO-injective with respect to the MPO constructed from $\cat_\g{1}$. They further satisfy the symmetry-enriched pulling through equations defined by the MPOs constructed from $\cat_\G$  
\begin{align}
	\raisebox{-.9cm}{$
	\begin{array}{c}
	\includeTikz{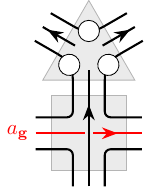}{}
	\end{array}
	$}
	&=
	\raisebox{.6cm}{$
	\begin{array}{c}
	\includeTikz{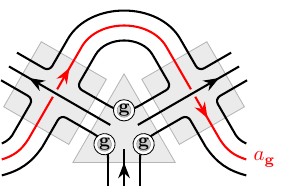}{}
	\end{array}
	$}
	,
\\
	\raisebox{-.6cm}{$
	\begin{array}{c}
	\includeTikz{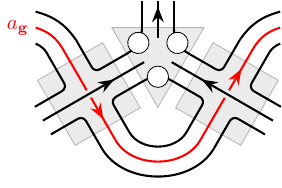}{}
	\end{array}
	$}
	&=
	\raisebox{.9cm}{$
	\begin{array}{c}
	\includeTikz{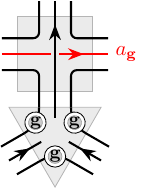}{}
	\end{array}
	$} 
	\, ,
\end{align}
where the shaded circle containing $\g{g}$ denotes the application of the on-site left regular group action
\begin{align}
			\raisebox{-.15cm}{$
			\begin{array}{c}
	\includeTikz{SETTCgtensor0}{}
	\end{array}
	$}
	:=
		\raisebox{-.15cm}{$
		\begin{array}{c}
	\includeTikz{SETTCgtensor1}{}
	\end{array}
	$}
	=
		\raisebox{-.15cm}{$
			\begin{array}{c}
	\includeTikz{SETTCgtensor2}{}
	\end{array}
	$}
 \,	.
\end{align}
Again a factor of the quantum dimension is included on the summed variable, which appears as a closed loop. These equations are implied by the pentagon equation, and the $\G$-grading of the fusion algebra for $\cat_\G$.  The two pulling through equations are related by a combined reflection and Hermitian conjugation. All other instances of the symmetry-enriched pulling through equation can be derived from the above by using Eqs.~\eqref{set:draggingMPOoffvirtualindex}, and \eqref{set:orientationreversalofSNlocaltensor}. 
 The symmetry-enriched pulling through equation~\cite{williamson2014matrix}, together with the properties of the MPOs, imply that the PEPS is symmetric under the on-site left regular representation of $\G$.

\subsubsection{Gauging the $\G$ symmetry}

The state gauging procedure defined in Ref.~\cite{Gaugingpaper} can be applied to the PEPS tensors of a fixed-point model, such as the symmetry-enriched string-nets, to produce PEPS tensors for the gauged model. This was explained in detail in Ref.~\cite{williamson2014matrix}, which focused on SPT tensor networks, but the results generalize straightforwardly. 
For a system of  $\mathbb{C}[\G]$ degrees of freedom on the vertices of a directed graph, in state $\ket{\psi}_V$, and a global symmetry given by a tensor product of the left regular representation on every vertex, the gauging map is given by 
\begin{align}
G\ket{\psi}_V := \prod_{v} \left( \frac{1}{|\G|} \sum_{\g{g}_v} L_v(\g{g}_v) \bigotimes_{e_v^+} L_{e_v^+}(\g{g}_v) \bigotimes_{e_v^-} R_{e_v^-}(\g{g}_v) \right) \ 
\ket{\psi}_V \bigotimes_{e} \ket{\g{1}}_e
\, ,
\end{align}
where $e_v^\pm$ is a neighboring edge that points away from (towards) vertex $v$. The resulting states $G\ket{\psi}_V$ are only defined on the subspace that satisfies the local gauge constraints, which does not have an immediately obvious tensor product structure. 

For the on-site left regular representation the following isometry disentangles the gauge constraints, and maps back to a tensor product Hilbert space 
\begin{align}
Y := \left(\bigotimes_{v} \bra{\g{+}}_{v} \right) \prod_{v} \, \prod_{e_v^+} \ctl{v}{e_v^+} \prod_{e_v^-} \ctr{v}{e_v^-}
\, ,
&&
\text{where}
&&
\ket{\g{+}}:=\sum_{\g{g}} \ket{\g{g}}
\, ,
 \end{align}
and  $CL\, (R)$ is the controlled left (right) multiplication. Note the $\ket{+}$ state above is not normalized. Hence $YG\ket{\psi}_V$ is a state on the edge degrees of freedom alone.

The composition of the global gauging projector followed by the disentangling isometry, $CG$, can be implemented directly on the local tensors 
\begin{align}
\bra{\g{+}}_{0}\bra{\g{+}}_{1}\bra{\g{+}}_{2} \ctl{0}{01} \ctl{0}{02} \ctl{1}{12} \ctr{1}{01} \ctr{2}{02} \ctr{2}{12}& \ket{\g{1}}_{01} \ket{\g{1}}_{12}\ket{\g{1}}_{23}
	\raisebox{.2cm}{$

	\, ,
\end{align}
where $0,\, 1,\, 2$ labels the closest $\mathbb{C}[\G]$ degree of of freedom, and $01,\, 12,\, 02$ labels the newly introduced $\mathbb{C}[\G]$ degree of of freedom on the relevant edge. 

Every edge degree of freedom of the tensor network now carries a redundant copy of the sector information i.e. $\ket{\defect{s}{g}}\ket{\g{g}}$. 
A further local isometry $W$ can be applied to each edge to disentangle and project out this redundancy
\begin{align}
W_{01} W_{12} W_{02}
	\begin{array}{c}
	\includeTikz{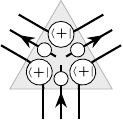}{}
	\end{array}%
	\, ,
\end{align}
where $W \ket{\defect{s}{g}}\ket{\g h} =(  \openone \otimes \bra{\g{1}} ) \ket{\defect{s}{g}}\ket{\conj{\g{g}} \g h}$. 
We find that this results in the string-net tensor for the full UFC $\cat_\G$, given in Eq.~\eqref{set:stringnetfixptpepstensors}, left.  A similar equation holds for the positively oriented tensor given in  Eq.~\eqref{set:stringnetfixptpepstensors}, right.  

In summary, for fixed point models 
\begin{itemize}
\item
The effect of gauging can be captured by simply projecting each local tensor onto the trivial representation of the on-site symmetry, in this case $\ket{\g{+}}$. This explains some of the results observed in Ref.~\cite{PhysRevB.94.155106}. 
\item
The inverse operation to gauging $\G$ corresponds to a $\repg$ anyon condensation. This can simply be implemented on the string-net tensors by copying the sectors of the virtual degrees of freedom to the physical index. This effectively ungauges the flat $\G$-connection defined by the $\G$-grading of the virtual indices. 
\item
Similarly, confining the $\G$-defects of the SET can be implemented directly on the local tensors by projecting the sector information onto the $\ket{\g{1}}$ state. 
\item
The inverse operation corresponds to allowing $\G$-defects and can be implemented on the local tensors by allowing the sector degree of freedom to fluctuate freely. 
\end{itemize}

These relations are summarized as follows
\begin{align}
\xymatrixcolsep{9pc}\xymatrix{
   \text{SNTN}(\cat)\ \ar@<+4pt>[r]^-{\text{Fluctuate sector d.o.f.}}  & \ \text{SESNTN}(\cat_\G)\  \ar@<+4pt>[r]^-{\text{Project sector d.o.f. onto ${\g{+}}$} 
   }  \ar@<+4pt>[l]^-{\text{Project sector d.o.f. onto $\g{1}$}}  & \ \text{SNTN}(\cat_\G) \ar@<+4pt>[l]^-{\text{Copy $\sector{a}$ to phys. index}}
}
\end{align}
where SNTN$(\cat)$ refers to the string-net tensor network constructed from the UFC $\cat$, and SESNTN$(\cat_\G)$ refers to the symmetry-enriched string-net tensor network.

\section{Ocneanu's tube algebra and emergent topological order from matrix product operators}
  \label{set:toptubealgsection}

In this section, we describe a construction known as Ocneanu's \emph{tube algebra}~\cite{tubealgebra,evans1998quantum,DrinfeldCenter,ocneanu1994chirality}. The derivation of the tube algebra from MPOs was first explained in Ref.~\cite{Bultinck2017183}, similar lattice constructions of the tube algebra have appeared in Refs.~\cite{Qalgebra,haah,aasen2017fermion,daveprep}. 
By block diagonalizing this tube algebra we find tensor network representations of the emergent superselection sectors. We are able to extract gauge invariant observable physical data  by using these representations to construct minimally entangled states, this includes the modular $S$ and $T$ matrices as well as the topological entanglement entropies of the superselection sectors. Furthermore we outline how the full emergent anyon theory, which is known to be the Drinfeld center (or double) $\double{\cat}$ of the MPO UFC $\cat$, can be constructed by finding the gauge variant fusion and braiding, $F$ and $R$ symbols. 
The tube algebra construction can be performed for the string-net examples tensors of the previous section.

\subsection{Definition of the tube algebra}

Throughout this section we work on the level of abstraction explained in Eq.~\eqref{set:abstractingtopictures1}, where the specifics of the underlying lattice are neglected. We use a diagrammatic notation in which grey shaded areas denote the PEPS tensor network, and black lines denote MPOs contracted with the virtual indices of the PEPS. 

The tube algebra is found by considering a puncture in the tensor network and calculating the action of the topological MPO symmetry on the virtual indices of the puncture. 
To this end we consider a cylinder (or twice punctured sphere) as there is a global constraint that forces the charge in a unique puncture on a sphere to be trivial.  
As explained in Eq.~\eqref{set:puncturedsphere} the MPO-injective tensor network states on the cylinder are found by closing the tensor network with an MPO. Hence it suffices to consider the action of the topological MPO symmetry algebra on the states
  \begin{align}
  \label{set:cylindertubeaction}

\, ,
\end{align}
where we have used Eqs.~\eqref{set:draggingMPOoffvirtualindex}, 
\eqref{set:SNMPOsplittingeqn}, 
\eqref{set:orientationreversalofSNlocaltensor}, 
and
\eqref{set:SNexamplelinebendingab}, or rather, the analogous equations for a general MPO algebra.  The dashed lines indicate periodic boundary conditions and the white triangles denote cup and cap matrices, see Eq.~\eqref{set:flags}.

Hence the basis elements that span the tube algebra are given by 
  \begin{align}
  \label{set:tubeconvention}
\tube{p q r}{s} = 
d_p^{\frac{1}{4}} d_r^{-\frac{1}{4}}
\begin{array}{c}
\includeTikz{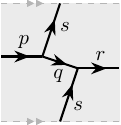}{\drawtube{p}{q}{r}{s}}
  \end{array}
  \, ,
\end{align}
corresponding to the minimal tensor network
  \begin{align}
  d_p^{\frac{1}{4}} d_r^{-\frac{1}{4}}
	\begin{array}{c}
	\includeTikz{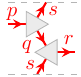}{}
	\end{array}
  \, , 
  \end{align}
we will often refer to these basis elements simply as tubes. 
We remark that in some cases this minimal tensor network may evaluate to $0$, this can be avoided by including one or more $s$ MPO tensors around the cylinder. 
There is some gauge freedom in the definition of these tubes which is similar to a 1-cochain. 
Eq.~\eqref{set:cylindertubeaction} can be rewritten as
\begin{align}
\tube{r r r}{\vac} = 
 \frac{ 1 }{d_s^2} 
 \sum_{p q} 
 (\tube{p q r}{s})^\dagger
 \tube{p p p}{\vac} 
\tube{p q r}{s}
 \, ,
 &&
 \text{and hence}
 &&
 \openone := \sum_r \tube{r r r}{\vac} = 
 \frac{ 1 }{\qd^2} 
 \sum_{p q r s} 
 (\tube{p q r}{s})^\dagger
 \tube{p q r}{s} 
 \, ,
\end{align}
where $\openone$ denotes the identity matrix on a relevant subspace corresponding to virtual indices of a cylinder PEPS.

 Multiplication of tubes is defined by the stacking operation
    \begin{align}
    \label{set:tubemult}
    \tube{p q r}{s} \tube{p'q'r'}{s'}
    & = 
    \delta_{r p'}
d_p^{\frac{1}{4}}       d_{r'}^{-\frac{1}{4}}
	\begin{array}{c}
	\includeTikz{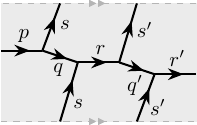}{
	\begin{tikzpicture}[scale=.1]
	\def\dx{-1};
	\def\ddx{1.5};
	\def\dy{.5};
	\def\p{16};
	\def\dz{1}
	\filldraw[UFCBackground](0,0)--(33,0)--(33,20)--(0,20)--(0,0);
	\draw[anyon](13,9+\dz)--(20,9+\dz) node[pos=0.55,above] {$r$};
	\draw[anyon](7,11+\dz)--(13,9+\dz) node[pos=0.45,below] {$q$};
	\draw[anyon](0,11+\dz)--(7,11+\dz) node[pos=0.55,above] {$p$};
	\draw[anyon](10,0)--(13,9+\dz)  node[pos=0.3,right] {$s$};
	\draw[anyon](7,11+\dz)--(10,20) node[pos=0.55,right] {$s$};
	\draw[draw=black!29,dashed](0,0)--(33,0) (33,20)--(0,20);
	\filldraw[draw=black!29,fill=black!29](\p,0)--(\p+\dx,-\dy)--(\p+\dx,\dy)--(\p,0);
	\filldraw[draw=black!29,fill=black!29](\p,20)--(\p+\dx,20-\dy)--(\p+\dx,20+\dy)--(\p,20);
	\filldraw[draw=black!29,fill=black!29](\p+\ddx,0)--(\p+\dx+\ddx,-\dy)--(\p+\dx+\ddx,\dy)--(\p+\ddx,0);
	\filldraw[draw=black!29,fill=black!29](\p+\ddx,20)--(\p+\dx+\ddx,20-\dy)--(\p+\dx+\ddx,20+\dy)--(\p+\ddx,20);
	\begin{scope}[xshift=13cm]
	\draw[anyon](13,9-\dz)--(20,9-\dz) node[pos=0.55,above] {$r'$};
	\draw[anyon](7,11-\dz)--(13,9-\dz) node[pos=0.45,below] {$q'$};
	\draw[anyon](10,0)--(13,9-\dz)  node[pos=0.3,right] {$s'$};
	\draw[anyon](7,11-\dz)--(10,20) node[pos=0.55,right] {$s'$};
	\end{scope}
		\end{tikzpicture}
	}
  \end{array}
  \\
  &
  = 
  \delta_{r p'} \sum_{q'' s''} \fs{s} \sqrt{\frac{d_s d_{s'}}{d_{s''}}} \rbend{q' \conj{s}}{q''} (\rbend{q s }{r })^{\dagger}
\shortf{s' q' \conj{s}}{q }{r }{ q''} {(\shortf{q'' s s'}{r' }{s''}{q' })}^* {(\shortf{s s' q''}{p }{q}{ s''})}^* \  \tube{r' q'' p }{s''} 
\, ,
    \end{align}
only tubes, $ \tube{pq'' r'}{s''} $,  satisfying $\delta_{s s'}^{s''} \neq 0$ have nonzero coefficients in this expansion. 
The tube algebra is a $C^*$ algebra as Hermitian conjugation is given by
    \begin{align}
    \label{set:tubeconj}
   (\tube{p q r}{s})^\dagger = 
	d_p^{\frac{1}{4}}	d_r^{-\frac{1}{4}}
	\begin{array}{c}
	\includeTikz{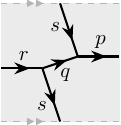}{
	\begin{tikzpicture}[scale=.1]
	\def\dx{-1};
	\def\ddx{1.5};
	\def\dy{.5};
	\def\p{5.5};
	\filldraw[UFCBackground](0,0)--(-20,0)--(-20,20)--(0,20)--(0,0);
	\draw[anyon](-20,9)--(-13,9) node[pos=0.55,above] {$r$};
	\draw[anyon](-13,9)--(-7,11) node[pos=0.65,below] {$q$};
	\draw[anyon](-7,11)--(0,11) node[pos=0.55,above] {$p$};
	\draw[anyon](-13,9)--(-10,0)  node[pos=0.7,left] {$s$};
	\draw[anyon](-10,20)--(-7,11) node[pos=0.45,left] {$s$};
	\begin{scope}[xshift=-20cm]
	\draw[draw=black!29,dashed](0,0)--(20,0) (20,20)--(0,20);
	\filldraw[draw=black!29,fill=black!29](\p,0)--(\p+\dx,-\dy)--(\p+\dx,\dy)--(\p,0);
	\filldraw[draw=black!29,fill=black!29](\p,20)--(\p+\dx,20-\dy)--(\p+\dx,20+\dy)--(\p,20);
	\filldraw[draw=black!29,fill=black!29](\p+\ddx,0)--(\p+\dx+\ddx,-\dy)--(\p+\dx+\ddx,\dy)--(\p+\ddx,0);
	\filldraw[draw=black!29,fill=black!29](\p+\ddx,20)--(\p+\dx+\ddx,20-\dy)--(\p+\dx+\ddx,20+\dy)--(\p+\ddx,20);
	\end{scope}
		\end{tikzpicture}
	}
  \end{array}
  =
  \sqrt{\frac{d_p}{d_r}}
\sum_{q'}  \rbend{ps}{q'} (\lbend{sq}{p})^\dagger \shortf{\conj{s} p s}{r}{q}{q'} \ \tube{p q' r}{\conj{s}}
\, .
    \end{align}

Each tube defines a state on the torus, which is identically zero when $p\neq r$ 
\begin{align}
\ket{\tube{p q r}{s}} :=
 \delta_{p r}
 \begin{array}{c}
\includeTikz{Torus}{\drawtorus{p}{q}{p}{s}}
\end{array}
=: \delta_{p r}
\ket{p s q}
.
\end{align}
As explained in Eq.~\eqref{set:torusclosure}, the above states span the ground space on the torus.

These states are not always independent, due to the relation induced by growing an $\omega_0$-loop over the whole torus and fusing it into the MPOs along the 1-skeleton 
\begin{align}
\label{set:toruslooprel}
\ket{\tube{r q r}{s}}
=
\frac{1}{\qd^2} \sum_{p q s} 
\left|
\tube{p q r}{s}
 \tube{rqr}{s} 
(\tube{p q r}{s})^\dagger
\right\rangle
.
 \end{align}

\subsection{Block diagonalizing the tube algebra with irreducible central idempotents}

We have demonstrated that the tube algebra is a $C^*$ algebra and hence can be block diagonalized. Since the tubes in Eq.~\eqref{set:tubeconvention} span the algebra we have
\begin{align}
D = \sum_{pqrs} \delta_{sq}^p \delta_{qs}^{r} = \sum_{\ici{a}} D_{{a}} \times D_{{a}}
\, ,
\end{align}
where the sum is over all blocks $\ici{a}$ of the tube algebra, $D$ is the dimension of the tube algebra, and $D_{{a}}$ is the dimension of the $\ici{a}$ block (which consists of $D_a\times D_a$-matrices). 

 The irreducible central idempotents (ICI) of the tube matrix algebra correspond to Hermitian projectors onto each irreducible block. 
 These ICIs determine the set of inequivalent topological sectors for a disc in the tensor network that may be used to fill in a puncture. 
A constructive algorithm to find these ICIs was explained in Ref.~\cite{Bultinck2017183}. Alternatively, several exact formulas exist when the input category has additional structure such as a modular braiding~\cite{ocneanu2001,koenig2010quantum,davepriv} or a $\mathcal{G}$-crossed modular braiding~\cite{Ren2024}. 

The irreducible central idempotents can be expressed as linear combinations of the tube basis elements
\begin{align}
\ici{a} &= \frac{1}{\qd^2}  \sum_{pqrs}  t^{pqrs}_{a} \tube{pqr}{s} \, ,
\\
	\begin{array}{c}
	\includeTikz{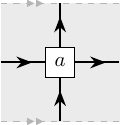}{
	\begin{tikzpicture}[scale=.1]
	\def\dx{-1};
	\def\ddx{1.5};
	\def\dy{.5};
	\def\p{5.5};
	\def\l{2.5};
	\filldraw[UFCBackground](0,0)--(20,0)--(20,20)--(0,20)--(0,0);
	\draw[anyon](10+\l,10)--(20,10) ;
	\draw[anyon](0,10)--(10-\l,10) ;
	\draw[anyon](10,0)--(10,10-\l) ;
	\draw[anyon](10,10+\l)--(10,20) ;
	\draw[draw=black!29,dashed](0,0)--(20,0) (20,20)--(0,20);
	\filldraw[draw=black!29,fill=black!29](\p,0)--(\p+\dx,-\dy)--(\p+\dx,\dy)--(\p,0);
	\filldraw[draw=black!29,fill=black!29](\p,20)--(\p+\dx,20-\dy)--(\p+\dx,20+\dy)--(\p,20);
	\filldraw[draw=black!29,fill=black!29](\p+\ddx,0)--(\p+\dx+\ddx,-\dy)--(\p+\dx+\ddx,\dy)--(\p+\ddx,0);
	\filldraw[draw=black!29,fill=black!29](\p+\ddx,20)--(\p+\dx+\ddx,20-\dy)--(\p+\dx+\ddx,20+\dy)--(\p+\ddx,20);
	\filldraw[draw=black,fill=white] (10-\l,10-\l)--(10-\l,10+\l)--(10+\l,10+\l)--(10+\l,10-\l)--cycle;
	\node at (10,10){$a$};
	\end{tikzpicture}
	}
  \end{array}
&=
\frac{1}{\qd^2} \sum_{pqrs} t^{pqrs}_{a}
\begin{array}{c}
\includeTikz{Tube}{}
\end{array}
\, ,
\end{align}
note the coefficients satisfy $ t^{pqrs}_{a} = \delta_{pr} t^{pqps}_{a}$ as only basis elements $\tube{pqr}{s}$ with $r=p$ can occur in the construction of central idempotents.  The ICIs are defined by the conditions, $\ici{a},\ici{b}\neq0$ 
\begin{align}
\ici{a}\, \ici{b} &= \delta_{ab}\, \ici{a}  
\, ,
&
\tube{pqr}{s} \ici{a}& = \ici{a} \tube{pqr}{s}
\, ,
&
\ici{a} &\neq \ici{b} + \ici{c} \, ,
\end{align}
for $\ici{a},\ici{b}$ ICIs and $\ici{c}$ a nonzero central idempotent. 
These idempotents realize the superselections sectors of the emergent anyon theory, or modular tensor category (MTC), which is the Drinfeld center $\double{\cat}$ of the input UFC $\cat$~\cite{DrinfeldCenter}. The crossing tensors $\ici{a}$, containing the weights $ t^{pqrs}_{a}$, can be directly interpreted as the the simple objects of this Drinfeld center MTC $\double{\cat}$. Furthermore, these crossing tensors yield a construction of the string operators in the Levin-Wen fixed-point models by using them to resolve an external string into the lattice~\cite{levin2005string,davepriv}.

There is always a unique ICI corresponding to the vacuum 
\begin{align}
\ici{\vac} = \sum_{s \in \cat} \frac{d_s}{\qd^2} \tube{\vac s \vac}{s}
\, ,
\end{align}
which is easily verified to be an ICI. 

Each irreducible central idempotent that projects onto a block, $a$, of the matrix algebra with dimension $D_a$ can be written as a sum of $D_a$ non-central irreducible idempotents that partition the identity within the block 
\begin{align}
\ici{a}= \sum_{i=1}^{D_a} (\ici{a})_{ii} \,
.
\end{align}

These non-central irreducible idempotents satisfy, $(\ici{a})_{ii}\neq 0 $ 
\begin{align}
(\ici{a})_{ii} (\ici{b})_{jj} &= \delta_{a b} \delta_{i j}  (\ici{a})_{ii} \, ,
&
(\ici{a})_{ii} &\neq (\ici{b})_{jj} + \ici{c}
\, ,
\end{align}
for $(\ici{b})_{jj},\ici{c}\neq0$ idempotents and $(\ici{b})_{jj}$ irreducible. 
There are also $D_a(D_a-1)$ off-diagonal, nilpotent elements $(\ici{a})_{ij}\neq 0$ in each degenerate block that satisfy
\begin{align}
\label{set:icihoppingterms}
(\ici{a})_{ij}(\ici{a})_{kl} &= \delta_{jk} (\ici{a})_{il}  
\, ,
& (\ici{a})_{ij}^\dagger &= (\ici{a})_{ji}
\, .
\end{align}
The irreducible idempotents are mixed by the conjugation action of the tube algebra $\tube{pqr}{s} (\ici{a})_{ii}  (\tube{pqr}{s})^\dagger$.  
In particular
\begin{align}
(\ici{a})= \sum_{j} (\ici{a})_{ji} (\ici{a})_{ii} (\ici{a})_{ji}^\dagger
\, .
\end{align}

We remark that the irreducible idempotents $(\ici{a})_{ii}$, together with the off-diagonal elements $(\ici{a})_{ij}$ within each block, define a basis for the tube algebra. Hence we can define the change of basis matrix
\begin{align}
(\ici{a})_{ij} = \frac{1}{\qd^2} \sum_{pqrs} t^{pqrs}_{a_{ij}} \tube{pqr}{s}
\, ,
\end{align}
which includes expressions for the off-diagonal elements as well as for the irreducible idempotents. 
Similarly we have the inverse change of basis matrix
\begin{align}
\tube{pqr}{s} = \sum_{a_{ij}} \conj{t}^{a_{ij}}_{pqrs} (\ici{a})_{ij}
\, ,
&&
\text{which implies}
&&
(\tube{pqr}{s})^\dagger = \sum_{a_{ij}} (\conj{t}^{a_{ij}}_{pqrs})^* (\ici{a})_{ji}
\, .
\end{align}
These coefficients satisfy
\begin{align}
\frac{1}{\qd^2} \sum_{pqrs}  {t}_{a_{ij}}^{pqrs} \conj{t}^{b_{kl}}_{pqrs} = \delta_{a b} \delta_{i k} \delta_{j l } 
\, ,
&&
\frac{1}{\qd^2} \sum_{a_{ij}}  \conj{t}^{a_{ij}}_{pqrs} {t}_{a_{ij}}^{xyzw} = \delta_{p x} \delta_{q y} \delta_{r z }  \delta_{s w}
\, .
\end{align}

The columns of the inverse change of basis matrix are orthogonal 
\begin{align}
\label{set:tubenorm}
\frac{1}{\qd^2} \sum_{pqrs}  \conj{t}^{a_{ij}}_{pqrs} (\conj{t}^{b_{kl}}_{pqrs})^* = \delta_{a b} \delta_{i k} \delta_{j l }  \frac{1}{\norm{a_{ij}}^2}
\, ,
\end{align}
for some positive weights $\norm{a_{ij}}$.

It follows from Eqs.~\eqref{set:cylindertubeaction} and \eqref{set:tubenorm} that 
\begin{align}
\label{set:cylinderiiaction}
\openone = \sum_{r}\tube{r r r}{\vac} 
= \frac{1}{\qd^2} \sum_{pqrs} (\tube{pqr}{s})^\dagger  \tube{pqr}{s}
= \sum_{a_{ij}}  \frac{1}{\norm{a_{ij}}^2} (\ici{a})_{ij}^\dagger (\ici{a})_{ij}  
\, .
\end{align}
Multiplying by $(\ici{a})_{ii}$ implies
\begin{align}
\label{set:icipartition}
\sum_j \frac{1}{\norm{a_{ij}}^2} =  1
\, ,
&&
\text{and hence}
&&
\openone= \sum_{p}\tube{ppp}{\vac} 
= \sum_{a} \ici{a}
\, ,
\end{align}
which is a partition of the identity, on the virtual indices of the cylinder tensor network, by ICIs.

\subsection{Minimally entangled states and modular matrices}

To diagnose the emergent topological order of the Drinfeld center $\double{\cat}$ described by the ICIs, we want to calculate gauge invariant observable quantities. In this section we describe how to extract the modular $S$ and $T$ matrices by using the ICIs to define minimally entangled states (MES)~\cite{PhysRevB.85.235151}. These modular matrices are useful invariants for the identification of an anyon theory, however they are not complete invariants even for nonchiral phases~\cite{mignard2017modular}.

Each irreducible idempotent defines a state on the torus 
\begin{align}
\ket{a_{ij}} := \delta_{ij} \frac{1}{\qd^2} \sum_{pqs} t^{pqps}_{a_{ii}} \ket{psq} 
\, .
\end{align}
Due to the relation
$
(\ici{a})_{ii}= (\ici{a})_{ij} (\ici{a})_{ji} 
\,
\sim 
\,
(\ici{a})_{ji}  (\ici{a})_{ij} =  (\ici{a})_{jj}$, 
all irreducible idempotents $(\ici{a})_{ii}$ within the same block give rise to the same state, i.e. $\ket{a_{ii}}=\ket{a_{jj}}$. 

In particular we take the following set of representatives as our definition 
\begin{align}
\ket{a} : = \frac{1}{D_a} \sum_{i} \ket{a_{ii}}
\, ,
\end{align}
which yields a basis of minimally entangled states with a definite anyonic flux $a$ threading the $y$-cycle of the torus. 
With this choice of normalization we have $\ket{a}=\ket{a_{ii}}$. 

Since the ground space on the torus is spanned by states of the form $\ket{psq}$, see Eq.~\eqref{set:torusclosure}, we recover the well known fact that the ground space dimension of the torus equals $|\double{\cat}|$ the number of emergent anyons or ICIs.

The modular $S$ and $T$ matrices are defined on the tubes as follows
\begin{align}
\label{set:snegsandt}
S( \tube{p q r}{s}) &:= \delta_{p r}
	\begin{array}{c}
	\includeTikz{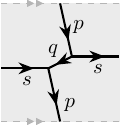}{
	\begin{tikzpicture}[scale=.1]
	\def\dx{-1};
	\def\ddx{1.5};
	\def\dy{.5};
	\def\p{5.5};
	\filldraw[UFCBackground](0,0)--(20,0)--(20,20)--(0,20)--(0,0);
		\draw[anyon](8,9)--(10,0) node[pos=0.7,right] {$p$};
		\draw[anyon](12,11)--(8,9) node[pos=0.8,above] {$q$};
		\draw[anyon](10,20)--(12,11) node[pos=0.45,right] {$p$};
		\draw[anyon](0,9)--(8,9)  node[pos=0.55,below] {$s$};
		\draw[anyon](12,11)--(20,11) node[pos=0.55,below] {$s$};
	\draw[draw=black!29,dashed](0,0)--(20,0) (20,20)--(0,20);
	\filldraw[draw=black!29,fill=black!29](\p,0)--(\p+\dx,-\dy)--(\p+\dx,\dy)--(\p,0);
	\filldraw[draw=black!29,fill=black!29](\p,20)--(\p+\dx,20-\dy)--(\p+\dx,20+\dy)--(\p,20);
	\filldraw[draw=black!29,fill=black!29](\p+\ddx,0)--(\p+\dx+\ddx,-\dy)--(\p+\dx+\ddx,\dy)--(\p+\ddx,0);
	\filldraw[draw=black!29,fill=black!29](\p+\ddx,20)--(\p+\dx+\ddx,20-\dy)--(\p+\dx+\ddx,20+\dy)--(\p+\ddx,20);
		\end{tikzpicture}
		}
	  \end{array}
   &
  T (\tube{pqr}{s}) &:=   \tube{p \vac p}{p} \tube{pqr}{s}
  \\
   & = \sum_{q'} \fs{s} \rbend{q' \conj{s}}{p} \rbend{s p}{q'} (\rbend{qs}{p})^{\dagger} (\lbend{ps}{q'})^\dagger {(\shortf{sp\conj{s}}{p}{q'}{q})}^* \ \tube{s q' s}{\conj{p}} 
     \, .
     &
\end{align}
For the normalized MES and ICIs they satisfy
\begin{align}
\label{set:sandtonidempotentsdefn}
  \ket{S(a)} &=  \sum_{b} S_{ab} \, \ket{b} \, ,
    &
    T (\ici{a}) &= \theta_a\,  \ici{a} \, .
\end{align}
The quantity $\theta_a\in\Uone$ is known as the topological spin of sector $\ici{a}$. The equation defining it follows from Schur's lemma and the observations 
\begin{align}
\label{set:Tmatrixcommutingwithtubes}
\sum_{a} \tube{a\vac a}{a} \, \tube{pqr}{s} = \tube{pqr}{s}  \sum_{a} \tube{a \vac a}{a} \, , && 
\text{and}
&&
\left(\sum_{a} \tube{a\vac a}{a}\right)^\dagger \left(\sum_{a} \tube{a\vac a}{a}\right)=\sum_{a}  \tube{aaa}{\vac} = \openone \, ,
\end{align}
which imply that $T$ acts as the identity times a phase on each irreducible block of the matrix algebra. 
The $S$ and $T$ matrices just defined match those derived from state overlaps of MES for fixed point models such as the string-nets. 

As a technical aside, note that $S$ was defined on MES --- rather than directly on ICIs --- as the relation $\ket{a_{ii}}=\ket{a} $ is necessary to find the correct $S$ matrix elements. 
An equivalent definition would be to look at the action of $S$ on ICIs up to equivalence by conjugation $(\ici{a})_{ii} \sim (\ici{a})_{ji} (\ici{a})_{ii} (\ici{a})_{ij}^\dagger = (\ici{a})_{jj}$. 
Another useful equivalent definition of the matrix elements for $S$ and $T$ is given by 
\begin{align}
\label{set:sandttraceformula}
   S_{ab} &=  \frac{D_b \tr[\ici{b}^\dagger S(\ici{a})]}{D_a \tr[\ici{b}^\dagger\ici{b}]} \, ,
    &
 \theta_a \delta_{a b} &= \frac{D_b \tr[\ici{b}^\dagger T(\ici{a})]}{D_a \tr[\ici{b}^\dagger\ici{b}]}
    \, ,
\end{align}
where the trace now ensures that we have the necessary relation 
\begin{align}
\tr[(\ici{a})_{ii}]=\tr[(\ici{a})_{jj}]=\frac{1}{D_a}\tr[\,\ici{a}\, ]
\, .
\end{align}
We remark that the formulas in Eq.~\eqref{set:sandttraceformula} where written with some unnecessary redundancies (such as the hermitian conjugation). We chose to present them in this way as it makes their generalization to the symmetry-enriched case more clear. The trace used in the above equation is the conventional matrix trace applied to elements of the tube algebra.

The $S$ and $T$ matrices constructed above are known to be unitary~\cite{muger2003subfactors,DrinfeldCenter} and hence the emergent theory  $\double{\cat}$ is modular. Furthermore, since all such theories can be realized by local commuting projector Hamiltonians --- via the string-net construction --- they have zero chiral central charge and do not support robust gapless edge modes.  Consequently $S$ and $T$ satisfy the modular relations
\begin{align}
(ST)^3=S^2
\, ,
&&
S^4=\openone
\, .
\end{align}
One can also verify that our definition of $S$ and $T$ must satisfy the above modular relations via direct manipulation of MPOs. 
Without using the previously known results it is clear that our definition of $T$ must be unitary. It should also be possible to directly show that our definition of $S$ is unitary. However we will not labor to reproduce this known result here.

One can find an explicit formula for $S$ in terms of matrix elements $t_{a_{ij}}^{pqrs},\conj{t}^{a_{ij}}_{pqrs}$ of the tube basis change matrix, and $\shortf{abc}{d}{e}{f}$ from the UFC data. 
This leads to a convenient formula for the quantum dimensions 
\begin{align}
\label{set:derivingqdfromtubes}
d_a = \sum_{p } \frac{d_p {t}_{a}^{ppp\vac} }{D_a} 
\, ,
\end{align}
which follows from
\begin{align}
\frac{d_a}{\qd^2}=S_{a\vac} = \frac{ \tr[\ici{\vac} S(\ici{a})]}{D_a \tr[\, \ici{\vac}\, ]} =  \frac{1}{\qd^2} \sum_{pqs}  {t}_{a}^{pqps} \frac{ \tr[\ici{\vac} S(\tube{pqp}{s}) ]}{D_a \tr[\, \ici{\vac}\, ]} 
= \sum_{p } \frac{d_p {t}_{a}^{ppp\vac} }{\qd^2 D_a} 
 \, .
\end{align}
Where we have used the relation $ S(\tube{pqp}{s}) \, \ici{\vac}= d_p  \delta_{s \vac} \delta_{p q}  \, \ici{\vac}$, which is implied by Eq.~\eqref{set:ICIomegazerosliding}.

\subsection{Topological entanglement entropy of the superselection sectors}
\label{set:teeofanyonsuperselectionsectors}

Our tensor network representation of the ICIs $\ici{a}$, that project onto each superselection sector, can be used to calculate the topological correction to the entanglement entropy for these sectors~\cite{KitaevPreskill,levinwenentanglement,PhysRevLett.108.196402}.  
 We first calculate the scaling of the rank of each sector, given by the trace of the relevant ICI, to find the topological correction. We then argue that this correction should be independent of the R\'enyi index for an RG fixed point model~\cite{topologicalrenyi} and, furthermore, that the value of topological entanglement entropy (TEE) should be robust throughout a gapped phase~\cite{KitaevPreskill}. 

Using Eq.~\eqref{set:derivingqdfromtubes}, we find that the topological correction to the $0$-R\'enyi entropy $\entropy{0}$ (i.e. the entanglement rank) for the $\ici{a}$ sector is given by 
\begin{align}
\gamma_a=\log (\qd_{\text{out}} ) - \log ( d_a )
\, ,
&&
\text{where}
&&
 \qd_{\text{out}}=\qd^2 
 \, ,
 \end{align}
$\qd_{\text{out}}$ is the total quantum dimension of the emergent theory.  
Note, this differs from the TEE of the MES, as reported in Ref.~\cite{PhysRevB.85.235151}, since we are considering a disk topology rather than a torus. 

For gapped, fixed point models --- such as the string-nets --- the 0-R\'enyi entropy is equal to the entanglement entropy as the Schmidt spectrum is flat. Hence, for fixed point models, the topological correction $\gamma_a$ we have calculated corresponds to the topological entanglement entropy (TEE) of the ${a}$ superselection sector~\cite{KitaevPreskill,levinwenentanglement}. Furthermore, Ref.~\cite{topologicalrenyi} argued that the result for the TEE calculated at a fixed point, via the 0-R\'enyi entropy, should hold throughout the gapped phase containing that fixed point model. 

The $0$-R\'enyi entropy $\entropy{0}$ of the $a$ superselection sector can be calculated by taking the trace of the ICI $\ici{a}$. To extract the topological correction we need to access the regime of asymptotic scaling in the size of the perimeter $L$. Hence we consider an ICI formed by an MPO of length $L$ contracted with the crossing tensor containing the weights $t_a^{pqps}$
\begin{align}
\label{set:hzerotracedefinition}
\entropy{0} = 
\log \Big (
\frac{1}{D_a \qd^2} \sum_{pqs} t_a^{pqps}
\ 
\tr
\ 
{\Big [ }
\raisebox{-.4cm}{$

$}
{\Big ] }
\Big )
\, .
\end{align}
The factor of $D_a^{-1}$ is included due to an argument presented in Ref.~\cite{Bultinck2017183} that for a pair of charges $a,\,\conj{a}$ on a sphere, each basis vector in the degenerate block appears with the same weight. To produce a correctly normalized state, this weight should be $D_a^{-1}$.

It was argued in Ref.~\cite{Bultinck2017183} that the trace of a projector MPO of length $L$ (assuming the existence of a unique unit element) is dominated by the leading eigenvector of the $\vac$ block, $\lambda_\vac > \lambda_s$ for $s\neq\vac$. This contributes a factor of $\lambda_\vac^L$ and picks out the elements of the crossing tensor $t_a^{ppp\vac}$ satisfying $s=0$. 
To be consistent with a faithful MPO representation of the UFC calculus, the traced crossing tensor must yield the quantum dimension of the closed loop it forms, i.e. $d_p$ for the entry of the crossing tensor with coefficient $t_a^{ppp\vac}$. 
Using Eq.~\eqref{set:derivingqdfromtubes} we find that $H_0$ scales as
\begin{align}
\label{set:TEEscalingfromtube}
\entropy{0} \approx \log(\lambda_\vac) L - \gamma_a
\, ,
&&
\text{where}
&&
\gamma_a = -\log \left( \sum_{p } \frac{d_p {t}_{a}^{ppp\vac} }{D_a \qd^2} \right) = \log(\qd_\text{out}) - \log(d_a)
\, .
\end{align}
This provides a general method to extract the TEE of superselection sectors directly on the lattice, including the well known vacuum TEE $\gamma_\vac = \log(\qd_\text{out})$.

\subsection{Fusion}
\label{set:icifusionsection}

The fusion spaces of the emergent theory are found by considering the tensor network on a thrice-punctured sphere. 
To construct these fusion spaces we make a slight change of notation for the tubes  
  \begin{align}
\begin{array}{c}
\includeTikz{Tube}{}
\end{array}%
  \, .
  \end{align}
Now the splitting space $V_c^{a b}$ is constructed from tensor networks  of the form 
\begin{align}
  \label{set:nabctube}
    	
\, ,
\end{align}
where
\begin{align}
     	\begin{array}{c}
	\includeTikz{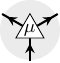}{}
  \end{array}%
\, .
\end{align}
There is an important subtlety in calculating the basis vectors of the spanning set, due to the local relation induced by growing an $\omega_0$-loop over the thrice punctured sphere and fusing it into the skeleton. 
Hence the splitting space is spanned by vertex tensors $\mu_{a b}^{c}$ in the support subspace of the following projector
\begin{align}
\label{set:nabcaction}
  \sum_{s i j k l p q}
\frac{ \qd^4}{d_s^2}
   	%
  \, ,
  \end{align}
    where $\mu_{ab}^{c}$ indicates $\mu \in \splittingspace{ab}{c}$. 
  More specifically, the relation is induced by creating an $\omega_0$-loop between $\ici{c}$ and $\ici{a},\ici{b}$ and fusing it into the edges around the $\mu_{a b}^{c}$ vertex. This also results in a correlated action by $(\ici{a})_{ij},(\ici{b})_{kl}$ and $(\ici{c})^\dagger_{pq}$ 
on the $\ici{a},\ici{b}$ and $\ici{c}$ sectors, respectively. These actions permute the internal states within each degenerate block.  

Considering  Eq.~\eqref{set:icihoppingterms}, and the fact that the $\omega_0$-loop is a projector, one finds that the action in Eq.~\eqref{set:nabcaction} is also a projector. 
By taking the trace of this projector we find $N_{ab}^c$, the dimension of its support subspace. After using the tube $S$-matrix to expand $\ici{a},\ici{b},\ici{c}$ in a complimentary tube basis one can derive the well known Verlinde formula~\cite{0305-4470-20-16-043,VERLINDE1988360}
\begin{align}
N_{ab}^{c} = \sum_{x} \frac{S_{ax} S_{bx} S_{\conj{c} x}}{S_{\vac x}}
\, ,
\end{align}
for example see Ref.~\cite{aasen2017fermion}. 
This formula holds for any modular theory, which includes all the theories we consider as they are Drinfeld centers.

In the case that $\ici{a},\ici{b},$ and $\ici{c}$ are nondegenerate, i.e. one dimensional, $c_\mu^{pqr}$ is a delta function that determines a unique fusion vertex, $\mu_{a b}^{c}$. 
This fusion vertex, $\mu_{a b}^{c}$, must satisfy
\begin{align}
\label{set:nabcproj}
  \sum_s
  \frac{\qd^4}{d_s^2}

  \, ,
  \end{align}
which allows one to absorb the $\ici{c}$ ICI into $\ici{a},\ici{b}$ and $\mu_{a b}^{c}$. 
More explicitly, for nondegenerate ICIs  $\ici{a},\ici{b},\ici{c}$, one has 
  \begin{align}
    	%
  \, ,
  \end{align}
  for the unique vertex $\mu_{a b}^{c}$ that satisfies Eq.~\eqref{set:nabcproj}. 

Eq.~\eqref{set:nabcaction} follows from the identity
    \begin{align}
    \label{set:ICIteleportation}
        \sum_s  \frac{d_s}{\qd^2}
	%
= \sum_{ij} (\ici{\widetilde a})_{ij}^\dagger \, \ici{a}\,  (\ici{{a}})_{ij}
\, ,
\end{align}
which  takes a slightly more illuminating form for a nondegenerate block
   \begin{align}
    \sum_s  \frac{d_s}{\qd^2}
    \label{set:nondegenICIteleportation}
	\begin{array}{c}
	\includeTikz{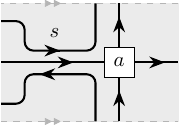}{	}
  \end{array}
= {\ici{\widetilde{a}}}^\dagger  \, \ici{a}
\, .
\end{align}
The special case of this equation for $\ici{a}=\ici{\vac}$, corresponds to sliding an MPO over an $\omega_0$-loop 
    \begin{align}
    \label{set:ICIomegazerosliding}
        \sum_s  
	%
  \, ,
  \end{align}
which can be intuitively understood as follows: the $\ici{\vac}$ ICI simply projects onto a disc of unaltered tensor network, through which the MPO is free to move.

The above identities are a consequence of the following slight generalization of Eq.~\eqref{set:cylinderiiaction}, which is proved in much the same way, 
\begin{align}
   \sum_{r s} \frac{d_s}{\qd^2}
%
\, ,
&
(\ici{\widetilde{a}})_{ij}^\dagger &= \frac{1}{\qd^2} \sum_{pqrs} (t^{pqrs}_{a_{ij}})^*\, (\widetilde{\mathcal{T}}_{pqr}^{s} )^\dagger .
\end{align}
We remark that Eq.~\eqref{set:nondegenICIteleportation} is somewhat analogous to state teleportation in quantum information, where $\ici{a}$ plays the role of the state being teleported and the delta condition arising from multiplication with the idempotent plays the role of the projective measurement.

Similar to Eq.~\eqref{set:ICIteleportation}, we have 
    \begin{align}
    \sum_s  \frac{d_s}{\qd^2}
    \label{set:ICIteleportationtwo}

= \sum_{ij}  (\ici{a})_{ij}^\dagger \, \ici{a} \,  (\ici{\widetilde{a}})_{ij}
\, ,
\end{align}
with analogous definitions of $\widetilde{\mathcal{T}}_{pqr}^{s}$ and $(\ici{\widetilde{a}})_{ij}$ to those in Eq.~\eqref{set:tildetanda}.

\subsubsection{Cups, caps and the Frobenius-Schur indicator}

The special case of pair fusion  (creation) to (from) the vacuum is captured by the cap (cup) tube
\begin{align}
\label{set:cupandcaptubes}
	%
  \, ,
\end{align}
respectively. Where $\ici{\bar{a}}$ is the ICI for which the above tubes are nonzero.  
The FS indicator can be found by composing a cup and a cap~\cite{davepriv}, which corresponds to unbending an anyon worldline
  \begin{align}
  \label{set:fsindtubeeqn}
	%
  =
  \fs{a}\, \ici{a}
  \, .
\end{align}
We remark that the the FS indicator only has an invariant meaning for self inverse particles, $\conj{a}=a$, in which case $\fs{a}=\pm 1$.

\subsubsection{$F$-symbol associators}

Assuming no block degeneracy, the $F$-symbols for the doubled theory, denoted $\mathcal{F}$, are given by
\begin{align}
\label{set:doublefsymboleqn}
 	%
  \, ,
  \end{align}
where, due to the no degeneracy assumption, $c_{pqt}^{\mu}$ and $\conj{c}^{qru}_{\rho}$ are delta functions. We remark that the above equation should be interpreted as part of a larger diagram, involving a sphere with four punctures, that have been projected onto idempotents $\ici{a},\ici{b},\ici{c}$ and $\ici{d}$ respectively, similar to the thrice punctured sphere depicted in Eq.~\eqref{set:nabctube}. The LHS of the above equation has an implicit $\ici{e}$ idempotent between the fusion vertices, which was absorbed into the $\mu$ vertex using Eq.~\eqref{set:nabcproj}. Similarly, the line connecting the fusion vertices on the RHS has been resolved into a sum over idempotents $\ici{f}$, using Eq.~\eqref{set:icipartition}, which have been absorbed into the $\rho$ vertex using Eq.~\eqref{set:nabcproj}. 

There is a similar, although somewhat more involved equation for $F$ for degenerate blocks.  
By considering a sphere with five punctures one can check that the emergent $F$-symbols satisfy the pentagon equation.

 \subsection{Braiding}

The exchange of superselection sectors can be implemented by a deformation of the tensor network
\begin{align}

  \, ,
  \end{align}
where we have used Eq.~\eqref{set:ICIteleportation} to find a tensor corresponding to the $R^{ab}$ matrix. 
We remark that there is a correlated action $(\ici{a})_{ij}$ on the internal states of the $\ici{a}$ block, if it is degenerate, which we do not explicitly depict.

\subsubsection{$R$-matrix}

If the $\ici{a}$ ICI is nondegenerate, there is no summation over $i,j$ and we simply have the above formula with $\ici{a}^\dagger$ in place of $(\ici{a})^\dagger_{ij}$. 
Hence for nondegenerate $\ici{a}$ and $\ici{b}$ we have a formula for the $R$ matrix
\begin{align}
\label{set:rabnondegen}
R^{ab}= 

  =
 \sum_{q}  \frac{(t_{a}^{s^a q \, s^a  s^b  })^*}{d_{s^b}}
  \, ,
\end{align}
where $s^a$ is defined by the equation $\ici{a} \tube{p p p}{\vac} = \delta_{p ,s^a}\,  \ici{a}$. 

The $R$ matrix can be resolved into sectors of definite charge to find the $\rmatrix{ab}{c}$ symbols. 
For nondegenerate ICIs $\ici{a},\ici{b},\ici{c}$, and the unique fusion vertices $\mu_{ba}^c$, and $\nu_{ab}^c$,  we have 
\begin{align}
      	%
	\, 
	,
\end{align}
which should be interpreted as part of a larger fusion diagram such as Eq.~\eqref{set:nabctube}. An explicit formula for $\rmatrix{ab}{c}$ can be derived from Eq.~\eqref{set:rabnondegen}. 
We remark that a similar line of reasoning applies to degenerate ICIs, but the analysis becomes more complicated.

  \section{Symmetry-enriched topological order from matrix product operators}
  \label{setsection:setmpos}

In Ref.~\cite{williamson2014matrix}, a formalism was introduced for the classification of symmetry-enriched topological order in two spatial dimensions using $\G$-graded matrix product operator algebras. 
The formalism was based on finding MPO representations ${\mpo{}{\g{g}}}$ of the physical symmetry group $\G$.
These MPOs correspond to $\G$-domain walls. Only unitary, on-site representations of finite groups were explicitly considered, but this turns out to capture the same SET phases without making the on-site assumption~\cite{etingof2009fusion} and there is a simple extension to also include anti-unitary elements~\cite{cheng2016exactly}. 

In this section we develop the theory of $\G$-graded MPO algebras, drawing on the previous results derived for MPO algebras in Ref.~\cite{Bultinck2017183}. 
We find that the $\G$-graded MPO algebras, with a given underlying $\g{1}$-sector MPO algebra, correspond to $\G$ extensions of that $\g{1}$-sector MPO algebra. 
We establish a close analogy between $\G$-graded MPO algebras and $\G$-graded UFCs, for which the extension problem has been thoroughly analyzed in Ref.~\cite{etingof2009fusion}. 
The $\G$-graded MPOs are used to define a symmetry-enriched pulling through condition for PEPS tensors that lead to symmetric tensor networks under the global symmetry. 
Finally we describe how calculations involving the $\G$-graded MPOs can be abstracted to a diagrammatic calculus that only keeps track of the relevant topological information.

\subsection{$\G$-graded MPO algebras}

We consider MPO representations of a finite group $\G$ consisting of translation invariant elements of the form 
\begin{align}
\mpo{L}{\g{g}} = \sum_{\substack{i_1 \dots i_L \\ j_1 \dots j_L}} 
\tr[ \Delta_{\g{g}} \mpotensor{\g{g} }{i_1 j_1} \mpotensor{\g{g}}{i_2 j_2} \cdots \mpotensor{\g{g}}{i_L j_L} ] 
\, \ket{i_1 \cdots i_L}\bra{j_1 \dots j_L}
\, ,
\end{align}
for arbitrary length $L$. The object $\mpotensor{\g{g} }{i j}$ is a $\mpobd_\g{g} \times \mpobd_\g{g}$ matrix for each value of $i,j\in \z_\pepsbd$, where $\chi_\g{g}$ is the bond dimension of the MPO and $\pepsbd$ corresponds to the bond dimension of a PEPS.  $\Delta_\g{g}$ is also a $\mpobd_\g{g} \times \mpobd_\g{g}$ matrix.  
In the conventional tensor network notation, we have
\begin{align}
(\mpotensor{\g{g} }{i j})_{\alpha \beta}
=
\, ,
\end{align}
where dashed lines are used to denote periodic boundary conditions.  

We restrict our attention to MPOs that can be brought into canonical form~\cite{MPSrepresentations,Cirac2017100} after insertion of $\Delta_\g{g}$. Hence the tensors have a block diagonal structure  
\begin{align}
\label{set:gblockstructure}
\mpotensor{\g{g} }{i j} = 
\bigoplus_{a\in \cat_{\g g}} \mpotensor{a }{i j} 
\, ,
&&
\Delta_\g{g} = 
\bigoplus_{a\in\cat_{\g g}} w_{a} \openone_{\chi_a}
\, ,
\end{align}
where $\mpotensor{a }{i j}$ are a set of $\chi_a\times\chi_a$ matrices that generate the full matrix algebra, $w_a$ are nonzero complex numbers, and $\cat_\g{g}$ refers to the collection of MPOs appearing in the decomposition of $\mpo{}{\g{g}}$. These MPOs are referred to as \emph{single block}, since they cannot be decomposed further. The full collection of single block MPOs in all sectors is denoted $\cat_\G$.  
We remark that $\Delta_\g{g}$ decomposes into a direct sum of multiples of the identity since it must commute with the full matrix algebra on each block, as the MPO is translationally invariant. 

In tensor network notation Eq.~\eqref{set:gblockstructure}, left, is written as
\begin{align}

=
(\mpotensor{\defect{a}{g} }{i j})_{\alpha \beta}
\, ,
\end{align}
and the shorthand notation $a_\g{g}$ indicates $\mpo{}{a}\in\cat_\g{g}$. We also use the shorthand $\sector{a}\in\G$ to indicate $\mpo{}{a}\in\cat_{\sector{a}}$, i.e. $\sector{\defect{a}{g}}=\g g$.  We remark that each $\defect{a}{g}$ MPO tensor is only nonzero within the relevant $\chi_a \times \chi_a$ block of the virtual index, hence a direct sum is implicit in the equation above, left. 

The \emph{single block} MPOs are then 
\begin{align}
\mpo{}{a_\g{g}}= 
%
\, ,
&&
\text{which satisfy}
&&
\mpo{}{\g{g}}= \sum_{a_\g{g}}  w_{a_\g{g}} \, \mpo{}{a_\g{g}}
\, .
\end{align}
Since the $\g g$ MPOs form a representation of $\G$ we have $\mpo{}{\g{g}} \mpo{}{\g{h}} = \mpo{}{\g{gh}}$. 
The MPO algebra must have a faithful $\G$-grading, i.e. $| \cat_{ \g g} | > 0$ for all $\g g$, otherwise application of the physical symmetry to a tensor network can annihilate the state. 
If $|\cat_{\g 1}|=1$ there is no underlying topological order, and the formalism recovers the cohomology classification of SPT phases~\cite{chen2013symmetry,czxmodel,levin2012braiding} in \std{2}, this case was thoroughly analyzed in Ref.~\cite{williamson2014matrix}. 
In general this representation yields a Hermitian MPO projector onto the symmetric subspace
\begin{align}
\label{set:mpogroupprojector}
\frac{1}{|\G|} \sum_{\g{g} \in \G} \mpo{L}{\g{g}}
\, ,
\end{align}
for all lengths $L$. 
In Ref.~\cite{Bultinck2017183} such MPO projectors where analyzed extensively, and it was found that many concepts from the theory of fusion categories naturally emerge. At an intuitive level, such MPO projectors correspond to a representation of a fusion category. 
In particular, the theory of MPOs and the fact that we have a representation, implies 
\begin{align}
\label{set:setmpoblockmult}
\mpo{}{a_\g{g}}\mpo{}{b_\g{h}} = \sum_{c \in \cat_\G} N_{\defect{a}{g} \defect{b}{h} }^{c}  \mpo{}{c}
=\sum_{c_\g{gh}\in \cat_{gh}} N_{\defect{a}{g} \defect{b}{h} }^{\defect{c}{gh}}  \mpo{}{\defect{c}{gh}}
\, ,
\end{align}
\begin{align}
\label{set:setmpoblockmult2}
\sum_{\defect{a}{g}\,\defect{b}{h} }  N_{\defect{a}{g}  \defect{b}{h}}^{\defect{c}{gh}} w_{\defect{a}{g}} w_{\defect{b}{h}} = w_{\defect{c}{gh}} 
\, ,
\end{align}
since $N_{\defect{a}{g} \defect{b}{h}}^{c}$ vanishes unless $c\in\cat_{\g{gh}}$. 
Remarkably it also implies a local version of this condition on the level of individual MPO tensors. That is, there exists a set of independent \emph{fusion tensors}, and their left inverses 
\begin{align}
\label{set:fusiontensors}
[X_{a_\g{g} b_\g{h}}^{c_\g{gh}\,\mu}]_{\alpha \beta}^{\gamma}=

\, ,
\end{align}
where the $\defect{c}{gh}$ line on the right indicates the identity on the $\defect{c}{gh}$ block of the virtual index. 
We remark that the fusion tensor are only defined up to an invertible gauge transformation on the degeneracy index,  $\mu$,
\begin{align}
X_{a_\g{g} b_\g{h}}^{c_\g{gh}\,\mu} \, \mapsto  \, 
\widetilde{X}_{a_\g{g} b_\g{h}}^{c_\g{gh}\,\nu} :=
\sum_\mu (Y_{ab}^c)_{\nu\mu} X_{a_\g{g} b_\g{h}}^{c_\g{gh}\,\mu} \, .
\end{align}

For a symmetric PEPS on a sphere, a second type of MPO arises that has a reversed orientation of the virtual bond. These MPO are not independent from those introduced earlier, and lead to an arrow reversing gauge transformation. Consider a sphere and partition the on-site site representation into an action on the northern and southern hemispheres, 
\begin{align}
U(\g g)=U_\text{N}(\g g) \otimes U_\text{S}(\g g)
\, ,
\end{align}
respectively. Applying $U_\text{S}(\g g)$ to a symmetric PEPS --- see Section~\ref{set:sepullthrupepstensors} --- leads to a domain wall along the equator, represented by $\mpo{}{\g{g}}$. Applying $U_\text{N}(\g{g})$ instead, leads to a different MPO
\begin{align}
\mpo{-}{\g{g}} &= \sum_{a_\g{g}} w_{a_{\g{g}}} \mpo{-}{a_\g{g}} 
\, ,
\end{align}
\begin{align}
\text{where}
&&
\mpo{-}{a_\g{g}}= 

\, .
\end{align}
This is simply because the boundary of the southern disc has the opposite orientation to the boundary of the northern disc. 
Hence, by applying $U_\text{N}(\g g) = ( \openone \otimes U_\text{S}(\conj{\g g})\, ) \, U(\g g)$ to a symmetric PEPS on the sphere we find 
\begin{align}
\mpo{-}{\g{g}}=\mpo{}{\conj{\g{g}}}
\, .
\end{align}
The theory of MPS implies that there is a unique label $\defect{\conj{a}}{\conj{\g{g}}}$, for each $\defect{a}{g}$, such that 
\begin{align}
\mpo{-}{\defect{a}{g}} = \mpo{}{\defect{\conj{a}}{\conj{\g{g}}}}
\, ,&&
\text{and}
\, , &&
w_{\defect{a}{g}} = w_{\defect{\conj{a}}{\conj{\g{g}}}}
\, ,
\end{align}
which also satisfies $\conj{\conj{a}}=a$. 
Furthermore, there exists a local gauge transformation 
\begin{align}
Z_a
=

\, .
\end{align}
Note the same gauge transformation must work for both orientations in a consistent MPO representation~\cite{williamson2014matrix}.
By applying Eq.~\eqref{set:mpoorientationreversal} twice, we find that $Z_a (Z_{\conj{a}}^{-1})^T$ must commute with the set of matrices $\mpotensor{a}{ij}$. Hence ${Z_a (Z_{\conj{a}}^{-1})^T=\fs{a} \openone_{\chi_a}}$, for some complex number $\fs{a}$. 
This implies 
\begin{align}
\label{set:fsindeqn}
	%
\, ,
\end{align}
 we also find $\fs{\conj{a}}=\fs{a}^{-1}$. The magnitude $|\fs{\conj{a}}|$ can be absorbed into the definition of $Z_a$ and the phase $\frac{\fs{a}}{|\fs{{a}}|}$ can also be absorbed unless $a=\conj{a}$, in which case we have $\fs{a}=\pm1$. It will be shown later that this quantity corresponds to the familiar Frobenius-Schur indicator from the theory of fusion categories.

\subsection{Classification via group extensions of MPO algebras}
\label{set:groupextensionclassification}

$\G$-graded matrix product operator algebras can be classified by $\G$ extensions of an underlying MPO algebra, that corresponds to the $\g{1}$-sector of the resulting graded algebra.

\subsubsection{Summary of the problem and comparison to previous work}

$\G$-graded MPO algebras $\cat_\G$ occur as group extensions of an ungraded MPO algebra $\catD$ that forms the $\g{1}$ sector. This is a generalization of the more familiar group extension problem, where $\catD$ plays the role of the normal subgroup, and $\G$ the quotient group. For the class of MPO algebras that correspond to fusion categories, which we focus on, one can apply  the thorough analysis of the group extension problem for fusion categories given in Ref.~\cite{etingof2009fusion}. 

The first step in the search for faithful $\G$ extensions of an MPO algebra $\catD$ can intuitively be thought of as a hunt for $\G$-graded MPO algebras $\cat_G$ that make 
\begin{align}
1 \rightarrow \catD \rightarrow \cat_\G \rightarrow \G \rightarrow 1
\, ,
\end{align}
into a short exact sequence. 
From left to right, the arrows correspond to: a map from unit to the $\vac$ MPO (defined more precisely below), an injection that includes $\catD$ as the trivial sector in $\cat_\G$, a  surjection that maps $\cat_\g{g}$ down to $\g{g}$, and a map from all elements of $\G$ to $1$, respectively. 

This extension problem includes the group extension problem as a special case in the following way: for $\catD=\vecg{N}{}$ and $\G=Q$, any group $G$ that is an extension of $Q$ by $N$ yields a solution $\cat_\G=\vecg{G}{}$. 
There are solutions beyond these group extensions, as demonstrated by the example $\catD=\vecg{\zt}{}$ and $\G=\zt$, for which we find solutions where $\cat_\G$ has Ising fusion rules, along with solutions that have $\zt \times \zt$ and $\z_4$ fusion rules. This example is explored further in Section~\ref{setsection:examples2}.

There is a potential obstruction to finding an associative set of $\G$-graded fusion rules which lies in the third cohomology group  $\cohomo{3}{\G}{\autoequiv{1}}$, explained in detail below. If this obstruction vanishes, we have a consistent set of $\G$-extended fusion rules. These form a torsor over $\cohomo{2}{\G}{\autoequiv{1}}$ as shifting by such a 2-cocycle leads to another consistent solution. 

A consistent $\G$-graded fusion algebra alone does not solve the extension problem. Since the fusion of open MPOs is only associative up to some boundary operator, we need to solve a categorification of the underlying algebra problem. Hence we must also check whether a consistent set of associators can be found, to relate the fusion of open MPOs in different orders. This amounts to finding $F$-symbols that solve the pentagon equation for the previously constructed $\G$-graded fusion algebra.  Any such solution can be realized by a $\G$-graded MPO algebra via the fixed point construction in Section~\ref{set:stringnetexample}. 
 It was shown in Ref.~\cite{etingof2009fusion} that there is a potential obstruction to finding a consistent  set of associators which lies in $\coho{4}$. We find that this obstruction vanishes for the class of MPOs considered here, and the resulting solutions form $\coho{3}$ torsors. That is, multiplying any solution by a 3-cocycle results in another --- possibly distinct --- solution.  Physically this corresponds to the fact that a classification of SET phases depends upon the choice of on-site symmetry with respect to which they are classified.

For the above examples, where $\catD=\vecg{N}{}$, $\G=Q$, and the fusion rules are given by $G$ --- a group extension of $Q$ by $N$ --- the potential $\coho{4}$ obstruction disappears as the trivial $F$-symbols are a valid solution to the pentagon equation that yields $\cat_\G=\vecg{G}{}$. 
Shifting by an element $[\alpha ] \in \coho{3}$ corresponds to twisting the extension by a 3-cocycle to obtain $\cat_\G = \vecg{G}{\alpha}$. 
More generally, for $\catD=\vecg{N}{\omega}$ with $[\omega ] \in \cohom{3}{N}$ and $\G=Q$ we find several valid extensions with $G$ fusion rules. These extensions are given by $\cat_\G=\vecg{G}{\hat \omega}$ for $[\hat \omega ] \in \cohom{3}{G}$ that satisfy $\hat \omega \big|_N \equiv \omega$.

This class of examples can be obtained by gauging normal subgroups of SPT phases, see Appendix~\ref{setappendix:gauging}. They capture the formalisms of Refs.~\cite{PhysRevB.65.165113,mesaros2013classification,PhysRevB.95.125107,garre2017symmetry,PhysRevB.95.235119} as follows:
\begin{itemize}
\item The unitary finite group case of Wen's projective symmetry group classification~\cite{PhysRevB.65.165113} is captured by restricting $\omega$ and $\hat \omega$ to be trivial. 
\item Similarly, Ref.~\cite{garre2017symmetry}  is captured by restricting $\omega$ and $\hat \omega$ to be trivial. 
\item Conversely, Ref.~\cite{mesaros2013classification} is captured by restricting to trivial direct product extensions $G\cong N \times Q$, but with arbitrary $\omega$ and $\hat \omega$. 
\item Interestingly, Ref.~\cite{PhysRevB.95.235119} is captured by restricting to $\vecg{G}{\hat \omega}$ --- with possibly nontrivial $\omega$ and $\hat \omega$ --- that are Morita equivalent to some $\vecg{A}{}$, for $A$ an abelian group. For further explanation see Sections \ref{set:algebragradingmposymmetrybreakingcondensation} and \ref{setsection:moritaequivegone}.
\item The unitary finite group case of Ref.~\cite{PhysRevB.95.125107} is captured with no further restrictions. 
\end{itemize}

It is clear that only considering extensions $\cat_\G$ whose fusion rules are given by a finite group is very restrictive and misses many interesting cases,  such as the Ising fusion rules needed to describe the EM duality enriched toric code, see Sections~\ref{set:tcextensioneg},~\ref{set:settceg}. Furthermore, all such MPOs can be generated by an on-site representation times a diagonal local unitary circuit, see Refs.~\cite{czxmodel,williamson2014matrix,PhysRevB.91.195134}, while the more general MPOs may require a linear depth unitary circuit, for instance the Ising MPO.

\subsubsection{MPO representations satisfying the group-zipper condition}

At this point we assume that a stronger version of the local fusion condition holds for the MPO group representation. This allows us to make contact with the theory of group extensions of fusion categories~\cite{etingof2009fusion}. The MPO representation of a $\G$ domain wall correspond to an invertible bimodule in the language of Ref.~\cite{etingof2009fusion}. 
We restrict to the case where the full group MPOs satisfy the \emph{group-zipper condition}: that there exist tensors, together with left inverses
\begin{align}
X_{\g{g}, \g{h}}=

	\, . 
\end{align}
These tensors define a left and right group action on the $\chi_g\times\chi_g$ matrix algebra of the $\g g$ virtual bond
\begin{align}
\label{set:hleftrightdwaction}
 \g{h} \circ M_{\g{g}} 
=
	%
\, .
\end{align}
The subset of matrices --- modulo multiples of the identity on each block --- that commute with $\mpotensor{\g g}{ij}$ for all $ij$, define the group of autoequivalences of the $\g g$ domain wall, $\autoequiv{g}$. 
The right and left actions in Eq.~\eqref{set:hleftrightdwaction} imply isomorphisms $\autoequiv{hg} \cong \autoequiv{g}\cong \autoequiv{gh}$, due to the zipper condition, and hence $\autoequiv{g} \cong \autoequiv{1}$. 
 It was shown in Ref.~\cite{etingof2009fusion} that $\autoequiv{1} \cong \aban$, the group of abelian anyons of the emergent topological order $\aban \subseteq \double{\cat_{\g 1}}$. 
  It is beyond the scope of this paper to rigorously establish that our definition of $\autoequiv{1}$ matches that of Ref.~\cite{etingof2009fusion}, but they are analogous. 
  
The equivalence $\autoequiv{1} \cong \aban$ can be understood physically by considering a topological order \double{\cat_{\g 1}} with a gapped boundary to vacuum $\cat_\g{1}$. Pushing abelian anyons from the bulk onto the boundary excitations induces an autoequivalence of $\cat_\g{1}$ for each element of $\aban$. This picture suggests that only the abelian anyons that do not condense on the boundary induce interesting autoequivalences of $\cat_\g{1}$, which lines up with the results of the example in Section~\ref{set:tcextensioneg}. 
 
The $X_{\g g ,\g h}$ tensors define a matrix
\begin{align}
T(\g{g},\g{h},\g{k}) = 

\, ,
\end{align}
that lies in $\autoequiv{ghk}$ and satisfies the 3-cocycle equation
\begin{align}
\g{g}_0 \circ T({\g{g}_1,\g{g}_2,\g{g}_3})\, T({\g{g}_0,\g{g}_1\g{g}_2,\g{g}_3})\, T({\g{g}_0,\g{g}_1,\g{g}_2}) \circ \g{g}_3 
=  T({\g{g}_0,\g{g}_1,\g{g}_2\g{g}_3}) \,  T({\g{g}_0\g{g}_1,\g{g}_2,\g{g}_3})
\, .
\end{align}
If we modify our choice of $X_{\g g, \g h}$, by an element $M_{\g{g},\g{h}}\in \autoequiv{gh}$, to $X'_{\g g, \g h}= X_{\g g, \g h} M_{\g{g},\g{h}}$ we find that 
\begin{align}
T'(\g{g},\g{h},\g{k}) = 
M_{\g{gh},\g{k}}^{-1} \, ( M_{\g{g},\g{h}}^{-1} \circ \g{k} ) \,
T(\g{g},\g{h},\g{k}) 
\, ( \g{g} \circ M_{\g{h},\g{k}} ) \, M_{\g{g},\g{hk}}
\, ,
\end{align}
which corresponds to a transformation by a coboundary.  

Hence $X_{\g g, \g h}$ yields an element $[T]\in\cohomo{3}{\G}{\autoequiv{1}}$, that is invariant under modifying the choice of $X_{\g g, \g h}$ by an element of $\autoequiv{gh}$. This element $[T]$ constitutes an obstruction to a consistent MPO representation of a set of $\G$ domain walls. 
Intuitively, the obstruction can be thought of as an anomalous anyonic charge appearing from the fusion of domain walls. 
It is unclear at this point whether an MPO group representation can be found with $[T]\neq 0$. If they were to be found, these MPOs would be relevant to the understanding of tensor network representations of SETs with anomalous symmetry fractionalization, such as those occurring on the boundary of a \std{3} higher form SPT~\cite{thorngren2015higher,fidkowski2015realizing}. 

Consistent fusion of the MPO representation requires $[T]=0$ and that $T$ has been transformed,  by a coboundary, into a multiple of the identity.
Tensor solutions $X_{\g g,\g h}$ that satisfy $[T]=0$, form $\cohomo{2}{\G}{\autoequiv{1}}$ torsors. That is, we can modify any solution $X_{\g g,\g h}$ by a 2-cocycle $M_{\g{g},\g{h}}\in \autoequiv{gh}$ satisfying
\begin{align}
 ( M_{\g{g},\g{h}} \circ \g{k} ) \, M_{\g{gh},\g{k}}
=
 ( \g{g} \circ M_{\g{h},\g{k}} ) \, M_{\g{g},\g{hk}}
 \, ,
\end{align}
 to find a new solution $X'_{\g g, \g h}= X_{\g g, \g h} M_{\g{g},\g{h}}$

Hence we have $T(\g g,\g h, \g k)=\alpha(\g g,\g h,\g k) \openone_{\g{ghk}}$, where we may take $\alpha(\g g,\g h,\g k)\in\Uone$ by normalizing the $X_{\g g,\g h}$ tensors.  
This yields an associator 
\begin{align}

\, .
\end{align}

Considering the reduction of $\g{g}_0 \g{g}_1 \g{g}_2 \g{g}_3$, one finds a relation 
\begin{align}
	%
	\, ,
\end{align}
leading to the phase
\begin{align}
\nu(\g{g}_0 , \g{g}_1 , \g{g}_2 , \g{g}_3) := 
\frac{ \alpha({\g{g}_1,\g{g}_2,\g{g}_3}) \alpha({\g{g}_0,\g{g}_1\g{g}_2,\g{g}_3}) \alpha({\g{g}_0,\g{g}_1,\g{g}_2})}
{\alpha({\g{g}_0,\g{g}_1,\g{g}_2\g{g}_3}) \alpha({\g{g}_0\g{g}_1,\g{g}_2,\g{g}_3})}
\, ,
\end{align}
which is a 4-cocycle. This constitutes an $\coho{4}$ obstruction~\cite{etingof2009fusion} to finding a solution to the pentagon equation for the MPO representation.
We find that this obstruction always vanishes for the MPOs we consider, due to the zipper condition we have assumed. It would be interesting to generalize our framework to capture a nontrivial $\coho{4}$ obstruction, which is relevant for the understanding of tensor network representations of anomalous SETs such as those occurring on the boundary of a \std{3} SPT~\cite{PhysRevX.3.041016,PhysRevB.90.245122,PhysRevX.5.041013}. 

Since this obstruction vanishes, the possible solutions to the pentagon equation form an $\coho{3}$ torsor. That is, any $F$-symbol solution may be multiplied by a 3-cocycle to obtain another solution which may or may not be distinct.

\subsection{ $\G$-graded MPO algebras satisfying the stronger zipper condition and $\G$-graded unitary fusion categories}

We move on by narrowing our focus to the class of MPOs that are relevant for the description of anomaly-free SET phases in (2+1)D. The properties thus introduced allow for an identification of these MPOs with $\G$-graded UFCs. 
In the process, we recount many results shown in Ref.~\cite{Bultinck2017183} and put them into the context of $\G$-graded MPO algebras, which are relevant for the description of domain walls in SET ordered states. 

From this point on we assume that the stronger \emph{zipper condition} holds for the single block MPOs, and the fusion tensors defined in Eq.~\eqref{set:fusiontensors}
\begin{align}

	\, .
\end{align}
The class of MPOs satisfying the zipper condition is sufficiently general to capture representations of all $\G$-graded fusion categories, via the fixed point construction explained in Section.~\ref{set:stringnetexample}.

The associativity of the product $(\mpo{}{\defect{a}{g}} \mpo{}{\defect{b}{h}}) \mpo{}{\defect{c}{k}}  = \mpo{}{\defect{a}{g}} (\mpo{}{\defect{b}{h}} \mpo{}{\defect{c}{k}})$ implies 
\begin{align}
\sum_{\defect{e}{gh}}
N_{\defect{a}{g} \defect{b}{h}}^{\defect{e}{gh}} N_{\defect{e}{gh} \defect{c}{k}}^{\defect{d}{ghk}}
=
\sum_{\defect{f}{hk}}
N_{\defect{a}{g} \defect{f}{hk}}^{\defect{d}{ghk}} N_{\defect{b}{h} \defect{c}{k}}^{\defect{f}{hk}}
\, .
\end{align}
The addition of the zipper condition further implies the existence of \emph{$F$-symbol} associators~\cite{Bultinck2017183}
\begin{align}

	\, ,
\end{align}
that satisfy the \emph{pentagon equation} 
\begin{align}
	%
\, ,
\\
	\sum_{\nu}\f{pcd}{e}{q \beta \gamma}{r \mu \nu}\f{abr}{e}{p \alpha \nu}{s \sigma \tau}=\sum_{x \kappa \lambda \eta}\f{abc}{q}{p \alpha \beta}{x \kappa \lambda }\f{axd}{e}{q \lambda \gamma}{s \eta \tau}\f{bcd}{s}{x \kappa \eta }{r \mu \sigma}
	\, .
\end{align}
Hence all obstructions vanish for MPO representations satisfying this zipper condition, and we have a consistent fusion theory. 
It was shown in Ref.~\cite{Bultinck2017183} that MPO algebras satisfying the zipper condition correspond closely to fusion categories. 
Hence the structure of $\G$-graded MPO algebras that satisfy the zipper condition correspond closely to $\G$-graded fusion categories. 

The classification of $\G$ SET phases for a given underlying topological order is given by $\G$-graded MPO algebras containing the MPO algebra of the underlying topological order  as $\cat_\g{1}$.
It was shown in Ref.~\cite{edie2017equivalences} that seemingly different $\G$-extensions of a UFC $\cat_\g{1}$ that has nontrivial monoidal auto-equivalences, can actually coincide. This leads to a possible overcounting of $\G$-extensions and we expect a similar result to hold for $\G$-graded MPO algebras. 
 There is a further subtlety in the classification of SET phases, as substantially different MPO algebras may give rise to the same emergent SET. We say that a pair of $\G$-graded MPO algebras are \emph{$\G$-graded Morita equivalent} if they lead to the same emergent SET. A method to calculate the emergent SET from the $\G$-graded MPO algebra is described in Section~\ref{setsection:dubes}. Hence, for the SET phase classification, one should consider $\G$ extensions of all representatives in the Morita equivalence class of the underlying topological order, up to $\G$-graded Morita equivalence. We remark that this is a looser equivalence relation than the graded monoidal equivalence considered in Ref.~\cite{edie2017equivalences}.

\subsubsection{Imposing further restrictions to isolate the $\G$-graded MPO algebras that describe SET orders}

Thus far we have seen the structure of associative fusion rules arise from MPO algebras. 
We proceed to restrict our attention to MPO algebras that satisfy a range of further conditions, derived from structures in a unitary fusion category. 

Firstly, we restrict to irreducible projector MPOs, i.e. those where $\mpo{}{\g{1}}$ cannot be written as a sum of inequivalent nonzero projector MPOs (PMPOs) 
\begin{align}
\mpo{}{\g{1}}\neq \text{PMPO} + \text{PMPO}'
\, .
\end{align}
 This excludes the possibility of cat state projectors that occur for spontaneous symmetry breaking phases. It corresponds to restricting to stable TQFTs, and we note the nonreducible case can be reconstructed by taking a sum of irreducible PMPOs. 

Next we further restrict our attention to MPO representations that satisfy $\mpo{\dagger}{\g{g}} = \mpo{}{\conj{\g g}}$. Equivalently, MPO representations that are unitary within the support subspace of a Hermitian unit MPO. This implies the MPO projector in Eq.~\eqref{set:mpogroupprojector} is also Hermitian. 
For a consistent unitary MPO representation we must have ${\mpo{\dagger}{\defect{a}{g}}=\mpo{}{\defect{\conj{a}}{\conj{\g{g}}}}=\mpo{-}{\defect{a}{g}}}$, that is, the conjugate particle defined by Hermitian conjugation must match that defined by orientation reversal. In this case the left handed tensor is defined by reflecting and complex conjugating the right handed tensor $(\mpotensor{a}{-})^{ij}_{\alpha \beta}=(\mpotensor{a}{ij})^{*}_{\beta \alpha}$. Graphically
\begin{align}
\begin{array}{c}
\includeTikz{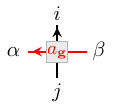}{}
\end{array}
=
\begin{array}{c}
\includeTikz{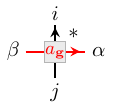}{
	\begin{tikzpicture}[scale=.05]
	\begin{scope}[decoration={markings,mark=at position 0.92 with {\arrow{stealth}}}]
	\draw[draw=red,line width=1pt,postaction=decorate] (-10,-15)-- (10,-15);
	\end{scope}
	\begin{scope}[decoration={markings,mark=at position 0.95 with {\arrow{stealth}}}]
	\draw[draw=black,line width=1pt,postaction=decorate]  (0,-24) -- (0,-6);		
	\node[transform shape,draw=black!29,fill=black!10!white!80,regular polygon,regular polygon sides=4,minimum size=10cm] at (0,-15){};			
	\node[anchor=west] at (2,-9){$*$};
	\node[anchor=west] at (10,-15){$\alpha$};
	\node[anchor=east] at (-9.5,-15){$\beta$};
	\node[anchor=south] at (0,-7){$i$};
	\node[anchor=north] at (0,-23.5){$j$};
	\node at (0,-15.25) {\small $ \red{a_\g{g}}$};
	\end{scope}
	\end{tikzpicture}
}
\end{array}
\, ,
\end{align}
i.e. complex conjugation is equivalent to reflection of the tensor.  This implies
\begin{align}
\label{set:somenidentity}
N_{a b}^{c} = N_{\conj{a} \conj{b}}^{\conj{c}}\, ,
\end{align}
and additionally we have $w_{\bar{a}}=w_a^*$. Moreover, it was shown in Ref.~\cite{Bultinck2017183} that the $w_a$ are real numbers and that $w_a>0$ follows from Eq.~\eqref{set:setmpoblockmult2}, and the fact that $N_{a b}^{c}$ has only nonnegative entries.

We further assume that the single block MPOs have been brought into a canonical form~\cite{MPSrepresentations,Cirac2017100} that admits unitary gauge transformation matrices $Z_a$ and isometric fusion matrices $X_{a b}^{c}$.

It was shown in Ref.~\cite{Bultinck2017183} that, for projector MPOs satisfying our assumptions, each block $a$ has an element $e_a$ that satisfies $N_{a \conj{a}}^{e_a}=1$.  We further assume that there is a unique, single block \emph{unit MPO} $\vac$ satisfying $N_{a \conj{a}}^{\vac}=1$ for all $a\in\cat$. It was shown~\cite{Bultinck2017183} that, with this assumption, $\conj{\vac}=\vac$ and $\fs{\vac}=1$. Furthermore,  $N_{a \conj{b}}^{\vac}=\delta_{ab}$ and 
\begin{align}
\label{set:someotherNidentity}
N_{ab}^{c} = N_{b \conj{c}}^{\conj{a}} = N_{\conj{c} a}^{\conj{b}}
\, ,
&& 
\text{which implies}
&&
 \delta_{a b} = N_{a\vac}^{b} = N_{\vac a}^{b} 
\, ,
\end{align}
and hence $\vac$ plays the role of the trivial element under fusion, as expected. 
We do not expect all irreducible projector MPOs to be unital, as the existence of a unit is posited as an assumption in the theory of fusion categories. 

Given the definition of the unit element, we require that the $F$-symbols have been brought into a gauge where 
\begin{align}
\delta_{a b}^{c} \delta_{\mu \nu}
=
\f{a\vac b}{c}{ a 0 \mu }{b 0 \nu } 
=
\f{ab\vac }{c}{ b \mu 0 }{a 0 \nu} 
=
\f{\vac ab}{c}{ b 0 \mu }{a \nu 0 } 
\, ,
\end{align}
where the first equality corresponds to the \emph{triangle equation} and the latter equalities follow by combining it with the pentagon equation.  
Then it was shown~\cite{Bultinck2017183} that the $F$-symbols can be brought into a compatible gauge satisfying 
\begin{align}
\f{a \conj{a} a}{a}{\vac 00}{\vac 00}
=
\frac{\fs{a}}{d_a}  
\, ,
\end{align}
with $d_a=d_{\conj{a}}>0$, and $\fs{a}$ matching the definition given in Eq.~\eqref{set:fsindeqn}. 

Consistent fusion of the MPOs within a tensor network requires that removing an infinitesimal loop of $\mpo{}{a}$ results in a weight $d_a$. A local sufficient condition for this is 
\begin{align}

\, ,
\end{align}
where $P$ corresponds to the projector onto the support subspace of the PEPS virtual index. 
This implies that $d_a$ matches the definition of quantum dimension via the Perron-Frobenius vector, i.e.
\begin{align}
\label{set:pfdeqn}
d_a d_b = \sum_c N_{a b}^{c} d_c 
\, .
\end{align}
For a consistent MPO representation of $\G$ domain walls within a tensor network, we furthermore require that an infinitesimal loop of domain wall $\G$ can be removed. This corresponds to a local sufficient condition 
\begin{align}
\label{set:agbubblepopping}
	%
\, ,
\end{align}
which implies 
\begin{align}
w_{\defect{a}{g}}=\frac{d_{\defect{a}{g}}}{\qd_\g{g}^2}
\, ,
&&
\text{where}
&&
\qd_{\g{g}}^2:= \sum_{\defect{a}{g}} d_{\defect{a}{g}}^2
\, .
\end{align}
Hence the MPO representation of $\G$ is given by 
\begin{align}
\mpo{}{\g g} = \sum_{\defect{a}{g}} \frac{d_\defect{a}{g}}{\qd_{\g{g}}^2} \, \mpo{}{\defect{a}{g}}
\, ,
\end{align}
in particular, $\mpo{}{\g{1}}$ corresponds to what is known as the $\omega_0$-projector for $\cat_{\g 1}$.  

At this point we renormalize the fusion and splitting tensors
\begin{align}
	\begin{array}{c}
	\includeTikz{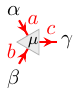}{}
	\end{array}
=
d_a^\frac{1}{4} d_b^\frac{1}{4} d_c^{-\frac{1}{4}} 
[X_{a_\g{g} b_\g{h}}^{c_\g{gh}\,\mu}]_{\alpha \beta}^{\gamma}
\, ,
&&
	\begin{array}{c}
	\includeTikz{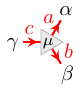}{}
	\end{array}
	=
d_a^\frac{1}{4} d_b^\frac{1}{4} d_c^{-\frac{1}{4}}  
([X_{a_\g{g} b_\g{h}}^{c_\g{gh}\,\mu}]_{\alpha \beta}^{\gamma})^{*}
	\, .
\end{align}
This is to ensure that removing a loop of $\mpo{}{a}\mpo{}{\conj{a}}$ fusing to the vacuum $\mpo{}{\vac}$, matches the removal of an $\mpo{}{a}$ loop within a tensor network
\begin{align}
	%
\, .
\end{align}

We focus on the class of MPOs admitting a set of unitary matrices $\lbend{ab}{c},\rbend{ab}{c}$ that satisfy \emph{pivotal identities} relating the fusion and splitting tensors 
\begin{align}
	%
	\, .
\end{align}
This requires $N_{ ab}^{c}=N_{\conj{a} c}^{b}$, which follows from Eqs.~\eqref{set:somenidentity} and \eqref{set:someotherNidentity}.  
These pivotal identities are related to the existence of a pivotal structure in the theory of fusion categories. 
We believe that the above pivotal identities can be derived for the class of MPOs we are considering, however a proof is beyond the scope of this paper. We remark that such a proof was given in Ref.~\cite{williamson2014matrix} for the SPT case where $|\cat_{\g{1}}|=1$. 

For the fixed point examples in Section~\ref{set:stringnetexample} we found 
\begin{align}
[\lbend{ab}{c}]_{\mu}^{\nu} &= \sqrt{\frac{d_a d_b}{d_c}} (\f{\conj{a} a b}{b}{\vac 0 0}{c \mu \nu})^* \, ,
& 
[\rbend{ab}{c}]_{\mu}^{\nu} &= \sqrt{\frac{ d_a d_b}{d_c}} \f{a b \conj{b}}{a}{c \mu \nu}{\vac 0 0} \, ,
\end{align} 
and we suspect that these formulas still hold for the full class of MPOs considered here.

\subsubsection{Implications of the $\G$-grading}

The existence of a $\G$-grading for the fusion structure implies further restrictions on the algebraic data of an MPO algebra, similar to results shown in Ref.~\cite{barkeshli2014symmetry}. 

Since each element $\defect{a}{g}$ satisfies $N_{a \conj{b}}^{\vac}=N_{a \vac}^{b}=N_{\vac a}^{b}=\delta_{a b}$, for all $\defect{b}{g}$ there must exist some $\defect{c}{1},c'_{\g 1}$ such that $N_{a \conj{b}}^{c}=N_{\conj{c} a}^{b}\neq 0$, and $N_{\conj{b} a}^{c'}=N_{a \conj{c'}}^{b}\neq 0$.  
That is, for any single block  $\mpo{}{ \defect{a}{g} }$, any other single block $\mpo{}{ \defect{b}{h} }$ appears in the product $\mpo{}{ \defect{a}{g}} \mpo{}{ \conj{c'_{\g 1}}}$ and also $ \mpo{}{ \defect{c}{1}} \mpo{}{ \defect{a}{g}}$ for some $\defect{c}{1},c'_{\g 1}$. 
However, we do not generally have $|\cat_{\g g}|=|\cat_{\g h}|$ for $\g g \neq \g h$.

The MPO algebras we consider are faithfully $\G$-graded, hence Eq.~\eqref{set:pfdeqn} implies that $\qd^2_0=\qd^2_{\g g}$ for all $\g g$. The argument for this is identical to that given in Ref.~\cite{barkeshli2014symmetry}. Consequently  we have
\begin{align}
\label{set:setqdrelation}
\qd^2_{\G}=|\G| \qd^2_{{\g 1}}
\, ,
&&
\text{where}
&&
\qd^2_{\G} = \sum_{\g g} \qd^2_{\g g}
\, .
\end{align}
Hence the MPO projector onto the symmetric subspace in Eq.~\eqref{set:mpogroupprojector} corresponds to the $\omega_0$-loop of $\cat_\G$, since 
\begin{align}
\frac{w_a}{|\G|}= \frac{d_a}{\qd^2_\G}
\, .
\end{align}

The $\omega_0$-MPO loop of $\cat_\g{1}$ satisfies an equation, generalizing Eq.~\eqref{set:ICIomegazerosliding}, that allows us to move a $\g g$-MPO through an $\mpo{}{\g{1}}$ loop by transforming it to an $\mpo{}{\g g}$ loop. That is, 
\begin{align}

\, ,
\end{align}
where we have used the zipper condition, the pivotal identities, and properties of the FS indicator. 
A similar equation holds for a $\g 1$ MPO of arbitrary length and with arbitrary orientations of the black indices. By applying orientation reversing gauge transformations, $Z_g$, on the open red indices and noting that reversal of the orientation of the $\g 1 $ MPO has no effect, $\mpo{-}{\g 1}=\mpo{}{\g 1}$, the equation can also bee seen to hold for all orientations of the red indices. Furthermore, the same argument applies for pulling through an open MPO from any contiguous subregion of a circle to its complement.

\subsection{SET tensor network states and the symmetry-enriched pulling through equation}
\label{set:sepullthrupepstensors}

The $\G$-graded MPO algebras that were just introduced yield representations of domain walls in SET tensor network states. We proceed to introduce a condition on the local tensors of a PEPS that ensures it is symmetric under an on-site representation of $\G$, and furthermore that the domain walls are described by a $\G$-graded MPO algebra.

The local tensors of a two dimensional PEPS on a directed trivalent lattice, dual to a triangulation with branching structure, are specified by 
\begin{align}
(\peps_+)^{ijk}_{p}
=

\, .
\end{align}
We assume that this PEPS is MPO-injective and hence describes a state with some underlying topological order. The MPO w.r.t. which the PEPS is MPO-injective corresponds to $\mpo{}{\g{1}}$ in the MPO representation of $\G$. 

The PEPS is symmetric under an on-site representation of $\G$ if it satisfies the \emph{symmetry-enriched pulling though equation}~\cite{williamson2014matrix}   
\begin{align}
\label{set:setpullthrough}
	%
	\, . 
\end{align}
Where we have used the following graphical notation for the on-site group action
\begin{align}
	\raisebox{-.1cm}{$
	%
	:=
\sum_{q}	[U_\g{g}]_{p}^{q} \, (\peps_+)^{ijk}_{q}
\, .
\end{align}
By the properties of the MPOs described above, Eq.~\eqref{set:setpullthrough} implies that all the similar pulling through equations, involving $\mpo{}{\g g}$ acting on a different choice of indices, also hold. 
While the tensors considered above can be used to build a PEPS on any triangulation with branching structure. Similar considerations apply for regular lattices, such as the square lattice, or more generally for locally planar directed graphs. 

Due to Eqs.~\eqref{set:agbubblepopping} and \eqref{set:sepullthru}, the symmetry-enriched pulling though equation is equivalent to the following group intertwiner property
\begin{align}

\, ,
\end{align}
which we expect to be more practical for numerical purposes.

\subsubsection{$\omega_0$-loop fixed-point tensors} 

The $\omega_0$-loop of $\cat_{\g{1}}$, $\mpo{}{\g 1}$, defines an MPO-injective PEPS tensor~\cite{Bultinck2017183} which we refer to as an MPO fixed-point PEPS tensor 
\begin{align}
\label{set:mpofixptpeps}
	%
\, .
\end{align}
The physical degree of freedom is identified with a subspace of the three MPO indices that enter the circle, given by the support of $\mpo{L=3}{\g 1}$. 
The on-site group action is given by $U_g=\mpo{L=3}{\g{g}}$, which is unitary on the support subspace of $\mpo{L=3}{\g{1}}$, corresponding to the local physical degree of freedom. The application of the group action to the PEPS tensors results in 
\begin{align}
	%
\, .
\end{align}
Hence it follows from Eq.~\eqref{set:sepullthru} that the MPO fixed-point PEPS tensors in Eq.~\eqref{set:mpofixptpeps} satisfy the symmetry-enriched pulling through equation~\eqref{set:setpullthrough}. 

We remark that this MPO fixed-point construction yields the same PEPS tensor for all $\G$-extensions of an underlying MPO algebra $\cat_{\g 1}$. This same tensor represents different SET phases, depending upon the choice of the on-site symmetry action. A similar phenomenon has previously been noted for the restricted case of SPT phases~\cite{czxmodel}.

\subsection{Abstracting the calculus of $\G$-graded MPO algebras to diagrams}

At this point we have developed the theory of $\G$-graded MPO algebras, and seen that they represent $\G$ domain walls of symmetric tensor networks. 
The characteristics of the domain wall MPOs within a tensor network are topological and do not have a strong dependence on local details, such as the number of sites within a certain region. 
In fact, the calculus of these MPOs within a tensor network can be abstracted further by not explicitly keeping track of the underlying lattice. This was described in detail in Eqs.~\eqref{set:abstractingtopictures1} and \eqref{set:abstractingtopictures2}. 

To carrying out this abstraction, we first identify the tensor network ground state with the vacuum. This corresponds to the empty diagram
\begin{align}

\, .
\end{align}
A loop on top of this background represents an MPO
\begin{align}
%
\, ,
\end{align}
and the application of the on-site symmetry to all sites within a region is depicted as
\begin{align}
%
\, .
\end{align}
When passing to the abstract diagrammatic calculus, we switched convention for the colour of MPO loops from red to black. 

All of the properties satisfied by the $\G$-graded MPO algebras that were discussed throughout this section can be carried over to the abstract diagrammatic calculus. 
This realizes all consistent fusion diagrams, modulo local relations from the $\G$-graded unitary fusion category, and planar isotopy. In particular, restricting to the $\g 1$-sector recovers the diagrammatic calculus~\cite{JOYAL199155,JOYAL199320,kitaev2006anyons} of the unitary fusion category $\cat_{\g{1}}$.  
The full $\cat_\G$ leads to a more general diagrammatic calculus, related to a 2-category, that was recently introduced in Ref.~\cite{barkeshli2016reflection}.

In this section we have adapted the theory of MPO algebras, presented in Ref.~\cite{Bultinck2017183}, to include a $\G$-grading. These MPOs form a representation of the domain walls of an SET ordered phase. Classifying the emergent SET order that arises from $\G$-graded MPO algebras corresponds to grouping them into graded Morita equivalence classes. Two $\G$-graded MPO algebras are equivalent if they give rise to the same symmetry-enriched $\G$-graded double, which is described by a UGxBFC. In the next section we will describe how to construct the symmetry-enriched double, and extract the gauge invariant physical data of the emergent SET. 

This contains the classification of the underlying emergent topological order as a subproblem. This emergent topological order is given by the Drinfeld double (or center) of the fusion category corresponding to the MPO algebra and hence is nonchiral. Moreover, this construction can realize all nonchiral topological orders~\cite{Davydov2013,davydov2013witt}. 
The classification is then given by collecting MPO algebras into Morita equivalence classes that lead to the same topological order~\cite{muger2003subfactors,DrinfeldCenter}. 

It was shown in Ref.~\cite{etingof2009fusion}, and later used in a condensed matter context by Refs.~\cite{cheng2016exactly,PhysRevB.94.235136}, that the symmetry-enriched double construction can realize all non-anomalous symmetry actions on nonchiral emergent anyons in an on-site manner. In the next section we will go further to construct the full SET on the lattice, including the theory of emergent symmetry defects.

  \section{The defect tube algebra and emergent symmetry-enriched topological order}
  \label{setsection:dubes}

In this section we generalize Ocneanu's tube algebra to an algebra of topological symmetries acting on each nontrivial monodromy $\g{g}$-defect, appearing at the termination point of a $\g{g}$-domain wall. 
We construct tensor network representations of the defect superselection sectors by block diagonalizing each $\g{g}$-sector of this defect tube algebra. 
Furthermore, we find tensor network representations of the projective group action upon each defect superselection sector. 
From these representations we extract physical data of the emergent theory, including the $\G$-crossed modular $S$ and $T$ matrices, the permutation action of the symmetry upon the defect sectors, the cohomology class of the projective representation carried by each defect sector, and the topological entanglement entropies of the defect sectors. 
We go on to describe a construction of the full $\G$-crossed modular U$\G$xBFC of the emergent SET order from the defect ICIs.  
In particular, this defect tube algebra construction can be applied to the symmetry-enriched string-net examples of Section~\ref{set:stringnetexample}.

\subsection{Definition of the dube algebra} 
  
Similar to the conventional tube algebra, the defect tube (\emph{dube}) algebra is found by considering the action of MPOs in the $\g{1}$-sector upon a puncture at the end of a $\g{g}$-domain wall. 
For this, we look at  a symmetry-twisted cylinder, which is equivalent to a domain wall on a sphere with a puncture at each of its end points.  
Again, for simplicity of presentation, we restrict our attention to the case of no fusion multiplicity. The results we find directly generalize to the case with nontrivial fusion multiplicity. 
  
By a slight generalisation of Eq.~\eqref{set:puncturedsphere} the $\g{g}$-twisted cylinder is spanned by tensor networks that have been closed with an MPO from the $\g{g}$-sector 
\begin{align}
\sum_{\defect{r}{g}\, \mu} c_\defect{r}{g}

\, .
\end{align}
We can repeat the analysis that lead to Eq.~\eqref{set:cylindertubeaction}  by considering the purely virtual MPO symmetries that act on the punctures, these come from the $\g{1}$-sector of the MPO algebra
  \begin{align}
  \label{set:Dubealgebrapunctureaction}
%
\, .
\end{align}
Elements of the dube algebra are defined in the same way as for the tube algebra, except they now split into sectors according to the label of the horizontal MPO  
  \begin{align}
\tube{\defect{p}{g} \defect{q}{g} \defect{r}{g}}{\defect{s}{1}} := 
d_p^{\frac{1}{4}} d_r^{-\frac{1}{4}}
%
  \, ,
\end{align}
corresponding to the minimal tensor network
\begin{align}
  d_p^{\frac{1}{4}} d_r^{-\frac{1}{4}}
	%
  \, .
  \end{align}
We often refer to elements of the dube algebra simply as dubes.
We remark that there is a gauge freedom in the definition of the dube elements, similar to a 1-cochain. 

 Hence Eq.~\eqref{set:Dubealgebrapunctureaction} can be rewritten as  
    \begin{align}
  \label{set:SETcylindertubeaction}
  \openone_{\g{g}}
  :=\sum_{\defect{r}{g}} \tube{\defect{r}{g} \defect{r}{g} \defect{r}{g}}{\vac} 
  = \frac{1}{\qd_\g{1}^2} \sum_{\substack{\defect{p}{g} \defect{q}{g} \\ \defect{r}{g} \defect{s}{1}}} (\tube{\defect{p}{g} \defect{q}{g} \defect{r}{g}}{\defect{s}{1}})^\dagger \tube{\defect{p}{g} \defect{q}{g} \defect{r}{g}}{\defect{s}{1}}
  \, ,
  \end{align}
where  $\openone_{\g{g}}$ is the identity on the virtual level of the $\g{g}$-twisted cylinder tensor network. 
  
 The definition of multiplication is given by stacking dubes, and is identical to the one given in Eq.~\eqref{set:tubemult}. Similarly, the definition of Hermitian conjugation is given in Eq.~\eqref{set:tubeconj}. 
 We remark that the $\G$-grading of the MPOs ensures that the dube algebra breaks into $|\G|$ orthogonal sectors, each of which is a $C^*$ algebra. 
 In particular, the $\g{1}$-sector of the dube algebra recovers the tube algebra  of $\cat_{\g{1}}$.

 \subsection{Block diagonalizing sectors of the dube algebra with ICIs}

Since the dube algebra breaks up into $|\G|$ decoupled $C^*$ algebras, we can separately block diagonalize each of these sectors. 
There is an exact formula for the ICIs and domain walls of a dube algebra given a $\mathcal{G}$-graded input category with a $\mathcal{G}$-crossed modular braiding, which generalizes the solution for the ICIs of a tube algebra given an input cateogry with a modular braiding, see the appendix of Ref.~\cite{Ren2024}. 
More generally, it is neccessary to calculate the block diagonalization of each $\g{g}$-sector of the dube algbera. 
 
 The sector generated by tubes of the form $\tube{\defect{p}{g} \defect{q}{g} \defect{r}{g}}{\defect{s}{1}}$ is referred to as the $\g{g}$-dube algebra. It has dimension 
\begin{align}
D_\g{g} = \sum_{\defect{p}{g} \defect{q}{g} \defect{r}{g} \defect{s}{1}} \delta_{\defect{s}{1} \defect{q}{g} }^\defect{p}{g} \delta_{\defect{q}{g} \defect{s}{1}}^{\defect{r}{g}} = \sum_{\ici{\defect{a}{g}}} D_{{\defect{a}{g}}} \times D_{{\defect{a}{g}}}
\, ,
\end{align}
 where the sum is over the irreducible blocks $\ici{\defect{a}{g}}$ of the $\g{g}$-dube algebra, and $D_{\defect{a}{g}}$ is the dimension of the  $\ici{\defect{a}{g}}$ block (consisting of $D_{\defect{a}{g}}\times D_{\defect{a}{g}}$-matrices). 
 
The ICIs of each $\g{g}$-sector are given by Hermitian projectors onto each irreducible $\ici{\defect{a}{g}}$ block.  
The ICIs of the $\g{g}$-sector determine the inequivalent superselection sectors for a disc filling a puncture at the termination point of a $\g{g}$-domain wall. 
The ICIs can be found constructively by following the approach in Ref.~\cite{Bultinck2017183}. 
Expressions for the defect ICIs of the $\G$-graded dube algebra for $\cat_{\G}$ can be derived from the ICIs of the tube algebra for $\cat_\G$, this is discussed further in Section~\ref{setsection:condensation}. 

The $\g g$-defect ICIs can be written as a linear combination of dubes from the $\g g$-sector 
 \begin{align}
\ici{\defect{a}{g}} &= \frac{1}{\qd_\g{1}^2}  \sum_{\substack{\defect{p}{g} \defect{q}{g} \\ \defect{r}{g} \defect{s}{1}}} t^{pqrs}_{\defect{a}{g}} \tube{\defect{p}{g} \defect{q}{g} \defect{r}{g}}{\defect{s}{1}} \, ,
\\
	\begin{array}{c}
	\includeTikz{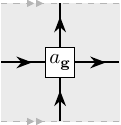}{
	\begin{tikzpicture}[scale=.1]
	\def\dx{-1};
	\def\ddx{1.5};
	\def\dy{.5};
	\def\p{5.5};
	\def\l{2.5};
	\filldraw[UFCBackground](0,0)--(20,0)--(20,20)--(0,20)--(0,0);
	\draw[anyon](10+\l,10)--(20,10) ;
	\draw[anyon](0,10)--(10-\l,10) ;
	\draw[anyon](10,0)--(10,10-\l) ;
	\draw[anyon](10,10+\l)--(10,20) ;
	\draw[draw=black!29,dashed](0,0)--(20,0) (20,20)--(0,20);
	\filldraw[draw=black!29,fill=black!29](\p,0)--(\p+\dx,-\dy)--(\p+\dx,\dy)--(\p,0);
	\filldraw[draw=black!29,fill=black!29](\p,20)--(\p+\dx,20-\dy)--(\p+\dx,20+\dy)--(\p,20);
	\filldraw[draw=black!29,fill=black!29](\p+\ddx,0)--(\p+\dx+\ddx,-\dy)--(\p+\dx+\ddx,\dy)--(\p+\ddx,0);
	\filldraw[draw=black!29,fill=black!29](\p+\ddx,20)--(\p+\dx+\ddx,20-\dy)--(\p+\dx+\ddx,20+\dy)--(\p+\ddx,20);
	\filldraw[draw=black,fill=white] (10-\l,10-\l)--(10-\l,10+\l)--(10+\l,10+\l)--(10+\l,10-\l)--cycle;
	\node at (10,10){$\defect{a}{g}$};
	\end{tikzpicture}
	}
  \end{array}
&=
\frac{1}{\qd_\g{1}^2} \sum_{\substack{\defect{p}{g} \defect{q}{g} \\ \defect{r}{g} \defect{s}{1}}} t^{pqrs}_{\defect{a}{g}}
\begin{array}{c}
\includeTikz{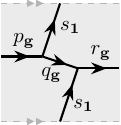}{}
\end{array}
\, ,
\end{align}
where the coefficients satisfy   $ t^{pqrs}_{\defect{a}{g}} =  \delta_{pr} t^{pqps}_{\defect{a}{g}} $, as only dubes $\tube{\defect{p}{g} \defect{q}{g} \defect{r}{g}}{\defect{s}{1}}$ with $\defect{p}{g}=\defect{r}{g}$ can appear in the ICIs.

The $\g{1}$-sector of the dube algebra recovers the tube algebra of the underlying fusion category $\cat_{\g{1}}$. Hence the emergent topological order is described by $\double{\cat_\g{1}}$, see Section~\ref{set:toptubealgsection}. 
The full set of $\g{g}$-sector ICIs lead to a description of the emergent SET order, which is denoted $\rdouble{\cat_\G}{\cat_\g{1}}$. The specifics of this construction, in particular how to extract the physical characteristics of the emergent theory, are described below.

In the mathematics literature $\rdouble{\cat_\G}{\cat_\g{1}}$ is known as a relative center.  It was shown in Ref.~\cite{gelaki2009centers} that the relative center $\rdouble{\cat_\G}{\cat_\g{1}}$ has a canonical $\G$-crossed braided structure. Our construction realizes this U$\G$xBFC, henceforth referred to as $\rdouble{\cat_\G}{\cat_\g{1}}$, on the lattice as a tensor network. 
Similar to the ungraded case, the crossing tensors of the defect ICIs $\defect{a}{g}$ can be directly interpreted as the simple objects of $\rdouble{\cat_\G}{\cat_\g{1}}$. 
We also expect the crossing tensors of these defect ICIs $\defect{a}{g}$ to yield a construction of the defect string operators in the symmetry-enriched Levin-Wen models~\cite{PhysRevB.94.235136,cheng2016exactly} by using them to resolve an external string into the lattice.

More generally, one can define the relative double $\rdouble{A}{\cat}$ for an arbitrary $\cat$-$\cat$-bimodule category $A$. There is an obvious extension of the tube algebra to this case, although it is unclear whether there is any notion of a modified braided structure on $\rdouble{A}{\cat}$ when $A$ is not invertible. In Ref.~\cite{Kitaev2012} a related extension of the tube algebra for a pair of $\cat$-modules was used to find boundary excitations.

For irreducible blocks of dimension $D_{\defect{a}{g}}>1$ the $\ici{\defect{a}{g}}$ ICI further decomposes into a sum of $D_{\defect{a}{g}}$ orthogonal irreducible idempotents $(\ici{\defect{a}{g}})_{ii}$. 
Furthermore, there are $D_{\defect{a}{g}}(D_{\defect{a}{g}} -1)$ nilpotent off diagonal elements $(\ici{\defect{a}{g}})_{ij}$ for $i\neq j$ 
\begin{align}
(\ici{\defect{a}{g}})_{ij} &= \frac{1}{\qd_\g{1}^2}  \sum_{\substack{\defect{p}{g} \defect{q}{g} \\ \defect{r}{g} \defect{s}{1}}}  t^{pqrs}_{(\defect{a}{g})_{ij}} \tube{\defect{p}{g} \defect{q}{g} \defect{r}{g}}{\defect{s}{1}} \, . 
\end{align}
These elements span the $D_{\defect{a}{g}}\times D_{\defect{a}{g}}$-matrix algebra of the $\defect{a}{g}$-block, hence the dubes in the $\g{g}$-sector can be expanded in this basis
\begin{align}
\tube{\defect{p}{g} \defect{q}{g} \defect{r}{g}}{\defect{s}{1}}
&= \frac{1}{\qd_\g{1}^2}  \sum_{(\ici{\defect{a}{g}})_{ij} }  \conj{t}_{pqrs}^{(\defect{a}{g})_{ij}}  
(\ici{\defect{a}{g}})_{ij} 
\, .
\end{align}
The columns of this change-of-basis matrices are orthogonal   
  \begin{align}
\label{set:SETtubenorm}
\frac{1}{\qd_{\g 1}^2} \sum_{\defect{p}{g} \defect{q}{g} \defect{r}{g} \defect{s}{1}}  \conj{t}^{(\defect{a}{g})_{ij}}_{pqrs} (\conj{t}^{(\defect{b}{h})_{kl}}_{pqrs})^* = \delta_{a b} \delta_{i k} \delta_{j l }  \frac{1}{\norm{(\defect{a}{g})_{ij}}^2}
\, ,
\end{align}
for some positive weights $\norm{(\defect{a}{g})_{ij}}$.

It follows from Eqs.~\eqref{set:SETcylindertubeaction} and \eqref{set:SETtubenorm} that 
\begin{align}
\label{set:SETcylinderiiaction}
\openone_\g{g} 
= \sum_{(\defect{a}{g})_{ij}}  \frac{1}{\norm{(\defect{a}{g})_{ij}}^2} (\ici{\defect{a}{g}})_{ij}^\dagger (\ici{\defect{a}{g}})_{ij}  
\, ,
\end{align}
where $\openone_\g{g} $ denotes the identity on the virtual level of the $\g g$-twisted cylinder. 
Multiplying by $(\ici{\defect{a}{g}})_{ii}$ implies
\begin{align}
\label{set:SETicipartition}
\sum_j \frac{1}{\norm{(\defect{a}{g})_{ij}}^2} =  1
\, ,
&&
\text{and hence}
&&
\openone_\g{g}
= \sum_{\defect{a}{g}} \ici{\defect{a}{g}}
\, ,
\end{align}
which is a partition of the identity on the virtual level of the $\g g$-twisted cylinder by the $\defect{a}{g}$ ICIs.

\subsection{Projective group actions on the defect ICIs}
  
Beyond the topological symmetries acting on each $\g{g}$-defect, there is also an action of each element of the physical symmetry group. 
This can be found by considering the action of the global symmetry on a $\g{g}$-twisted cylinder tensor network  
  \begin{align}

\, ,
\label{set:groupactionongtwistedcylinder}
\end{align}
where the darker shaded region denotes the application of the on-site physical action of $\g{h}\in\G$. 
  
The action of $\g{h}$ on a $\g{g}$-defect involves tube elements of the form 
\begin{align}
\tube{\defect{p}{$\act{h}$g} \defect{q}{g\conj{\g{h}}} \defect{r}{g}}{\defect{s}{h}} := 
d_p^{\frac{1}{4}} d_r^{-\frac{1}{4}}
%
  \, . 
\end{align}
These tubes match the set of elements of the tube algebra for $\cat_\G$, however they are interpreted differently in light of the $\G$-graded structure.

Hence Eq.~\eqref{set:groupactionongtwistedcylinder} can be written as 
  \begin{align}
  \label{set:SETcylindertubehgroupaction}
 \globalu{h}[ \openone_{\g{g}}]
  = \frac{1}{\qd_\g{1}^2} \sum_{\substack{\defect{p}{$\act{h}$g} \defect{q}{g\conj{\g{h}}} \\ \defect{r}{g}  \defect{s}{h}}} (\tube{\defect{p}{$\act{h}$g} \defect{q}{g\conj{\g{h}}} \defect{r}{g}}{\defect{s}{h}} )^\dagger \tube{\defect{p}{$\act{h}$g} \defect{q}{g\conj{\g{h}}} \defect{r}{g}}{\defect{s}{h}} 
  \, ,
  \end{align}
where $\globalu{h}$ denotes the action of the global symmetry $\g{h}\in\G$ to the physical level of the tensor network.

The action of an $\g h$ domain wall on an ICI $\ici{\defect{a}{g}}$ is defined by projecting the collection of tubes with $s\in\cat_{\g{h}}$ onto the ICI
\begin{align}
\label{set:hdwdefnonicig}
\dw{h}{\defect{a}{g}}
:=
K \left( \sum_{\substack{\defect{p}{$\act{h}$g} \defect{q}{g\conj{\g{h}}} \\ \defect{r}{g}  \defect{s}{h}}}  c_{\defect{a}{g}}^{pqrs} \,  \tube{\defect{p}{$\act{h}$g} \defect{q}{g\conj{\g{h}}} \defect{r}{g}}{\defect{s}{h}}  \right) \,  \ici{\defect{a}{g}}
\, ,
\end{align} 
for arbitrary constants $c_{\defect{a}{g}}^{pqrs}\neq 0$, and some normalization $K\neq0$ that depends on the $c_{\defect{a}{g}}^{pqrs}$.
For simplicity of presentation we ignore some subtleties that may arise in the case of a degenerate block, the analysis presented can still be applied to the degenerate case with some care.  
We remark that the RHS of the above equation may be $0$ for certain fine-tuned choices of $  c_{\defect{a}{g}}^{pqrs}$, we intend it to hold for generic choices, i.e. after adding a random perturbation to any given fine-tuned values. 

The normalization is fixed, up to a multiplicative phase factor, by the unitarity requirement 
\begin{align}
\label{set:aghdwnorm}
(\dw{h}{\defect{a}{g}})^\dagger \dw{h}{\defect{a}{g}} = \ici{\defect{a}{g}}
\, .
\end{align}
The domain walls are only defined up to a multiplicative 1-cochain $\varepsilon_{\defect{a}{g}}^{\g{h}}\in\Uone$. We always work with a choice of 1-cochain such that $\dw{1}{\defect{a}{g}}=\ici{\defect{a}{g}}$ and  $\dw{\g{g}}{\vac}=\mpo{}{\g{g}}$.

The domain walls on each sector can be expanded in the tube basis 
 \begin{align}
\dw{h}{\defect{a}{g}} &= \frac{1}{\qd_\g{1}^2}  
\sum_{\substack{\defect{p}{$\act{h}$g} \defect{q}{g\conj{\g{h}}} \\ \defect{r}{g} \defect{s}{h}}} 
t^{pqrs}_{\dw{h}{\defect{a}{g}}} 
  \tube{\defect{p}{$\act{h}$g} \defect{q}{g\conj{\g{h}}} \defect{r}{g}}{\defect{s}{h}} 
 \, ,
\\
	\begin{array}{c}
	\includeTikz{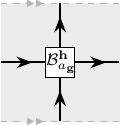}{
	\begin{tikzpicture}[scale=.1]
	\def\dx{-1};
	\def\ddx{1.5};
	\def\dy{.5};
	\def\p{5.5};
	\def\l{2.5};
	\filldraw[UFCBackground](0,0)--(20,0)--(20,20)--(0,20)--(0,0);
	\draw[anyon](10+\l,10)--(20,10) ;
	\draw[anyon](0,10)--(10-\l,10) ;
	\draw[anyon](10,0)--(10,10-\l) ;
	\draw[anyon](10,10+\l)--(10,20) ;
	\draw[draw=black!29,dashed](0,0)--(20,0) (20,20)--(0,20);
	\filldraw[draw=black!29,fill=black!29](\p,0)--(\p+\dx,-\dy)--(\p+\dx,\dy)--(\p,0);
	\filldraw[draw=black!29,fill=black!29](\p,20)--(\p+\dx,20-\dy)--(\p+\dx,20+\dy)--(\p,20);
	\filldraw[draw=black!29,fill=black!29](\p+\ddx,0)--(\p+\dx+\ddx,-\dy)--(\p+\dx+\ddx,\dy)--(\p+\ddx,0);
	\filldraw[draw=black!29,fill=black!29](\p+\ddx,20)--(\p+\dx+\ddx,20-\dy)--(\p+\dx+\ddx,20+\dy)--(\p+\ddx,20);
	\filldraw[draw=black,fill=white] (10-\l,10-\l)--(10-\l,10+\l)--(10+\l,10+\l)--(10+\l,10-\l)--cycle;
	\node at (10.2,10){\footnotesize $\dw{h}{\defect{a}{g}}$};
	\end{tikzpicture}
	}
  \end{array}
&=
\frac{1}{\qd_\g{1}^2} \sum_{\substack{\defect{p}{$\act{h}$g} \defect{q}{g\conj{\g{h}}} \\ \defect{r}{g} \defect{s}{h}}} t^{pqrs}_{\dw{h}{\defect{a}{g}}}
\begin{array}{c}
\includeTikz{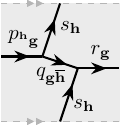}{}
\end{array}
\, .
\end{align}
Due to Eqs.~\eqref{set:SETicipartition}, and~\eqref{set:aghdwnorm} we have
\begin{align}
\label{set:dwidpartition}
\globalu{h}[\openone_\g{g} ] = \sum_{\defect{a}{g}} (\dw{h}{\defect{a}{g}})^\dagger \dw{h}{\defect{a}{g}} 
\, ,
\end{align}
which corresponds to the intertwining of a physical $\g{h}$ action to the virtual level by the $\g{g}$-twisted cylinder tensor network.

\subsubsection{Permutation action on the defects}

The action of the domain wall on the set of defects may result in a nontrivial permutation $\rho_h(\defect{a}{g})=\act{h}(\ici{\defect{a}{g}})=\ici{\defect{(\act{h}a)}{$\act{h}$g}}$.
Which can be found as follows 
\begin{align}
\label{set:dwgroupactiondefinition}
\act{h}(\ici{\defect{a}{g}}) := \dw{h}{\defect{a}{g}} (\dw{h}{\defect{a}{g}})^\dagger  
\, ,
&&
\text{or equivalently}
&&
\dw{h}{\defect{a}{g}} \,  \ici{\defect{a}{g}} =\act{h}(\ici{\defect{a}{g}}) \, \dw{h}{\defect{a}{g}} 
\, .
\end{align}
To reiterate, the permuted defect satisfies $\act{h}(\ici{\defect{a}{g}})\in \cat_{\act{h}\g{g}}$, and even in the case $\g{gh}=\g{hg}$ there may still be a nontrivial permutation action of $\g{h}$ upon the defects within sector $\g{g}$, captured by $\ici{\defect{(\act{h} a)}{g}}$. 
In particular, the potentially nontrivial permutation action of the global symmetry $\G$ upon the emergent anyons (or $\g{1}$-defects) can be extracted. 

By definition~\eqref{set:hdwdefnonicig}, we have  $ \dw{h}{\act{k}(\defect{a}{g})} \dw{k}{\defect{a}{g}} = C \, \dw{hk}{\defect{a}{g}}$, for some constant $C$. Hence $\rho$ is a homomorphism from $\G$ to the permutation group on $\cat_\G$
\begin{align}
\act{h}(\act{k}(\ici{\defect{a}{g}}))=\act{hk}(\ici{\defect{a}{g}})
\, .
\end{align}
We remark that the action of the domain walls commutes with the $T$ matrix, see Eqs.~\eqref{set:snegsandt}, and~\eqref{set:gcrossedmodularmatrixdefinition}. Hence the topological spin of all superselection sectors in an orbit of the permutation action must match.

We refer to the set of all group elements that do not permute a given defect $\defect{a}{g}$ as its centralizer
\begin{align}
\centralizer{\defect{a}{g}}
:= 
\{ \g{h}\in\G \, | \, \act{h}\defect{a}{g}=\defect{a}{g}
\}
\, ,
\end{align}
which is a subgroup of $\G$. Note $\g{g}\in\centralizer{\defect{a}{g}}$ as the $T$ matrix, see Eq.~\eqref{set:Tmatrixcommutingwithtubes},  commutes with $\defect{a}{g}$.

The orbit of a defect $\defect{a}{g}$ under the $\G$-action is denoted by
\begin{align}
\orbit{\defect{a}{g}}
:=
\{  \act{h}\defect{a}{g}  \, | \,  \g{h} \in \G 
\}
\, ,
\end{align}
and a set of representatives for each right coset of $\centralizer{\defect{a}{g}}$ is written as $\G / \centralizer{\defect{a}{g}}$, hence
\begin{align}
\orbit{\defect{a}{g}}=
\{  \act{h}\defect{a}{g}  \, | \,  \g{h} \in \G / \centralizer{\defect{a}{g}}
\}
\, .
\end{align}

\subsubsection{2-cocycle of the projective representation on a defect}

We can go beyond the analysis of the previous section to work out precise composition rules of the domain walls. Since 
\begin{align}
\left( \sum_{\defect{a}{g}} (\dw{hk}{\defect{a}{g}})^\dagger \dw{h}{\act{k}(\defect{a}{g})} \dw{k}{\defect{a}{g}} \right) \ici{ \defect{b}{g}}
=
\ici{ \defect{b}{g}} \left( \sum_{\defect{a}{g}} (\dw{hk}{\defect{a}{g}})^\dagger \dw{h}{\act{k}(\defect{a}{g})} \dw{k}{\defect{a}{g}} \right) 
\, ,
\\
\left( \sum_{\defect{a}{g}} (\dw{hk}{\defect{a}{g}})^\dagger \dw{h}{\act{k}(\defect{a}{g})} \dw{k}{\defect{a}{g}} \right) ^\dagger
\left( \sum_{\defect{a}{g}} (\dw{hk}{\defect{a}{g}})^\dagger \dw{h}{\act{k}(\defect{a}{g})} \dw{k}{\defect{a}{g}} \right) 
=\openone_\g{g}
\, ,
\end{align}
by an application of Schur's lemma we have
\begin{align}
\label{set:dwtwococycleeqn}
\dw{h}{\act{k}(\defect{a}{g})} \dw{k}{\defect{a}{g}} = \factsys{a}{\g{h},\g{k}} \dw{hk}{\defect{a}{g}}
\, ,
\end{align}
for some phase $ \factsys{a}{\g h, \g k} \in \Uone$.  Due to the associativity of matrix multiplication, $\eta_a$ must satisfy the twisted 2-cocycle equation
\begin{align}
\label{set:etatwococycleeqn}
\factsys{\act{k}a}{\g f ,\g h}\factsys{a}{\g{fh},\g k}
=
\factsys{a}{\g f, \g{hk}} \factsys{a}{\g h, \g k}
\, .
\end{align}
Since $\dw{k}{\defect{a}{g}} $ is only defined up to a multiplicative 1-chain,  $\factsys{a}{\g h, \g k}$ is only defined up to a 2-coboundary and hence each defect $\defect{a}{g}$ carries a $\rho$-twisted projective representation in cohomology class $[\eta_a]\in\twistedcoho$. 
Furthermore, each defect $\defect{a}{g}$ transforms under a conventional projective representation of its own centralizer $\centralizer{\defect{a}{g}}$ with cohomology class ${[\eta_a]\in\cohom{2}{\centralizer{a}}}$.  
 We assume these cocycles have been normalized such that 
\begin{align}
\factsys{a}{\g h, \g 1} =
\factsys{a}{\g 1, \g h}
=
1
\, .
\end{align}
We remark that a previous work demonstrated how to extract this symmetry fractionalization label for anyons, or $\g{1}$-defects, using tensor networks~\cite{PhysRevB.90.045142}.

All defects in a $\G$-orbit $\act{h}\defect{a}{g} \in \orbit{\defect{a}{g}}$  have isomorphic centralizer groups ${\centralizer{\defect{a}{g}}\cong \centralizer{\act{h}\defect{a}{g}}}$, via the obvious isomorphism $\g{k}\mapsto \act{h}\g{k}$. 
It was shown in Ref.~\cite{barkeshli2014symmetry} that the 2-cocycle of the projective rep of $\centralizer{\act{h}a}$ acting on $\act{h}\defect{a}{g}$ is related to the projective rep of $\centralizer{a}$ acting on $\defect{a}{g}$ by a coboundary
\begin{align}
\eta_{\act{h}a}(\act{h}\g{x},\act{h}\g{y})= d \varepsilon_a^{\g{h}}(\g{x},\g{y})
\,  \eta_a(\g{x},\g{y})
\, ,
&&
\text{where}
&&
\varepsilon_a^{\g{h}}(\g{x}) := \frac{\eta_a(\conj{\g{h}},\act{h}\g{x})}{\eta_a(\g{x},\g{k})}
\, ,
\end{align}
hence $[\eta_{\act{h}a}(\act{h}\g{x},\act{h}\g{y})]=[\eta_a(\g{x},\g{y})]$ give the same element in both $\cohom{2}{\centralizer{a}}$ and $\cohom{2}{\centralizer{\act{h}a}}$. 

Furthermore, this gives a canonical isomorphism between the $\eta_{\act{h} a}$-projective irreps of $\centralizer{\act{h} a}$ and the $\eta_a$-projective irreps of $\centralizer{a}$, for a chosen representative $a\in [a]$, as follows 
\begin{align}
\label{set:isomofprojirreps}
\pirrep{\act{h}a}{\act{h}\mu}{\act{h}k} = \varepsilon_{a}^{\g{h}}(\g{k}) 
\, \pirrep{a}{\mu}{k}
\, .
\end{align}
Where $\pi_a^\mu$ is an irreducible projective representation of $\centralizer{a}$ with cocycle $\eta_a$, labeled by $\mu$.

\subsection{Symmetry-twisted minimally entangled states and $\G$-crossed modular matrices}
\label{set:stmesandsandtmatrices}

The representations of $\g{h}$ on the $\g g$-defects $\dw{h}{\defect{a}{g}}$ yield all the $(\g{g},\g{h})$ symmetry-twisted, minimally entangled states~\cite{barkeshli2014symmetry} on the torus   
 \begin{align}
 \label{set:defdomainwallstate}
\ket{\dw{h}{\defect{a}{g}}}
 = 
 \frac{1}{D_a \qd_\g{1}^2}  \sum_{{\defect{p}{g} \defect{q}{g\conj{\g{h}}} \defect{s}{h}}} t^{pqps}_{\dw{h}{\defect{a}{g}}} 
 \ket{\defect{p}{g} \, \defect{s}{h} \, \defect{q}{g\conj{\g{h}}} }
 \, .
\end{align}
Only tubes $ \tube{\defect{p}{$\act{h}$g} \defect{q}{g\conj{\g{h}}} \defect{r}{g}}{\defect{s}{h}} $ with $p=r$ lead to nonzero states, and hence nonzero symmetry-twisted sectors must satisfy $\g{gh}=\g{hg}$.  
Furthermore, only group elements with nonzero matrix trace $\tr[ \dw{h}{\defect{a}{g}} ]\neq0$ lead to nonzero states. 
By using Eq.~\eqref{set:dwgroupactiondefinition}, and the orthogonality of defect ICIs, this implies that nonzero symmetry-twisted MES must satisfy $\act{h}{\defect{a}{g}}=\defect{a}{g}$. 
Hence the dimension of the $(\g{g},\g{h})$-sector equals the number of $\g{h}$-invariant $\g{g}$-defects.

Similarly, one can use the crossing tensors from the domain wall tensor networks $\dw{h}{\defect{a}{g}}$ --- containing the weights  $t^{pqps}_{\dw{h}{\defect{a}{g}}}$ --- to construct general symmetry-twisted states on arbitrary oriented 2-manifolds, i.e. higher genus tori. 
Here we focus on the torus, as it already allows us to construct the $\G$-crossed $S$ and $T$ matrices that generate a representation of the modular group. 

The action of the global, on-site symmetry upon the symmetry-twisted MES can be calculated by using Eq.~\eqref{set:dwidpartition} 
\begin{align}
\label{set:mpoloopsymtwistedsectorgroupaction}
\globalu{k} \ket{\dw{h}{\defect{a}{g}}} 
= 
\ket{ \dw{k}{\defect{a}{g}} \dw{h}{\defect{a}{g}} (\dw{k}{\defect{a}{g}})^\dagger } 
=
\frac{\factsys{a}{\g k , \g h } \factsys{\act{k}a}{\g{kh} , \conj{\g{k}}}}{\factsys{a}{\conj{\g{k}}, \g{k} }}
\ket{\dw{$\act{k}$h}{\act{k}\defect{a}{g}}} 
\, .
\end{align}
Where, by  Eq.~\eqref{set:dwtwococycleeqn}, and  Eq.~\eqref{set:etatwococycleeqn}, we have 
$\factsys{a}{\conj{\g{k}}, \g{k} } (\dw{k}{\defect{a}{g}})^\dagger= \dw{$\conj{\g{k}}$}{\act{k}\defect{a}{g}}$. 

For calculational purposes it is often convenient to derive the permutation action on the defects $\act{k}\defect{a}{g}$ directly from the MES. This can be done by intertwining the global $\globalu{k}$ action on the physical indices to $\mpo{}{\g{k}}$ on the virtual indices, and fusing it into the 1-skeleton of the torus. 

For $\g{k}\in\centralizer{h}$, by Eq.~\eqref{set:aghdwnorm},  we have 
 \begin{align}
\globalu{k} \ket{\dw{h}{\defect{a}{g}}} 
=
\eta_{\defect{a}{g}}^{\g{h}}(\g k) \ket{\dw{h}{\act{k}\defect{a}{g}}} 
\, ,
&&
\text{where}
&&
\eta_{\defect{a}{g}}^{\g{h}}(\g k) = \frac{\factsys{a}{\g k , \g h }}{\factsys{a}{\g h , \g k }}
 \, ,
 \end{align}
 which is the slant product. 
 This phase is only gauge invariant for $\g{k} \in \centralizer{\dw{h}{\defect{a}{g}}}$, which is defined to be the elements of $\centralizer{\defect{a}{g}}$ that also commute with $\g{h}$. In that case we find a group homomorphism $\eta_{\defect{a}{g}}^{\g{h}}: \centralizer{\dw{h}{\defect{a}{g}}} \rightarrow \Uone$, due to the 2-cocycle Equation~\eqref{set:etatwococycleeqn}.

The $\G$-crossed modular $S$ and $T$ matrices are defined by the same action on tubes as the topological modular matrices, see Eq.~\eqref{set:snegsandt}. They satisfy 
\begin{align}
\label{set:gcrossedmodularmatrixdefinition}
\ket{S(\dw{h}{a_{\g{g}}})} &= \sum_{b_{\g{h}} }S_{a_{\g{g}} b_{\g{h}}}  \, \ket{\dw{{\conj{\text{g}}}}{b_{\g{h}}}}
\, ,    &
    \ket{T (\dw{h}{a_{\g{g}}} ) } &=  \theta_{\dw{h}{a_{\g{g}}}} \ket{\dw{gh}{a_\g{g}} }
\,    ,
\end{align}
where
\begin{align}
   S_{\defect{a}{g} \defect{b}{h}} &=  \frac{D_b \tr[ ( \dw{{\conj{\text{g}}}}{b_{\g{h}}})^\dagger S(\dw{h}{a_{\g{g}}}) ]}{D_a \tr[ ( \dw{{\conj{\text{g}}}}{b_{\g{h}}})^\dagger \dw{{\conj{\text{g}}}}{b_{\g{h}}} ]} \, ,
    &
 \theta_{\dw{h}{a_{\g{g}}}} \delta_{a, b} &= \frac{D_b \tr[ ( \dw{{\conj{\text{g}}}}{b_{\g{h}}})^\dagger T(\dw{h}{a_{\g{g}}}) ]}{D_a \tr[ ( \dw{{\conj{\text{g}}}}{b_{\g{h}}})^\dagger \dw{{\conj{\text{g}}}}{b_{\g{h}}} ]}
    \, ,
\end{align}
and we have used that $\act{g}\defect{a}{g}=\defect{a}{g}$, since the $T$ matrix tube commutes with $\defect{a}{g}$, see Eq.~\eqref{set:Tmatrixcommutingwithtubes}.  The trace used above corresponds to the conventional matrix trace. 

This includes the special case of the topological spin of defect superselection sectors 
\begin{align}
    \ket{T (\defect{a}{g} )} &=  \theta_{\defect{a}{g}} \ket{ \dw{g}{a_\g{g}}}
    \, ,
\end{align}
and due to Eq.~\eqref{set:dwgroupactiondefinition} we have 
\begin{align}
\theta_{\dw{h}{a_{\g{g}}}} = \factsys{a}{\g{g},\g h} \theta_{\defect{a}{g}} 
\, .
\end{align}

The $\G$-crossed modular matrices are broken down into sectors, outside of which they are identically zero. That is, 
\begin{align}
\braket{ \dw{y}{\defect{b}{x}} \, | \,S(\dw{h}{a_{\g{g}}})} =: \delta_{\g{x},\g{h}} \delta_{\g{y},\conj{\g{g}}} S^{(\g{g},\g{h})}_{a_{\g{g}} b_{\g{h}}}
\, ,
&&
\braket{ \dw{y}{\defect{b}{x}} \, | \, T(\dw{h}{a_{\g{g}}})} =: \delta_{\g{x},\g{g}} \delta_{\g{y},{\g{gh}}} \delta_{a,b} T^{(\g{g},\g{h})}_{a_{\g{g}} a_{\g{g}}}
\, .
\end{align}

Since $S$ and $T$ are clearly invertible by definition, the dimension of the $(\g{g},\g{h})$ and $(\g{h},\g{g})$ sectors must match. This implies the number of $\g{h}$-invariant $\g{g}$-defects equals the number of $\g{g}$-invariant $\g{h}$-defects, where $\g{gh}=\g{hg}$.  
In particular, the number of $\g{g}$-defects equals the number of $\g{g}$-invariant anyons ($\g{1}$-defects). 
We also have that the $(\g{g},\g{h})$ and $(\g{g},\g{gh})$ sectors must have the same dimension.

Similar to the case without symmetry, the $S$ and $T$ matrices are unitary. This is because the emergent theory describes a modular non-chiral topological order that has been enriched by a non-anomalous symmetry, and remarkably it has been shown that $\G$-extensions of modular theories are $\G$-crossed modular~\cite{kirillov2004g,etingof2005fusion,barkeshli2014symmetry}. 

Unitarity of the $T$ matrix follows directly from our definition, and we expect that unitarity of $S$ can also be derived in similar fashion to the ungraded tube algebra. However, we will not give such a proof here.  

The $\G$-crossed $S$ and $T$ matrices satisfy modular relations, similar to the ungraded case. For technical reasons~\cite{barkeshli2014symmetry} we first rescale $S$ by a complex phase $\splitact{h}{a,\conj{a};\vac}\in\Uone$ that arises from the group action on the fusion space, see Eq.~\eqref{set:DefectnabcU}, 
\begin{align}
\widetilde{S}_{a_{\g{g}} b_{\g{h}}} := [\splitact{h}{a,\conj{a};\vac}]^* S_{a_{\g{g}} b_{\g{h}}} 
\, .
\end{align}
Since the underlying topological order is a Drinfeld center $\double{\cat_\g{1}}$, the chiral central charge is trivial, and hence we have the modular relations
\begin{align}
(\widetilde{S}T)^3=\widetilde{S}^2
\, ,
&&
\widetilde{S}^4=\openone
\, .
\end{align}
One can also verify these formulas by direct manipulation of MPOs, using the definitions of $S$ and $T$ given in Eq.~\eqref{set:snegsandt}.

We remark that $S_{a_{\g{g}} b_{\g{h}}}$  and $ \theta_{\dw{h}{a_{\g{g}}}}$ are not gauge-invariant, although they do contain gauge invariant data~\cite{barkeshli2014symmetry}. In particular, the $S$ and $T$ matrices of the underlying topological order are given by $S^{(\g{1},\g{1})}$ and $T^{(\g{1},\g{1})}$. 

Furthermore, with the normalizations we have fixed for the $\dw{1}{\defect{a}{g}},\dw{g}{\vac}$ domain walls, we can extract the normalized quantum dimensions of the defects using a similar approach to that which led to Eq.~\eqref{set:derivingqdfromtubes}. The result
\begin{align}
\label{set:derivingdefectqdfromtubes}
d_\defect{a}{g} = \sum_{\defect{p}{g} } \frac{d_\defect{p}{g} {t}_{\defect{a}{g}}^{\defect{p}{g}\defect{p}{g}\defect{p}{g}\vac} }{D_\defect{a}{g}} 
\, ,
\end{align}
follows from 
\begin{align}
\frac{d_\defect{a}{g}}{\qd_\g{1}^2}=S_{\defect{a}{g}\vac} = \frac{ \tr[(\dw{$\conj{\g{g}}$}{\vac})^\dagger S(\ici{\defect{a}{g}})]}{D_a \tr[\, (\dw{$\conj{\g{g}}$}{\vac})^\dagger \dw{$\conj{\g{g}}$}{\vac}\, ]} =  \frac{1}{\qd_\g{1}^2} \sum_{pqs}  {t}_{\defect{a}{g}}^{\defect{p}{g}\defect{q}{g}\defect{p}{g}\defect{s}{1}} \frac{ \tr[(\dw{$\conj{\g{g}}$}{\vac})^\dagger S(\tube{pqp}{s}) ]}{D_a \tr[\, \ici{\vac}\, ]} 
= \sum_{\defect{p}{g} } \frac{d_p {t}_{a}^{ppp\vac} }{\qd_\g{1}^2 D_a} 
 \, ,
\end{align}
where we have used $ S(\tube{\defect{p}{g}\defect{q}{g}\defect{p}{g}}{\defect{s}{1}}) \, \ici{\vac}= d_p \,  \delta_{s \vac} \delta_{p q}  \, \dw{$\conj{\g{g}}$}{\vac}$, which is implied by  Eq.~\eqref{set:DefectICIomegazerosliding}.

Since we have a $\G$-extension of $\double{\cat_\g{1}}$, and the total quantum dimension of the $\g{1}$-defect topological superselection sectors is given by $\qd_\text{out}=\qd_{\g{1}}^2$, the total quantum dimension of all defects is given by $\qd_{\G,\text{out}}=\sqrt{|\G|} \qd_\g{1}^2$.

\subsection{Topological entanglement entropy of the defect superselection sectors}

The topological entanglement entropy (TEE) of the defect superselection sectors can be calculated via the topological correction to the 0-R\'enyi entropy, similarly to the purely topological case considered in Section~\ref{set:teeofanyonsuperselectionsectors}. 
The 0-R\'enyi entropy for the $\defect{a}{g}$ defect superselection sector is calculated by taking the trace of the $\ici{\defect{a}{g}}$ ICI, of varying MPO length $L$, similar to Eq.~\eqref{set:hzerotracedefinition}. 

By using Eq.~\eqref{set:derivingdefectqdfromtubes} we recover a similar result to Eq.~\eqref{set:TEEscalingfromtube}
\begin{align}
\label{set:icidefecttee}
\entropy{0} \approx \log(\lambda_\vac) L - \gamma_{\defect{a}{g}}
\, ,
\end{align}
where
\begin{align}
\gamma_{\defect{a}{g}} = -\log \left( \sum_{\defect{p}{g} } \frac{d_\defect{p}{g} {t}_{\defect{a}{g}}^{\defect{p}{g}\defect{p}{g}\defect{p}{g}\vac}  }{D_\defect{a}{g} \qd_\g{1}^2} \right) = \log(\qd_\text{out}) - \log(d_\defect{a}{g})
\, ,
\end{align}
and $\qd_\text{out}=\qd_\g{1}^2$ is the quantum dimension of the underlying emergent topological order.

We argued in Section~\ref{set:teeofanyonsuperselectionsectors} that the topological correction to the entropy is independent of the R\'enyi index for an RG fixed point model, and that the topological entanglement entropy is a robust quantity throughout a gapped phase of matter. The same arguments apply for the topological correction to the entropy of a defect superselection sector, and hence the quantity  $\gamma_{\defect{a}{g}}$ in  Eq.~\eqref{set:icidefecttee}  does indeed correspond to the TEE of a defect. 
Hence Eq.~\eqref{set:icidefecttee} provides a general method to calculate the TEE of defect superselection sectors directly on the lattice~\cite{PhysRevLett.111.220402}.

  \subsection{$\G$-graded fusion}

  The fusion structure of the defect ICIs proceeds almost identically to the purely topological case, explained in Section~\ref{set:icifusionsection}. We briefly summarize the points of similarity, before moving on to discuss the interplay of fusion with the group action and domain walls, which are aspects that do not occur in the topological case. 
 
We use a similar notation to that established in Section~\ref{set:icifusionsection}
  \begin{align}

  \, ,
  \end{align}
where the group elements on the far right indicate the sector of the strings appearing in the definition of $\ici{\defect{a}{g}}$.

The fusion of defects $\ici{\defect{a}{g}}$ and $\ici{\defect{b}{h}}$ to sector $\ici{\defect{c}{k}}$ is captured by 
  \begin{align}
  \label{set:defectnabctube}
    	%
\, ,
  \end{align}
are fusion vertex tensors that lead to nonzero tensor networks. 
Due to the structure of the defect ICIs,  $c^{\defect{p}{g} \defect{q}{h} \defect{r}{k}}_\mu$ contains the constraint $\delta_{gh,k}$ and hence the fusion is $\G$-graded. 
  
The fusion space $\splittingspace{\defect{a}{g} \defect{b}{h}}{\defect{c}{gh}}$ is spanned by states $\mu_{ab}^{c}$ in the support subspace of the projector described in Eq.~\eqref{set:nabcaction}.  This is found using the same approach as in the purely topological case,  by creating an $\mpo{}{\g{1}}$ loop projector between $\ici{\defect{c}{gh}}$ and $\ici{\defect{a}{g}}, \, \ici{\defect{b}{h}}$ and fusing it into the edges surrounding $\mu_{a b}^{c}$. 

By taking the trace of the projector in Eq.~\eqref{set:nabcaction} we find $N_{\defect{a}{g} \defect{b}{h}}^{\defect{c}{gh}}$, the dimension of its support subspace. 
After using the $\G$-crossed $S$-matrix to expand $\ici{\defect{a}{g}}, \,\ici{\defect{b}{h}}, \,\ici{\defect{c}{gh}}$ in a complimentary basis one can derive the $\G$-crossed Verlinde formula, recently introduced in Ref.~\cite{barkeshli2014symmetry} 
\begin{align}
N_{\defect{a}{g} \defect{b}{h}}^{\defect{c}{gh}} = \sum_{\defect{x}{1}} \frac{S_{\defect{a}{g}\defect{x}{1}} S_{\defect{b}{h}\defect{x}{1}} S^*_{\defect{c}{gh} \defect{x}{1}}}{S_{\vac \defect{x}{1}}} \factsys{x}{\conj{\g{h}},\conj{\g{g}}}
\, .
\end{align}
This holds for any $\G$-crossed modular theory, which includes all the theories in our formalism as they are symmetry-enriched Drinfeld centers.

Further properties of defect fusion that follow from a direct application of the methods described in Section~\ref{set:icifusionsection} are briefly summarized below:  
  \begin{itemize}
  \item
If the ICIs $\ici{\defect{a}{g}}, \,\ici{\defect{b}{h}}, \,\ici{\defect{c}{gh}}$ are all nondegenerate, there is a unique fusion vertex $\mu_{ab}^{c}$ and one can absorb the $\ici{c}$ ICI into $\ici{a},\, \ici{b}$ and $\mu$,  similar to Eq.~\eqref{set:absorbingcintoabmu}. 
  \item  
  The antiparticle $\ici{\conj{\defect{a}{g}}}$ for each $\ici{\defect{a}{g}}$ is found using the same approach as in the topological case. That is, by finding the ICI $\ici{\conj{\defect{a}{g}}}$ that leads to nonzero cup and cap tubes, as defined in Eq.~\eqref{set:cupandcaptubes}. The $\G$-grading of the fusion implies that $\conj{\defect{a}{g}}$ lies in the $\conj{\g{g}}$-sector. 
  \item
  The FS indicator $\fs{\defect{a}{g}}$ can then be calculated by multiplying a cup and cap tube as in Eq.~\eqref{set:fsindtubeeqn}. 
  This is only gauge invariant for $\conj{\defect{a}{g}}=\defect{a}{g}$, which requires $\g{g}^2=1$, and implies $\fs{\defect{a}{g}}=\pm 1$. 
  \item
The $F$-symbols of the emergent theory are calculated using the same approach as in the topological case, see Eq.~\eqref{set:doublefsymboleqn}. 
  \end{itemize}

Similar to Eq.~\eqref{set:dwidpartition} we have 
    \begin{align}
    \label{set:defectICIteleportation}
        \sum_{\defect{s}{h}}  \frac{d_s}{\qd_\g{1}^2}

= (\tdw{h}{\defect{a}{g}})^\dagger \dw{h}{\defect{a}{g}}
\, 
\end{align}

For the special case $\defect{a}{g}=\vac$ we recover a version of Eq.~\eqref{set:sepullthru} for tubes 
    \begin{align}
    \label{set:DefectICIomegazerosliding}
        \sum_{\defect{s}{g}}
	%
  \, ,
  \end{align}
which corresponds to sliding a $\g{g}$ domain wall over an $\mpo{}{\g{1}}$ loop, thus transforming it into an $\mpo{}{\g{g}}$ loop. This is similar to Eq.~\eqref{set:sepullthru}. 

Eq.~\eqref{set:defectICIteleportation} follows from the following slight  generalization of Eq.~\eqref{set:dwidpartition} 
\begin{align}
\sum_{\g{g},\g{h}}   \sum_{\defect{r}{g} \defect{s}{h}} \frac{d_s}{\qd_\g{1}^2}
%
=  \frac{1}{\qd_\g{1}^2} \sum_{\g{g},\g{h}} \sum_{\substack{\defect{p}{$\act{h}$g} \defect{q}{g\conj{\g{h}}} \\ \defect{r}{g} \defect{s}{h}}} (\widetilde{\mathcal{T}}_{\defect{p}{$\act{h}$g} \defect{q}{g\conj{\g{h}}} \defect{r}{g}}^{\defect{s}{h}} )^\dagger  \tube{\defect{p}{$\act{h}$g} \defect{q}{g\conj{\g{h}}} \defect{r}{g}}{\defect{s}{h}} 
=  \sum_{\g{g},\g{h}} (\tdw{h}{\defect{a}{g}})^\dagger \dw{h}{\defect{a}{g}}
\, ,
\end{align}
where
\begin{align}
(\widetilde{\mathcal{T}}_{\defect{p}{$\act{h}$g} \defect{q}{g\conj{\g{h}}} \defect{r}{g}}^{\defect{s}{h}} )^\dagger &=  d_p^{\frac{1}{4}}  d_r^{-\frac{1}{4}}

\, ,
&
 (\tdw{h}{\defect{a}{g}})^\dagger &= \frac{1}{\qd_\g{1}^2} \sum_{\substack{\defect{p}{$\act{h}$g} \defect{q}{g\conj{\g{h}}} \\ \defect{r}{g} \defect{s}{h}}} (t^{pqrs}_{\dw{h}{\defect{a}{g}}})^*\, (\widetilde{\mathcal{T}}_{\defect{p}{$\act{h}$g} \defect{q}{g\conj{\g{h}}} \defect{r}{g}}^{\defect{s}{h}} )^\dagger .
\end{align}
We also have the similar result
    \begin{align}
    \sum_{\defect{s}{h}}  \frac{d_s}{\qd_\g{1}^2}
	%
= (\dw{h}{\defect{a}{g}})^\dagger \, \tdw{h}{\defect{a}{g}} 
\, .
\end{align}

\subsubsection{Group action on the fusion space}

The action of the on-site representation of $\G$ on the fusion pair-of-pants tensor network can be calculated as follows
  \begin{align}
  \label{set:Uactiondefectnabctube}
    	%
  \, .
\end{align}
Where the darker shaded region indicates the on-site action of $\g{k}$ on the fusion tensor network, and 
\begin{align}
\label{set:DefectnabcU}
    	%
  \, ,
  \end{align}
  defines the group action $\splitact{k}{a,b;c}$ on the fusion space.

We denote the pair-of-pants tensor network in Eq.~\eqref{set:defectnabctube} by $\fusiontube{\defect{a}{g}\defect{b}{h}}{\defect{c}{gh}}$, hence 
\begin{align}
\globalu{k}[\fusiontube{\defect{a}{g}\defect{b}{h}}{\defect{c}{gh}}] = (\dw{k}{c})^\dagger \, \fusiontube{\act{k}\defect{a}{g} \act{k}\defect{b}{h}}{\act{k}\defect{c}{gh}} \, \dw{k}{a} \otimes  \splitact{k}{a,b;c}  \otimes \dw{k}{b}
\, .
\end{align}
Since $\globalu{k}$ forms a representation and $\dw{k}{a}$ forms a projective representation we find that $\splitact{k}{a,b;c}$ forms a projective representation 
\begin{align}
 \splitact{x}{\act{y}a,\act{y}b;\act{y}c} \splitact{y}{a,b;c} = \ufactsys{\g{x},\g{y}}{a,b;c}   \splitact{xy}{a,b;c} 
 \, .
\end{align}
Where
\begin{align}
\ufactsys{\g{x},\g{y}}{a,b;c} = \frac{\factsys{c}{\g{x},\g{y}}}{\factsys{a}{\g{x},\g{y}}\factsys{b}{\g{x},\g{y}}}
\, ,
\end{align}
is a $\Uone$ phase that satisfies the twisted 2-cocycle equation
\begin{align}
\ufactsys{\g{x},\g{yz}}{a,b;c} \, \ufactsys{\g{y},\g{z}}{a,b;c}  = \ufactsys{\g{x},\g{y}}{\act{z}a,\act{z}b;\act{z}c} \, \ufactsys{\g{xy},\g{z}}{a,b;c}
\, .
\end{align} 
For $\g{x},\, \g{y} \in \centralizer{a,b,c}$ we find $[\ufactsys{}{a,b;c}]\in \cohom{2}{\centralizer{a,b,c}}$ as it is only defined up to a coboundary, corresponding to rephasing $\splitact{k}{a,b;c}$.

Eq.~\eqref{set:Uactiondefectnabctube} included an action of the symmetry within the $\ici{a},\,\ici{b},$ and $\ici{c}$ superselection sectors. 
To this end we consider closing the pair-of-pants tensor network via defect tensors $\peps_a^-,\, \peps_b^-, \, \peps_c$, where the $-$ indicates that the boundary of the puncture is negatively oriented. The defect tensors  must transform under the on-site physical symmetry  as follows 
\begin{align}
\globalu{k}[\peps_a]=\peps_{\act{k}a} \dw{k}{a}
\, , &&
\globalu{k}[\peps^-_a]= (\dw{k}{a})^\dagger \peps^-_{\act{k}a }
\, ,
\end{align}
if they are to correctly capture the behaviour of the defect $\defect{a}{g}$.  
For the special case $\peps_a=\ici{a}$, we have $\globalu{k}[\peps_a]=\dw{k}{a}$.

Including the action of the physical symmetry upon the defect superselection sectors yields the following modification of Eq.~\eqref{set:Uactiondefectnabctube} 
\begin{align}
\globalu{k}[\peps_c \, \fusiontube{\defect{a}{g}\defect{b}{h}}{\defect{c}{gh}} \, \peps_a^- \otimes \openone \otimes \peps_b^-] 
&= \globalu{k}[\peps_c] (\dw{k}{c})^\dagger \, \fusiontube{\act{k}\defect{a}{g} \act{k}\defect{b}{h}}{\act{k}\defect{c}{gh}} \, \dw{k}{a} \, \globalu{k}[\peps_a^-] \otimes  \splitact{k}{a,b;c}  \otimes \dw{k}{b} \, \globalu{k}[\peps_b^-]
\\
&= \peps_{\act{k}c} \, \fusiontube{\act{k}\defect{a}{g} \act{k}\defect{b}{h}}{\act{k}\defect{c}{gh}} \, \peps^-_{\act{k}a} \otimes  \splitact{k}{a,b;c}  \otimes \peps^-_{\act{k}b}
\, .
\end{align}
That is, the physical action of the symmetry compensates the action on each defect sector appearing in Eq.~\eqref{set:Uactiondefectnabctube}.

 \subsection{$\G$-crossed braiding}

The process of exchanging defects differs from the exchange of anyons, as we must now also move the domain walls attached to the defects, which requires the application of the physical symmetry 
\begin{align}

  \, .
  \end{align}
  During the exchange step we have applied the physical symmetry to the dark shaded region between the initial and final position of the $\textbf{h}$ defect line, and to the superselection sector $\defect{a}{g}$, which resulted in the application of $(\dw{h}{a})^\dagger$ to the ICI $\ici{\defect{a}{g}}$.

For nondegenerate $\ici{\defect{a}{g}}$ and $\ici{\defect{b}{h}}$ we have
\begin{align}
\label{set:Defectrabnondegen}
R^{\act{h}\defect{a}{g}  \defect{b}{h}}:= 
    	%
  =
 \sum_{\defect{q}{g$\conj{\g{h}}$}}  \frac{(t_{\dw{h}{a}}^{s^{\act{h}{a}} q \, s^a  s^b } )^*}{d_{s^b}}
  \, ,
\end{align}
where $s_a$ is defined by the equation $\ici{\defect{a}{g}} \tube{p p p}{\vac} = \delta_{p s^a} \, \ici{\defect{a}{g}}$.

The $\G$-crossed $R$-symbols $  \rmatrix{{\act{h}\defect{a}{g} \defect{b}{h}}}{\defect{c}{hg}}$ are defined by resolving the $R$-matrix into defect superselection sectors of definite topological charge. For nondegenerate defect ICIs $\ici{\defect{a}{g}},\, \ici{\defect{b}{h}},\, \ici{\defect{c}{hg}},$  the fusion vertices $\mu_{b a}^{c}$ and $\nu_{\act{h}a b}^{c}$ are unique, and we have 
\begin{align}

	\, 
	,
\end{align}
as part of a larger fusion pair-of-pants tensor network  $\fusiontube{\act{h}\defect{a}{g}\defect{b}{h}}{\defect{c}{gh}}$.  An explicit expression for $\rmatrix{{\act{h}\defect{a}{g} \defect{b}{h}}}{\defect{c}{hg}}$ can be derived directly from Eq.~\eqref{set:Defectrabnondegen}. 
These results can be generalized to degenerate defect ICIs by following a  similar  line of reasoning.

  \subsection{Shifting by 2 and 3 cocycles}
  
In Section~\ref{set:groupextensionclassification} we found that consistent fusion rules for a $\G$-graded MPO algebra correspond to an $\cohomo{2}{\G}{\autoequiv{1}}$ torsor, while $F$-symbol associators that satisfy the pentagon equation correspond to $\coho{3}$-torsors. 
Hence one might consider the effect on the emergent theory, of shifting the MPO algebra by an element of $\cohomo{2}{\G}{\autoequiv{1}}$ or $\coho{3}$. 
Shifting the fusion rules by an element of $\cohomo{2}{\G}{\autoequiv{1}}$ can totally change the structure of the $\g{g}$-defect tube algebras for $\g{g}\neq\g{1}$. Beyond the $\g{1}$-sector, which remains the same, there is no obvious relation between the emergent theories before and after the shift. This is demonstrated in Example~\ref{set:tcextensioneg}.

Shifting the $F$-symbol associators by an element of $\coho{3}$ is more subtle, as it preserves the set of emergent defects, but it may change some of their properties. Except for the underlying emergent topological order, corresponding to the $\g{1}$-sector, which is left invariant. 
Specifically, we consider shifting by a nontrivial cocycle $[\alpha]\in \coho{3}$ 
\begin{align}
\shortf{\defect{a}{g} \defect{b}{h} \defect{c}{k}}{d}{e}{ f} \mapsto  \alpha(\g{g},\g{h},\g{k}) \, \shortf{\defect{a}{g} \defect{b}{h} \defect{c}{k}}{d}{e}{ f}
\, ,
\end{align}
which leads to an obvious change in the $F$-symbols of the emergent theory, via Eq.~\eqref{set:doublefsymboleqn}. 
The effect on the $\G$-crossed $R$ matrices can also be derived. 

The $\G$-crossed modular matrices are modified as follows
\begin{align}
   S_{\defect{a}{g} \defect{b}{h}} \mapsto     \fs{\g{g}}^* \,   \alpha^{\g h}(\g  g ,  \conj{\g g})^* \frac{\alpha(\g h , \g g \conj{\g h}, \g h)}{\alpha(\conj{\g{g}}, \g h \g g, \conj{\g{g}} )}  \,  S_{\defect{a}{g} \defect{b}{h}}
\, ,
&&
 \theta_{\dw{h}{a_{\g{g}}}}  \mapsto   \alpha(\g g, \g h, \g g)^* \frac{\alpha(\g h , \g g \conj{\g h}, \g h)}{\alpha( \g {gh} ,  \conj{\g h}, \g {gh})} \, \theta_{\dw{h}{a_{\g{g}}}}
\, .
\end{align}

The effects of the cocycle shift on the physically observable data are more apparent after gauging the symmetry, see Section~\ref{setsection:gauging}. These effects are largely a consequence of the following transformations. 

The Frobenius-Schur indicator transforms as follows
  \begin{align}
  \fs{\defect{a}{g}} \mapsto    \alpha(\g{g},\conj{\g{g}},\g{g})\,  \fs{\defect{a}{g}}
  \, .
  \end{align}
The 2-cocycle of the projective representation acting on the $\defect{a}{g}$ defect transforms as
\begin{align}
\factsys{a}{\g{h},\g{k}} \mapsto 
\omega_{\g g}(\g{h},\g{k})
\,\factsys{a}{\g{h},\g{k}}
\, ,
\end{align}
 where 
\begin{align}
\label{set:equationwiththeslantproduct}
\omega_{\g g}(\g{h},\g{k}) &=
\alpha^{\g{g}}(\g{h},\g{k}) \, 
d \varepsilon^{\g g}(\g h)
\, ,
&
   \alpha^{\g{g}}(\g{h},\g{k}) &= 
   \frac{\alpha(\g h,\act{k}\g{g},\g{k})}{\alpha(\act{hk}\g{g},\g{h},\g{k}) \alpha(\g{h},\g{k},\g{g})}
   \, ,
\\
   d \varepsilon^{\g g} (\g{h},\g{k}) &= 
   \frac{\varepsilon^{\g g}(\g k) \varepsilon^{\act{k}\g g}(\g h,\g k) }{ \varepsilon^{\g g}(\g{hk}) }
\, ,
   &
\varepsilon^{\g g}(\g h) &= \alpha(\g h,\g g \conj{\g{h}}, \g h)
   \, .
\end{align}
Hence the group action on the symmetry-twisted MES becomes 
\begin{align}
\eta_{\defect{a}{g}}^{\g{h}}(\g k) \mapsto  \alpha^{ \g{g},\g{h}}(\g k) \eta_{\defect{a}{g}}^{\g{h}}(\g k)
\, ,
&&
\text{where}
&&
\alpha^{ \g{g},\g{h}}(\g k) = \frac{\alpha^{\g g}(\g{k},\g{h}) }{\alpha^{\g g}(\g{h},\g{k})} 
\, ,
\end{align}
and $\g{k}\in\centralizer{\g{g},\g{h}}$, with $\g{gh}=\g{hg}$. 
Finally the group action on the fusion vertices transforms as follows   
  \begin{align}
  \splitact{k}{\defect{a}{g},\defect{b}{h};\defect{c}{gh}} \mapsto  \omega_{\g k}(\g{g},\g{h})  \splitact{k}{\defect{a}{g},\defect{b}{h};\defect{c}{gh}}
  \, ,
  \end{align}
  for $\g{k} \in \centralizer{\defect{a}{g},\defect{b}{h},\defect{c}{gh}}$.

We remark that shifting by a 3-cocycle cannot, in general, be absorbed by a change of gauge of the MPO data. 
For instance it is apparent that the projective representations acting on the defects are shifted when the 2-cocycle $\alpha^{\g g}$ lies in a nontrivial cohomology class. 
Hence one must assess whether shifting by a 3-cocycle actually changes the emergent UGxBFC on a case by case basis.  
In Section~\ref{setsection:condensation} we present some examples for which a 3-cocycle shift can be absorbed by a change of gauge, and some for which it cannot.

    \section{ Gauging a global symmetry}
  \label{setsection:gauging}

In this section, we describe the effect of applying the state gauging procedure --- developed in Refs.~\cite{Gaugingpaper,williamson2014matrix} --- to SET tensor networks. 
This corresponds to equivariantization in the category theory language used in Refs.~\cite{turaev2000homotopy,kirillov2004g,etingof2009fusion,gelaki2009centers}. 
Our analysis focuses on the transformation of the $\G$-graded MPO symmetry algebra. Gauging effectively promotes the nontrivial $\g{g}$-sectors into purely topological symmetries. 
From this we derive formulas for the the ICIs of the gauged topological order. 
These gauged ICIs lead to expressions for the resulting $S$ and $T$ matrices, as well as the fusion degeneracies and quantum dimensions. 
Before addressing the general case, we focus on a simple example that captures many features of the gauging procedure.

      \subsection{Example: Gauging SPT orders}

We  begin by applying the gauging procedure to the simplest class of examples where a unitary, on-site SPT order is gauged to a twisted quantum double (Dijkgraaf-Witten) topological order~\cite{levin2012braiding,williamson2014matrix,qdouble,DijkgraafWitten}.

\subsubsection{MPO description of an SPT order}

We consider a $\G$-graded MPO algebra with $|\cat_\g{1}|=1$. It was shown in Ref.~\cite{williamson2014matrix} that this corresponds to a non-chiral, invertible topological order. Furthermore it was shown that $|\cat_\g{g}|=1,\, \forall \g{g}\in\G,$ for a faithful MPO representation with $|\cat_\g{1}|=1$. 

In this case the single block MPOs are simply labeled by group elements $\mpo{}{\g g}$, and their fusion is given by the group product $\mpo{}{\g g}\mpo{}{\g h}=\mpo{}{\g{gh}}$. 
Consequently, the associators are simply $\Uone$ phases ${\shortf{\g{g},\g{h},\g{k}}{\g{ghk},}{\g{gh},}{\g{hk}}=\alpha(\g{g},\g{h},\g{k})}$ that satisfy the 3-cocycle equation 
\begin{align}
\alpha(\g{f},\g{g},\g{h}) \alpha(\g{f},\g{gh},\g{k}) \alpha(\g{g},\g{h},\g{k}) = \alpha(\g{fg},\g{h},\g{k}) \alpha(\g{f},\g{g},\g{hk})
\, ,
\end{align}
due to the pentagon equation.  
The algebraic data of the MPO algebra is described by the UFC $\vecg{\G}{\alpha}$, considered as a $\G$-graded category. The classification of such MPO algebras recovers the third cohomology classification of SPTs~\cite{czxmodel,chen2012symmetry,PhysRevB.90.235137}. In this case the $\G$-graded Morita equivalence class is given by a representative, up to a 1-cochain, since each sector consists of only one single block MPO. 
Fixed point representatives of these SPT phases are captured by a special case of the $\G$-graded string-net example, see Section~\ref{set:stringnetexample}.

The significance of the fact that $F$-symbol solutions are torsors can be understood in this case as follows: the classification of SPT phases is dependent upon the precise choice of physical symmetry, with respect to which the classification is done. As an explicit example, the well known CZX model~\cite{czxmodel} is in the trivial phase w.r.t. an on-site $X^{\otimes N}$ symmetry, but is in a nontrivial phase for the more complicated on site symmetry chosen in Ref.~\cite{czxmodel}.

Since $|\cat_\g{g}|=1$ the dube algebras are all one dimensional. There is a single defect ICI for each sector, which is simply given by $\ici{\g{g}}=\tube{\g{g}, \g{g}, \g{g}}{\g{1}}$.

When considering the group action on the defects we use the following choice of gauge 
\begin{align}
\label{set:sptexampledwgaugechoice}
\dw{h}{\g g}=\alpha^*(\g h ,\g g \conj{\g{h}}, \g h ) \tube{\act{h}\g{g}, \g{g}\conj{\g{h}} , \g{g}}{\g{h}} = : \tube{\g{g}}{\g{h}}
\, .
\end{align} 

The permutation action is simply given by group conjugation, i.e. $\act{h}\ici{\g{g}}=\ici{\g{hg\conj{\g h}}}$. 

Multiplication of domain walls is given by
\begin{align}
\dw{h}{\act{k}\g g} \dw{k}{\g g} = \frac{\alpha(\g h, \act{k} \g g, \g k)}{ \alpha(\act{hk}\g g, \g h, \g k) \alpha(\g h, \g k, \g g)} \dw{hk}{\g g}
\, ,
&& 
\text{hence}
&&
\factsys{\g{g}}{\g h , \g k} = \alpha^{\g{g}}(\g h , \g k )
\, .
\end{align}

The group action of $\g{k}\in\centralizer{h}$ on the MES is given by 
\begin{align}
\globalu{k} \ket{\dw{h}{\g{g}}} 
=
\alpha^{\g g , \g h}(\g k) \ket{\dw{h}{\act{k}\g{g}}} 
\, ,
&&
\text{where}
&&
\alpha^{\g g , \g h}(\g k) = \frac{\alpha^{\g g}(\g k , \g h)}{\alpha^{\g g}(\g h , \g k)}
\, .
\end{align}

The $\G$-crossed modular matrices are given by 
\begin{align}
\braket{ \dw{y}{\g{x}} \, | \,S(\dw{h}{\g{g}})} &= \delta_{\g{x},\g{h}} \delta_{\g{y},\conj{\g{g}}} \delta_{\g{g},\act{h}\g{g}} \,   \fs{\g{g}}^* \,   \alpha^{\g h}(\g  g ,  \conj{\g g})^*
\, ,
\\
\braket{ \dw{y}{\g{x}} \, | \, T(\dw{h}{\g{g}})} &= \delta_{\g{x},\g{g}} \delta_{\g{y},{\g{gh}}} \delta_{\g{g},\act{h}\g{g}}\, \alpha(\g g, \g h, \g g)^* 
\, .
\end{align}

Fusion of the defect ICIs is again given by the group multiplication rule, and it is multiplicity free. Hence the group action on the fusion space is simply a $\Uone$ phase, which is only defined up to multiplication with a 1-cocycle. One can verify that the following is a valid choice~\cite{Bultinck2017183} 
\begin{align}
\splitact{k}{\g{g} , \g{h} ; \g {gh}} = \alpha^{\g{k}}(\g g , \g h)
\, .
\end{align}
Even though $U$ is a phase in this case, it cannot generally be absorbed into the gauge of the projective representation $\dw{h}{\g g}$ as it is a function of three group variables. 

 The $F$-symbols of the defects are given by the 3-cocycle $\alpha$ of the underlying MPO algebra. Hence the FS indicators of the defects are given by $\fs{\g g}=\alpha(\g g , \conj{\g g} , \g g )$.

\subsubsection{Gauging the global symmetry of an SPT order}

 It was shown in Ref.~\cite{williamson2014matrix} that applying the lattice gauging procedure introduced in Ref.~\cite{Gaugingpaper},  for the full symmetry group $\G$, results in a tensor network with an MPO algebra where the $\G$-grading is essentially forgotten. That is, the $1$-sector of the resulting MPO algebra is given by the full $\G$-graded MPO algebra of the ungauged model.  The algebraic data describing this is again the UFC $\vecg{\G}{\alpha}$, but without including the $\G$-grading. Hence the emergent topological order is $\double{\vecg{\G}{\alpha}}$, which is a twisted quantum double model. 
  See Section~\ref{set:stringnetexample} for a description of the gauging map applied to the fixed point PEPS tensors.

After gauging, the former domain walls $\dw{h}{\g g}$ become purely virtual operators that do not require a physical group action. These MPOs can now be thought of as fluctuating in the vacuum given by the tensor network representation of the ground state. 

This produces a tube algebra for $\vecg{\G}{\alpha}$ from the dube algebra of the $\G$-graded $\vecg{\G}{\alpha}$, together with the former domain walls $\dw{h}{\g g}$. 
Hence all the sectors in a conjugacy class $\conjclass{\g g}$ of the dube algebra are coupled by the domain wall tubes $\dw{h}{\g g}$ and result in a $|\conjclass{\g g}| \times |\G|$ block of the gauged tube algebra.  

An individual defect $\g{g}$ supports a projective representation of $\centralizer{\g g}$ with 2-cocycle $\alpha^{\g g}$. This contributes a block of size $|\centralizer{\g g}|$ to the gauged tube algebra, which is block-diagonalized by idempotents that project onto the projective irreps $\pi^\mu_{\g g}$ 
\begin{align}
\gaugedii{\g g}{ \mu} := \frac{d_\mu}{|\centralizer{\g g}|}
\sum_{\g{h}\in\centralizer{\g g} } \cirrep{\g g}{\mu}{h} \, \dw{h} {\g g}
\, ,
\end{align}
where $\cirrep{\g g}{\mu}{h} := \tr [\pi^\mu_{\g g}(\g h)]^*$ is a complex conjugated character.  
Each $\gaugedii{\g g}{ \mu}$ projects onto a block of size $d_\mu$, and since 
\begin{align}
\sum_{\mu} d_\mu^2 = |\centralizer{\g g}|
\, ,
\end{align}
we see that the decomposition is complete. 

These irreducible idempotents can be extended to the full $|\conjclass{\g g}| \times |\G|$ block, by summing over the orbit of $(\g g, \mu)$ under conjugation 
\begin{align}
\label{set:fullorbiticiforgaugethy}
\ici{\gauged{\g g}{\mu}} =\sum_{\g{k}\in \G / \centralizer{\g g }} \gaugedii{\act{k}\g g}{\act{k} \mu}
\, ,
\end{align}
where the canonical isomorphism between projective irreps, given in Eq.~\eqref{set:isomofprojirreps}, is used to define $ \gaugedii{\act{k}\g g}{\act{k} \mu}$.
The resulting $\ici{\gauged{\g g}{\mu}}$ are ICIs that project onto $d_\mu \times |\conjclass{\g g}|$ dimensional blocks, and by counting one can verify that they give a complete decomposition.

After gauging, some states on the torus vanish. As discussed in Ref.~\cite{williamson2014matrix} only those states $\ket{\dw{h}{\g g}}$ satisfying $\alpha^{\g g , \g h} \equiv 1 $ survive the gauging process.

The gauged $S$ and $T$ can be derived from the $\G$-crossed modular matrices, leading to the formulas 
\begin{align}
  S^{\double{\vecg{\G}{\alpha}}}_{\gauged{{\g{g}}}{\mu} \gauged{{\g{h}}}{\nu}} &=  \frac{1}{|\G|} 
  \sum_{\substack{\g{r}\in \G/\centralizer{\g g}\\ \g{s}\in \G/\centralizer{\g h}}} 
\delta_{\act{r}\g{g}\act{s}\g{h},\act{s}\g{h}\act{r}\g{g}} \,  \act{r}\cirrep{\g g}{\mu}{\act{s}h}\, \act{s} \irrep{ \g h}{\nu}{\act{r}\conj{\text{g}}}  \, \fs{\,\act{r}\g{g}}^* \, \alpha ^{\act{s} \g h}(\act{r}\g g , \act{r} \conj{\g{g}})
  \, ,
    \\
    T^{\double{\vecg{\G}{\alpha}}}_{\gauged{{\g{g}}}{\mu} \gauged{{\g{h}}}{\nu}} &=   \delta_{ {\gauged{{\g{g}}}{\mu} \gauged{{\g{h}}}{\nu}}}
    \frac{\irrep{\g g}{\mu}{g}}{\irrep{\g g}{\mu}{1}} 
    ,
\end{align}
where $ \act{r}\chi_{\g g}^{\mu} := \chi_{\act{r} \g g}^{ \act{k} \mu}$. 
From the special case of the $S$ matrix with $  \gauged{{\g{h}}}{\nu}= \gauged{\g{1}}{1}$, we find 
\begin{align}
d_\gauged{{\g{g}}}{\mu} = \frac{|\G| d_\mu}{|\centralizer{\g g}|}
\, .
\end{align}

The fusion multiplicities can be derived from $S$ via the Verlinde formulas, but they can also be obtained directly by considered the intertwining property of a defect pair-of-pants tensor network 
\begin{align}
\dw{k}{\g{gh}}
\globalu{k}[\fusiontube{\g g, \g h}{\g {gh}}] 
= 
\fusiontube{\g g, \g h}{\g {gh}} 
\, \alpha^{\g{k}}(\g g , \g h)  \, \dw{k}{g} \otimes \dw{k}{h} \, ,
\end{align}
for $\g{k}\in\centralizer{\g g,\g h} $. After gauging there is no global group action $\globalu{k}$ remaining, hence the projector onto the $\pi^\kappa_{\g {gh}}(\g k)$  projective rep is intertwined by the gauged pair-of-pants tensor network as follows 
\begin{align}
\frac{d_\kappa}{|\centralizer{\g {gh}}|}
\sum_{\g{k}\in\centralizer{\g g,\g h} } \cirrep{\g {gh}}{\kappa}{k} \, \dw{k} {\g {gh}}
\ \mapsto \
\frac{d_\kappa}{|\centralizer{\g {gh}}|}
\sum_{\g{k}\in\centralizer{\g g,\g h} } \cirrep{\g{gh}}{\kappa}{k} \, \alpha^{\g{k}}(\g g , \g h)\, \dw{k}{\g g} \otimes \dw{k}{\g h}
\, ,
\end{align}
which projects onto the diagonal  $\pi^\kappa_{\g {gh}}(\g k)\big|_{\g{k}\in\centralizer{\g g , \g h}}$ irrep within ${ \left((\alpha^{\g{k}}(\g g , \g h)  \,  \pi^\mu_{\g {g}}(\g k) \otimes \pi^\nu_{\g {h}}(\g k)  \right)\big|_{\g{k}\in\centralizer{\g g , \g h}}   }$. 
To calculate the fusion multiplicity one must also sum over conjugacy classes $\conjclass{g},\, \conjclass{h},$ and $\conjclass{gh}$.  This is implemented by summing over conjugates $(\act{x}\g{g},\act{y}\g{h})$, where $\g{x}\in\G /\centralizer{g}, \, \g{y} \in \G/ \centralizer{h}$, modulo left multiplication $(\g{zx},\g{zy})$, since that leads to conjugation $\act{z}(\act{x}\g{g}\act{y}\g{h})$ which remains within $\conjclass{\act{x}g\act{y}h}$.  The set of elements $\G /\conjclass{g}\times \G/ \conjclass{h}$, modulo left multiplication, is known to be isomorphic to the double coset $\centralizer{g} \backslash \G / \centralizer{h} $.

After including the summations  over conjugacy classes we find that the fusion multiplicity is given by
\begin{align}
N_{{\gauged{\g g}{\mu}}   {\gauged{\g h}{\nu}}}^{{\gauged{\g {k}}{ \kappa  }}}
= 
\, &\sum_{ \substack{(\g x , \g y ) \in \centralizer{g} \backslash \G / \centralizer{h} \\  \g{z} \in \G / \centralizer{k}}}
\delta_{\act{x}\g {g} \act{y}\g h, \act{z}\g{k}} 
\nonumber \\
&m\left (\act{z}\pi^\kappa_{\g {k}}( \cdot)\Big|_{\centralizer{\act{x}\g g , \act{y}\g h}} , \ \alpha^{({\cdot})}(\act{x}\g g , \act{y}\g h)\Big|_{\centralizer{\act{x}\g g , \act{y}\g h}}    \act{x}\pi^\mu_{\g {g}}(\cdot)\Big|_{\centralizer{\act{x}\g g , \act{y}\g h}}  \otimes \, \act{y}\pi^\nu_{\g {h}}(\cdot) \Big|_{\centralizer{\act{x}\g g , \act{y}\g h}}  \right)
\, ,
\end{align}
where 
\begin{align}
\label{set:definitionofmfunction}
m(\pi,\rho) = \dim \left[
\text{Hom}_\G(\pi,\rho)
\right]
\, ,
\end{align}
 counts the dimension of the space of intertwiners between the $\pi$ and $\rho$ reps. For the special case of $\pi$ an irrep. it counts the number of times $\pi$ appears in the irrep decomposition of $\rho$. For a further discussion of the $m$ function, and how it can be computed in terms of projective characters, see Ref.~\cite{barkeshli2014symmetry}. 
 In Ref.~\cite{Bultinck2017183} this multiplicity was evaluated explicitly in terms of projective characters
 \begin{align}
N_{{\gauged{\g g}{\mu}}   {\gauged{\g h}{\nu}}}^{{\gauged{\g {k}}{ \kappa  }}}
= 
\frac{1}{|\G|} \sum_{\g x \in \G/\centralizer{\g g}}
\, \sum_{\g y \in \G/\centralizer{\g h }}
\, \sum_{\g z \in \G/\centralizer{\g {k}}}
\, 
\delta_{\act{x}\g {g} \act{y}\g h, \act{z}\g{k}} 
 \sum_{\g w \in \centralizer{\act{x}\g g , \act{y}\g h}} 
 \, \alpha^{\g{w}}(\act{x}\g g , \act{y} \g h)  \,  \act{x}\cirrep{\g g}{\mu}{w} \,  \act{y}\cirrep{\g h}{\nu}{w} \, \act{z}\irrep{\g {k}}{\kappa}{w}
 \, .
  \end{align}

\subsection{The general case}

Gauging a global symmetry can be implemented explicitly on a tensor network state using the prescription of Ref.~\cite{Gaugingpaper}. It was shown in Ref.~\cite{williamson2014matrix} that gauging a symmetric tensor network results in the removal of the $\G$-grading from the MPO algebra. That is, the $\g{1}$-sector of the gauged MPO algebra consists of all sectors of the ungauged $\G$-graded MPO algebra.  
The application of the gauging map to the fixed point SET string-net tensors is described explicitly in Section~\ref{set:stringnetexample}.

Since the $\g{1}$-sector of the gauged MPO algebra is $\cat_\G$, the emergent superselection sectors are given by $\double{\cat_{\G}}$ and can be constructed via the tube algebra. 
The total quantum dimension is given by $\qd_{\text{out}}=\qd_\G^2=|\G| \qd_\g{1}^2$, see Eq.~\eqref{set:setqdrelation}. 

Rather than calculating the gauged ICIs directly with the tube algebra, we can derive them from the defect ICIs $\ici{\defect{a}{g}}$ and the domain walls $\dw{h}{\defect{a}{g}}$ of the dube algebra. In this section, and the next, we demonstrate that the problem of finding the gauged ICIs is equivalent to finding the ungauged defect ICIs and domain walls. 
This provides a solution for the ICIs of the tube algebra for an input category with a $\mathcal{G}$-crossed modular braiding by combining the solution for the ICIs and domain walls of the dube algebra from Ref.~\cite{Ren2024} with the gauging procedure explained below. 

In Ref.~\cite{gelaki2009centers} the authors show that $\double{\cat_{\G}}$ is canonically equivalent to a $\G$-equivariantization of the relative center $\rdouble{\cat_\G}{\cat_\g{1}}$. Physically $\G$-equivariantization corresponds to gauging a global $\G$ symmetry, and hence our results can be understood as a tensor network realization of the theorem in Ref.~\cite{gelaki2009centers}. 

After gauging, the former domain walls $\dw{h}{\defect{a}{g}}$ are now included in the tube algebra. 
Hence the sectors of the dube algebra within a conjugacy class $\conjclass{\g g}$ are coupled by the domain walls  $\dw{h}{\defect{a}{g}}$, and contribute a block to the resulting tube algebra.

The projective representation of $\centralizer{\g g}$, acting on each $\defect{a}{g}$ defect, was constructed in Section~\ref{setsection:dubes}. Idempotents of the gauged theory are given by Hermitian projectors onto projective irreps $\pi^\mu_{\defect{a}{g}}$ with 2-cocycle $\eta_\defect{a}{g}$ 
\begin{align}
\gaugedii{\defect{a}{g}}{ \mu} := \frac{d_\mu}{|\centralizer{\defect{a}{g}}|}
\sum_{\g{h}\in\centralizer{\defect{a}{g}} } \cirrep{\defect{a}{g}}{\mu}{h} \, \dw{h} {\defect{a}{g}}
\, .
\end{align}
The projector $\gaugedii{\defect{a}{g}}{ \mu}$ correspond to the identity on a block of dimension $d_\mu \times D_\defect{a}{g}$. 

The ICIs are then found by summing over the orbit of the defect $\defect{a}{g}$ under the group action 
\begin{align}
\label{set:gaugedtubeformula}
\ici{\gauged{\defect{a}{g}}{\mu}} =\sum_{\g{k}\in \G / \centralizer{\defect{a}{g} }} \gaugedii{\act{k}\defect{a}{g}}{ \act{k} \mu}
\, ,
\end{align}
which project onto blocks of dimension $d_\mu \times D_\defect{a}{g} \times |\G|/ | \centralizer{\defect{a}{g}}|$.  
Similar to the SPT example~\cite{williamson2014matrix}, only states satisfying $\eta_{\defect{a}{g}}^{\g{h}} \equiv 1$ lead to nonzero gauged states.

The $S$ and $T$ matrices of the gauged theory can be calculated in terms of the $\G$-crossed modular matrices by using Eqs.~\eqref{set:sandttraceformula},\eqref{set:gcrossedmodularmatrixdefinition},\eqref{set:gaugedtubeformula} 
\begin{align}
   S_{ab}
    &=
     \sum_{\substack{\g{r}\in \G/\centralizer{a}\\ \g{s}\in \G/\centralizer{b}}} \sum_{\substack{ \g{f}\in\centralizer{\act{s}\defect{a}{g}} \\ \g{k}\in\centralizer{\act{s}\defect{b}{h}} }} 
     \frac{D_{[\defect{b}{h}, \nu]} \,  d_\mu d_\nu \, \act{r}\cirrep{\defect{a}{g}}{\mu}{f} \,  \act{s}\irrep{\defect{b}{h}}{\nu}{k} \,
\tr[(\dw{k} {\act{s}\defect{b}{h}})^\dagger S( \dw{f} {\act{r}\defect{a}{g}})]}{D_{[\defect{a}{g}, \mu]}  \,  |\centralizer{\defect{a}{g}}| \,  |\centralizer{\defect{b}{h}}| \, \tr \ici{\gauged{\defect{b}{h}}{\mu}} } 
   \\
   &=
    \sum_{\substack{\g{r}\in \G/\centralizer{a}\\ \g{s}\in \G/\centralizer{b}}}
     \frac{D_\defect{b}{h}\,
     d_\nu^2 \, \act{r}\cirrep{\defect{a}{g}}{\mu}{\act{s}\g{h}} \,  \act{s}\irrep{\defect{b}{h}}{\nu}{\act{r}\conj{\g g}} \,
\tr[(\dw{\act{r}\conj{\g{g}}} {\act{s}\defect{b}{h}})^\dagger S( \dw{\act{s}{\g{h}}} {\act{r}\defect{a}{g}})]}{ D_\defect{a}{g} \,  |\centralizer{\defect{b}{h}}|^2 \, \tr \ici{\gauged{\defect{b}{h}}{\mu}} } 
   \\
   &=
    \sum_{\substack{\g{r}\in \G/\centralizer{a}\\ \g{s}\in \G/\centralizer{b}}}
     \frac{
      \act{r}\cirrep{\defect{a}{g}}{\mu}{\act{s}\g{h}} \,  \act{s}\irrep{\defect{b}{h}}{\nu}{\act{r}\conj{\g g}} \,
S_{\act{r}\defect{a}{g}\,\act{s}\defect{b}{h}} }{  |\G|  } 
   \, .
    \end{align}
Where we have used that $\tr[(\dw{k} {\defect{b}{h}})^\dagger S( \dw{f} {\defect{a}{g}})]$ contains the delta condition $\delta_{\g{k},\conj{\g{g}}}\delta_{\g{f},\g{h}}$, and 
\begin{align}
\tr \ici{\gauged{\defect{b}{h}}{\nu}}
= 
\sum_{\g{s}\in \G / \centralizer{\defect{b}{h} }} \frac{d_\nu}{|\centralizer{\defect{b}{h}}|}
\sum_{\g{k}\in\centralizer{\act{s}\defect{b}{h}} } \act{s}\cirrep{\defect{b}{h}}{\nu}{k} \, \tr( \dw{k} {\act{s}\defect{b}{h}} )
=
\frac{|\G| d_\nu^2}{|\centralizer{\defect{b}{h}}|^2}  \tr( \ici{\defect{b}{h}})
\, ,
\end{align}
 which follows from $\cirrep{\defect{b}{h}}{\nu}{1}=d_\nu$, $\tr( \dw{k} {\defect{b}{h}} )=\delta_{\g{k},\g{1}} \tr( \dw{1} {\defect{b}{h}} )$ and $ \dw{1} {\act{s}\defect{b}{h}}=\ici{\defect{b}{h}}$.

Since it was shown in Eq.~\eqref{set:sandtonidempotentsdefn} that the $T$ matrix simply results in a phase, it can be calculated by keeping track of the relative phase of the underlying defect ICI $\ici{\defect{a}{g}}$
\begin{align}
T ( \ici{\gauged{\defect{a}{g}}{\mu}} ) 
&=  \left( \sum_{\g{k}, \defect{s}{k}} \tube{s \vac s}{s} \right) \frac{d_\mu}{|\centralizer{\defect{a}{g}}|} \left(  \cirrep{\defect{a}{g}}{\mu}{1} \, \ici{\defect{a}{g}} 
+ \dots \right)
\\
&=
\frac{d_\mu}{|\centralizer{\defect{a}{g}}|} \left(  \cirrep{\defect{a}{g}}{\mu}{1} \frac{\irrep{\defect{a}{g}}{\mu}{g}}{\irrep{\defect{a}{g}}{\mu}{1}} \left(\frac{\irrep{\defect{a}{g}}{\mu}{g}}{\irrep{\defect{a}{g}}{\mu}{1}}\right)^*  \, \theta_{\defect{a}{g}} \, \dw{g} {\defect{a}{g}} + \dots \right)
\\
&=
\frac{\cirrep{\defect{a}{g}}{\mu}{g}}{\irrep{\defect{a}{g}}{\mu}{1}} \, \theta_{\defect{a}{g}} \,
\ici{\gauged{\defect{a}{g}}{\mu}} 
\, ,
\end{align}
where we have used that $\frac{\irrep{\defect{a}{g}}{\mu}{g}}{\irrep{\defect{a}{g}}{\mu}{1}}\in\Uone$. 

Hence we have~\cite{barkeshli2014symmetry}
\begin{align}
\label{set:gaugedsetmodularmatrices}
  S^{\double{\cat_\G}}_{\gauged{a_{\g{g}}}{\mu} \gauged{b_{\g{h}}}{\nu}} &=  \frac{1}{|\G|} 
  \sum_{\substack{\g r \in \G/\centralizer{a}\\ \g s \in \G/\centralizer{b}}} 
  \act{r}\cirrep{a}{\mu}{\act{s}h}\, \act{s} \irrep{b}{\nu}{\act{r}\conj{\text{g}}}  \, S_{\act{r}a_{\g{g}} \act{s}b_{\g{h}}}
  \, ,
\\
    T^{\double{\cat_\G}}_{\gauged{a_{\g{g}}}{\mu} \gauged{b_{\g{h}}}{\nu}} &=   \delta_{ {\gauged{a_{\g{g}}}{\mu} \gauged{b_{\g{h}}}{\nu}}}
    \frac{\irrep{a}{\mu}{g}}{\irrep{a}{\mu}{1}} \theta_a
    \, .
\end{align}
These quantities are gauge invariant, as the variation of $ \theta_{\defect{a}{g}}$ and $S_{\act{r}a_{\g{g}} \act{s}b_{\g{h}}}$ under a 2-coboundary is canceled by that of $\irrep{\defect{a}{g}}{\mu}{g}$. In particular, the topological spins of the anyons ($\g 1$-defects), after gauging, remain the same. 
Setting $  \gauged{b_{\g{h}}}{\nu}= \gauged{\vac}{0}$ in Eq.~\eqref{set:gaugedsetmodularmatrices} we find the quantum dimensions of the gauged superselection sectors to be 
\begin{align}
d_{\gauged{a_{\g{g}}}{\mu}}= \frac{|\G| d_\mu d_{\defect{a}{g}}}{|\centralizer{\defect{a}{g}}|}
\, .
\end{align} 

The fusion multiplicities can be derived from the $S$ matrix, calculated in Eq.~\eqref{set:gaugedsetmodularmatrices}, by using the Verlinde formula. It is also possible to derive them directly, by using the intertwining property of the pair-of-pants tensor network
\begin{align}
\dw{k}{\defect{c}{gh}} \, 
\globalu{k}[\fusiontube{\defect{a}{g},\defect{b}{h}}{\defect{c}{gh}} ] 
= 
\fusiontube{\defect{a}{g},\defect{b}{h}}{\defect{c}{gh}} 
\, \dw{k}{\defect{a}{g}} \otimes  \splitact{k}{\defect{a}{g},\defect{b}{h};\defect{c}{gh}}  \otimes \dw{k}{\defect{b}{h}} 
\, ,
\end{align}
for $\g k \in \centralizer{\defect{a}{g},\defect{b}{h},\defect{c}{gh}}$. 
After gauging, there is no longer a physical group action, and hence the projector onto the $\pi^\kappa_{\defect{c}{gh}}$  rep is intertwined by the gauged pair-of-pants fusion tube as follows 
\begin{align}
\frac{d_\kappa}{|\centralizer{\defect{c} {gh}}|}
\sum_{\g{k}\in\centralizer{\defect{a}{g},\defect{b}{h},\defect{c}{gh}}} \cirrep{\defect{c}{gh}}{\kappa}{k} \, \dw{k} {\defect{c} {gh}}
\mapsto 
\frac{d_\kappa}{|\centralizer{\defect{c} {gh}}|}
\sum_{\g{k}\in\centralizer{\defect{a}{g},\defect{b}{h},\defect{c}{gh}} } \cirrep{\defect{c}{gh}}{\kappa}{k} \, \dw{k}{\defect{a}{g}} \otimes  \splitact{k}{\defect{a}{g},\defect{b}{h};\defect{c}{gh}}  \otimes \dw{k}{\defect{b}{h}} 
\, .
\end{align}
This projects onto the $\pi^\kappa_{\defect{c}{gh}}\big|_{\centralizer{\defect{a}{g},\defect{b}{h},\defect{c}{gh}}}$ rep within $\left( \pi^\mu_{\defect{a}{g}}\,  \otimes \splitact{$(\, \cdot\, )$}{\defect{a}{g},\defect{b}{h};\defect{c}{gh}} \,  \otimes \pi^\nu_{\defect{b}{h}} \right)\big|_{\centralizer{\defect{a}{g},\defect{b}{h},\defect{c}{gh}}}$. 
To calculate the fusion multiplicity one must also include a summation over defect orbits $[ \defect{a}{g}],$ $[ \defect{b}{h}],$ $[ \defect{c}{gh}]$. 
We implement this by summing over pairs $(\act{x}\defect{a}{g},\act{y}\defect{b}{h})$, for $ \g{x} \in \G / \centralizer{\defect{a}{g}}, \, \g{y} \in \G / \centralizer{\defect{b}{h}}$, modulo left multiplication $(\g{zx},\g{zy})$, since that leads to defects $\act{z}(\act{x}\defect{a}{g}\times\act{y}\defect{b}{h})$ in the same orbit. Hence
\begin{align}
&N_{{\gauged{\defect{a}{ g}}{\mu}}   {\gauged{\defect{b}{ h}}{\nu}}}^{{\gauged{\defect{c}{k}}{ \kappa  }}}
= \sum_{ \substack{(\g x , \g y ) \in \centralizer{\defect{a}{g}} \backslash \G / \centralizer{\defect{b}{h}} \\  \g{z} \in \G / \centralizer{\defect{c}{k}}}}
\delta_{\act{x}\defect{a} {g} \act{y}\defect{b}{ h}}^{ \act{z}\defect{c}{k}} 
\nonumber \\
& 
m\left (\act{z}\pi^\kappa_{\defect{c}{k}}\big|_{\centralizer{\act{x}\defect{a}{g},\act{y}\defect{b}{h},\act{z}\defect{c}{k}}} ,
\act{x}\pi^\mu_{\defect{a}{g}}\big|_{\centralizer{\act{x}\defect{a}{g},\act{y}\defect{b}{h},\act{z}\defect{c}{k}}}  \,  
\otimes \splitact{$(\, \cdot\, )$}{\act{x}\defect{a}{g},\act{y}\defect{b}{h};\act{z}\defect{c}{k}}\big|_{\centralizer{\act{x}\defect{a}{g},\act{y}\defect{b}{h},\act{z}\defect{c}{k}}}  \,  
\otimes \act{y}\pi^\nu_{\defect{b}{h}}\big|_{\centralizer{\act{x}\defect{a}{g},\act{y}\defect{b}{h},\act{z}\defect{c}{k}}}  \right)
\, ,
\nonumber
\end{align}
where $m$ is the projective rep multiplicity counting function defined in Eq.~\eqref{set:definitionofmfunction}.

    \section{ Anyon condensation phase transitions dual to gauging}
  \label{setsection:condensation}

In this section, we study the $\repg$ anyon condensation phase transition induced by breaking a $\G$-graded MPO symmetry down to the $\g{1}$-sector subalgebra~\cite{PhysRevB.95.235119,garre2017symmetry}. This corresponds to de-equivariantization in the category theory language used in Ref.~\cite{turaev2000homotopy,kirillov2004g,etingof2009fusion,gelaki2009centers}. 
These phase transition are dual to those induced by gauging a global $\G$ symmetry, which were studied in the previous section. 
We recount how symmetry breaking perturbations to a local tensor lead to tensor network representations of an anyon condensate~\cite{chen2010tensor,shadows,shukla2016boson}. Furthermore, we present a decomposition of these perturbations into topological charge sectors that identify which anyons are condensed by a given perturbation. 
A procedure to extract the defect superselection sectors and domain walls of the condensed theory from the topological superselection sectors of the original theory is explained. This allows one to identify which anyons split, which are identified, and which are confined during the condensation phase transition, as well as the full U$\G$xBFC description of the resulting SET. 
We also outline how different --- Morita equivalent --- MPO algebras, that lead to the same emergent topological order, can be used to find different anyon condensation phase transitions. It is conjectured that this approach will recover all  possible boson condensation transitions from a given nonchiral topological order.

\subsection{MPO symmetry breaking perturbations and anyon condensation} 

We first explain how anyon condensation phase transitions, induced by a reduction of the topological MPO symmetry algebra, can be analyzed with superselection sector ICIs. The reduction of the topological MPO symmetry algebra can either originate from explicit symmetry breaking~\cite{chen2010tensor,shukla2016boson,garre2017symmetry}, due to a non-symmetric perturbation to the local tensor, or from spontaneous symmetry breaking~\cite{PhysRevB.95.235119,shadows,marien2016condensation}, where the local tensor is varied symmetrically until the boundary theory undergoes a phase transition.  

For the case of explicit symmetry breaking, it was argued in Ref.~\cite{shukla2016boson} that perturbations to the local tensor can be classified according to the MPO symmetry they respect. 
The perturbations that are not even symmetric under the vacuum $\mpo{}{\vac}$ do not change the tensor network, while perturbations respecting $\mpo{}{\vac}$ lead to fluctuating anyons on top of the tensor network ground state vacuum. The full MPO algebra symmetry of the unperturbed tensor network implies that the fluctuating anyons satisfy a global charge constraint.  

The perturbations respecting $\mpo{}{\vac}$ can be further divided into sectors corresponding to the subset of irreducible idempotents of the tube algebra that come attached to a vacuum string $\mpo{}{\vac}$.  
The set of ICI sectors on which a perturbation has support identify the emergent anyons that it condenses. 

We denote the set of irreducible idempotents that can be realized at the end of a vacuum $\mpo{}{\vac}$ string by $\{\ici{a_0}\}$. Hence we have 
\begin{align}
\ici{a_0} = \frac{1}{\qd^2}  \sum_{s}  t^{\vac s \vac s}_{a_0} \, \tube{\vac s\vac}{s}
= \frac{1}{\qd^2}  \sum_{s}  t^{\vac s\vac s}_{a_0} \, \mpo{}{s} \, .
\end{align}
Each $\ici{a_0}$ corresponds to an internal state of a topological superselection sector satisfying $\ici{a} = (\ici{a_0} + \dots)\, $. 
In particular the vacuum sector $\ici{\vac}$ always appears in this set of idempotents that occur at the end of  $\mpo{}{\vac}$. 

Due to Eq.~\eqref{set:icipartition} any tensor perturbation $\pepspert{}$ can be partitioned into a sum over $\ici{a_0}$ sectors 
\begin{align}
\label{set:decompositionoflocaltensorpert}
\pepspert{}=\pepspert{\times} + \sum_{a_0} \pepspert{a_0} 
\, ,
\end{align}
where $\pepspert{\times}$ is orthogonal to $\mpo{}{\vac}$ and hence leads to a zero state, while $\pepspert{a_0}$ satisfies $\pepspert{a_0} \ici{a_0} = \pepspert{a_0}$, and hence lies in the $\ici{a}$ charge sector. The perturbations  $\pepspert{\vac}$, satisfying $\pepspert{\vac}\, \ici{\vac} = \pepspert{\vac}$, are symmetric under the full MPO algebra. 

Let $\ket{\Psi[\peps^{(i)}]}$ denote the tensor network resulting from the contraction of tensors $\peps^{(i)}$ on every site $i$ of some lattice. We assume the local tensor is symmetric under the full MPO algebra i.e. $\peps \, \ici{\vac}=\peps$.  
A perturbed tensor network is given by 
\begin{align}
\label{set:perturbedtensornetworkanyoncondensate}
\ket{\Psi[(\peps+ \varepsilon_\times \pepspert{\times}+\sum_{a_0} \varepsilon_{a_0} \pepspert{a_0})^{(i)}]}
= \ket{\Psi[\peps]} &+
\sum_{a_0} \varepsilon_{a_0} \sum_{v} \ket{\Psi[\peps^{(i\neq v)},\pepspert{a_0}^{(v)}]}
\nonumber \\
&+\sum_{a_0,b_0}\varepsilon_{a_0}\varepsilon_{b_0}\sum_{u,v} \ket{\Psi[\peps^{(i\neq u, v)},\pepspert{a_0}^{(u)},\pepspert{b_0}^{(v)}]}
\nonumber \\
&+ \dots
\, ,
\end{align} 
where  $i,u,v$ are lattice sites. We remark that the first order perturbations, aside from $ \pepspert{\vac}$, generically contribute a zero state for a closed manifold due to a global charge constraint. 
The resulting tensor network states can be interpreted as anyon condensates, as they involve a fluctuation of the charges $a_0$ with $\varepsilon_{a_0}\neq 0$ on top of the tensor network ground state vacuum. 

We expect the case of spontaneous MPO symmetry breaking to follow a similar picture. That is, once the phase transition has been crossed the physical state becomes a cat (GHZ) state given by a sum of explicit MPO symmetry breaking tensor networks, each of which can be interpreted as an anyon condensate.

\subsection{Extracting the SET order that results from $\repg$ anyon condensation}
  
Due to the additional structure appearing in the ICIs of a tube algebra derived from a graded MPO algebra $\cat_\G$, it is possible to infer the defect ICIs and domain walls of the $\cat_\G$ dube algebra from the anyon ICIs of the $\cat_\G$ tube algebra. This procedure allows us to calculate the effect of condensing a  bosonic $\repg$ subtheory of $\double{\cat_\G}$ that induces a phase transition to $\rdouble{\cat_\G}{\cat_\g{1}}$. While the anyon theory that results from this condensation transition is known~\cite{barkeshli2014symmetry} to be $\double{\cat_\g{1}}$, which can simply be constructed by finding the ICIs of the $\cat_\g{1}$ tube algebra, we go beyond this to explicitly derive the full SET defect theory $\rdouble{\cat_\G}{\cat_\g{1}}$. In particular we calculate which ICIs split, which become identified, and which become confined defect ICIs during the condensation phase transition~\cite{PhysRevB.79.045316}. The resulting dube ICIs and domain walls can be used to construct the full U$\G$xBFC description of the resulting SET, by following the procedure explained in Section~\ref{setsection:dubes}.

We assume that the MPO algebra $\cat_\G$ of the model under consideration admits  a $\G$-grading. Hence there is a group element assigned to each of the single block MPOs $\sector{a}\in\G$.  
The analysis of Refs.~\cite{chen2010tensor,shukla2016boson} can be extended to the case where the full virtual MPO symmetry $\cat_\G$ is broken down to a nontrivial residual MPO symmetry, given by the $\g{1}$-sector subalgebra $\cat_\g{1}$.
 This corresponds to an anyon condensation transition from $\double{\cat_\G}$ to a nontrivial topological phase $\double{\cat_\g{1}}$. 
To calculate the relationship between the $\double{\cat_\G}$ theory and the condensate $\rdouble{\cat_\G}{\cat_\g{1}}$ we examine the implications of a $\G$-grading for the $\cat_\G$ tube algebra ICIs. 
For an ICI of the $\cat_\G$ tube algebra 
\begin{align}
\ici{a} &= \frac{1}{\qd^2}  \sum_{pqs}  t^{pqps}_{a} \, \tube{pqp}{s} \, ,
\end{align}
we can separate out sectors according to the group elements $\sector{p}$ and $\sector{s}$ 
\begin{align}
(\ici{a})_{\g{g}}^{\g{h}} &:= \frac{1}{\qd^2}  \sum_{pqs} \delta_{\sector{p},\g{g} } \, \delta_{\sector{s},\g{h}} \,  t^{pqps}_{a} \, \tube{pqp}{s} \, ,
\end{align}
we must have $\g{gh}=\g{hg}$ for a nonzero $(\ici{a})_{\g{g}}^{\g{h}}$.  

Each ICI $\ici{a}$ can be written as a sum of subidempotents
\begin{align}
\ici{a} &= \sum_{\g{g}\in \conjclass{}}\, \defect{(\ici{a})}{g}
\, ,
&&
\text{where}
&&
\defect{(\ici{a})}{g} :=
\sum_{\g{h}\in \centralizer{\g{g}}} (\ici{a})_{\g{g}}^{\g{h}}  \, 
\end{align}
for some conjugacy class $\conjclass{}$. 

Since the MPO symmetry has been broken down to $\cat_\g{1}$, the MPOs from the $\cat_\g{g}$ sectors for $\g{g}\neq \g{1}$ now correspond to immobile domain walls in the tensor network. Hence, tubes $\tube{pqr}{s}$ with $\sector{s}\neq 1$ should be dropped from the tube algebra. Furthermore, the remaining tubes $\tube{pqr}{s}$ generate a $\G$ dube algebra, that is graded by the sector $\sector{p}=\sector{q}=\sector{r}$.  
By dropping the immobile domain walls from the topological symmetry algebra, the  subidempotents of an ICI $\ici{a}$ are mapped as follows 
\begin{align}
\defect{(\ici{a})}{g}
\, \mapsto \, 
  (\ici{a})_{\g{g}}^{\g{1}}
 \, .
\end{align}
The resulting idempotents $ (\ici{a})_{\g{g}}^{\g{1}}$ are attached to an MPO string in the $\g g$-sector. For $\g{g}\neq\g{1}$ the idempotents correspond to confined defects, as they are connected to a physically observable $\g{g}$-domain wall that is detected by the local terms of the PEPS parent Hamiltonian~\cite{GarciaVerstraeteWolfCirac08}, and hence the energy penalty of a defect pair scales with their separation. 

For a nontrivial conjugacy class $|\conjclass{}|>1$ the ICI $\ici{a}$ splits into multiple defects 
\begin{align}
\ici{a} \,  \mapsto \, \{ \,  K \, (\ici{a})_{\g{g}}^{\g{1}} \, \}_{\g g \in \conjclass{}}
\, , 
\end{align}
for some positive constant normalization $K>0$ such that $ K (\ici{a})_{\g{g}}^{\g{1}}$ is a projector. 
We can also calculate which ICIs condense into the same defect, by looking for all $\ici{a}$ and $\ici{b}$ that satisfy
\begin{align}
(\ici{a})_{\g{g}}^{\g{1}} = \widetilde K \, (\ici{b})_{\g{g}}^{\g{1}} \neq 0
\, ,
\end{align}
for some positive constant $\widetilde K$. 

The centralizer $\centralizer{\defect{a}{g}}$ of a defect $\defect{a}{g}$ is given by the set of group elements $\g{h}$ that lead to a nonzero $(\ici{a})_{\g{g}}^{\g{h}}$, and the orbit of a defect $[{(\ici{{a}})}^{\g{1}}_\g{g}]$ is given by the set of defects $  (\ici{a})_{\act{k}\g{g}}^{\g{1}}$ that are nonzero.

The $(\ici{a})_{\g{1}}^{\g{1}}$ idempotents give the anyon ICIs of the condensed topological order. The procedure explained above allows one to determine the splitting and condensation relations between the anyons of $\double{\cat_\G}$ and those of $\double{\cat_\g{1}}$.  
In particular, we can calculate the anyons that condense into the vacuum
\begin{align}
\ici{\defect{\vac}{1}} =   \sum_{s} \delta_{\sector{s},\g{1}} \, \frac{d_s}{\qd_\g{1}^2} \, \tube{\vac s \vac}{s}
\, ,
\end{align}
by looking for those $\ici{a}$ that satisfy $(\ici{a})_{\g{1}}^{\g{1}}= \frac{1}{|\G|} \ici{\defect{\vac}{1}}$.  
For condensation induced by symmetry breaking down to the $\g{1}$-sector of a $\G$-graded MPO algebra, the ICIs that condense to vacuum are precisely given by $\ici{\gauged{\defect{0}{1}}{\mu}}$, see Eq.~\eqref{set:gaugedtubeformula}. These superselection sectors correspond to bosonic $\G$-charges, which generate a $\repg$ subtheory of $\double{\cat_\G}$ that condenses to yield the resulting SET $\rdouble{\cat_\G}{\cat_\g{1}}$.

Hence symmetry breaking perturbations to the local tensor can be divided into irreps, or $\G$-charges, 
\begin{align}
\pepspert{}= \varepsilon_\times \pepspert{\times} + \sum_{\mu} \varepsilon_\mu \pepspert{\mu} 
\, ,
\end{align}
similar to Eq.~\eqref{set:decompositionoflocaltensorpert}. 
When a tensor network is built from a symmetric local tensor $\peps$ with such a perturbation $\pepspert{}$ --- for $\varepsilon_\mu\neq 0$ --- the $\G$-charges fluctuate and induce a $\repg$ anyon condensation, see Eq.~\eqref{set:perturbedtensornetworkanyoncondensate}.  

To construct a tensor network for the condensed SET, where the symmetry is realized on-site, we can modify the local tensor with a perturbation that is correlated to a newly introduced physical group variable  
\begin{align}
\peps \, \mapsto \,  \sum_{\g{g}} \ket{\conj{\g{g}}} \otimes (\peps + \pepspert{0} \, \mpo{}{\g{g}} )
\, .
\end{align}
One can verify that this tensor has the correct symmetry transformation under a physical group action, given by the left regular representation, acting on the newly introduced variable. This corresponds to ungauging a $\G$-gauge symmetry, or equivalently gauging a 1-form symmetry~\cite{PhysRevB,Gaiotto2015,kapustin2013higher}. 
We remark that the perturbations $\pepspert{0} \, \mpo{}{\g{g}}$ can be constructed from a linear combination of the $\pepspert{\mu}$ perturbations.

Beyond the defect ICIs, we can also extract the $\G$ domain walls of the SET resulting from $\repg$ condensation
\begin{align}
{\dw{\defect{a}{g}}{h}} = K \, (\ici{a})_{\g{g}}^{\g{h}}
\, ,
\end{align}
where $\g{gh}=\g{hg}$, and $K>0$ is a normalization fixed by the equation $({\dw{\defect{a}{g}}{h}})^\dagger{\dw{\defect{a}{g}}{h}}=\ici{\defect{a}{g}}$. 
These domain walls can be used to extract the second cohomology label of each defect ${[\eta_a]\in\twistedcoho}$  that describes its projective transformation under the global symmetry. 

Furthermore, the domain walls define the full set of symmetry-twisted MES on the torus $\ket{\dw{\defect{a}{g}}{h}}$, from which the $\G$-crossed $S$ and $T$ matrices can be extracted, see Section~\ref{set:stmesandsandtmatrices}.  
To calculate the permutation action on the defects, and symmetry twisted states, one can simply fuse an $\mpo{}{\g k}$ loop into the skeleton of the torus
\begin{align}
 \ket{\mpo{}{\g k}\circ \dw{h}{\defect{a}{g}}} 
\, \mapsto \,
\frac{\factsys{a}{\g k , \g h } \factsys{\act{k}a}{\g{kh} , \conj{\g{k}}}}{\factsys{a}{\conj{\g{k}}, \g{k} }}
\ket{\dw{$\act{k}$h}{\act{k}\defect{a}{g}}} 
\, ,
\end{align}
see Eq.~\eqref{set:mpoloopsymtwistedsectorgroupaction}.

\subsection{Using Morita equivalences to find all anyon condensation phase transitions}
\label{set:algebragradingmposymmetrybreakingcondensation}

The concept of Morita equivalence plays an important role in the search for all possible boson condensation transitions in the framework we have developed. 
Two MPO algebras $\cat,\, \widetilde\cat$ are Morita equivalent, denoted $\cat\sim \widetilde\cat$, precisely when they lead to the same emergent anyons~\cite{muger2003subfactors,DrinfeldCenter}, $\double{\cat}\cong\double{\widetilde\cat}$. That is, when the physical properties derived from the ICIs of the tube algebras match. 

Different, Morita equivalent, MPO algebras that lead to the same emergent topological order can exhibit different $\G$-gradings and different $\g{1}$-sectors $\cat_\g{1}$.  Hence, by considering such different presentations of the same emergent topological order we can find different $\repg$ condensation transitions from $\double{\cat}$. 
An example of this phenomenon is given in Section~\ref{setsection:examples2}. 

By Ref.~\cite{etingof2009fusion} all $\repg$ condensation transitions from a given MTC $\double{\cat}$, which are dual to gauging an on-site $\G$ symmetry on a nonchiral topological order, can be realized by considering Morita equivalent MPO algebras that lead to the same $\double{\cat}$ emergent topological order. 
To find Morita equivalent representations more explicitly  would require a framework for invertible domain walls between different MPO algebras. This is more general than the $\G$-extension problem considered in Section~\ref{set:groupextensionclassification}, which only considered invertible domain walls between the same MPO algebra. We plan to address this in future work.

The process we have outlined for calculating the effects of anyon condensation, by breaking the MPO symmetry down to a nontrivial subalgebra, is straightforwardly generalized beyond the case of a $\G$-grading.  The most general case corresponds to a ``grading'' by an algebra, note it is trivial that every MPO symmetry is ``graded'' in this way by its own fusion algebra.  

Motivated by the $\G$-graded case, \emph{we conjecture} that all possible boson condensation phase transitions from a nonchiral topological phase $\double{\cat}$ can be realized by symmetry breaking 
down to the $\g{1}$-sector of some algebra-``graded'' 
 MPO symmetry algebra in the $\double{\cat}$ Morita equivalence class. 
 Further results in this direction will be presented in a forthcoming work~\cite{MPOanyoncondprep}.

Beyond computing the effects of anyon condensation, the MPO symmetry breaking procedure leads to a simple method to derive the ICIs of a UFC $\cat$ from the ICIs of a modular extension of $\cat$. This is because $\cat$ naturally forms a subcategory of any modular extension of $\cat$. 
We expect this to prove useful as explicit formulas exist for the ICIs of a modular theory, making them particularly easy to find~\cite{koenig2010quantum,davepriv}.

\section{Further examples}
\label{setsection:examples2}

In this section, we present several examples that demonstrate different aspects of the constructions explained throughout the paper. 
First, the $\zt$-extension problem for $\vecg{\zt}{}$ is solved explicitly. 
Then, a Morita equivalent description of $\double{\vecg{\z_{4}}{}}$ is used to study a $\rep{\zt}$ condensation phase transition to the doubled semion model $\double{\vecg{\zt}{\omega_{\text{I}}}}$, induced by MPO symmetry breaking. 
Finally, we study a condensation phase transition from $\double{\vecg{S_3}{}}$ to $\double{\vecg{\z_{3}}{}}$ which increases the number of anyon types.

\subsection{ $\zt$-extension of $\vecg{\zt}{}$ and symmetry-enriched toric codes}
\label{set:tcextensioneg}

The $\zt$-extension problem for an underlying UFC $\vecg{\zt}{}=\{\vac,\psi\}$ provides a simple and illustrative example of the extension procedure described in Section~\ref{set:groupextensionclassification}. 
It corresponds to classifying anomaly-free $\zt$ SET orders, where the underlying topological order is the toric code $\double{\vecg{\zt}{}}$. 
One of these extensions corresponds to the EM duality-enriched toric code, which was analyzed in detail in Section~\ref{setsection:example1}.

An extension is calculated in two stages, firstly we search for consistent $\zt$-graded fusion rules containing $\vecg{\zt}{}$ as the trivial sector. Note the $\zt=\{\g{1},\g{x}\}$ action on the objects of $\vecg{\zt}{}$ must be trivial, as there is only a single nontrivial object.
Since the total quantum dimension of each $\g{g}$-sector must be the same we have $\qd_\g{x}^2=\qd^2_\g{1}=2$. Hence the $\g{x}$-sector can either consist of a single object $\sigma$ with quantum dimension $\sqrt{2}$, or a pair of inequivalent objects $\sigma_+,\sigma_-$ both with quantum dimension $1$. 
We find consistent fusion rules for both of these cases, for which the potential obstructions in $H^3(\zt,\zt^2)$ disappear. 

In the former case the fusion rules must be
\begin{align}
\sigma \times \sigma = \vac + \psi \,  ,
\end{align}
and the action of the $H^2$ torsor, permuting $\vac$ and $\psi$, is trivial. Hence the unique choice is given by the Ising fusion rules. 

In the later case there is a trivial solution to the fusion rules 
\begin{align}
\sigma_\pm \times \sigma_\pm = \vac\, ,
&& 
\sigma_\pm \times \sigma_\mp = \psi \, ,
\end{align}
which corresponds to $\zt\times\zt$ fusion rules. 

In this case the action of the $H^2$ torsor changes the fusion rules to an inequivalent choice
\begin{align}
\sigma_\pm \times \sigma_\pm = \psi \, ,
&& 
\sigma_\pm \times \sigma_\mp = \vac \, ,
\end{align}
which corresponds to $\z_4$ fusion rules. This exhausts the different choices of consistent fusion rules. 

The possible obstruction at the second stage also vanishes since $H^4(\zt,\Uone)=0$. For each choice of fusion rule we can now look for solutions to the pentagon equation. These are $H^3(\zt,\Uone)$ torsors, which can be realized through multiplication of a representative $F$-symbol with the nontrivial 3-cocycle function $\alpha(\g{g},\g{h},\g{k})=(-1)^{\g{g}\g{h}\g{k}}$. That is
\begin{align}
\shortf{\defect{a}{g} \defect{b}{h} \defect{c}{k}}{d}{e}{f} \mapsto \alpha(\g g,\g h,\g k)\shortf{\defect{a}{g} \defect{b}{h} \defect{c}{k}}{def}
\, ,
\end{align}
note the only nontrivial value of $\alpha$ is $\alpha(\g{x},\g{x},\g{x})=-1$.

Solutions to the pentagon equation for each choice of graded fusion rules are listed below, where we use the notation of Ref.~\cite{propitius,DEWILDPROPITIUS1997297} for the 3-cocycles $\omega_{\text{I}},\, \omega_{\text{II}}$. 
\begin{itemize}
\item[Ising]
\begin{itemize}
\item
For Ising fusion rules, the $F$-symbols in Eq.~\eqref{set:isingfsymbols} solve the pentagon equation. In this case the FS indicator of $\sigma$ is given by $\fs{\sigma}=1$. As discussed in Section~\ref{set:isingexample} this corresponds to the EM duality enriched toric code that gauges to the doubled Ising model, $\text{Ising}^{(1)} \boxtimes  \overline{\text{Ising}}^{(1)}$. 
\item
Multiplication with the 3-cocycle $\alpha(\sigma,\sigma,\sigma)=-1$ flips the phase of $F^{\sigma \sigma \sigma}_{\sigma \vac\vac}$ and hence yields $\fs{\sigma}=-1$. 
This corresponds to the EM duality enriched toric code with a pair of self-inverse defects with negative FS indicators. Gauging this leads to the Drinfeld center of the modified Ising model with $\fs{\sigma}=-1$, $\text{Ising}^{(3)} \boxtimes  \overline{\text{Ising}}^{(3)}$. 
\end{itemize}
\item[$\zt^2$]
\begin{itemize}
\item
For $\zt\times\zt$ fusion rules, there is a trivial solution $F^{ghk}_{abc}=\delta_{ghk,a}\delta_{gh,b}\delta_{hk,c}$ which corresponds to a toric code tensor producted with a trivial paramagnet. This gauges to two copies of the toric code.  
\item
Multiplying by $\alpha(\sigma_\pm,\sigma_\pm,\sigma_\pm)=-1$ leads to a $\zt$-extension of $\vecg{\zt}{}$ with two self inverse defects that have negative FS indicator $\fs{\sigma_+}=\fs{\sigma_-}=-1$. This corresponds to a symmetry-enriched toric code that gauges to $\double{\vecg{\zt\times\zt}{\omega_{\text{I}^{(2)}}}}$. 
\item 
There is another distinct solution to the pentagon equation with these fusion rules, corresponding to a 3-cocycle $\omega_{\text{I}^{(1)}} \omega_{\text{II}}$, which gives a toric code tensor producted with a nontrivial $\zt$ SPT. This gauges to a toric code tensor producted with a doubled semion model. 
\\
Multiplying by $\alpha(\sigma_\pm,\sigma_\pm,\sigma_\pm)=-1$ in this case corresponds to relabeling $\sigma_\pm$, and hence yields the same solution. 
\end{itemize}
The general $F$-symbol solutions for $\zt\times\zt$ fusion rules are given by 3-cocycles representatives of the elements $H^3(\zt\times\zt,\Uone)=\zt^3$, which are generated by $\omega_{\text{I}^{(1)}}, \, \omega_{\text{I}^{(2)}}, \, \omega_{\text{II}}$. For a fixed choice of $\zt$ subgroup, corresponding to $\cat_\g{1}$, there are four inequivalent 3-cocycles with a trivial restriction. Two of these 3-cocycles are captured by the first two cases above, while the other two are both captured by the third case. 
\item[$\z_4$]
\begin{itemize}
\item
For $\z_4$ fusion rules, again there is the trivial solution, which corresponds to a toric code with two pairs of topologically distinct defects that are mutual inverses. This gauges to the $\z_4$ quantum double model $\double{\vecg{\z_4}{}}$. 
\item
Multiplying by $\alpha(\sigma_\pm,\sigma_\pm,\sigma_\pm)=-1$ corresponds to a different symmetry-enriched toric code with two pairs of mutual inverse defects. This gauges to the $\omega_{\text{I}}^2$-twisted $\z_4$ quantum double model $\double{\vecg{\z_4}{\omega_{\text{I}}^2}}$. 
\end{itemize}
The general $F$-symbol solutions for $\z_4$ fusion rules are given by 3-cocycles representatives of the elements $H^3(\z_4,\Uone)=\z_4$. There are two 3-cocycles that have trivial restriction to the $\zt$ subgroup. These correspond to the two cases above. 
\end{itemize}

We remark that similar treatments of this example have appeared in Refs.~\cite{cheng2016exactly,PhysRevB.94.235136}.

\subsection{Morita equivalent MPO algebras ${\normalfont \vecg{\z_{4}}{}\sim \vecg{\zt\times\zt}{\omega_{\text{II}}}}$ and condensation to the doubled semion model}
\label{setsection:moritaequivegone}

In Ref.~\cite{PhysRevB.95.235119} a condensation phase transition from $\double{\vecg{\z_4}{}}$ to the doubled semion model $\vecg{\zt}{\omega_{\text{I}}}$ was described in terms of a boundary SPT phase under the MPO symmetry. Here we revisit this condensation with our framework, and by using a Morita equivalent description of the topological order $\double{\vecg{\z_4}{}} \cong \double{\vecg{\zt\times\zt}{\omega_{\text{II}}}}$ we find a reduction of the MPO symmetry that induces the same $\rep{\zt}$ condensation phase transition without any reference to a boundary SPT. This exemplifies the general fact, that all $\repg$ condensation phase transitions can be realized through $\G$-graded MPO symmetry breaking. 
Hence any condensation phase transition corresponding to a boundary SPT phase in the language of Ref.~\cite{PhysRevB.95.235119}, can be induced by symmetry breaking in some Morita equivalent MPO algebra using our framework. 
Our approach has the advantage that it captures all $\repg$ condensation phase transition, while no such proof was given in Ref.~\cite{PhysRevB.95.235119}.

There is a well known Morita equivalence $\vecg{\z_{4}}{}\sim \vecg{\zt\times\zt}{\omega_{\text{II}}}$, see Ref.~\cite{propitius}, and hence 
${ \double{\vecg{\z_{4}}{}}\cong \double{\vecg{\zt\times\zt}{\omega_{\text{II}}}} }$. 
Therefore, to describe an emergent $\double{\vecg{\z_{4}}{}}$ topological order we may use an MPO algebra $\vecg{\zt\times\zt}{\omega_{\text{II}}}$. 

We first construct the tube algebra for $\vecg{\zt\times\zt}{\omega_{\text{II}}}$ and find its ICIs, from which one can verify the equivalence ${ \double{\vecg{\z_{4}}{}}\cong \double{\vecg{\zt\times\zt}{\omega_{\text{II}}}} }$.  Since the fusion rules are Abelian and all group elements are self inverse, the tubes appearing are of the form $\tube{g, g+h, g}{h}$ and we use the shorthand notation $\tube{g}{h}:= \alpha(h,g+h,h) \tube{g ,g+h ,g}{h}$, established in Eq.~\eqref{set:sptexampledwgaugechoice}. 
This tube algebra has multiplication rules
\begin{align}
\tube{f}{h}\tube{g}{k} = \delta_{f,g} \,  \alpha^{g}(h,k) \tube{g}{h + k}
\, ,
\end{align}
where $\alpha^{g}$ is the slant product defined in Eq.~\eqref{set:equationwiththeslantproduct}. The 3-cocycle is given by $\alpha\equiv\omega_{\text{II}}$ where
\begin{align}
\omega_{\text{II}}(A,B,C)=(-1)^{a_0b_1 c_1} \, , && \text{and}&&  A=(a_0,a_1)
\, .
\end{align}
Using the ICI formula for twisted gauge theory, given in Eq.~\eqref{set:fullorbiticiforgaugethy}, we find 
\begin{align}
\bgroup		
\setlength\tabcolsep{.09cm}
\begin{tabular}{ | c | c c c c c c c c c c c c c c c c  | } 
\hline
Anyons 
 & \tube{00}{00}   & \tube{00}{01}  & \tube{00}{10}  & \tube{00}{11} & \tube{01}{00} & \tube{01}{01}  & \tube{01}{10}  & \tube{01}{11}  & \tube{10}{00}   & \tube{10}{01}  & \tube{10}{10}  & \tube{10}{11} &  \tube{11}{00}   & \tube{11}{01}  & \tube{11}{10}  & \tube{11}{11}
\\ \hline
$(00,++)$
 & $1$  & $1$ & $1$ & $1$ & & &  & & & & & & & & &
 \\ 
$(00,+-)$
 &  $1$ & $-1$ & $1$ & $-1$ & & &  & & & & & & & & &
 \\ 
 $(00,-+)$
 & $1$  & $1$ & $-1$ & $-1$ & & &  & & & & & & & & &
 \\ 
 $(00,--)$
 & $1$ & $-1$  & $-1$ & $1$ & & &  & & & & & & & & &
 \\ 
$(01,++)$
 &  &  & & & $1$ & $1$ & $1$  & $1$ & & & & & & & &
 \\ 
$(01,+-)$
 &  &  & & & $1$ & $-1$ & $1$  & $-1$ & & & & & & & &
 \\ 
 $(01,-+)$
 &  &  & & & $1$ & $1$ & $-1$  & $-1$ & & & & & & & &
 \\ 
 $(01,--)$
 &  &  & & & $1$ & $-1$ & $-1$  & $1$ & & & & & & & &
 \\ 
$(10,++)$
 &  &  & & & & &  & & $1$ & $-i$ & $1$ & $-i$ & & & &
 \\ 
$(10,+-)$
 &  &  & & & & &  & & $1$ & $i$ & $1$ & $i$ & & & &
 \\ 
 $(10,-+)$
 &  &  & & & & &  & & $1$ & $-i$ & $-1$ & $i$ & & & &
 \\ 
 $(10,--)$
 &  &  & & & & &  & & $1$ & $i$ & $-1$ & $-i$ & & & &
 \\ 
$(11,++)$
 &  &  & & & & &  & & & & & & $1$ & $-i$ & $1$ & $-i$
 \\ 
$(11,+-)$
 &  &  & & & & &  & & & & & &  $1$ & $i$ & $1$ & $i$
 \\ 
 $(11,-+)$
 &  &  & & & & &  & & & & & &  $1$ & $-i$ & $-1$ & $i$
 \\ 
 $(11,--)$
 &  &  & & & & &  & & & & & & $1$ & $i$ & $-1$ & $-i$
 \\ 
\hline
\end{tabular}
\egroup
\end{align}
 the table gives the elements  $t_{i}^{j}$ of the ICIs and should be read as follows
\begin{align}
\ici{a_i} = \frac{1}{\qd^2} \sum_{j} t_{i}^{j} (\tube{h}{g})_j ,
\end{align}
where $\qd^2=4$ in this case.

There are several $\zt$-gradings of $\vecg{\zt\times\zt}{\omega_{\text{II}}}$ that we could consider for $\cat_{\g{1}} \oplus \cat_{\g{x}} $
\begin{align}
\{00,01\}\oplus \{ 10,11\}, \, && \{00,10\}\oplus \{ 01,11\}, \, && \{00,11\}\oplus \{ 01,10\}
\, .
\end{align}
 The first two are equivalent and lead to a condensation to the toric code phase, while the third leads to a condensation to the doubled semion phase. This is because the restriction of the 3-cocycle $\omega_{\text{II}}$ to the $\{00,11\}$ subgroup lies in the nontrivial cohomology class of ${\cohom{3}{\zt}=\zt}$. 
We remark that shifting by a $\omega_I$ cocycle on any of the $\zt$ subgroups rearranges which of the above $\zt$-gradings corresponds to the doubled semion phase.

We choose the $ \{00,11\}\oplus \{ 01,10\}$ grading for the MPO symmetry algebra. Then breaking the MPO symmetry down to $\cat_{\g{1}}$ leads to a $\rep{\zt}$ condensation to a symmetry enriched doubled semion phase described in the table below. 
\begin{align}
\label{set:zfourdsemeg}

\end{align}
The domain walls for the nontrivial element $\g{x}$ are given by
\begin{align}
	%
\end{align}
where a normalization by $\qd_\g{1}^2=2$ is implicit.  For a more detailed guide to reading the tables, see Eq.~\eqref{set:defectisingexample}. 

Denote the $\zt$ generator on defect  $a$ by $X_a$, then $X_{a}^2=1$ for $a\in\{\vac, s\conj{s}, \sigma_0, \sigma_1\}$ and  $X_{a}^2=-1$ for $a\in\{ s , \conj{s}, \sigma_2, \sigma_3\}$. We remark, in this case, the projective phase $X_{a}^2=-1$ can be removed by a choice of coboundary. It is important, however, to keep track of the choice when calculating the effect of gauging the symmetry.

This example demonstrates that the topological data obtained by gauging an SET is not a faithful label. 
In particular one finds the same gauged data for the symmetry-enriched doubled semion, described in Eq.~\eqref{set:zfourdsemeg}, and the symmetry-enriched toric code, described as the first case under $\z_4$ in Section~\ref{set:tcextensioneg}.  For this example, the gauge invariant $S^{(\g{1},\g{1})},T^{(\g{1},\g{1})}$ matrices of the SETs clearly distinguish the two SET phases, as they have different underlying topological orders. 

We remark that even when the underlying topological order is the same, the gauged data still does not provide a faithful label. A well known example of this occurs for the trivial $D_4$ SPT and the type-III $\zt^3$ SPT, since ${\double{\vecg{D_4}{}}\cong \double{\vecg{\zt^3}{\omega_{\text{III}}}}}$ and the underlying topological orders are both trivial~\cite{propitius,PhysRevB.91.195134,bridgeman2017anomalies}. In this case the SPTs are clearly distinguished by their different symmetry groups.

  \subsection{ A phase transition from $\double{ \vecg{S_3}{} }$ to $\double{\vecg{\z_3}{}}$ }

For a final example, we use the $\zt$-grading of $\vecg{S_3}{}$ to study a phase transition to $\double{\vecg{\z_3}{}}$. We remark that one can start from the Morita equivalent $\rep{S_3}$ and follow a similar approach to study a phase transition to $\double{\rep{\z_2}}$ that is not induced by a $\repg$ boson condensation. We plan to explore this further in a future work~\cite{MPOanyoncondprep}.

We work with the following presentation of $S_3$ 
    \begin{align}
\left \langle s,r  \ |\ s^2=r^3=(sr)^2=1 \right\rangle
\, ,
  \end{align}
hence the set of group elements is given by $\{ 1,\, r,\, \conj{r},\, s,\, sr,\, s\conj{r} \}$. 
Furthermore, we denote the irreducible representations of $S_3$ by $\{\vac,\,\psi,\,\pi\}$, correspond to the trivial, sign, and two-dimensional irrep, respectively.
  
The fusion rules of $\vecg{S_3}{}$ are given by $S_3$ multiplication, and the $F$-symbols are trivial.   
  The anyons of $\double{\vecg{S_3}{}}$ can be labeled by a conjugacy class, together with an irreducible representation of the centralizer of a representative from the conjugacy class. We use the notation $\gauged{g}{\rho}$ to indicate the conjugacy class $\conjclass{g}$ and an irrep $\rho$ of the centralizer $\centralizer{g}$. 
  A formula for the ICIs is given in Eq.~\eqref{set:fullorbiticiforgaugethy}.   
\begin{align}
	\bgroup		
\setlength\tabcolsep{.04cm}
\begin{tabular}{ | c | c c c c c c c c c c c c c c c c c c | } 
\hline
Anyons 
 & \tube{1}{1} & \tube{1}{r}  & \tube{1}{\conj{r}} & \tube{1}{s} & \tube{1}{sr} & \tube{1}{s\conj{r}} & \tube{r}{1} & \tube{r}{r} & \tube{r}{\conj{r}} & \tube{\conj{r}}{1} & \tube{\conj{r}}{r} & \tube{\conj{r}}{\conj{r}} & \tube{s}{1} & \tube{s}{s} & \tube{sr}{1} & \tube{sr}{sr}& \tube{s\conj{r}}{1} & \tube{s\conj{r}}{s\conj{r}} 
\\ \hline
$\gauged{1}{\vac}$
 & $1$  & $1$ & $1$ & $1$ & $1$ & $1$ &  & & & & & & & & & & &
\\
$\gauged{1}{\psi}$
 & $1$  & $1$ & $1$ & $-1$ & $-1$  & $-1$ &  & & & & & & & & & & &
\\
$\gauged{1}{\pi}$
 & $4$  & $-2$ & $-2$ &  & & &  & & & & & & & & & & &
\\
$\gauged{r}{1}$
 &  &  &  &  & & & $2$  & $2$ & $2$ &  $2$ & $2$ & $2$ & & & & & &
\\
$\gauged{r}{e^{\frac{2 \pi i}{3}}}$
 &  &  &  &  & & & $2$  & $2e^{-\frac{2 \pi i}{3}}$ & $2e^{\frac{2 \pi i}{3}}$ &  $2$ & $2e^{\frac{2 \pi i}{3}}$ & $2e^{-\frac{2 \pi i}{3}}$ & & & & & &
\\
$\gauged{r}{e^{-\frac{2 \pi i}{3}}}$
 &  &  &  &  & & & $2$  & $2e^{\frac{2 \pi i}{3}}$ & $2e^{-\frac{2 \pi i}{3}}$ &  $2$ & $2e^{-\frac{2 \pi i}{3}}$ & $2e^{\frac{2 \pi i}{3}}$ & & & & & &
 \\
$\gauged{s}{+}$
 &  &  &  &  & & &  & & & & & &  $3$  & $3$ & $3$ & $3$ & $3$ & $3$
\\
$\gauged{s}{-}$
 &  &  &  &  & & &  & & & & & &  $3$  & $-3$ & $3$ & $-3$ & $3$ & $-3$
\\
\hline
\end{tabular}
\egroup
\end{align}
Note the implicit normalization of $\qd^2=6$ in the above table.

The full two dimensional $\ici{\gauged{1}{\pi}}$, $\ici{\gauged{r}{1}}$, $\ici{\gauged{r}{e^{\frac{2 \pi i}{3}}}}$, and $\ici{\gauged{r}{e^{-\frac{2 \pi i}{3}}}}$ blocks are given by
\begin{align}
\ici{\gauged{1}{\pi}}_{00} &=  \frac{1}{3} \left( \tube{1}{1} +  e^{-\frac{2 \pi i}{3}}\tube{1}{r} + e^{\frac{2 \pi i}{3}}\tube{1 }{\conj{r}}  \right)
& \ici{\gauged{1}{\pi}}_{01} &=  \frac{1}{3} \left( \tube{1}{s} +  e^{\frac{2 \pi i}{3}}\tube{1}{ sr} + e^{-\frac{2 \pi i}{3}}\tube{1 }{s \conj{r}}   \right)
\nonumber \\ 
\ici{\gauged{1}{\pi}}_{10} &=  \frac{1}{3} \left( \tube{1}{s} +  e^{-\frac{2 \pi i}{3}}\tube{1}{sr} + e^{\frac{2 \pi i}{3}}\tube{1 }{s\conj{r}}  \right)
&  \ici{\gauged{1}{\pi}}_{11} &=  \frac{1}{3} 
\left( \tube{1}{1} +  e^{\frac{2 \pi i}{3}}\tube{1}{r} + e^{-\frac{2 \pi i}{3}}\tube{1 }{\conj{r}}  \right)   \, ,
\\
\ici{\gauged{r}{1}}_{00} &=  \frac{1}{3} \left( \tube{r}{1} + \tube{r}{r} + \tube{r }{\conj{r}} \right)
& \ici{\gauged{r}{1}}_{01} &= \frac{1}{3}\left( \tube{\conj{r}}{s} +  \tube{\conj{r}}{sr} + \tube{\conj{r}}{s\conj{r}} \right)
\nonumber \\ 
\ici{\gauged{r}{1}}_{10} &=  \frac{1}{3} \left( \tube{r}{s} + \tube{r}{sr} + \tube{r }{s\conj{r}}  \right)
&  \ici{\gauged{r}{1}}_{11} &=  \frac{1}{3} \left(\tube{\conj{r}}{1} + \tube{\conj{r}}{r} + \tube{\conj{r} }{\conj{r}}  \right)
   \, ,
\\
\ici{\gauged{r}{e^{\pm\frac{2 \pi i}{3}}}}_{00} &=   \frac{1}{3} \left( \tube{r}{1} +  e^{\mp\frac{2 \pi i}{3}}\tube{r}{r} + e^{\pm\frac{2 \pi i}{3}}\tube{r }{\conj{r}}  \right)
& \ici{\gauged{r}{e^{\pm\frac{2 \pi i}{3}}}}_{01} &=  \frac{1}{3} \left( \tube{\conj{r}}{s} +   e^{\pm\frac{2 \pi i}{3}}\tube{\conj{r}}{sr} + e^{\mp\frac{2 \pi i}{3}}\tube{\conj{r}}{s\conj{r}} \right) 
\nonumber \\ 
\ici{\gauged{r}{e^{\pm\frac{2 \pi i}{3}}}}_{10} &=  \frac{1}{3} \left( \tube{r}{s} +  e^{\mp\frac{2 \pi i}{3}}\tube{r}{sr} + e^{\pm\frac{2 \pi i}{3}}\tube{r }{s\conj{r}} \right)
&  \ici{\gauged{r}{e^{\pm\frac{2 \pi i}{3}}}}_{11} &=  \frac{1}{3} \left( \tube{\conj{r}}{1} + e^{\pm\frac{2 \pi i}{3}}\tube{\conj{r}}{r} + e^{\mp\frac{2 \pi i}{3}}\tube{\conj{r} }{\conj{r}}  \right)  
\,  .
\end{align}
While the three dimensional $\ici{\gauged{s}{+}}$ and $\ici{\gauged{s}{-}}$ blocks are given by  
\begin{align}
\ici{\gauged{s}{\pm}}_{00} &= \frac{1}{2} \left( \tube{s}{1} \pm \tube{s}{s}  \right)
& \ici{\gauged{s}{\pm}}_{01} &= \frac{1}{2} \left( \tube{sr}{\conj{r}} \pm \tube{sr}{s\conj{r}}  \right)
& \ici{\gauged{s}{\pm}}_{02} &= \frac{1}{2} \left( \tube{s\conj{r}}{{r}} \pm \tube{s\conj{r}}{sr}  \right)
\nonumber \\
\ici{\gauged{s}{\pm}}_{10} &= \frac{1}{2} \left( \tube{s}{r} \pm \tube{s}{s\conj{r}}  \right)
&  \ici{\gauged{s}{\pm}}_{11} &=    \frac{1}{2} \left( \tube{sr}{1} \pm \tube{sr}{sr}  \right)
 & \ici{\gauged{s}{\pm}}_{12} &=    \frac{1}{2} \left( \tube{s\conj{r}}{\conj{r}} \pm \tube{s\conj{r}}{s}  \right)
\nonumber \\
\ici{\gauged{s}{\pm}}_{20} &= \frac{1}{2} \left( \tube{s}{\conj{r}} \pm \tube{s}{sr}  \right)
&  \ici{\gauged{s}{\pm}}_{21} &=    \frac{1}{2} \left( \tube{sr}{r} \pm \tube{sr}{s}  \right)
& \ici{\gauged{s}{\pm}}_{22} &=   \frac{1}{2} \left( \tube{s\conj{r}}{1} \pm \tube{s\conj{r}}{s\conj{r}}  \right)
\, .
\end{align}

$\vecg{S_3}{}$ admits a $\zt$-grading as follows $\{ 1,\, r,\, \conj{r}\} \oplus \{ s,\, sr,\, s\conj{r} \}$. Breaking the MPO algebra down to the $\g{1}$-sector induces a $\rep{\zt}$ condensation phase transition to $\double{\vecg{\z_3}{}}$. 
\begingroup
\renewcommand*{\arraystretch}{1.5}
	\setlength\tabcolsep{.02cm}
\begin{align}

\end{align}
\endgroup
The three dimensional  $\ici{\Delta}$ block is given by 
\begin{align}
\ici{\Delta}_{00} &= \tube{s}{1}
& \ici{\Delta}_{01} &=\tube{sr}{\conj{r}}
& \ici{\Delta}_{02} &= \tube{s\conj{r}}{r}
\nonumber \\
\ici{\Delta}_{10} &= \tube{s}{r}
&  \ici{\Delta}_{11} &=   \tube{sr}{1}
 & \ici{\Delta}_{12} &=  \tube{s\conj{r}}{\conj{r}}
\nonumber \\
\ici{\Delta}_{20} &= \tube{s}{\conj{r}}
&  \ici{\Delta}_{21} &=   \tube{sr}{{r}}
& \ici{\Delta}_{22} &=  \tube{s\conj{r}}{1}
\, .
\end{align}
The reader may find it amusing that this condensation transition actually increases the number of anyons from eight to nine.

The action of the nontrivial group element $\g{x}$ upon the defect superselection sectors is given by 
\begingroup
\renewcommand*{\arraystretch}{1.5}
	\setlength\tabcolsep{.13cm}
\begin{align}

\end{align}
\endgroup
where there is an implicit normalization by $\qd_\g{1}^2=3$.

From these actions we can extract the permutation action upon the defects, summarized below. 
\begingroup
\renewcommand*{\arraystretch}{1.5}
	\setlength\tabcolsep{.045cm}
\begin{align}
	%
\end{align}
\endgroup

\section{ Discussion and conclusions}
\label{setsection:conclusion}

In this work, we have established a description of emergent symmetry-enriched topological order in tensor network states in terms of $\G$-graded matrix product operator algebras. A classification of these $\G$-graded MPO algebras was given in terms of $\G$-extensions of an underlying topological MPO symmetry algebra. 
An extension of Ocneanu's tube algebra to nontrivial defect sectors was established, from which the physical data of the emergent SET order was extracted.  
This induced a $\G$-graded Morita equivalence relation on the $\G$-graded MPO algebras, relating those which led to the same emergent SET order. 
We described the effect that gauging the global $\G$ symmetry had upon the MPO symmetry algebra, and found the relationship between the emergent SET and the topological order that results from gauging. 
The dual $\repg$ anyon condensation process was also described in terms of the MPO symmetry algebra. 

The results of Sections~\ref{set:toptubealgsection},~\ref{setsection:setmpos},~\ref{setsection:dubes},~\ref{setsection:gauging}, and~\ref{setsection:condensation} can be summarized succinctly with a diagram 
\begin{align}
\xymatrixcolsep{8pc}\xymatrixrowsep{3.5pc}\xymatrix{
\ \  \ \ {\cat_\g{1}} \ \ \ar@<+4pt>[r]^-{\G\text{-extension}} \ar@<+4pt>[d]_-{\text{Double}}  & \ \ \ \ {\cat}_\G \ \  \ar@<+4pt>[l]^-{\text{Restrict to $\cat_\g{1}$} }   \ar@<+4pt>[d]_-{\text{$\G$-Double}}  \ar@<+4pt>[dr]^-{\text{Double}} &
 \\
   \double{\cat_\g{1}}\ \ar@<+4pt>[r]^-{\text{Add } \G \text{ defects}}  & \ \rdouble{\cat_\G}{\cat_\g{1}}\ \ar@<+4pt>[r]^-{\text{Gauge } \G \text{ symmetry}}  \ar@<+4pt>[l]^-{\text{Confine } \G \text{ defects}}  &\  \double{\cat_\G} \, . \ar@<+4pt>[l]^-{\text{Condense Rep}(\G)}
}
\end{align}
Where  $\cat_\g{1}$ is an MPO algebra (or UFC), $\cat_\G$ is a $\G$-graded MPO algebra (or $\G$-graded UFC), $ \double{\cat_\g{1}}$ and $\double{\cat_{\G}}$ are emergent topological orders (or MTCs), and $\rdouble{\cat_\G}{\cat_\g{1}}$ is an emergent SET (or $\G$-crossed MTC). 
The theory of $\G$-graded MPO algeras $\cat_\G$ was described in Section~\ref{setsection:setmpos}, the double construction was described in Section~\ref{set:toptubealgsection}, $\G$-Double refers to the symmetry-enriched relative double construction described in Section~\ref{setsection:dubes}, the gauging procedure was described in Section~\ref{setsection:gauging}, and the condensation procedure was described in Section~\ref{setsection:condensation}.

In summary, the results developed in this paper provide a comprehensive toolbox for the study of anomaly-free, nonchiral, SET orders under an on-site unitary group representation, on a lattice in two spatial dimensions. 
The tools thus developed were brought to bear on numerous examples: including an EM duality enriched toric code in Section~\ref{setsection:example1}, the symmetry-enriched string-nets in Section~\ref{set:stringnetexample}, symmetry-protected orders in Section~\ref{setsection:gauging}, and several further examples in Section~\ref{setsection:examples2}.

Moving forward, the work reported here has generated a number of outstanding questions and uncovered a number of promising paths toward future research. 

For instance, it would be interesting --- and extremely relevant --- to incorporate anti-unitary symmetries such as time-reversal into our formalism. We remark symmetry-enriched string-net models with time reversal symmetry have been described in Refs.~\cite{Gaugingtime,cheng2016exactly,barkeshli2016reflection}.  
Another immediate generalization would be the inclusion of spatial symmetries and lattice defects, for which tensor networks are naturally suited. We anticipate an approach similar to that presented in Ref.~\cite{thorngren2016gauging} would prove fruitful.  

It would also be interesting to adapt our approach to systems with constituent fermion degrees of freedom. It is known that the fermionic parity symmetry can be gauged, and that this is dual to condensing an emergent fermion in the gauged system, which is made up of bosonic constituent degrees of freedom~\cite{walkerfermions,Bhardwaj2017,kapustin2017fermionic,Wan2017,aasen2017fermion}. The fermion parity defects can be interpreted as singularities in a lattice spin structure~\cite{doi:10.1142/S0217751X16450445,PhysRevB.94.115115,PhysRevB.94.115127,NewFermionPEPSPaper2017}.

A question that we have not answered is how to construct local isometry circuits between different symmetry-enriched phases with the same underlying topological order, which are expected to exist following the framework of Ref.~\cite{chen2010local}. We remark that such circuits are simple to construct for the special case of SPT phases, see Refs.~\cite{PhysRevB.90.235137,williamson2014matrix} for example.

The $\G$-graded MPO algebras we have studied can be viewed as a construction of invertible gapped domain walls between a $\double{\cat_\g{1}}$ topological order and itself. More specifically, these are gapped domain walls between a specific representative of the $\double{\cat_\g{1}}$ Morita equivalence class, given by the MPO algebra $\cat_\g{1}$, and itself. A natural extension would be to consider invertible gapped domain walls between arbitrary representative MPO algebras from the $\double{\cat_\g{1}}$ Morita equivalence class. Such a gapped domain wall, between representative UFCs $\cat_A$ and $\cat_B$, would correspond to an invertible $\cat_A$-$\cat_B$-bimodule~\cite{etingof2015tensor,Kitaev2012}. Developing the theory of these bimodules should allow us to understand the Morita equivalence relation more explicitly at the level of the local tensors of an MPO algebra.

Another extension would be to consider noninvertible gapped domain walls, which correspond to $\mathcal A$-``graded'' MPO algebras, where $\mathcal A$ is also an algebra. It was discussed in Section~\ref{set:algebragradingmposymmetrybreakingcondensation} how an $\mathcal A$-``grading'' can be used to calculate the effects of an anyon condensation phase transition induced by breaking the MPO algebra down to the $\g{1}$-sector. We conjectured that this captures all anyon condensation phase transitions between nonchiral topological orders in two spatial dimensions. Results in this direction will be presented in a future work~\cite{MPOanyoncondprep}.
 
 It would be interesting to develop an $\mathcal{A}$-extension procedure, generalizing $\G$-extension, dual to these anyon condensations
  and to understand the connection between the $\mathcal{A}$ algebra and the \emph{algebra object} approach to anyon condensation~\cite{Kitaev2012,walkerfermions}.  
We expect $\mathcal A$-graded UFCs that admit a braiding can be used to generalize the Hamiltonian construction of topological phases presented in Ref.~\cite{williamson2016hamiltonian}. 

The even more general case of  $\cat_A$-$\cat_B$-bimodules for arbitrary UFCs $\cat_A$ and $\cat_B$ (not necessarily Morita equivalent), describes all gapped domain walls between $\double{\cat_A}$ and $\double{\cat_B}$. This captures the gapped boundary conditions of $\double{\cat_A}$ via the special case where $\cat_B$ is trivial, $\cat_B=\{ \vac \}$, note the general case can be recovered by using the folding trick. We remark that a systematic study of this problem for string-net models was given in Ref.~\cite{Kitaev2012}.

The tools we have developed also allow one to find and construct all possible transversal gates on topological quantum codes in two spatial dimensions. 
We remark that the results of Ref.~\cite{etingof2009fusion} imply all locality preserving gates on lattice topological quantum codes, that have an anomaly-free action on the superselection sectors of the topological order, can in fact be realized transversally. This may require one to use a different underlying model with the same emergent topological order. The symmetry-enriched string-nets~\cite{cheng2016exactly,PhysRevB.94.235136} suffice to realize all such transversal gates in local commuting projector topological quantum codes. We plan to explain this in more detail in a follow up work~\cite{SETtransgates}. 
It may also be interesting to interpret the computations in Refs.~\cite{edie2017brauer,edie2017field}, of symmetry groups for anyon theories that are doubles of $ADE$ fusion categories, in terms of transversal gate sets.

The boundary phases of an SET tensor network, and their phase transitions, can be classified in terms of the $\G$-graded MPO symmetry algebra. These phases are classified w.r.t.~a symmetry given by the MPO algebra $\cat_\g{1}$, and $\G$ plays the role of a group of dualities between the phases. Conversely, the framework developed in Section~\ref{setsection:setmpos} allows one to construct the possible MPO algebras of dualities between the one dimensional $\cat_\g{1}$ phases. 
This uncovers a beautiful connection to the results of Refs.~\cite{1751-8121-49-35-354001,daveprep}. 
The MPO dualities can be used to find phase transitions, some of which are described by emergents conformal field theories (CFTs)~\cite{fuchs2002tft,PhysRevLett.93.070601,FROHLICH2007354,1751-8121-47-45-452001}. At the phase transition the dualities become symmetries and topological superselection sectors in the CFT can be  constructed using the approach described in Section~\ref{set:toptubealgsection}, see Ref.~\cite{bridgeman2017anomalies} for example.  
The \std{1} quantum Euclidean path integrals, or $2$D classical partition functions, of the boundary phases can be realized using a generalized strange correlator~\cite{PhysRevLett.112.247202}, given by an overlap between a product state and the SET tensor network state. Generalized strange correlators derived from the symmetry-enriched string-nets can be used to construct many familiar lattice statistical mechanics models~\cite{1751-8121-49-35-354001,daveprep}, including the Ising model and $\z_3$ Potts model.  
Stable renormalization group fixed points on the boundary correspond to gapped phases and can be interpreted as gapped boundary conditions, while unstable gapless fixed points correspond to CFTs. This connection is studied further in Ref.~\cite{Vanhove2018,Vanhove2022}.

It would be very interesting to find tensor networks that are symmetric under an anomalous symmetry in two spatial dimensions. Such anomalous symmetries can be expressed as tensor network operators, that arise at the boundary of a (generalized) SPT phase in three spatial dimensions~\cite{PhysRevX.3.041016,PhysRevB.90.245122,PhysRevX.5.041013,thorngren2015higher,fidkowski2015realizing}. We anticipate that such a tensor network could be found by following a similar approach to Ref.~\cite{bridgeman2017anomalies}. We plan to study this in the future. 

\medskip

\textbf{Note added.} This work first appeared on the arXiv preprint server in November 2017. Since then, a number of relevant follow up works have appeared. This includes work on matrix product operator algebras~\cite{Molnar2022,Ruiz-de-Alarcn2022,Liu2025} and dualities~\cite{Thorngren2019,Aasen2020b,Lootens2021b,Lootens2022,Bridgeman2022,Haller2023,Cuiper2025,Jones2026}, string-net realizations~\cite{Lootens2021,Lootens2021a,Schatz2024,Lin2024,Barter2019,Lan2025}, and adaptive quantum circuit implementations~\cite{Tantivasadakarn2021,Bravyi2022,Lootens2023,Ren2024,Tantivasadakarn2025a}.

\section*{Acknowledgements}
We thank Dave Aasen, Jacob Bridgeman, Meng Cheng, Jutho Haegeman, Micha{\"e}l Mari{\"e}n, Ryan Thorngren and Zhenghan Wang for many useful discussions. 

\section*{Author contributions}
All authors contributed extensively to this research and the writing of this manuscript. 

\bibliographystyle{quantum}
\bibliography{Thesis_Refs.bib}

\begin{thebibliography}{100}

\bibitem{landau1965course}
Lev~Davidovich Landau and Evgenii~Mikhailovich Lifshitz.
\newblock ``{Course of theoretical physics}''.
\newblock Pergamon Press. ~(1965).

\bibitem{segal1988definition}
G.~B. Segal.
\newblock ``The definition of conformal field theory''.
\newblock \href{https://dx.doi.org/10.1007/978-94-015-7809-7_9}{Pages 165--171}.
\newblock Springer Netherlands. Dordrecht~(1988).

\bibitem{witten1988topological}
Edward Witten.
\newblock ``Topological quantum field theory''.
\newblock \href{https://dx.doi.org/10.1007/BF01223371}{Commun. Math. Phys. {\bf 117}, 353--386}~(1988).

\bibitem{atiyah1988topological}
Michael Atiyah.
\newblock ``Topological quantum field theories''.
\newblock \href{https://dx.doi.org/10.1007/BF02698547}{Inst. Hautes Études Sci. Publ. Math. {\bf 68}, 175--186}~(1988).

\bibitem{wegner1971duality}
Franz~J Wegner.
\newblock ``{Duality in generalized Ising models and phase transitions without local order parameters}''.
\newblock \href{https://dx.doi.org/10.1063/1.1665530}{J. Math. Phys. {\bf 12}, 2259--2272}~(1971).

\bibitem{kosterlitz1973ordering}
John~Michael Kosterlitz and David~James Thouless.
\newblock ``{Ordering, metastability and phase transitions in two-dimensional systems}''.
\newblock \href{https://dx.doi.org/10.1088/0022-3719/6/7/010}{J. Phys. C: Solid State Phys. {\bf 6}, 1181}~(1973).

\bibitem{PhysRevB.40.7387}
X.-G. Wen.
\newblock ``{Vacuum degeneracy of chiral spin states in compactified space}''.
\newblock \href{https://dx.doi.org/10.1103/PhysRevB.40.7387}{Phys. Rev. B {\bf 40}, 7387--7390}~(1989).

\bibitem{einarsson}
Torbj\"orn Einarsson.
\newblock ``{Fractional statistics on a torus}''.
\newblock \href{https://dx.doi.org/10.1103/physrevlett.64.1995}{Phys. Rev. Lett. {\bf 64}, 1995--1998}~(1990).

\bibitem{doi:10.1142/S0217979290000139}
X.-G. Wen.
\newblock ``{Topological Orders in Rigid States}''.
\newblock \href{https://dx.doi.org/10.1142/S0217979290000139}{Int. J. Mod. Phys. B {\bf 04}, 239--271}~(1990).

\bibitem{PhysRevLett.50.1153}
F.~D.~M. Haldane.
\newblock ``{Nonlinear Field Theory of Large-Spin Heisenberg Antiferromagnets: Semiclassically Quantized Solitons of the One-Dimensional Easy-Axis N\'eel State}''.
\newblock \href{https://dx.doi.org/10.1103/PhysRevLett.50.1153}{Phys. Rev. Lett. {\bf 50}, 1153--1156}~(1983).

\bibitem{gu2009tensor}
Zheng-Cheng Gu and Xiao-Gang Wen.
\newblock ``{Tensor-entanglement-filtering renormalization approach and symmetry-protected topological order}''.
\newblock \href{https://dx.doi.org/10.1103/PhysRevB.80.155131}{Phys. Rev. B {\bf 80}, 155131}~(2009).
\newblock  \href{http://arxiv.org/abs/0903.1069}{arXiv:0903.1069}.

\bibitem{pollmann2010entanglement}
Frank Pollmann, Ari~M Turner, Erez Berg, and Masaki Oshikawa.
\newblock ``{Entanglement spectrum of a topological phase in one dimension}''.
\newblock \href{https://dx.doi.org/10.1103/PhysRevB.81.064439}{Phys. Rev. B {\bf 81}, 064439}~(2010).
\newblock  \href{http://arxiv.org/abs/0910.1811}{arXiv:0910.1811}.

\bibitem{chen2013symmetry}
Xie Chen, Zheng-Cheng Gu, Zheng-Xin Liu, and Xiao-Gang Wen.
\newblock ``{Symmetry protected topological orders and the group cohomology of their symmetry group}''.
\newblock \href{https://dx.doi.org/10.1103/PhysRevB.87.155114}{Phys. Rev. B {\bf 87}, 155114}~(2013).
\newblock  \href{http://arxiv.org/abs/1106.4772}{arXiv:1106.4772}.

\bibitem{turaev2000homotopy}
Vladimir Turaev.
\newblock ``Homotopy field theory in dimension 3 and crossed group-categories''.
\newblock preprint,~(2000).
\newblock  \href{http://arxiv.org/abs/math/0005291}{arXiv:math/0005291}.

\bibitem{kirillov2004g}
Alexander Kirillov~Jr.
\newblock ``On $g$--equivariant modular categories''.
\newblock preprint,~(2004).
\newblock  \href{http://arxiv.org/abs/math/0401119}{arXiv:math/0401119}.

\bibitem{kitaev2006anyons}
Alexei Kitaev.
\newblock ``Anyons in an exactly solved model and beyond''.
\newblock \href{https://dx.doi.org/10.1016/j.aop.2005.10.005}{Ann. Phys. {\bf 321}, 2 -- 111}~(2006).
\newblock  \href{http://arxiv.org/abs/cond-mat/0506438}{arXiv:cond-mat/0506438}.

\bibitem{drinfeld2010braided}
Vladimir Drinfeld, Shlomo Gelaki, Dmitri Nikshych, and Victor Ostrik.
\newblock ``On braided fusion categories {I}''.
\newblock \href{https://dx.doi.org/10.1007/s00029-010-0017-z}{Selecta Mathematica {\bf 16}, 1--119}~(2010).
\newblock  \href{http://arxiv.org/abs/0906.0620}{arXiv:0906.0620}.

\bibitem{etingof2009fusion}
Pavel Etingof, Dmitri Nikshych, and Victor Ostrik.
\newblock ``Fusion categories and homotopy theory''.
\newblock preprint,~(2009).
\newblock  \href{http://arxiv.org/abs/0909.3140}{arXiv:0909.3140}.

\bibitem{bombin2010topological}
H.~Bombin.
\newblock ``Topological order with a twist: Ising anyons from an abelian model''.
\newblock \href{https://dx.doi.org/10.1103/PhysRevLett.105.030403}{Phys. Rev. Lett. {\bf 105}, 030403}~(2010).
\newblock  \href{http://arxiv.org/abs/1004.1838}{arXiv:1004.1838}.

\bibitem{hung2013quantized}
Ling-Yan Hung and Xiao-Gang Wen.
\newblock ``Quantized topological terms in weak-coupling gauge theories with a global symmetry and their connection to symmetry-enriched topological phases''.
\newblock \href{https://dx.doi.org/10.1103/PhysRevB.87.165107}{Phys. Rev. B {\bf 87}, 165107}~(2013).
\newblock  \href{http://arxiv.org/abs/1212.1827}{arXiv:1212.1827}.

\bibitem{mesaros2013classification}
Andrej Mesaros and Ying Ran.
\newblock ``Classification of symmetry enriched topological phases with exactly solvable models''.
\newblock \href{https://dx.doi.org/10.1103/PhysRevB.87.155115}{Phys. Rev. B {\bf 87}, 155115}~(2013).
\newblock  \href{http://arxiv.org/abs/1212.0835}{arXiv:1212.0835}.

\bibitem{Barkeshli2013a}
Maissam Barkeshli, Chao~Ming Jian, and Xiao~Liang Qi.
\newblock ``Theory of defects in abelian topological states''.
\newblock \href{https://dx.doi.org/10.1103/PhysRevB.88.235103}{Physical Review B - Condensed Matter and Materials Physics{\bf 88}}~(2013).

\bibitem{barkeshli2014symmetry}
Maissam Barkeshli, Parsa Bonderson, Meng Cheng, and Zhenghan Wang.
\newblock ``Symmetry, defects, and gauging of topological phases''.
\newblock preprint,~(2014).
\newblock  \href{http://arxiv.org/abs/1410.4540}{arXiv:1410.4540}.

\bibitem{tarantino2015symmetry}
Nicolas Tarantino, Netanel~H Lindner, and Lukasz Fidkowski.
\newblock ``Symmetry fractionalization and twist defects''.
\newblock \href{https://dx.doi.org/10.1088/1367-2630/18/3/035006}{New J. Phys. {\bf 18}, 035006}~(2016).
\newblock  \href{http://arxiv.org/abs/1506.06754}{arXiv:1506.06754}.

\bibitem{teo2015theory}
Jeffrey~CY Teo, Taylor~L Hughes, and Eduardo Fradkin.
\newblock ``Theory of twist liquids: gauging an anyonic symmetry''.
\newblock \href{https://dx.doi.org/10.1016/j.aop.2015.05.012}{Ann. Phys. {\bf 360}, 349--445}~(2015).
\newblock  \href{http://arxiv.org/abs/1503.06812}{arXiv:1503.06812}.

\bibitem{PhysRevB.94.235136}
Chris Heinrich, Fiona Burnell, Lukasz Fidkowski, and Michael Levin.
\newblock ``Symmetry-enriched string nets: Exactly solvable models for set phases''.
\newblock \href{https://dx.doi.org/10.1103/PhysRevB.94.235136}{Phys. Rev. B {\bf 94}, 235136}~(2016).
\newblock  \href{http://arxiv.org/abs/1606.07816}{arXiv:1606.07816}.

\bibitem{cheng2016exactly}
Meng Cheng, Zheng-Cheng Gu, Shenghan Jiang, and Yang Qi.
\newblock ``Exactly solvable models for symmetry-enriched topological phases''.
\newblock \href{https://dx.doi.org/10.1103/PhysRevB.96.115107}{Phys. Rev. B {\bf 96}, 115107}~(2017).
\newblock  \href{http://arxiv.org/abs/1606.08482}{arXiv:1606.08482}.

\bibitem{beigi2011quantum}
Salman Beigi, Peter~W. Shor, and Daniel Whalen.
\newblock ``The quantum double model with boundary: Condensations and symmetries''.
\newblock \href{https://dx.doi.org/10.1007/s00220-011-1294-x}{Commun. Math. Phys. {\bf 306}, 663--694}~(2011).
\newblock  \href{http://arxiv.org/abs/1006.5479}{arXiv:1006.5479}.

\bibitem{PhysRevLett.111.220402}
Benjamin~J. Brown, Stephen~D. Bartlett, Andrew~C. Doherty, and Sean~D. Barrett.
\newblock ``Topological entanglement entropy with a twist''.
\newblock \href{https://dx.doi.org/10.1103/PhysRevLett.111.220402}{Phys. Rev. Lett. {\bf 111}, 220402}~(2013).
\newblock  \href{http://arxiv.org/abs/1303.4455}{arXiv:1303.4455}.

\bibitem{yoshida2015gapped}
Beni Yoshida.
\newblock ``Gapped boundaries, group cohomology and fault-tolerant logical gates''.
\newblock \href{https://dx.doi.org/10.1016/j.aop.2016.12.014}{Ann. Phys. {\bf 377}, 387 -- 413}~(2017).
\newblock  \href{http://arxiv.org/abs/1509.03626}{arXiv:1509.03626}.

\bibitem{PhysRevA.91.012305}
Fernando Pastawski and Beni Yoshida.
\newblock ``Fault-tolerant logical gates in quantum error-correcting codes''.
\newblock \href{https://dx.doi.org/10.1103/PhysRevA.91.012305}{Phys. Rev. A {\bf 91}, 012305}~(2015).
\newblock  \href{http://arxiv.org/abs/1408.1720}{arXiv:1408.1720}.

\bibitem{PhysRevLett.110.170503}
Sergey Bravyi and Robert K\"onig.
\newblock ``Classification of topologically protected gates for local stabilizer codes''.
\newblock \href{https://dx.doi.org/10.1103/PhysRevLett.110.170503}{Phys. Rev. Lett. {\bf 110}, 170503}~(2013).
\newblock  \href{http://arxiv.org/abs/1206.1609}{arXiv:1206.1609}.

\bibitem{doi:10.1063/1.4939783}
Michael~E. Beverland, Oliver Buerschaper, Robert K\"onig, Fernando Pastawski, John Preskill, and Sumit Sijher.
\newblock ``Protected gates for topological quantum field theories''.
\newblock \href{https://dx.doi.org/10.1063/1.4939783}{J. Math. Phys. {\bf 57}, 022201}~(2016).
\newblock  \href{http://arxiv.org/abs/1409.3898}{arXiv:1409.3898}.

\bibitem{PhysRevB.91.245131}
Beni Yoshida.
\newblock ``Topological color code and symmetry-protected topological phases''.
\newblock \href{https://dx.doi.org/10.1103/PhysRevB.91.245131}{Phys. Rev. B {\bf 91}, 245131}~(2015).
\newblock  \href{http://arxiv.org/abs/1503.07208}{arXiv:1503.07208}.

\bibitem{bridgeman2017tensor}
Jacob~C Bridgeman, Stephen~D Bartlett, and Andrew~C Doherty.
\newblock ``Tensor networks with a twist: Anyon-permuting domain walls and defects in peps''.
\newblock preprint,~(2017).
\newblock  \href{http://arxiv.org/abs/1704.04221}{arXiv:1704.04221}.

\bibitem{webster2017locality}
Paul Webster and Stephen~D Bartlett.
\newblock ``Locality-preserving logical operators in topological stabiliser codes''.
\newblock preprint~(2017).
\newblock  \href{http://arxiv.org/abs/1709.00020}{arXiv:1709.00020}.

\bibitem{VerstraeteMurgCirac2008}
F~Verstraete, J.~I. Cirac, and V~Murg.
\newblock ``Matrix product states, projected entangled pair states, and variational renormalization group methods for quantum spin systems''.
\newblock \href{https://dx.doi.org/10.1080/14789940801912366}{Adv. Phys. {\bf 57}, 143--224}~(2009).
\newblock  \href{http://arxiv.org/abs/0907.2796}{arXiv:0907.2796}.

\bibitem{Orus2014}
Rom{\'{a}}n Or{\'{u}}s.
\newblock ``{A practical introduction to tensor networks: Matrix product states and projected entangled pair states}''.
\newblock \href{https://dx.doi.org/10.1016/j.aop.2014.06.013}{Ann. Phys. {\bf 349}, 117--158}~(2014).
\newblock  \href{http://arxiv.org/abs/1306.2164}{arXiv:1306.2164}.

\bibitem{TNReview}
Jacob~C Bridgeman and Christopher~T Chubb.
\newblock ``Hand-waving and interpretive dance: an introductory course on tensor networks''.
\newblock \href{https://dx.doi.org/10.1088/1751-8121/aa6dc3}{J. Phys. A {\bf 50}, 223001}~(2017).
\newblock  \href{http://arxiv.org/abs/1603.03039}{arXiv:1603.03039}.

\bibitem{PhysRevLett.69.2863}
Steven~R. White.
\newblock ``{Density matrix formulation for quantum renormalization groups}''.
\newblock \href{https://dx.doi.org/10.1103/PhysRevLett.69.2863}{Phys. Rev. Lett. {\bf 69}, 2863--2866}~(1992).

\bibitem{PhysRevLett.75.3537}
Stellan \"Ostlund and Stefan Rommer.
\newblock ``{Thermodynamic Limit of Density Matrix Renormalization}''.
\newblock \href{https://dx.doi.org/10.1103/PhysRevLett.75.3537}{Phys. Rev. Lett. {\bf 75}, 3537--3540}~(1995).
\newblock  \href{http://arxiv.org/abs/cond-mat/9503107}{arXiv:cond-mat/9503107}.

\bibitem{0295-5075-43-4-457}
J.~Dukelsky, M.~A. Mart{\'i}n-Delgado, T.~Nishino, and G.~Sierra.
\newblock ``{Equivalence of the variational matrix product method and the density matrix renormalization group applied to spin chains}''.
\newblock \href{https://dx.doi.org/10.1209/epl/i1998-00381-x}{EPL {\bf 43}, 457}~(1998).
\newblock  \href{http://arxiv.org/abs/cond-mat/9710310}{arXiv:cond-mat/9710310}.

\bibitem{PhysRevLett.59.799}
Ian Affleck, Tom Kennedy, Elliott~H. Lieb, and Hal Tasaki.
\newblock ``{Rigorous results on valence-bond ground states in antiferromagnets}''.
\newblock \href{https://dx.doi.org/10.1103/PhysRevLett.59.799}{Phys. Rev. Lett. {\bf 59}, 799--802}~(1987).

\bibitem{Fannes92}
M.~Fannes, B.~Nachtergaele, and R.~F. Werner.
\newblock ``{Finitely correlated states on quantum spin chains}''.
\newblock \href{https://dx.doi.org/10.1007/BF02099178}{Commun. Math. Phys. {\bf 144}, 443--490}~(1992).

\bibitem{0295-5075-24-4-010}
A.~Klümper, A.~Schadschneider, and J.~Zittartz.
\newblock ``{Matrix Product Ground States for One-Dimensional Spin-1 Quantum Antiferromagnets}''.
\newblock EPL {\bf 24}, 293~(1993).
\newblock  \href{http://arxiv.org/abs/cond-mat/9307028}{arXiv:cond-mat/9307028}.

\bibitem{MPSrepresentations}
D.~P{\'e}rez-Garc{\'i}a, F.~Verstraete, M.~M. Wolf, and J.~I. Cirac.
\newblock ``Matrix product state representations''.
\newblock Quantum Info. Comput. {\bf 7}, 401--430~(2007).
\newblock  \href{http://arxiv.org/abs/quant-ph/0608197}{arXiv:quant-ph/0608197}.

\bibitem{1Done}
Xie Chen, Zheng-Cheng Gu, and Xiao-Gang Wen.
\newblock ``{Classification of gapped symmetric phases in one-dimensional spin systems}''.
\newblock \href{https://dx.doi.org/10.1103/PhysRevB.83.035107}{Phys. Rev. B {\bf 83}, 035107}~(2011).
\newblock  \href{http://arxiv.org/abs/1008.3745}{arXiv:1008.3745}.

\bibitem{SchuchGarciaCirac11}
Norbert Schuch, David P\'erez-Garc\'ia, and Ignacio Cirac.
\newblock ``{Classifying quantum phases using matrix product states and projected entangled pair states}''.
\newblock \href{https://dx.doi.org/10.1103/PhysRevB.84.165139}{Phys. Rev. B {\bf 84}, 165139}~(2011).
\newblock  \href{http://arxiv.org/abs/1010.3732}{arXiv:1010.3732}.

\bibitem{Cirac2017100}
J.I. Cirac, D.~P{\'e}rez-Garc{\'i}a, N.~Schuch, and F.~Verstraete.
\newblock ``{Matrix product density operators: Renormalization fixed points and boundary theories}''.
\newblock \href{https://dx.doi.org/10.1016/j.aop.2016.12.030}{Ann. Phys. {\bf 378}, 100 -- 149}~(2017).
\newblock  \href{http://arxiv.org/abs/1606.00608}{arXiv:1606.00608}.

\bibitem{doi:10.1143PTP.105.409}
Tomotoshi Nishino, Yasuhiro Hieida, Kouichi Okunishi, Nobuya Maeshima, Yasuhiro Akutsu, and Andrej Gendiar.
\newblock ``{Two-Dimensional Tensor Product Variational Formulation}''.
\newblock \href{https://dx.doi.org/10.1143/PTP.105.409}{Progr. Theor. Phys. {\bf 105}, 409}~(2001).
\newblock  \href{http://arxiv.org/abs/cond-mat/0011103}{arXiv:cond-mat/0011103}.

\bibitem{peps}
F.~{Verstraete} and J.~I. {Cirac}.
\newblock ``Renormalization algorithms for quantum-many body systems in two and higher dimensions''~(2004).
\newblock  \href{http://arxiv.org/abs/cond-mat/0407066}{arXiv:cond-mat/0407066}.

\bibitem{Ginjectivity}
N.~Schuch, I.~Cirac, and D.~Perez-Garcia.
\newblock ``Peps as ground states: Degeneracy and topology''.
\newblock \href{https://dx.doi.org/10.1016/j.aop.2010.05.008}{Ann. Phys. {\bf 325}, 2153 -- 2192}~(2010).
\newblock  \href{http://arxiv.org/abs/1001.3807}{arXiv:1001.3807}.

\bibitem{Buerschaper14}
Oliver Buerschaper.
\newblock ``Twisted injectivity in projected entangled pair states and the classification of quantum phases''.
\newblock \href{https://dx.doi.org/10.1016/j.aop.2014.09.007}{Ann. Phys. {\bf 351}, 447 -- 476}~(2014).
\newblock  \href{http://arxiv.org/abs/1307.7763}{arXiv:1307.7763}.

\bibitem{MPOpaper}
Mehmet~B. \ifmmode \mbox{\c{S}}\else \c{S}\fi{}ahino\ifmmode~\breve{g}\else \u{g}\fi{}lu, D.~Williamson, N.~Bultinck, M.~Mari{\"e}n, J.~Haegeman, N.~Schuch, and F.~Verstraete.
\newblock ``{Characterizing topological order with matrix product operators}''.
\newblock preprint,~(2014).
\newblock  \href{http://arxiv.org/abs/1409.2150}{arXiv:1409.2150}.

\bibitem{Bultinck2017183}
N.~Bultinck, M.~Mari{\"e}n, D.J. Williamson, Mehmet~B. \ifmmode \mbox{\c{S}}\else \c{S}\fi{}ahino\ifmmode~\breve{g}\else \u{g}\fi{}lu, J.~Haegeman, and F.~Verstraete.
\newblock ``{Anyons and matrix product operator algebras}''.
\newblock \href{https://dx.doi.org/10.1016/j.aop.2017.01.004}{Ann. Phys. {\bf 378}, 183 -- 233}~(2017).
\newblock  \href{http://arxiv.org/abs/1511.08090}{arXiv:1511.08090}.

\bibitem{ribbons}
J.C. Bridgeman, S.~T. Flammia, and D.~Poulin.
\newblock ``{Detecting Topological Order with Ribbon Operators}''.
\newblock \href{https://dx.doi.org/10.1103/PhysRevB.94.205123}{Phys. Rev. B {\bf 94}, 205123}~(2016).
\newblock  \href{http://arxiv.org/abs/1603.02275}{arXiv:1603.02275}.

\bibitem{williamson2016fermionic}
Dominic~J Williamson, Nick Bultinck, Jutho Haegeman, and Frank Verstraete.
\newblock ``{Fermionic Matrix Product Operators and Topological Phases of Matter}''.
\newblock preprint,~(2016).
\newblock  \href{http://arxiv.org/abs/1609.02897}{arXiv:1609.02897}.

\bibitem{NewFermionPEPSPaper2017}
Nick Bultinck, Dominic~J. Williamson, Jutho Haegeman, and Frank Verstraete.
\newblock ``Fermionic projected entangled-pair states and topological phases''.
\newblock \href{https://dx.doi.org/10.1088/1751-8121/aa99cc}{J. Phys. A}~(2017).
\newblock  \href{http://arxiv.org/abs/1707.00470}{arXiv:1707.00470}.

\bibitem{muger2003subfactors}
Michael M\"uger.
\newblock ``{From subfactors to categories and topology I: Frobenius algebras in and Morita equivalence of tensor categories}''.
\newblock \href{https://dx.doi.org/10.1016/S0022-4049(02)00247-5}{J. of Pure Appl. Algebr. {\bf 180}, 81 -- 157}~(2003).
\newblock  \href{http://arxiv.org/abs/math/0111204}{arXiv:math/0111204}.

\bibitem{levin2005string}
Michael~A. Levin and Xiao-Gang Wen.
\newblock ``String-net condensation: A physical mechanism for topological phases''.
\newblock \href{https://dx.doi.org/10.1103/PhysRevB.71.045110}{Phys. Rev. B {\bf 71}, 045110}~(2005).
\newblock  \href{http://arxiv.org/abs/cond-mat/0404617}{arXiv:cond-mat/0404617}.

\bibitem{stringnet1}
Oliver Buerschaper, Miguel Aguado, and Guifr\'e Vidal.
\newblock ``Explicit tensor network representation for the ground states of string-net models''.
\newblock \href{https://dx.doi.org/10.1103/PhysRevB.79.085119}{Phys. Rev. B {\bf 79}, 085119}~(2009).
\newblock  \href{http://arxiv.org/abs/0809.2393}{arXiv:0809.2393}.

\bibitem{stringnet2}
Zheng-Cheng Gu, Michael Levin, Brian Swingle, and Xiao-Gang Wen.
\newblock ``Tensor-product representations for string-net condensed states''.
\newblock \href{https://dx.doi.org/10.1103/PhysRevB.79.085118}{Phys. Rev. B {\bf 79}, 085118}~(2009).
\newblock  \href{http://arxiv.org/abs/0809.2821}{arXiv:0809.2821}.

\bibitem{PhysRevLett.114.106803}
Shuo Yang, Thorsten~B. Wahl, Hong-Hao Tu, Norbert Schuch, and J.~Ignacio Cirac.
\newblock ``Chiral projected entangled-pair state with topological order''.
\newblock \href{https://dx.doi.org/10.1103/PhysRevLett.114.106803}{Phys. Rev. Lett. {\bf 114}, 106803}~(2015).
\newblock  \href{http://arxiv.org/abs/1411.6618}{arXiv:1411.6618}.

\bibitem{PhysRevB.92.205307}
J.~Dubail and N.~Read.
\newblock ``{Tensor network trial states for chiral topological phases in two dimensions and a no-go theorem in any dimension}''.
\newblock \href{https://dx.doi.org/10.1103/PhysRevB.92.205307}{Phys. Rev. B {\bf 92}, 205307}~(2015).
\newblock  \href{http://arxiv.org/abs/1307.7726}{arXiv:1307.7726}.

\bibitem{PhysRevLett.111.236805}
T.~B. Wahl, H.-H. Tu, N.~Schuch, and J.~I. Cirac.
\newblock ``{Projected Entangled-Pair States Can Describe Chiral Topological States}''.
\newblock \href{https://dx.doi.org/10.1103/PhysRevLett.111.236805}{Phys. Rev. Lett. {\bf 111}, 236805}~(2013).
\newblock  \href{http://arxiv.org/abs/1308.0316}{arXiv:1308.0316}.

\bibitem{williamson2014matrix}
Dominic~J. Williamson, Nick Bultinck, Michael Mari\"en, Mehmet~B. \ifmmode \mbox{\c{S}}\else \c{S}\fi{}ahino\ifmmode~\breve{g}\else \u{g}\fi{}lu, Jutho Haegeman, and Frank Verstraete.
\newblock ``Matrix product operators for symmetry-protected topological phases: Gauging and edge theories''.
\newblock \href{https://dx.doi.org/10.1103/PhysRevB.94.205150}{Phys. Rev. B {\bf 94}, 205150}~(2016).
\newblock  \href{http://arxiv.org/abs/1412.5604}{arXiv:1412.5604}.

\bibitem{gelaki2009centers}
Shlomo Gelaki, Deepak Naidu, and Dmitri Nikshych.
\newblock ``Centers of graded fusion categories''.
\newblock \href{https://dx.doi.org/10.2140/ant.2009.3.959}{Algebra \& Number Theory {\bf 3}, 959--990}~(2009).
\newblock  \href{http://arxiv.org/abs/0905.3117}{arXiv:0905.3117}.

\bibitem{chang2015enriching}
Liang Chang, Meng Cheng, Shawn~X Cui, Yuting Hu, Wei Jin, Ramis Movassagh, Pieter Naaijkens, Zhenghan Wang, and Amanda Young.
\newblock ``On enriching the {L}evin-{W}en model with symmetry''.
\newblock \href{https://dx.doi.org/10.1088/1751-8113/48/12/12FT01}{J. Phys. A {\bf 48}, 12FT01}~(2015).
\newblock  \href{http://arxiv.org/abs/1412.6589}{arXiv:1412.6589}.

\bibitem{ocneanu1994chirality}
Adrian Ocneanu.
\newblock ``Chirality for operator algebras''.
\newblock Subfactors (Kyuzeso, 1993)Pages 39--63~(1994).

\bibitem{tubealgebra}
David Evans and Yasuyuki Kawahigashi.
\newblock ``{On Ocneanu's theory of asymptotic inclusions for subfactors, topological quantum field theories and quantum doubles}''.
\newblock \href{https://dx.doi.org/10.1142/s0129167x95000468}{Int. J. Math. {\bf 6}, 205--228}~(1995).

\bibitem{evans1998quantum}
David~E Evans and Yasuyuki Kawahigashi.
\newblock ``Quantum symmetries on operator algebras''.
\newblock Volume 147.
\newblock Clarendon Press Oxford. ~(1998).

\bibitem{wilson1974confinement}
Kenneth~G. Wilson.
\newblock ``Confinement of quarks''.
\newblock \href{https://dx.doi.org/10.1103/PhysRevD.10.2445}{Phys. Rev. D {\bf 10}, 2445--2459}~(1974).

\bibitem{kogut1975hamiltonian}
John Kogut and Leonard Susskind.
\newblock ``Hamiltonian formulation of wilson's lattice gauge theories''.
\newblock \href{https://dx.doi.org/10.1103/PhysRevD.11.395}{Phys. Rev. D {\bf 11}, 395--408}~(1975).

\bibitem{levin2012braiding}
Michael Levin and Zheng-Cheng Gu.
\newblock ``{Braiding statistics approach to symmetry-protected topological phases}''.
\newblock \href{https://dx.doi.org/10.1103/PhysRevB.86.115109}{Phys. Rev. B {\bf 86}, 115109}~(2012).
\newblock  \href{http://arxiv.org/abs/1202.3120}{arXiv:1202.3120}.

\bibitem{Gaugingpaper}
Jutho Haegeman, Karel Van~Acoleyen, Norbert Schuch, J~Ignacio Cirac, and Frank Verstraete.
\newblock ``{Gauging quantum states: from global to local symmetries in many-body systems}''.
\newblock \href{https://dx.doi.org/10.1103/PhysRevX.5.011024}{Phys. Rev. X {\bf 5}, 011024}~(2015).
\newblock  \href{http://arxiv.org/abs/1407.1025}{arXiv:1407.1025}.

\bibitem{PhysRevB.79.045316}
F.~A. Bais and J.~K. Slingerland.
\newblock ``Condensate-induced transitions between topologically ordered phases''.
\newblock \href{https://dx.doi.org/10.1103/PhysRevB.79.045316}{Phys. Rev. B {\bf 79}, 045316}~(2009).
\newblock  \href{http://arxiv.org/abs/0808.0627}{arXiv:0808.0627}.

\bibitem{qdouble}
A.Yu. Kitaev.
\newblock ``Fault-tolerant quantum computation by anyons''.
\newblock \href{https://dx.doi.org/10.1016/S0003-4916(02)00018-0}{Ann. Phys. {\bf 303}, 2 -- 30}~(2003).
\newblock  \href{http://arxiv.org/abs/quant-ph/9707021}{arXiv:quant-ph/9707021}.

\bibitem{etingof2005fusion}
Pavel Etingof, Dmitri Nikshych, and Viktor Ostrik.
\newblock ``{On fusion categories}''.
\newblock \href{https://dx.doi.org/10.4007/annals.2005.162.581}{Ann. Math.Pages 581--642}~(2005).
\newblock  \href{http://arxiv.org/abs/math/0203060}{arXiv:math/0203060}.

\bibitem{etingof2015tensor}
Pavel Etingof, Shlomo Gelaki, Dmitri Nikshych, and Victor Ostrik.
\newblock ``Tensor categories''.
\newblock \href{https://dx.doi.org/10.1090/surv/205}{Volume 205}.
\newblock American Mathematical Soc. ~(2015).

\bibitem{bakalov2001lectures}
Bojko Bakalov and Alexander~A Kirillov.
\newblock ``{Lectures on tensor categories and modular functors}''.
\newblock \href{https://dx.doi.org/10.1090/ulect/021}{Volume~21}.
\newblock American Mathematical Soc. ~(2001).

\bibitem{Nussinov06102009}
Zohar Nussinov and Gerardo Ortiz.
\newblock ``Sufficient symmetry conditions for topological quantum order''.
\newblock \href{https://dx.doi.org/10.1073/pnas.0803726105}{Proceedings of the National Academy of Sciences {\bf 106}, 16944--16949}~(2009).
\newblock  \href{http://arxiv.org/abs/cond-mat/0605316}{arXiv:cond-mat/0605316}.

\bibitem{NUSSINOV2009977}
Zohar Nussinov and Gerardo Ortiz.
\newblock ``A symmetry principle for topological quantum order''.
\newblock \href{https://dx.doi.org/10.1016/j.aop.2008.11.002}{Annals of Physics {\bf 324}, 977 -- 1057}~(2009).
\newblock  \href{http://arxiv.org/abs/cond-mat/0702377}{arXiv:cond-mat/0702377}.

\bibitem{doi:10.1080/00018732.2011.619814}
Emilio Cobanera, Gerardo Ortiz, and Zohar Nussinov.
\newblock ``The bond-algebraic approach to dualities''.
\newblock \href{https://dx.doi.org/10.1080/00018732.2011.619814}{Advances in Physics {\bf 60}, 679--798}~(2011).
\newblock  \href{http://arxiv.org/abs/1103.2776}{arXiv:1103.2776}.

\bibitem{1751-8121-49-35-354001}
David Aasen, Roger S~K Mong, and Paul Fendley.
\newblock ``{Topological Defects on the Lattice I: The Ising model}''.
\newblock \href{https://dx.doi.org/10.1088/1751-8113/49/35/354001}{J. Phys. A {\bf 49}, 354001}~(2016).
\newblock  \href{http://arxiv.org/abs/1601.07185}{arXiv:1601.07185}.

\bibitem{JOYAL199155}
André Joyal and Ross Street.
\newblock ``{The geometry of tensor calculus, I}''.
\newblock \href{https://dx.doi.org/10.1016/0001-8708(91)90003-P}{Advances in Mathematics {\bf 88}, 55 -- 112}~(1991).

\bibitem{JOYAL199320}
A.~Joyal and R.~Street.
\newblock ``Braided tensor categories''.
\newblock \href{https://dx.doi.org/10.1006/aima.1993.1055}{Advances in Mathematics {\bf 102}, 20 -- 78}~(1993).

\bibitem{turaev1992state}
V.G. Turaev and O.Y. Viro.
\newblock ``State sum invariants of 3-manifolds and quantum 6j-symbols''.
\newblock \href{https://dx.doi.org/10.1016/0040-9383(92)90015-A}{Topology {\bf 31}, 865 -- 902}~(1992).

\bibitem{barrett1996invariants}
John Barrett and Bruce Westbury.
\newblock ``Invariants of piecewise-linear 3-manifolds''.
\newblock \href{https://dx.doi.org/10.1090/S0002-9947-96-01660-1}{Trans. Am. Math. Soc. {\bf 348}, 3997--4022}~(1996).
\newblock  \href{http://arxiv.org/abs/hep-th/9311155}{arXiv:hep-th/9311155}.

\bibitem{walker1991witten}
Kevin Walker.
\newblock ``On {W}itten’s 3-manifold invariants \& {TQFT}s''.
\newblock preprints, available at http://canyon23.net/math~(1991,2006).
\newblock  url:~\url{http://canyon23.net/math}.

\bibitem{freedman2008picture}
Michael Freedman, Chetan Nayak, Kevin Walker, and Zhenghan Wang.
\newblock ``On picture $(2+1)$-{TQFT}s''.
\newblock \href{https://dx.doi.org/10.1142/9789812819116_0002}{Pages 19--106}.
\newblock WORLD SCIENTIFIC. ~(2011).
\newblock  \href{http://arxiv.org/abs/0806.1926}{arXiv:0806.1926}.

\bibitem{koenig2010quantum}
Robert Koenig, Greg Kuperberg, and Ben~W. Reichardt.
\newblock ``{Quantum computation with Turaev-Viro codes}''.
\newblock \href{https://dx.doi.org/10.1016/j.aop.2010.08.001}{Ann. Phys. {\bf 325}, 2707 -- 2749}~(2010).
\newblock  \href{http://arxiv.org/abs/1002.2816}{arXiv:1002.2816}.

\bibitem{GarciaVerstraeteWolfCirac08}
D.~P\'erez-Garc\'ia, F.~Verstraete, M.~M. Wolf, and J.~I. Cirac.
\newblock ``Peps as unique ground states of local hamiltonians''.
\newblock Quantum Info. Comput. {\bf 8}, 650--663~(2008).
\newblock  \href{http://arxiv.org/abs/0707.2260}{arXiv:0707.2260}.

\bibitem{higherdto}
Mehmet~Burak Sahinoglu, Michael Walter, and Dominic~J Williamson.
\newblock ``A tensor network framework for topological order in any dimension''.
\newblock preprint, available at https://sites.google.com/site/dominicjw/papers~(2015).
\newblock  url:~\url{https://sites.google.com/site/dominicjw/papers}.

\bibitem{mignard2017modular}
Micha{\"e}l Mignard and Peter Schauenburg.
\newblock ``Modular categories are not determined by their modular data''.
\newblock preprint,~(2017).
\newblock  \href{http://arxiv.org/abs/1704.04221}{arXiv:1704.04221}.

\bibitem{DrinfeldCenter}
Michael Muger.
\newblock ``{From subfactors to categories and topology II: The quantum double of tensor categories and subfactors}''.
\newblock \href{https://dx.doi.org/10.1016/S0022-4049(02)00248-7}{J. Pure Appl. Algebr. {\bf 180}, 159 -- 219}~(2003).
\newblock  \href{http://arxiv.org/abs/math/0111205}{arXiv:math/0111205}.

\bibitem{barkeshli2016reflection}
Maissam Barkeshli, Parsa Bonderson, Chao-Ming Jian, Meng Cheng, and Kevin Walker.
\newblock ``Reflection and time reversal symmetry enriched topological phases of matter: path integrals, non-orientable manifolds, and anomalies''.
\newblock preprint,~(2016).
\newblock  \href{http://arxiv.org/abs/1612.07792}{arXiv:1612.07792}.

\bibitem{cui2017state}
Shawn~X Cui and Zhenghan Wang.
\newblock ``State sum invariants of three manifolds from spherical multi-fusion categories''.
\newblock preprint,~(2017).
\newblock  \href{http://arxiv.org/abs/1702.07113}{arXiv:1702.07113}.

\bibitem{PhysRevB}
Dominic~J. Williamson.
\newblock ``Fractal symmetries: Ungauging the cubic code''.
\newblock \href{https://dx.doi.org/10.1103/PhysRevB.94.155128}{Phys. Rev. B {\bf 94}, 155128}~(2016).
\newblock  \href{http://arxiv.org/abs/1603.05182}{arXiv:1603.05182}.

\bibitem{WalterOgburn1999}
R.~Walter~Ogburn and John Preskill.
\newblock ``Topological quantum computation''.
\newblock \href{https://dx.doi.org/10.1007/3-540-49208-9_31}{Pages 341--356}.
\newblock Springer Berlin Heidelberg. Berlin, Heidelberg~(1999).

\bibitem{freedman2003topological}
Michael Freedman, Alexei Kitaev, Michael Larsen, and Zhenghan Wang.
\newblock ``Topological quantum computation''.
\newblock \href{https://dx.doi.org/10.1090/S0273-0979-02-00964-3}{Bull. Am. Math. Soc. {\bf 40}, 31--38}~(2003).
\newblock  \href{http://arxiv.org/abs/quant-ph/0101025}{arXiv:quant-ph/0101025}.

\bibitem{nayak2008non}
Chetan Nayak, Steven~H. Simon, Ady Stern, Michael Freedman, and Sankar Das~Sarma.
\newblock ``Non-abelian anyons and topological quantum computation''.
\newblock \href{https://dx.doi.org/10.1103/RevModPhys.80.1083}{Rev. Mod. Phys. {\bf 80}, 1083--1159}~(2008).
\newblock  \href{http://arxiv.org/abs/0707.1889}{arXiv:0707.1889}.

\bibitem{wang2010topological}
Zhenghan Wang.
\newblock ``Topological quantum computation''.
\newblock \href{https://dx.doi.org/10.1090/cbms/112}{Volume 112}.
\newblock American Mathematical Soc. ~(2010).

\bibitem{moore1988polynomial}
Gregory Moore and Nathan Seiberg.
\newblock ``Polynomial equations for rational conformal field theories''.
\newblock \href{https://dx.doi.org/10.1016/0370-2693(88)91796-0}{Phys. Lett. B {\bf 212}, 451 -- 460}~(1988).

\bibitem{Moore1989}
Gregory Moore and Nathan Seiberg.
\newblock ``Classical and quantum conformal field theory''.
\newblock \href{https://dx.doi.org/10.1007/BF01238857}{Commun. Math. Phys. {\bf 123}, 177--254}~(1989).

\bibitem{mac2013categories}
Saunders Mac~Lane.
\newblock ``Categories for the working mathematician''.
\newblock \href{https://dx.doi.org/10.1007/978-1-4612-9839-7}{Volume~5}.
\newblock Springer. ~(2013).

\bibitem{luo2016structure}
Zhu-Xi Luo, Ethan Lake, and Yong-Shi Wu.
\newblock ``The structure of fixed-point tensor network states characterizes the patterns of long-range entanglement''.
\newblock \href{https://dx.doi.org/10.1103/PhysRevB.96.035101}{Phys. Rev. B {\bf 96}, 035101}~(2017).
\newblock  \href{http://arxiv.org/abs/1611.01140}{arXiv:1611.01140}.

\bibitem{PhysRevB.94.155106}
Carlos Fern\'andez-Gonz\'alez, Roger S.~K. Mong, Olivier Landon-Cardinal, David P\'erez-Garc\'{\i}a, and Norbert Schuch.
\newblock ``Constructing topological models by symmetrization: A projected entangled pair states study''.
\newblock \href{https://dx.doi.org/10.1103/PhysRevB.94.155106}{Phys. Rev. B {\bf 94}, 155106}~(2016).
\newblock  \href{http://arxiv.org/abs/1608.00594}{arXiv:1608.00594}.

\bibitem{Qalgebra}
Tian Lan and Xiao-Gang Wen.
\newblock ``{Topological quasiparticles and the holographic bulk-edge relation in $(2+1)$-dimensional string-net models}''.
\newblock \href{https://dx.doi.org/10.1103/PhysRevB.90.115119}{Phys. Rev. B {\bf 90}, 115119}~(2014).
\newblock  \href{http://arxiv.org/abs/1311.1784}{arXiv:1311.1784}.

\bibitem{haah}
J.~{Haah}.
\newblock ``{An invariant of topologically ordered states under local unitary transformations}''.
\newblock \href{https://dx.doi.org/10.1007/s00220-016-2594-y}{Commun. Math. Phys. {\bf 342}, 771--801}~(2016).
\newblock  \href{http://arxiv.org/abs/1407.2926}{arXiv:1407.2926}.

\bibitem{aasen2017fermion}
David Aasen, Ethan Lake, and Kevin Walker.
\newblock ``Fermion condensation and super pivotal categories''.
\newblock preprint~(2017).
\newblock  \href{http://arxiv.org/abs/1709.01941}{arXiv:1709.01941}.

\bibitem{daveprep}
David Aasen, Paul Fendley, and Roger S.~K. Mong.
\newblock ``{Topological Defects on the Lattice: Dualities and Degeneracies}''~(2020).
\newblock  \href{http://arxiv.org/abs/2008.08598}{arXiv:2008.08598}.

\bibitem{ocneanu2001}
Adrian Ocneanu.
\newblock ``Operator algebras, topology and subgroups of quantum symmetry--- construction of subgroups of quantum groups''.
\newblock Taniguchi Conference on Mathematics Nara 1998, Adv. Stud. Pure Math., {\bf 31}, 235--263~(2001).

\bibitem{davepriv}
David {Aasen}.
\newblock ``Lessons from the tube master''.
\newblock private communication~(2017).

\bibitem{Ren2024}
Yuanjie Ren, Nathanan Tantivasadakarn, and Dominic~J. Williamson.
\newblock ``Efficient preparation of solvable anyons with adaptive quantum circuits''.
\newblock \href{https://dx.doi.org/10.1103/b9hf-gx4f}{Physical Review X{\bf 15}}~(2025).

\bibitem{PhysRevB.85.235151}
Yi~Zhang, Tarun Grover, Ari Turner, Masaki Oshikawa, and Ashvin Vishwanath.
\newblock ``Quasiparticle statistics and braiding from ground-state entanglement''.
\newblock \href{https://dx.doi.org/10.1103/PhysRevB.85.235151}{Phys. Rev. B {\bf 85}, 235151}~(2012).
\newblock  \href{http://arxiv.org/abs/1111.2342}{arXiv:1111.2342}.

\bibitem{KitaevPreskill}
Alexei Kitaev and John Preskill.
\newblock ``Topological entanglement entropy''.
\newblock \href{https://dx.doi.org/10.1103/PhysRevLett.96.110404}{Phys. Rev. Lett. {\bf 96}, 110404}~(2006).
\newblock  \href{http://arxiv.org/abs/hep-th/0510092}{arXiv:hep-th/0510092}.

\bibitem{levinwenentanglement}
Michael Levin and Xiao-Gang Wen.
\newblock ``{Detecting Topological Order in a Ground State Wave Function}''.
\newblock \href{https://dx.doi.org/10.1103/PhysRevLett.96.110405}{Phys. Rev. Lett. {\bf 96}, 110405}~(2006).
\newblock  \href{http://arxiv.org/abs/cond-mat/0510613}{arXiv:cond-mat/0510613}.

\bibitem{PhysRevLett.108.196402}
Xiao-Liang Qi, Hosho Katsura, and Andreas W.~W. Ludwig.
\newblock ``General relationship between the entanglement spectrum and the edge state spectrum of topological quantum states''.
\newblock \href{https://dx.doi.org/10.1103/PhysRevLett.108.196402}{Phys. Rev. Lett. {\bf 108}, 196402}~(2012).
\newblock  \href{http://arxiv.org/abs/1103.5437}{arXiv:1103.5437}.

\bibitem{topologicalrenyi}
S.~Flammia, A.~Hamma, T.~Hughes, and X.-G. Wen.
\newblock ``{Topological entanglement Renyi entropy and reduced density matrix structure}''.
\newblock \href{https://dx.doi.org/10.1103/PhysRevLett.103.261601}{Phys. Rev. Lett. {\bf 103}, 261601}~(2009).
\newblock  \href{http://arxiv.org/abs/0909.3305}{arXiv:0909.3305}.

\bibitem{0305-4470-20-16-043}
V~Pasquier.
\newblock ``Operator content of the ade lattice models''.
\newblock \href{https://dx.doi.org/10.1088/0305-4470/20/16/043}{Journal of Physics A: Mathematical and General {\bf 20}, 5707}~(1987).

\bibitem{VERLINDE1988360}
Erik Verlinde.
\newblock ``Fusion rules and modular transformations in 2d conformal field theory''.
\newblock \href{https://dx.doi.org/10.1016/0550-3213(88)90603-7}{Nuclear Physics B {\bf 300}, 360 -- 376}~(1988).

\bibitem{czxmodel}
Xie Chen, Zheng-Xin Liu, and Xiao-Gang Wen.
\newblock ``{Two-dimensional symmetry-protected topological orders and their protected gapless edge excitations}''.
\newblock \href{https://dx.doi.org/10.1103/PhysRevB.84.235141}{Phys. Rev. B {\bf 84}, 235141}~(2011).
\newblock  \href{http://arxiv.org/abs/1106.4752}{arXiv:1106.4752}.

\bibitem{PhysRevB.65.165113}
Xiao-Gang Wen.
\newblock ``Quantum orders and symmetric spin liquids''.
\newblock \href{https://dx.doi.org/10.1103/PhysRevB.65.165113}{Phys. Rev. B {\bf 65}, 165113}~(2002).
\newblock  \href{http://arxiv.org/abs/cond-mat/0107071}{arXiv:cond-mat/0107071}.

\bibitem{PhysRevB.95.125107}
Shenghan Jiang and Ying Ran.
\newblock ``Anyon condensation and a generic tensor-network construction for symmetry-protected topological phases''.
\newblock \href{https://dx.doi.org/10.1103/PhysRevB.95.125107}{Phys. Rev. B {\bf 95}, 125107}~(2017).
\newblock  \href{http://arxiv.org/abs/1611.07652}{arXiv:1611.07652}.

\bibitem{garre2017symmetry}
Jos{\'e} Garre-Rubio, Sofyan Iblisdir, and David P{\'e}rez-Garc{\'\i}a.
\newblock ``Symmetry reduction induced by anyon condensation: a tensor network approach''.
\newblock preprint,~(2017).
\newblock  \href{http://arxiv.org/abs/1702.08759}{arXiv:1702.08759}.

\bibitem{PhysRevB.95.235119}
Kasper Duivenvoorden, Mohsin Iqbal, Jutho Haegeman, Frank Verstraete, and Norbert Schuch.
\newblock ``Entanglement phases as holographic duals of anyon condensates''.
\newblock \href{https://dx.doi.org/10.1103/PhysRevB.95.235119}{Phys. Rev. B {\bf 95}, 235119}~(2017).
\newblock  \href{http://arxiv.org/abs/1702.08469}{arXiv:1702.08469}.

\bibitem{PhysRevB.91.195134}
Juven~C. Wang, Luiz~H. Santos, and Xiao-Gang Wen.
\newblock ``{Bosonic anomalies, induced fractional quantum numbers, and degenerate zero modes: The anomalous edge physics of symmetry-protected topological states}''.
\newblock \href{https://dx.doi.org/10.1103/PhysRevB.91.195134}{Phys. Rev. B {\bf 91}, 195134}~(2015).
\newblock  \href{http://arxiv.org/abs/1403.5256}{arXiv:1403.5256}.

\bibitem{thorngren2015higher}
Ryan Thorngren and Curt von Keyserlingk.
\newblock ``{Higher {SPT}'s and a generalization of anomaly in-flow}''.
\newblock preprint,~(2015).
\newblock  \href{http://arxiv.org/abs/1511.02929}{arXiv:1511.02929}.

\bibitem{fidkowski2015realizing}
Lukasz Fidkowski and Ashvin Vishwanath.
\newblock ``Realizing anomalous anyonic symmetries at the surfaces of 3d gauge theories''.
\newblock preprint,~(2015).
\newblock  \href{http://arxiv.org/abs/1511.01502}{arXiv:1511.01502}.

\bibitem{PhysRevX.3.041016}
Lukasz Fidkowski, Xie Chen, and Ashvin Vishwanath.
\newblock ``Non-abelian topological order on the surface of a 3d topological superconductor from an exactly solved model''.
\newblock \href{https://dx.doi.org/10.1103/PhysRevX.3.041016}{Phys. Rev. X {\bf 3}, 041016}~(2013).
\newblock  \href{http://arxiv.org/abs/1305.5851}{arXiv:1305.5851}.

\bibitem{PhysRevB.90.245122}
F.~J. Burnell, Xie Chen, Lukasz Fidkowski, and Ashvin Vishwanath.
\newblock ``Exactly soluble model of a three-dimensional symmetry-protected topological phase of bosons with surface topological order''.
\newblock \href{https://dx.doi.org/10.1103/PhysRevB.90.245122}{Phys. Rev. B {\bf 90}, 245122}~(2014).
\newblock  \href{http://arxiv.org/abs/1302.7072}{arXiv:1302.7072}.

\bibitem{PhysRevX.5.041013}
Xie Chen, F.~J. Burnell, Ashvin Vishwanath, and Lukasz Fidkowski.
\newblock ``Anomalous symmetry fractionalization and surface topological order''.
\newblock \href{https://dx.doi.org/10.1103/PhysRevX.5.041013}{Phys. Rev. X {\bf 5}, 041013}~(2015).
\newblock  \href{http://arxiv.org/abs/1403.6491}{arXiv:1403.6491}.

\bibitem{edie2017equivalences}
Cain Edie-Michell.
\newblock ``Equivalences of graded categories''.
\newblock preprint~(2017).
\newblock  \href{http://arxiv.org/abs/1711.00645}{arXiv:1711.00645}.

\bibitem{Davydov2013}
Alexei Davydov, Dmitri Nikshych, and Victor Ostrik.
\newblock ``On the structure of the witt group of braided fusion categories''.
\newblock \href{https://dx.doi.org/10.1007/s00029-012-0093-3}{Selecta Mathematica {\bf 19}, 237--269}~(2013).
\newblock  \href{http://arxiv.org/abs/1109.5558}{arXiv:1109.5558}.

\bibitem{davydov2013witt}
Alexei Davydov, Michael M{\"u}ger, Dmitri Nikshych, and Victor Ostrik.
\newblock ``The witt group of non-degenerate braided fusion categories''.
\newblock \href{https://dx.doi.org/10.1515/crelle.2012.014}{Journal f{\"u}r die reine und angewandte Mathematik {\bf 2013}, 135--177}~(2013).
\newblock  \href{http://arxiv.org/abs/1009.2117}{arXiv:1009.2117}.

\bibitem{Kitaev2012}
Alexei Kitaev and Liang Kong.
\newblock ``Models for gapped boundaries and domain walls''.
\newblock \href{https://dx.doi.org/10.1007/s00220-012-1500-5}{Commun. Math. Phys. {\bf 313}, 351--373}~(2012).
\newblock  \href{http://arxiv.org/abs/1104.5047}{arXiv:1104.5047}.

\bibitem{PhysRevB.90.045142}
Ching-Yu Huang, Xie Chen, and Frank Pollmann.
\newblock ``Detection of symmetry-enriched topological phases''.
\newblock \href{https://dx.doi.org/10.1103/PhysRevB.90.045142}{Phys. Rev. B {\bf 90}, 045142}~(2014).
\newblock  \href{http://arxiv.org/abs/1312.3093}{arXiv:1312.3093}.

\bibitem{DijkgraafWitten}
Robbert Dijkgraaf and Edward Witten.
\newblock ``Topological gauge theories and group cohomology''.
\newblock \href{https://dx.doi.org/10.1007/BF02096988}{Commun. Math. Phys. {\bf 129}, 393--429}~(1990).

\bibitem{chen2012symmetry}
Xie Chen, Zheng-Cheng Gu, Zheng-Xin Liu, and Xiao-Gang Wen.
\newblock ``Symmetry-protected topological orders in interacting bosonic systems''.
\newblock \href{https://dx.doi.org/10.1126/science.1227224}{Science {\bf 338}, 1604--1606}~(2012).
\newblock  \href{http://arxiv.org/abs/1301.0861}{arXiv:1301.0861}.

\bibitem{PhysRevB.90.235137}
Dominic~V. Else and Chetan Nayak.
\newblock ``{Classifying symmetry-protected topological phases through the anomalous action of the symmetry on the edge}''.
\newblock \href{https://dx.doi.org/10.1103/PhysRevB.90.235137}{Phys. Rev. B {\bf 90}, 235137}~(2014).
\newblock  \href{http://arxiv.org/abs/1409.5436}{arXiv:1409.5436}.

\bibitem{chen2010tensor}
Xie Chen, Bei Zeng, Zheng-Cheng Gu, Isaac~L. Chuang, and Xiao-Gang Wen.
\newblock ``Tensor product representation of a topological ordered phase: Necessary symmetry conditions''.
\newblock \href{https://dx.doi.org/10.1103/PhysRevB.82.165119}{Phys. Rev. B {\bf 82}, 165119}~(2010).
\newblock  \href{http://arxiv.org/abs/1003.1774}{arXiv:1003.1774}.

\bibitem{shadows}
Jutho Haegeman, V~Zauner, N~Schuch, and Frank Verstraete.
\newblock ``Shadows of anyons and the entanglement structure of topological phases''.
\newblock \href{https://dx.doi.org/10.1038/ncomms9284}{Nat. Commun.{\bf 6}}~(2015).
\newblock  \href{http://arxiv.org/abs/1410.5443}{arXiv:1410.5443}.

\bibitem{shukla2016boson}
Sujeet~K Shukla, M~Burak {\c{S}}ahino{\u{g}}lu, Frank Pollmann, and Xie Chen.
\newblock ``Boson condensation and instability in the tensor network representation of string-net states''.
\newblock preprint,~(2016).
\newblock  \href{http://arxiv.org/abs/1610.00608}{arXiv:1610.00608}.

\bibitem{marien2016condensation}
Micha{\"e}l Mari{\"e}n, Jutho Haegeman, Paul Fendley, and Frank Verstraete.
\newblock ``Condensation-driven phase transitions in perturbed string nets''.
\newblock preprint,~(2016).
\newblock  \href{http://arxiv.org/abs/1607.05296}{arXiv:1607.05296}.

\bibitem{Gaiotto2015}
Davide Gaiotto, Anton Kapustin, Nathan Seiberg, and Brian Willett.
\newblock ``{Generalized global symmetries}''.
\newblock \href{https://dx.doi.org/10.1007/JHEP02(2015)172}{J. High Energy Phys. {\bf 2015}, 172}~(2015).
\newblock  \href{http://arxiv.org/abs/1412.5148}{arXiv:1412.5148}.

\bibitem{kapustin2013higher}
Anton Kapustin and Ryan Thorngren.
\newblock ``{Higher symmetry and gapped phases of gauge theories}''.
\newblock preprint,~(2013).
\newblock  \href{http://arxiv.org/abs/1309.4721}{arXiv:1309.4721}.

\bibitem{MPOanyoncondprep}
Dominic {J. Williamson \emph{et al.}}
\newblock ``Matrix product operator symmetry-breaking and anyon condensation''.
\newblock in preparation~(2017).

\bibitem{propitius}
M.~{de Wild Propitius}.
\newblock ``{Topological interactions in broken gauge theories}''.
\newblock PhD thesis.
\newblock University of Amsterdam.
\newblock ~(1995).
\newblock  \href{http://arxiv.org/abs/hep-th/9511195}{arXiv:hep-th/9511195}.

\bibitem{DEWILDPROPITIUS1997297}
Mark de~Wild~Propitius.
\newblock ``{(Spontaneously broken) Abelian Chern-Simons theories}''.
\newblock \href{https://dx.doi.org/10.1016/S0550-3213(97)00005-9}{Nucl. Phys. B {\bf 489}, 297 -- 359}~(1997).
\newblock  \href{http://arxiv.org/abs/hep-th/9606029}{arXiv:hep-th/9606029}.

\bibitem{bridgeman2017anomalies}
Jacob~C. Bridgeman and Dominic~J. Williamson.
\newblock ``Anomalies and entanglement renormalization''.
\newblock \href{https://dx.doi.org/10.1103/PhysRevB.96.125104}{Phys. Rev. B {\bf 96}, 125104}~(2017).
\newblock  \href{http://arxiv.org/abs/1703.07782}{arXiv:1703.07782}.

\bibitem{Gaugingtime}
Xie Chen and Ashvin Vishwanath.
\newblock ``Towards gauging time-reversal symmetry: A tensor network approach''.
\newblock \href{https://dx.doi.org/10.1103/PhysRevX.5.041034}{Phys. Rev. X {\bf 5}, 041034}~(2015).
\newblock  \href{http://arxiv.org/abs/1401.3736}{arXiv:1401.3736}.

\bibitem{thorngren2016gauging}
Ryan Thorngren and Dominic~V Else.
\newblock ``Gauging spatial symmetries and the classification of topological crystalline phases''.
\newblock preprint,~(2016).
\newblock  \href{http://arxiv.org/abs/1612.00846}{arXiv:1612.00846}.

\bibitem{walkerfermions}
Kevin Walker.
\newblock ``Codimension-1 defects, categorified group actions, and condensing fermions''.
\newblock talk at IPAM workshop~(Jan. 26-30, 2015).
\newblock  url:~\url{http://www.ipam.ucla.edu/abstract/?tid=12400&pcode=STQ2015}.

\bibitem{Bhardwaj2017}
Lakshya Bhardwaj, Davide Gaiotto, and Anton Kapustin.
\newblock ``State sum constructions of spin-tfts and string net constructions of fermionic phases of matter''.
\newblock \href{https://dx.doi.org/10.1007/JHEP04(2017)096}{J. High Energy Phys. {\bf 2017}, 96}~(2017).
\newblock  \href{http://arxiv.org/abs/1605.01640}{arXiv:1605.01640}.

\bibitem{kapustin2017fermionic}
Anton Kapustin and Ryan Thorngren.
\newblock ``Fermionic {SPT} phases in higher dimensions and bosonization''.
\newblock preprint,~(2017).
\newblock  \href{http://arxiv.org/abs/1701.08264}{arXiv:1701.08264}.

\bibitem{Wan2017}
Yidun Wan and Chenjie Wang.
\newblock ``Fermion condensation and gapped domain walls in topological orders''.
\newblock \href{https://dx.doi.org/10.1007/JHEP03(2017)172}{J. High Energy Phys. {\bf 2017}, 172}~(2017).
\newblock  \href{http://arxiv.org/abs/1607.01388}{arXiv:1607.01388}.

\bibitem{doi:10.1142/S0217751X16450445}
Davide Gaiotto and Anton Kapustin.
\newblock ``Spin {TQFT}s and fermionic phases of matter''.
\newblock \href{https://dx.doi.org/10.1142/S0217751X16450445}{Int. J. Mod. Phys. A {\bf 31}, 1645044}~(2016).
\newblock  \href{http://arxiv.org/abs/1505.05856}{arXiv:1505.05856}.

\bibitem{PhysRevB.94.115115}
Nicolas Tarantino and Lukasz Fidkowski.
\newblock ``Discrete spin structures and commuting projector models for two-dimensional fermionic symmetry-protected topological phases''.
\newblock \href{https://dx.doi.org/10.1103/PhysRevB.94.115115}{Phys. Rev. B {\bf 94}, 115115}~(2016).
\newblock  \href{http://arxiv.org/abs/1604.02145}{arXiv:1604.02145}.

\bibitem{PhysRevB.94.115127}
Brayden Ware, Jun~Ho Son, Meng Cheng, Ryan~V. Mishmash, Jason Alicea, and Bela Bauer.
\newblock ``Ising anyons in frustration-free majorana-dimer models''.
\newblock \href{https://dx.doi.org/10.1103/PhysRevB.94.115127}{Phys. Rev. B {\bf 94}, 115127}~(2016).
\newblock  \href{http://arxiv.org/abs/1605.06125}{arXiv:1605.06125}.

\bibitem{chen2010local}
Xie Chen, Zheng-Cheng Gu, and Xiao-Gang Wen.
\newblock ``Local unitary transformation, long-range quantum entanglement, wave function renormalization, and topological order''.
\newblock \href{https://dx.doi.org/10.1103/PhysRevB.82.155138}{Phys. Rev. B {\bf 82}, 155138}~(2010).
\newblock  \href{http://arxiv.org/abs/1004.3835}{arXiv:1004.3835}.

\bibitem{williamson2016hamiltonian}
Dominic~J Williamson and Zhenghan Wang.
\newblock ``Hamiltonian models for topological phases of matter in three spatial dimensions''.
\newblock \href{https://dx.doi.org/10.1016/j.aop.2016.12.018}{Ann. Phys. {\bf 377}, 311 -- 344}~(2017).
\newblock  \href{http://arxiv.org/abs/1606.07144}{arXiv:1606.07144}.

\bibitem{SETtransgates}
Dominic~J. {Williamson \emph{et al.}}
\newblock ``Symmetry-enriched topological order and fault-tolerant logic gates''.
\newblock in preparation~(2017).

\bibitem{edie2017brauer}
Cain Edie-Michell.
\newblock ``{The Brauer-Picard groups of the $ADE$ fusion categories}''.
\newblock preprint~(2017).
\newblock  \href{http://arxiv.org/abs/1709.04721}{arXiv:1709.04721}.

\bibitem{edie2017field}
Cain Edie-Michell and Scott Morrison.
\newblock ``{A field guide to categories with $A_n$ fusion rules}''.
\newblock preprint~(2017).
\newblock  \href{http://arxiv.org/abs/1710.07362}{arXiv:1710.07362}.

\bibitem{fuchs2002tft}
J{\"u}rgen Fuchs, Ingo Runkel, and Christoph Schweigert.
\newblock ``{TFT construction of RCFT correlators I: Partition functions}''.
\newblock \href{https://dx.doi.org/10.1016/S0550-3213(02)00744-7}{Nucl. Phys. B {\bf 646}, 353--497}~(2002).
\newblock  \href{http://arxiv.org/abs/hep-th/0204148}{arXiv:hep-th/0204148}.

\bibitem{PhysRevLett.93.070601}
J\"urg Fr\"ohlich, J\"urgen Fuchs, Ingo Runkel, and Christoph Schweigert.
\newblock ``Kramers-wannier duality from conformal defects''.
\newblock \href{https://dx.doi.org/10.1103/PhysRevLett.93.070601}{Phys. Rev. Lett. {\bf 93}, 070601}~(2004).
\newblock  \href{http://arxiv.org/abs/cond-mat/0404051}{arXiv:cond-mat/0404051}.

\bibitem{FROHLICH2007354}
Jürg Fröhlich, Jürgen Fuchs, Ingo Runkel, and Christoph Schweigert.
\newblock ``Duality and defects in rational conformal field theory''.
\newblock \href{https://dx.doi.org/10.1016/j.nuclphysb.2006.11.017}{Nucl. Phys. B {\bf 763}, 354 -- 430}~(2007).
\newblock  \href{http://arxiv.org/abs/hep-th/0607247}{arXiv:hep-th/0607247}.

\bibitem{1751-8121-47-45-452001}
Roger S~K Mong, David~J Clarke, Jason Alicea, Netanel~H Lindner, and Paul Fendley.
\newblock ``Parafermionic conformal field theory on the lattice''.
\newblock \href{https://dx.doi.org/10.1088/1751-8113/47/45/452001}{J. Phys. A {\bf 47}, 452001}~(2014).
\newblock  \href{http://arxiv.org/abs/1406.0846}{arXiv:1406.0846}.

\bibitem{PhysRevLett.112.247202}
Yi-Zhuang You, Zhen Bi, Alex Rasmussen, Kevin Slagle, and Cenke Xu.
\newblock ``Wave function and strange correlator of short-range entangled states''.
\newblock \href{https://dx.doi.org/10.1103/PhysRevLett.112.247202}{Phys. Rev. Lett. {\bf 112}, 247202}~(2014).
\newblock  \href{http://arxiv.org/abs/1312.0626}{arXiv:1312.0626}.

\bibitem{Vanhove2018}
Robijn Vanhove, Matthias Bal, Dominic~J. Williamson, Nick Bultinck, Jutho Haegeman, and Frank Verstraete.
\newblock ``{Mapping Topological to Conformal Field Theories through strange Correlators}''.
\newblock \href{https://dx.doi.org/10.1103/PhysRevLett.121.177203}{Phys. Rev. Lett. {\bf 121}, 177203}~(2018).
\newblock  \href{http://arxiv.org/abs/1801.05959}{arXiv:1801.05959}.

\bibitem{Vanhove2022}
Robijn Vanhove, Laurens Lootens, Hong-Hao Tu, and Frank Verstraete.
\newblock ``Topological aspects of the critical three-state potts model''.
\newblock \href{https://dx.doi.org/10.1088/1751-8121/ac68b1}{Journal of Physics A: Mathematical and Theoretical{\bf 55}}~(2022).

\bibitem{Molnar2022}
Andras Molnar, Alberto~Ruiz de~Alarcón, José Garre-Rubio, Norbert Schuch, J.~Ignacio Cirac, and David Pérez-García.
\newblock ``Matrix product operator algebras i: representations of weak hopf algebras and projected entangled pair states''~(2022).
\newblock  url:~\url{http://arxiv.org/abs/2204.05940}.

\bibitem{Ruiz-de-Alarcn2022}
Alberto~Ruiz de~Alarcón, José Garre-Rubio, András Molnár, and David Pérez-García.
\newblock ``Matrix product operator algebras ii: Phases of matter for 1d mixed states''~(2022).

\bibitem{Liu2025}
Yuhan Liu, Andras Molnar, Xiao-Qi Sun, Frank Verstraete, Kohtaro Kato, and Laurens Lootens.
\newblock ``Trading mathematical for physical simplicity: Bialgebraic structures in matrix product operator symmetries''~(2025).
\newblock  url:~\url{http://arxiv.org/abs/2509.03600}.

\bibitem{Thorngren2019}
Ryan Thorngren and Yifan Wang.
\newblock ``Fusion category symmetry i: Anomaly in-flow and gapped phases''~(2019).
\newblock  url:~\url{http://arxiv.org/abs/1912.02817}.

\bibitem{Aasen2020b}
David Aasen, Paul Fendley, and Roger S.~K. Mong.
\newblock ``Topological defects on the lattice: Dualities and degeneracies''~(2020).
\newblock  url:~\url{http://arxiv.org/abs/2008.08598}.

\bibitem{Lootens2021b}
Laurens Lootens, Clement Delcamp, Gerardo Ortiz, and Frank Verstraete.
\newblock ``Dualities in one-dimensional quantum lattice models: symmetric hamiltonians and matrix product operator intertwiners''.
\newblock \href{https://dx.doi.org/10.1103/PRXQuantum.4.020357}{PRX Quantum {\bf 4}, 020357}~(2021).

\bibitem{Lootens2022}
Laurens Lootens, Clement Delcamp, and Frank Verstraete.
\newblock ``Dualities in one-dimensional quantum lattice models: topological sectors''.
\newblock \href{https://dx.doi.org/10.1103/PRXQuantum.5.010338}{PRX Quantum {\bf 5}, 010338}~(2022).

\bibitem{Bridgeman2022}
Jacob~C. Bridgeman, Laurens Lootens, and Frank Verstraete.
\newblock ``Invertible bimodule categories and generalized schur orthogonality''~(2022).

\bibitem{Haller2023}
Lukas Haller, Wen~Tao Xu, Yu~Jie Liu, and Frank Pollmann.
\newblock ``Quantum phase transition between symmetry enriched topological phases in tensor-network states''.
\newblock \href{https://dx.doi.org/10.1103/PhysRevResearch.5.043078}{Physical Review Research {\bf 5}, 043078}~(2023).

\bibitem{Cuiper2025}
Bram Vancraeynest-De Cuiper, José Garre-Rubio, Frank Verstraete, Kevin Vervoort, Dominic~J. Williamson, and Laurens Lootens.
\newblock ``From gauging to duality in one-dimensional quantum lattice models''~(2025).
\newblock  url:~\url{http://arxiv.org/abs/2509.22051}.

\bibitem{Jones2026}
Corey Jones, Kylan Schatz, and Dominic~J. Williamson.
\newblock ``Quantum cellular automata and categorical dualities of spin chains''.
\newblock \href{https://dx.doi.org/10.1007/s00220-026-05571-y}{Communications in Mathematical Physics 2026 407:4 {\bf 407}, 66--}~(2026).

\bibitem{Lootens2021}
Laurens Lootens, Jürgen Fuchs, Jutho Haegeman, Christoph Schweigert, and Frank Verstraete.
\newblock ``Matrix product operator symmetries and intertwiners in string-nets with domain walls''.
\newblock \href{https://dx.doi.org/10.21468/SCIPOSTPHYS.10.3.053}{SciPost Physics{\bf 10}}~(2021).

\bibitem{Lootens2021a}
Laurens Lootens, Bram Vancraeynest-De Cuiper, Norbert Schuch, and Frank Verstraete.
\newblock ``Mapping between morita-equivalent string-net states with a constant depth quantum circuit''.
\newblock \href{https://dx.doi.org/10.1103/PhysRevB.105.085130}{Physical Review B{\bf 105}}~(2022).

\bibitem{Schatz2024}
Kylan Schatz.
\newblock ``Boundary symmetries of (2+1)d topological orders''~(2024).
\newblock  url:~\url{https://arxiv.org/abs/2408.10832v1}.

\bibitem{Lin2024}
Chien~Hung Lin and Fiona~J. Burnell.
\newblock ``Anyon condensation in string-net models''.
\newblock \href{https://dx.doi.org/10.1103/PhysRevB.110.115127}{Physical Review B {\bf 110}, 115127}~(2024).

\bibitem{Barter2019}
Daniel Barter, Jacob~C. Bridgeman, and Corey Jones.
\newblock ``{Domain Walls in Topological Phases and the Brauer–Picard Ring for Vec(Z/pZ)}''.
\newblock \href{https://dx.doi.org/10.1007/s00220-019-03338-2}{Communications in Mathematical Physics 2019 369:3 {\bf 369}, 1167--1185}~(2019).

\bibitem{Lan2025}
Tian Lan.
\newblock ``Tube category, tensor renormalization and topological holography''.
\newblock \href{https://dx.doi.org/10.1007/s00220-025-05383-6}{Communications in Mathematical Physics 2025 406:8 {\bf 406}, 196--}~(2025).

\bibitem{Tantivasadakarn2021}
Nathanan Tantivasadakarn, Ryan Thorngren, Ashvin Vishwanath, and Ruben Verresen.
\newblock ``Long-range entanglement from measuring symmetry-protected topological phases''.
\newblock \href{https://dx.doi.org/10.1103/PhysRevX.14.021040}{Physical Review X{\bf 14}}~(2021).

\bibitem{Bravyi2022}
Sergey Bravyi, Isaac Kim, Alexander Kliesch, and Robert Koenig.
\newblock ``Adaptive constant-depth circuits for manipulating non-abelian anyons''~(2022).
\newblock  url:~\url{http://arxiv.org/abs/2205.01933}.

\bibitem{Lootens2023}
Laurens Lootens, Clement Delcamp, Dominic Williamson, and Frank Verstraete.
\newblock ``Low-depth unitary quantum circuits for dualities in one-dimensional quantum lattice models''.
\newblock \href{https://dx.doi.org/10.1103/PhysRevLett.134.130403}{Physical Review Letters{\bf 134}}~(2023).

\bibitem{Tantivasadakarn2025a}
Nathanan Tantivasadakarn, Xinyu Liu, and Xie Chen.
\newblock ``Sequential circuit as generalized symmetry on lattice''~(2025).
\newblock  url:~\url{http://arxiv.org/abs/2507.22394}.

\bibitem{Cui2016}
Shawn~X. Cui, C{\'e}sar Galindo, Julia~Yael Plavnik, and Zhenghan Wang.
\newblock ``On gauging symmetry of modular categories''.
\newblock \href{https://dx.doi.org/10.1007/s00220-016-2633-8}{Communications in Mathematical Physics {\bf 348}, 1043--1064}~(2016).
\newblock  \href{http://arxiv.org/abs/1510.03475}{arXiv:1510.03475}.

\end{thebibliography}

\appendix

\numberwithin{equation}{section}

\section{Sequential gauging on the lattice}\label{setappendix:gauging}

Given a finite group $\G$, with normal subgroup $N\leq \G$ and quotient group $\G/N\cong Q$, i.e. a short exact sequence of groups
\begin{align}
1\rightarrow N \rightarrow \G \rightarrow Q \rightarrow 1
\, ,
\end{align}
it has been shown at the level of anyon theories~\cite{Cui2016,barkeshli2014symmetry} that sequentially gauging the subgroup $N$, which results in a theory with $Q$ global symmetry, followed by gauging the remaining $Q$ symmetry yields the same result as directly gauging $\G$. Here we demonstrate how to perform this on the lattice using the state gauging procedure developed in Refs.~\cite{Gaugingpaper,williamson2014matrix}. A subtlety occurs, in that the remaining symmetry $Q$ is no longer on-site after the initial gauging step for $N$. We remedy this with an additional local unitary circuit that restores the on-site nature of $Q$, allowing us to proceed with gauging it. 
For simplicity of presentation we consider a right regular representation of $\G$ and gauging maps that arise from topologically trivial flat connections~\cite{williamson2014matrix}. 

We begin by picking a tensor product structure $\mathbb{C}[Q]\otimes\mathbb{C}[N]$ on the Hilbert space $\mathbb{C}[\G]$ (note these are generally not isomorphic as group algebras). To this end we pick a section $s:Q\rightarrow \G$ such that  $s(\g{1})=\g{1}$ and $s(\conj{\g{q}})= \conj{s(\g{q})}$. Then we can uniquely decompose any element of $\G$ as $\g{g}=s(\g{q})\g{n}$ for some $\g{q}\in Q ,\g{n}\in N$.  Under multiplication we have 
\begin{align}
s(\g{q}_1)\g{n}_1s(\g{q}_2)\g{n}_2=s(\g{q}_1\g{q}_2) {c}(\g{q}_1,\g{q}_2) \g{n}_1^{\conj{\g{q}}_2}\g{n}_2 \, ,
\end{align}
 where we use the short hand notation $\g{n}^{\g{q}}:=s(\g{q})\,\g{n}\, \conj{s(\g{q})}$ for the action of $Q$ on $N$.  The function ${c:Q\times Q \rightarrow N}$ satisfies the following twisted 2-cocycle equation
\begin{align}
c(\g{q}_0\g{q}_1,\g{q}_2) c(\g{q}_0,\g{q}_1)^{\conj{\g{q}}_2}
= 
c(\g{q}_0,\g{q}_1\g{q}_2) c(\g{q}_1,\g{q}_2) \, .
\end{align}
Our choice of $s$ implies $c(\g{1},\g{q})=c(\g{q},\g{1})=1=c(\g{q},\conj{\g{q}})=c(\conj{\g{q}},\g{q})$. We remark that the group $\G$ reduces to a semidirect product when $c$ is trivial, and is a central extension when $N$ is abelian and $\g{n}^{\g{q}}=\g{n}$ for all $\g{n},\,\g{q}$. 

With our choice of tensor product structure the right regular multiplication on a vertex $v$ decomposes as follows 
\begin{align}
\text{R}_{s(\g{q})\g{n}} = (\text{R}_\g{q} \otimes \openone_N)  \text{CL}[c(q_v,\conj{\g{q}})]_{Q,N} (\openone_Q \otimes \text{L}_\g{q}\text{R}_\g{q}) (\openone_Q \otimes \text{R}_\g{n}) \, ,
\end{align}
where $\text{CL}[c(q_v,\conj{\g{q}})]_{Q,N}$ indicates a left multiplication on the $N$ qudit by the element  $c(q_v,\conj{\g{q}})$ which is controlled by the $Q$ qudit, whose state is denoted $q_v$. We remark the combined action $ \text{L}_\g{q}\text{R}_\g{q}$ on the $N$ qubit denotes the linear extension of the map $n_v \mapsto n_v^{\g{q}}$ which preserves the $\mathbb{C}[N]$ Hilbert space.

In a Hilbert space given by a tensor product of qudits living on the vertices of a graph embedded in a two dimensional manifold, the gauging map for a finite group $\G$, which acts
via $\bigotimes\limits_v R_{\g{g}}$, and a  state $\ket{\psi}$  is given by 
\begin{align}
G_\G\ket{\psi} := \prod_{v} \left( \frac{1}{|\G|} \sum_{\g{g}_v} R_v(\g{g}_v) \bigotimes_{e_v^+} L_{e_v^+}(\g{g}_v) \bigotimes_{e_v^-} R_{e_v^-}(\g{g}_v) \right) \ 
\ket{\psi} \bigotimes_{e} \ket{\g{1}}_e
\, ,
\end{align}
where $\ket{\g{1}}_e$ are newly introduced $\mathbb{C}[\G]$ gauge variables on the edges and $e_v^\pm$ denotes a neighboring edge that points away from (towards) vertex $v$. Here the summation is over all configurations of $\G$ elements assigned to vertices $\g{g}_v$.  We remark that the local qudit Hilbert space on each vertex need not be precisely $\mathbb{C}[\G]$ it only needs to transform under the right regular action of $\G$. 
The resulting states $G\ket{\psi}$ are only defined on the subspace that satisfies the local gauge constraints on every vertex
\begin{align}
P_v^\G:=\frac{1}{|\G|}\sum\limits_\g{g} \text{R}_\g{g} \bigotimes\limits_{e_v^+} \text{L}_\g{g} \bigotimes\limits_{e_v^-} \text{R}_\g{g}\, ,
\end{align}
i.e.~$P_v^\G G\ket{\psi}=G\ket{\psi}$. We remark that this subspace does not have an immediately obvious tensor product structure. 

For the on-site right regular action the following isometry disentangles the gauge constraints and projects them out, thereby mapping back to a tensor product Hilbert space 
\begin{align}
Y_\G := \left(\bigotimes_{v} \bra{\g{+}}_{v} \right) \prod_{v} \,  \prod_{e_v^+} \text{CL}_{v,e_v^+}  \prod_{e_v^-} \text{CR}_{v,e_v^-}
\, ,
&&
\text{where}
&&
\ket{\g{+}}:=\sum_{\g{g}} \ket{\g{g}}
\, ,
 \end{align}
and  $CL\, (R)$ is the controlled left (right) multiplication. 
 
To sequentially gauge a symmetry we will require an additional local unitary circuit that restores the on-site nature of the global symmetry $Q$ after the initial $N$-gauging and disentangling steps, this is given by
\begin{align}
T _Q:= \prod_e \text{CL}\left[c({q}_{v_e^+},\conj{q}_{v_e^-})\right]_{(v_e^+,v_e^-),e} \text{CL}_{v_e^-,e}  \text{CR}_{v_e^-,e}
\, ,
 \end{align}
 where $v_e^{\pm}$ indicates the vertex that edge $e$ points away from (towards). In the above equation $\text{CL}\left[c({q}_{v_e^+},\conj{q}_{v_e^-})\right]_{(v_e^+,v_e^-),e}$ indicates  a left multiplication on an $e$-edge $N$ qudit by the element  $c({q}_{v_e^+},\conj{q}_{v_e^-})$ which is controlled by the neighbouring vertex $Q$ qudits, whose states are denoted $q_{v_e^{\pm}}$. 
 We remark that the combined action of left and right $Q$-controlled multiplication $\text{CL}_{v_e^-,e}  \text{CR}_{v_e^-,e}$ induces an action of $Q$ on $N$ which preserves the $\mathbb{C}[N]$ edge Hilbert space.

The full sequential gauging procedure involves five steps. 1: gauging the $N$ subgroup symmetry. 2: disentangling the $N$ gauge constraints. 3: making the remaining $Q$ symmetry on-site. 4: gauging the $Q$ symmetry. 5: disentangling the $Q$ gauge constraints. These steps are summarized in the following table
\begin{align}
\begin{tabular}{| c | c | c | c | c |} 
\hline 
\multirow{2}{*}{Step} & \multirow{2}{*}{State} & \multirow{2}{*}{Hilbert space} & \multirow{2}{*}{Global symmetry} & Gauge
\\
& & & & constraints
\\ \hline
 0 & $\ket{\psi}$ & $\mathbb{H}_{\G\text{-matter}}$  & $\bigotimes\limits_v \text{R}_\g{g}$ & none
\\
{1} & {$G_N\ket{\psi}$} &  {GIS[$ \mathbb{H}_{\G\text{-matter}}\otimes \mathbb{H}_{N\text{-gauge}}$]} & $\bigotimes\limits_v \text{R}_{s(\g{q})} \bigotimes\limits_e \text{L}_\g{q} \text{R}_\g{q}$ & {$P_v^N$, $\forall v$}
\\
 2 & $Y_N G_N \ket{\psi}$ &  $\mathbb{H}_{Q\text{-matter}}\otimes \mathbb{H}_{N\text{-gauge}}$ & 
 see Eq.~\eqref{set:nononsiteQsym}
   & none
\\
 3 & $T_Q Y_N G_N \ket{\psi}$ & $\mathbb{H}_{Q\text{-matter}}\otimes \mathbb{H}_{N\text{-gauge}}$ & $\bigotimes\limits_v R_\g{q}$  & none
\\
 4 & $G_Q T_Q Y_N G_N \ket{\psi}$ & GIS[$ \mathbb{H}_{Q\text{-matter}}\otimes \mathbb{H}_{\G\text{-gauge}}$] & none & $P_v^Q$, $\forall v$
 \\
 5 & $Y_\G G_\G \ket{\psi}$ & $\mathbb{H}_{\G\text{-gauge}}$ & none & none
 \\
\hline
\end{tabular}
\end{align}
where $G_N$ denotes the gauging map applied only for the normal subgroup $N$ symmetry. 
Similarly $Y_N$ denotes the disentangling circuit applied only for the normal subgroup elements, in particular the controlled left and right multiplications involved in $Y_N$ are only supported on the $\mathbb{C}[N]$ vertex degrees of freedom. 
The gauge and matter Hilbert spaces above are given by 
\begin{align}
\mathbb{H}_{\G\text{-matter}}:=\bigotimes\limits_v \mathbb{C}[\G] \, , && \text{ and } && \mathbb{H}_{\G\text{-gauge}}:=\bigotimes\limits_e \mathbb{C}[\G]\, . 
\end{align}
In steps 1 and 4 GIS denote the gauge invariant subspace, i.e. the simultaneous $+1$ eigenspace of all gauge constraints. 
We remark that the global symmetry in step 1 only multiplies up to elements of the gauge group. 
The global symmetry in step 2 is given by 
\begin{align}
\label{set:nononsiteQsym}
\bigotimes\limits_v \text{R}_{\g{q}} \prod\limits_v \prod\limits_{e_v^-} \text{CR}\left[c({q}_v,\conj{\g{q}})\right]_{v,e_v^-} \prod\limits_{e_v^+}\text{CL}\left[c({q}_v,\conj{\g{q}})\right]_{v,e_v^+} \bigotimes\limits_e \text{L}_\g{q} \text{R}_\g{q} \, ,
\end{align}
which one can easily verify is not on-site.  
The state achieved in step 5 is $Y_Q G_Q T_Q Y_N G_N \ket{\psi}$ which is equal to the state achieved by directly gauging and disentangling the full $\G$ symmetry $Y_\G G_\G \ket{\psi}$. 

The effect that gauging a normal subgroup $N$ symmetry has upon a $\G$-graded MPO algebra $\cat_\G$ is to convert it into a $Q$-graded MPO algebra $\cat_Q$ whose $\g{q}$-sector is given by the collection of MPOs from the sectors of $\cat_\G$ that fall within the $\g{q}N$-coset,
\begin{align}
\cat_\g{q}:=\bigoplus_{\g{g}\in\g{q}N} \cat_\g{g} \, .
\end{align}
Further gauging the $Q$ symmetry amounts to forgetting the $Q$-grading of this MPO algebra $\cat_Q$. This clearly results in the same ungraded MPO algebra as gauging the full $\G$ symmetry in the first place, which amounts to simply forgetting the $\G$-grading of the MPO algebra $\cat_\G$.

\end{document}